\documentclass[%
 reprint,
 floatfix,
longbibliography,
 amsmath,amssymb,
 aps, physrev,
]{revtex4-2}

\usepackage{graphicx}% Include figure files
\usepackage{dcolumn}% Align table columns on decimal point
\usepackage{bm}% bold math
\usepackage{braket}
\usepackage{appendix}
\usepackage{esvect}
\usepackage{esint}
\usepackage{natbib}
\usepackage{hyperref}
\usepackage[shortlabels]{enumitem}

\usepackage{dsfont}
\usepackage{verbatim}
\begin{document}
\author{Basil M. Smitham} 
\email{bsmitham@princeton.edu}
\affiliation{Department of Electrical and Computer Engineering, Princeton University, Princeton, New Jersey 08540, USA}
\author{Andrew A. Houck}
\email{aahouck@princeton.edu}
\affiliation{Department of Electrical and Computer Engineering, Princeton University, Princeton, New Jersey 08540, USA}

\title{Reciprocal lumped-element superconducting circuits: quantization, decomposition, and model extraction}

\date{\today}

\begin{abstract}
In this work, we introduce new methods for the quantization, decomposition, and extraction (from electromagnetic simulations) of lumped-element circuit models for superconducting quantum devices. Our flux-charge symmetric procedures center on the network matrix, which encodes the connectivity of a circuit's inductive loops and capacitive nodes. First, we use the network matrix to demonstrate a simple algorithm for circuit quantization, giving novel predictions for the Hamiltonians of circuits with both Josephson junctions and quantum phase slip wires. We then show that by performing pivoting operations on the network matrix, we can decompose a superconducting circuit model into its simplest equivalent ``fundamental" form, in which the harmonic degrees of freedom are separated out from the Josephson junctions and phase slip wires. Finally, we illustrate how to extract an exact, transformerless circuit model from electromagnetic simulations of a device's hybrid admittance/impedance response matrix, by matching the lumped circuit's network matrix to the network topology of the physical layout. Overall, we provide a toolkit of intuitive methods that can be used to construct, analyze, and manipulate superconducting circuit models.
\end{abstract}

\maketitle

\section{\label{sec:Introduction}Introduction}

Lumped-element circuit models are an essential tool in the analysis of superconducting quantum devices. They describe a continuous physical layout using a discrete set of circuit variables, which can be promoted to quantum operators. For lossless systems that obey electromagnetic reciprocity, circuit models may consist of linear elements like capacitors and inductors, and nonlinear quantum tunneling elements such as Josephson junctions (Cooper pair tunneling) \cite{josephson_possible_1962} and phase slip wires (magnetic fluxon tunneling) \cite{mooij_superconducting_2006}.

Superconducting circuits can be studied using a variety of techniques, which normally employ flux and charge variables instead of voltages and currents \cite{michel_h_devoret_quantum_1995, burkard_multilevel_2004, burkard_circuit_2005, ulrich_dual_2016}. Though charge variables can sometimes be eliminated in favor or fluxes (or vice versa), frameworks that treat fluxes and charges symmetrically provide the most powerful tools for analyzing circuits. Recent advances by Parra-Rodriguez and Egusquiza \cite{egusquiza_algebraic_2022, parra-rodriguez_geometrical_2024} as well as Osborne et al. \cite{osborne_symplectic_2024, osborne_flux-charge_2024} have shown that flux-charge symmetric methods are essential for quantizing circuits with both Josephson junctions and quantum phase slip wires.

In this work, we introduce a modified flux-charge symmetric framework for lumped-element superconducting circuit analysis (Section \ref{sec:LumpedEquationsOfMotion}). Our approach demonstrates novel predictions in circuit quantization (Section \ref{sec:Quantization})---and also provides new techniques for circuit decomposition (Section \ref{sec:Decomposition}) and lumped model extraction from electromagnetic simulations (Section \ref{sec:Synthesis}). In our method, we impose physical constraints that ensure our lumped-element models exhibit a natural tree-cotree structure, with a capacitive spanning tree and inductive spanning cotree. Under these restrictions, the topology of a circuit is characterized by its network matrix $\mathbf{\Omega}$ \cite{schrijver_theory_1998, wolsey_integer_2014}, which encodes the connectivity between the system's capacitive and inductive components---and serves as a central focus of this work.

First, in Section \ref{sec:LumpedEquationsOfMotion}, we outline our physically constrained superconducting circuit models and their equations of motion. We discuss how they can be represented in tree-cotree notation, as well as how they are characterized by their network matrix---which functions similarly to the symplectic forms presented in \cite{osborne_symplectic_2024, parra-rodriguez_geometrical_2024}. The network matrix appears in the system's equations of motion, which we construct for each capacitive node and each inductive loop of the system. We then show how basis changes on the underlying node flux and loop charge variables correspond to transformations of the network matrix---allowing us to manipulate it into equivalent forms through row and column operations.

 In Section \ref{sec:Quantization}, we demonstrate a straightforward circuit quantization algorithm, which generates novel predictions for the Hamiltonians of systems possessing both Josephson junctions and phase slip wires. This procedure starts by using row and column operations to reduce the network matrix to an identity submatrix, with zeros elsewhere. We then perform a canonical transformation and utilize the system's ``integrated equations of motion" to classify its modes according to whether their charge and flux conjugate variables have extended, discrete, or compact spectra. In particular, we hypothesize that certain circuits with both Josephson junctions and phase slips \cite{le_doubly_2019} possess doubly-discrete charge and flux conjugate pairs of variables that drop out of the final Hamiltonians. We remark that our approach treats time-dependent external flux in a symmetric manner to time-dependent external charge, and provide a generalization of the results of \cite{you_circuit_2019} in Appendix \ref{subsec:ZeroCapInd}.

In Section \ref{sec:Decomposition} of this manuscript, we discuss how to decompose superconducting circuits into equivalent forms using changes of basis on the network matrix. To do this, we transform our system to the edge (or branch) basis, in which the transformed ``edge" network matrix $\mathbf{\Omega}_E$ \cite{schrijver_theory_1998,wolsey_integer_2014} encodes all the topological information of the system. We further discuss how the edge network matrix relates to our tree-cotree circuit notation, where a capacitor/Josephson junction spanning tree is connected to an inductor/phase slip cotree. We then describe how to a apply a set of ``structure-preserving transformations" to a circuit's edge network matrix, in order to simplify the circuit to an equivalent ``fundamental form." We give visual interpretations of these transformations, showing how they separate out the free and harmonic modes from the junction and phase slip degrees of freedom. We then demonstrate how these decomposed forms could be used to classify superconducting circuit models by their network structure, which can aid in efforts to enumerate classes of circuit Hamiltonians \cite{weissler_enumeration_2024}.

The final portion of this work (Section \ref{sec:Synthesis}) uses the network matrix to systematically extract lumped circuit models from electromagnetic simulation data of superconducting devices. This process is a type of ``black-box quantization," of which there have been several paradigmatic approaches \cite{nigg_black-box_2012, solgun_blackbox_2014, labarca_toolbox_2024}. Here we demonstrate a novel flux-charge symmetric approach to the problem, using the hybrid admittance/impedance response matrix \cite{newcomb_linear_1966, anderson_network_2013}, calculated from the reciprocal, lossless electromagnetic simulation of a device layout. We detail how to synthesize an exact lumped circuit model that emulates this response, by (in part) matching the circuit's network matrix to the network topology of the device's layout. In addition, we synthesize the AC poles of the hybrid matrix with auxiliary LC oscillators. In this way, the models we generate through our circuit extraction procedure (starting with an electromagnetic simulation) mirror those we obtain from our decomposition procedure (starting with a lumped-element model). We note that the extracted circuits have only capacitors and inductors as their linear elements, with no transformers.

We note that the Appendices of this paper provide a detailed, standalone account of our results.
 
\section{Lumped superconducting equations of motion} \label{sec:LumpedEquationsOfMotion}

\subsection{Physical model} \label{subsec:PhysicalModel}
\begin{figure}
    \centering
    \includegraphics[width=1\linewidth]{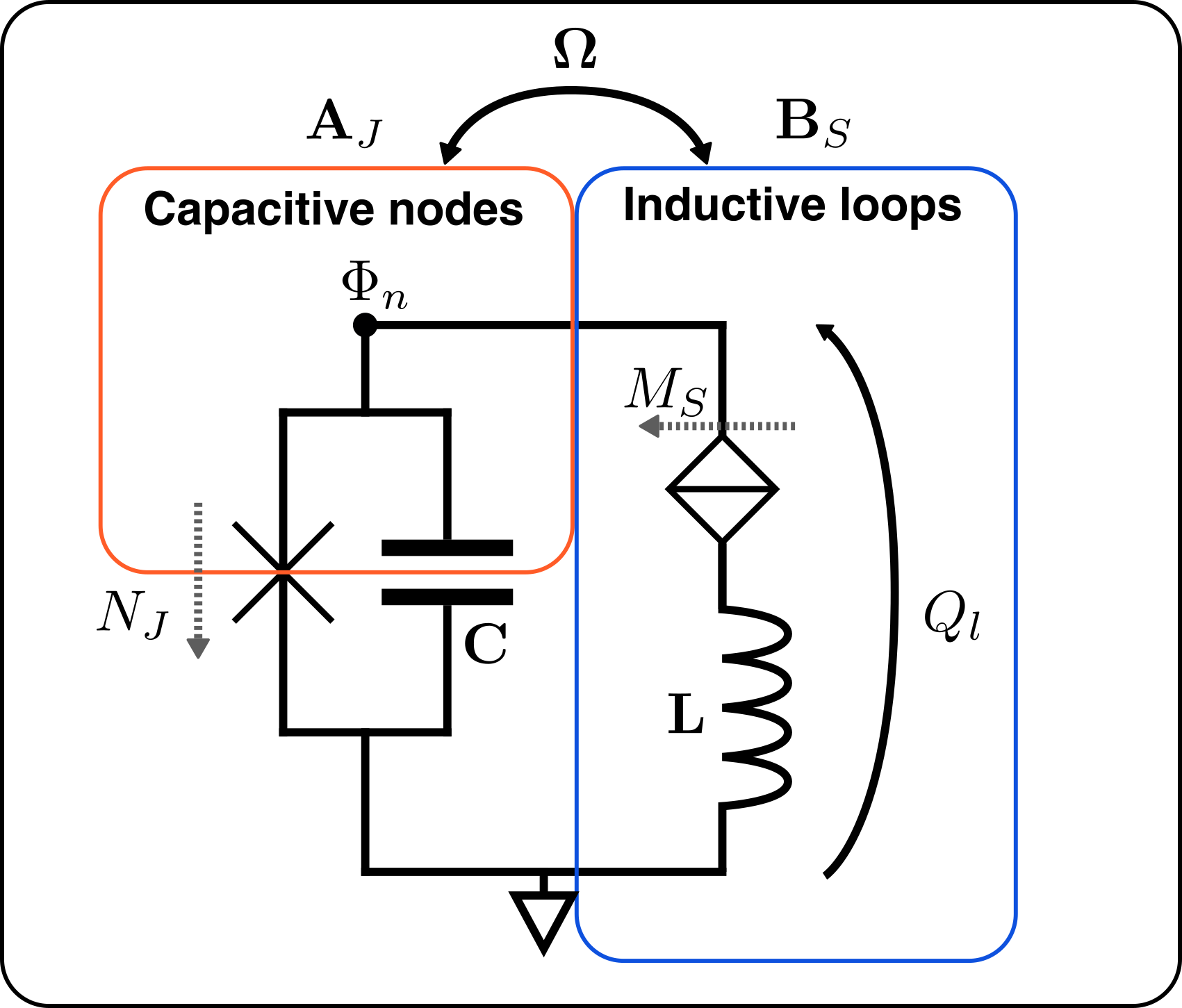}
    \caption{Overview of the flux-charge symmetric model. Circuits are characterized by both node flux variables $\Phi_n$ and loop charge variables $Q_l$. Capacitive networks possess nodal capacitance matrix $\mathbf{C}$ and inductive networks have loop inductance matrix $\mathbf{L}$. Josephson tunneling elements describe the tunneling of $N_J$ Cooper pairs between nodes enumerated by incidence matrix $\mathbf{A}_J$. Quantum phase slip elements describe the tunneling of $M_S$ magnetic flux quanta between loops, parametrized by loop matrix $\mathbf{B}_S$. The network matrix $\mathbf{\Omega}$ encodes the inter-connectivity between capacitive nodes and inductive loops.}
    \label{fig:ModelOverview}
\end{figure}

Before discussing quantization, decomposition, and model extraction, we provide an overview of the flux-charge symmetric model we use to analyze reciprocal lumped-element superconducting circuits. The circuit elements represent linear capacitance, linear inductance, Cooper pair tunneling (Josephson junction), and fluxon tunneling (phase slip wire). We construct vector-valued equations of motion for the capacitive nodes and inductive loops, in both ``standard" and ``integrated" forms. A set of heuristic derivations and physical interpretations of these models is given in Appendices \ref{sec:AppendixA} and \ref{sec:AppendixB}. Throughout, we employ the convention of assigning positive charge $2e$ to the charge carriers (Cooper pairs), and magnetic flux $\Phi_0$ to the flux carriers (fluxons).

Fig. \ref{fig:ModelOverview} gives an outline of our circuit framework, which employs similar notation to that used in \cite{osborne_symplectic_2024, osborne_flux-charge_2024}. Fundamentally, the graph circuit model is broken up into a set of interconnected nodes (or islands) and loops. Each capacitive node is characterized by a node flux variable $\Phi_n$, whose time derivative is the node voltage $V_n = \dot{\Phi}_n$ \cite{michel_h_devoret_quantum_1995}, and each inductive loop by a loop charge $Q_l$ \cite{ulrich_dual_2016}, whose time derivative is the loop current $I_l = \dot{Q}_l$. Appendix \ref{subsec:Vectorized} gives further background on this model, which employs linear-algebraic graph theory to model superconducting circuits \cite{michel_h_devoret_quantum_1995, vool_introduction_2017, burkard_multilevel_2004, burkard_circuit_2005, kerman_efficient_2020, parra-rodriguez_canonical_2019, egusquiza_algebraic_2022, parra-rodriguez_geometrical_2024, osborne_symplectic_2024, osborne_flux-charge_2024}.

In the constitutive relations for the linear portion of the system, the node fluxes $\vv{\Phi}_n$ are linked to node charges $\vv{Q}_n$ through the capacitance matrix $\mathbf{C}$, while the loop charges $\vv{Q}_l$ are related to loop fluxes $\vv{\Phi}_l$ by the inductance matrix $\mathbf{L}$ (discussed in Appendix \ref{subsec:CapacitanceInductanceMatrices}). These equations can be written in a vector form as \cite{kerman_efficient_2020}:
\begin{align}
\vv{Q}_n &= \mathbf{C}\dot{\vv{\Phi}}_n + \vv{Q}_\text{ext} \label{eq:CapacitiveConstitutiveFlux}\\
\vv{\Phi}_l &= \mathbf{L}\dot{\vv{Q}}_l + \vv{\Phi}_\text{ext} \label{eq:InductiveConstitutiveCharge}
\end{align}
The diagonal entries of $\mathbf{C}$/$\mathbf{L}$ represent the total self-capacitance/inductance, and the off-diagonals the coupling capacitance/inductance (or mutual inductance). We show how these matrices can be constructed from branch capacitors and inductors in Appendix \ref{sec:BranchCapacitorsAndInductors}---while emphasizing $\mathbf{C}$ and $\mathbf{L}$ as the more fundamental objects. Observe also that, in our framework, capacitance is only defined at nodes connected to the capacitive sub-network, and inductance is only defined in loops containing inductors---helping ensure that the capacitance and inductance matrices are invertible.
 
As detailed in Appendix \ref{subsec:ExternalChargeAndFlux}, $\vv{Q}_\text{ext}$ and $\vv{\Phi}_\text{ext}$ stand for external charge and external flux, respectively---and arise from either applied bias or noise. Note that in this framework, external flux is treated in a symmetric manner to external charge, and is assigned fundamentally to the inductive loop.

Pairs of capacitive nodes and inductive loops can also be connected through nonlinear quantum tunneling (Appendix \ref{subsec:ChargeAndFluxTunneling}). As illustrated in Fig. \ref{fig:ModelOverview}, the tunneling current flowing out of an node can be written as $I_J = 2e \dot{N}_J$, where $N_J$ is the number of charge carriers that have left the node since time $t_0$. Similarly, the voltage generated by fluxon tunneling through the loop is given by $V_S = -\Phi_0 \dot{M}_S$, where $M_S$ is the number of magnetic flux quanta that have tunneled into the loop. The presence of charge tunneling through a capacitor is denoted by the cross-shaped Josephson junction symbol, and that of fluxon-tunneling through the loop by the diamond shaped icon for quantum phase slip wire. We thus make the standard physical assumption that charge tunneling is intrinsically paired with parallel capacitance and flux tunneling with series inductance \cite{egusquiza_algebraic_2022} (depicted in Appendix Fig. \ref{fig:JJPS}).

The constitutive relation of a Josephson junction is given by $I_J = I_0 \sin \left(\frac{2 \pi}{\Phi_0}\Phi_J \right)$ \cite{josephson_possible_1962, tinkham_introduction_1996}, and the analogous one for a phase slip element by $V_S = V_0 \sin \left(\frac{2 \pi}{2e} Q_S \right) $\cite{mooij_superconducting_2006}. $I_0$ and $V_0$ are constants that characterize the tunneling strengths, and $\Phi_J$ is the capacitive flux across the Josephson junction, while $Q_S$ is the inductive charge along the wire.

The junction and phase slip constitutive equations can also be put in a vector form, for the case of multiple nodes and loops:
\begin{align}
\vv{I}_J &= \mathbf{I}_0 \sin \left(\frac{2 \pi}{\Phi_0} \mathbf{A}_J^T \vv{\Phi}_n \right) \label{eq:JunctionConstitutive}\\
\vv{V}_S &= \mathbf{V}_0 \sin \left(\frac{2 \pi}{2e} \mathbf{B}_S^T \vv{Q}_l \right) \label{eq:PhaseSlipConstitutive}
\end{align}
In these equations, $\mathbf{I}_0 = \text{diag}\left(\vv{I}_0 \right)$ and $\mathbf{V}_0 = \text{diag}\left(\vv{V}_0 \right)$ represent the set of Cooper pair and fluxon tunneling amplitudes, respectively. The sine functions are applied element-wise to the vector arguments, expressed in terms of junction incidence matrix $\mathbf{A}_J$ and phase slip loop matrix $\mathbf{B}_S$ (Appendix \ref{subsec:IncidenceLoopMatrices}).

$\mathbf{A}_J$ is the junction incidence matrix, which describes how the $J$ Josephson junction edges connect between the capacitive nodes of the circuit. The incidence matrix $\mathbf{A}$ of a directed graph has rows that correspond to graph nodes and columns that correspond to edges, such that \cite{strang_introduction_2016, bapat_graphs_2014}:
\begin{align} 
A_{ij} =
\begin{cases} 
      1 & \text{edge $j$ enters node $i$} \\
      -1 & \text{edge $j$ leaves node $i$} \\
      0 & \text{edge $j$ not at node $i$}
\end{cases}
\end{align}
When the row corresponding to the ground node is removed, $\mathbf{A}_J$ has at most one $-1$ and one $+1$ per column, with the rest of the entries being $0$. The vector of branch fluxes that appears in the sine argument in equation \ref{eq:JunctionConstitutive} is given by $\vv{\Phi}_J = \mathbf{A}_J^T \vv{\Phi}_n$---an expression of Kirchhoff's voltage law for capacitive nodes. By the structure of the incidence matrix, each branch flux will be the difference of two node fluxes. We make an additional physics-based assumption in the model, which is that there are no Josephson junction-only loops in the circuit. This is to say that each loop of junctions has some inductance, represented by an inductor that also appears in that loop. This condition implies that $\mathbf{A}_J$ has full column rank, equal to the number of tunneling junctions in the circuit.

Analogously, for a system with $S$ phase slip elements, the loop matrix $\mathbf{B}_S$ details how the phase slip tunneling branches connect between the inductive loops of the circuit. Here, two loops have tunneling between them if they share a phase slip edge. The loop matrix $\mathbf{B}$ of a directed graph has rows that correspond to loops and columns that correspond to edges, such that \cite{strang_introduction_2016,bapat_graphs_2014}:
\begin{align}
B_{ij} =
\begin{cases} 
      1 & \text{edge $j$ within loop $i$ --- same direction} \\
      -1 & \text{edge $j$ within loop $i$ --- opposite direction} \\
      0 & \text{edge $j$ not in loop $i$}
\end{cases}
\end{align}
We assign another physically-motivated condition for the phase slip loop matrix, which is that any node that is connected to multiple phase slip wires must also connect to a capacitor. This means that $\mathbf{B}_S$ will have full column rank, equal to the number of fluxon tunneling/phase slip segments. Also, if the loops are faces of planar graphs, each of these segments connects a pair of loops, such that the matrix again has at most one $-1$ and one $+1$ entry in each column. The sine argument here is applied to a vector of inductive branch charges $\vv{Q}_S = \mathbf{B}_S^T \vv{Q}_l$. This equation is analogous to Kirchhoff's current law for inductive loops.

The final ingredient needed to construct the equations of motion is the node-loop network matrix $\mathbf{\Omega}$ \cite{egusquiza_algebraic_2022,osborne_symplectic_2024,osborne_flux-charge_2024}. This matrix has rows that represent nodes and columns that represent loops, such that:
\begin{align}
\Omega_{ij} =
\begin{cases} 
      1 & \text{node $i$ touches $1$ inductive}  \\ & \text{edge in loop $j$ (tip)} \\
      -1 & \text{node $i$ touches $1$ inductive} \\ & \text{edge in loop $j$ (tail)} \\
      0 & \text{node $i$ touches even number} \\ & \text{of inductive edges in loop $j$}
\end{cases}
\end{align}
The analysis and manipulation of this object will play a fundamental role in the remainder of this work. An example circuit and its corresponding $\mathbf{A}_J$, $\mathbf{B}_S$, and $\mathbf{\Omega}$ matrices are shown in Fig. \ref{fig:TopologicalMatricesTreeCotree}(a).

In Appendix \ref{sec:AppendixB} (culminating in Appendix \ref{subsec:Vectorized}), we detail how to construct vector-valued equations of motion for superconducting circuits in complementary standard and integrated forms. Here we present a summary. The standard form is derived from the lumped superconducting analogues of the electromagnetic continuity equation and Faraday's law (or Kirchhoff's current law and voltage law):
\begin{align}
\dot{\vv{Q}}_n - \vv{I}_n &= \vv{0}_n \label{eq:ContinuityInitial} \\
\dot{\vv{\Phi}}_l + \vv{V}_l  &= \vv{0}_l \label{eq:FaradayInitial}
\end{align}

Here, the node charges ${\vv{Q}}_n$ and loop fluxes ${\vv{\Phi}}_l$ are defined as in Eqs.  \ref{eq:CapacitiveConstitutiveFlux} and \ref{eq:InductiveConstitutiveCharge}. The total current flowing into each node is the sum of the incident Josephson tunneling currents and net inductive loop currents. Correspondingly, the voltage drop around each loop is the sum of all the loop's phase slip voltage drops plus the voltage drops across the loop's capacitive nodes. These relations are expressed as:
\begin{align}
\vv{I}_n &= -\mathbf{A}_J \vv{I}_J + \mathbf{\Omega} \dot{\vv{Q}}_l \label{eq:NetNodeCurrent} \\
\vv{V}_l &=  \mathbf{B}_S \vv{V}_S  + \mathbf{\Omega}^T \dot{\vv{\Phi}}_n \label{eq:NetLoopVoltage}
\end{align}

Combining Eqs. \ref{eq:CapacitiveConstitutiveFlux}, \ref{eq:ContinuityInitial}, and  \ref{eq:NetNodeCurrent}---as well as Eqs. \ref{eq:InductiveConstitutiveCharge}, \ref{eq:FaradayInitial}, and \ref{eq:NetLoopVoltage}---we obtain the system's equations of motion in standard form:
\begin{align}
\mathbf{C}\ddot{\vv{\Phi}}_n + \dot{\vv{Q}}_{\text{ext}}+ \mathbf{A}_J \vv{I}_J - \mathbf{\Omega} \dot{\vv{Q}}_l &= \vv{0}_n \label{eq:ContinuityStandardForm} \\
\mathbf{L}\ddot{\vv{Q}}_l + \dot{\vv{\Phi}}_{\text{ext}}  + \mathbf{B}_S \vv{V}_S  + \mathbf{\Omega}^T \dot{\vv{\Phi}}_n  &= \vv{0}_l  \label{eq:FaradayStandardForm}
\end{align}
where $\vv{I}_J$ and $\vv{V}_S$ are defined in terms of $\vv{\Phi}_n$ and $\vv{Q}_l$ (respectively) in Eqs. \ref{eq:JunctionConstitutive} and \ref{eq:PhaseSlipConstitutive}. Here, the node flux variables $\vv{\Phi}_n$ and the loop charges $\vv{Q}_l$ represent the system's dynamical degrees of freedom, which each generate a canonically conjugate pair of operators when the system is quantized in Section \ref{sec:Quantization} \cite{michel_h_devoret_quantum_1995, osborne_symplectic_2024, parra-rodriguez_geometrical_2024}.

A central result shown in Appendix \ref{subsec:Vectorized} is that these equations can alternatively be expressed in a time-integrated form:
\begin{align}
\mathbf{C}\dot{\vv{\Phi}}_n + \vv{Q}_{\text{ext}}+ \mathbf{A}_J \left[ 2e \vv{N}_J \right]- \mathbf{\Omega} \vv{Q}_l &= 2e \vv{N}_0  \label{eq:ContinuityIntegratedForm} \\
\mathbf{L}\dot{\vv{Q}}_l + \vv{\Phi}_{\text{ext}}  - \mathbf{B}_S  \left[\Phi_0 \vv{M}_S \right]  + \mathbf{\Omega}^T \vv{\Phi}_n  &= \Phi_0 \vv{M}_0 \label{eq:FaradayIntegratedForm}
\end{align}
Here $\vv{N}_0$ is an integer-valued vector representing the number of initial charge carriers on each node at time $t_0$, and $\vv{M}_0$ is the analogous vector of magnetic fluxons enclosed in each loop at that time. Note that this expression keeps track of which quantities will become integer-valued when the circuit is quantized ($\vv{N}_J$, $\vv{N}_0$, $\vv{M}_S$, and $\vv{M}_0$). This form of the equations is used to eliminate free modes and back-solve for integer-valued canonical variables in the process of circuit quantization.

\subsection{Change of basis} \label{subsec:ChangeOfBasisMain}

Throughout this work, linear changes of basis are employed to manipulate the equations of motion (discussed further in Appendix \ref{subsec:ChangeOfBasis2}). Performing one set of transformations generates a quantization algorithm (Section \ref{sec:Quantization}) while carrying out another decomposes the circuit model to a simplified equivalent form (Section \ref{sec:Decomposition}). We often write the basis changes to $\vv{\Phi}_n$ and $\vv{Q}_l$  with the following notation:
\begin{align}
\vv{\Phi}_n &\rightarrow \mathbf{U}^{T^{-1}} \vv{\Phi}_n \\
\vv{Q}_l &\rightarrow \mathbf{W}^{T^{-1}} \vv{Q}_l
\end{align}

These transformations act on the system's matrices as \cite{chitta_computer-aided_2022}:
\begin{align}
\mathbf{C} &\rightarrow \mathbf{U} \mathbf{C} \mathbf{U}^T \\
\mathbf{L} &\rightarrow \mathbf{W} \mathbf{L} \mathbf{W}^T \\
\mathbf{A}_J &\rightarrow \mathbf{U} \mathbf{A}_J \\
\mathbf{B}_S &\rightarrow \mathbf{W} \mathbf{B}_S \\
\mathbf{\Omega} &\rightarrow \mathbf{U} \mathbf{\Omega} \mathbf{W}^T 
\end{align}

The other vectorial quantities transform as:
\begin{align}
\vv{Q}_\text{ext} &\rightarrow \mathbf{U} \vv{Q}_\text{ext}\\
\vv{\Phi}_\text{ext} &\rightarrow \mathbf{W} \vv{\Phi}_\text{ext} \\
\vv{N}_0 &\rightarrow \mathbf{U} \vv{N}_0\\
\vv{M}_0 &\rightarrow \mathbf{W} \vv{M}_0 
\end{align}

Thus $\mathbf{U}$ simultaneously performs row operations on $\mathbf{A}_J$ and $\mathbf{\Omega}$, while $\mathbf{W}$ performs row operations on $\mathbf{B}_S$ and column operations on $\mathbf{\Omega}$. Note that if $\mathbf{U}$ and $\mathbf{W}$ are integer-valued (as will often be the case), then the integer-valued natures of $\mathbf{A}_J$, $\mathbf{B}_S$, $\mathbf{\Omega}$, $\vv{N}_0$, $\vv{M}_0$ are preserved under transformation.

\subsection{Tree-cotree notation} \label{subsec:TreeCotreeNotation}

\begin{figure*}
    \centering
    \includegraphics[width=1\linewidth]{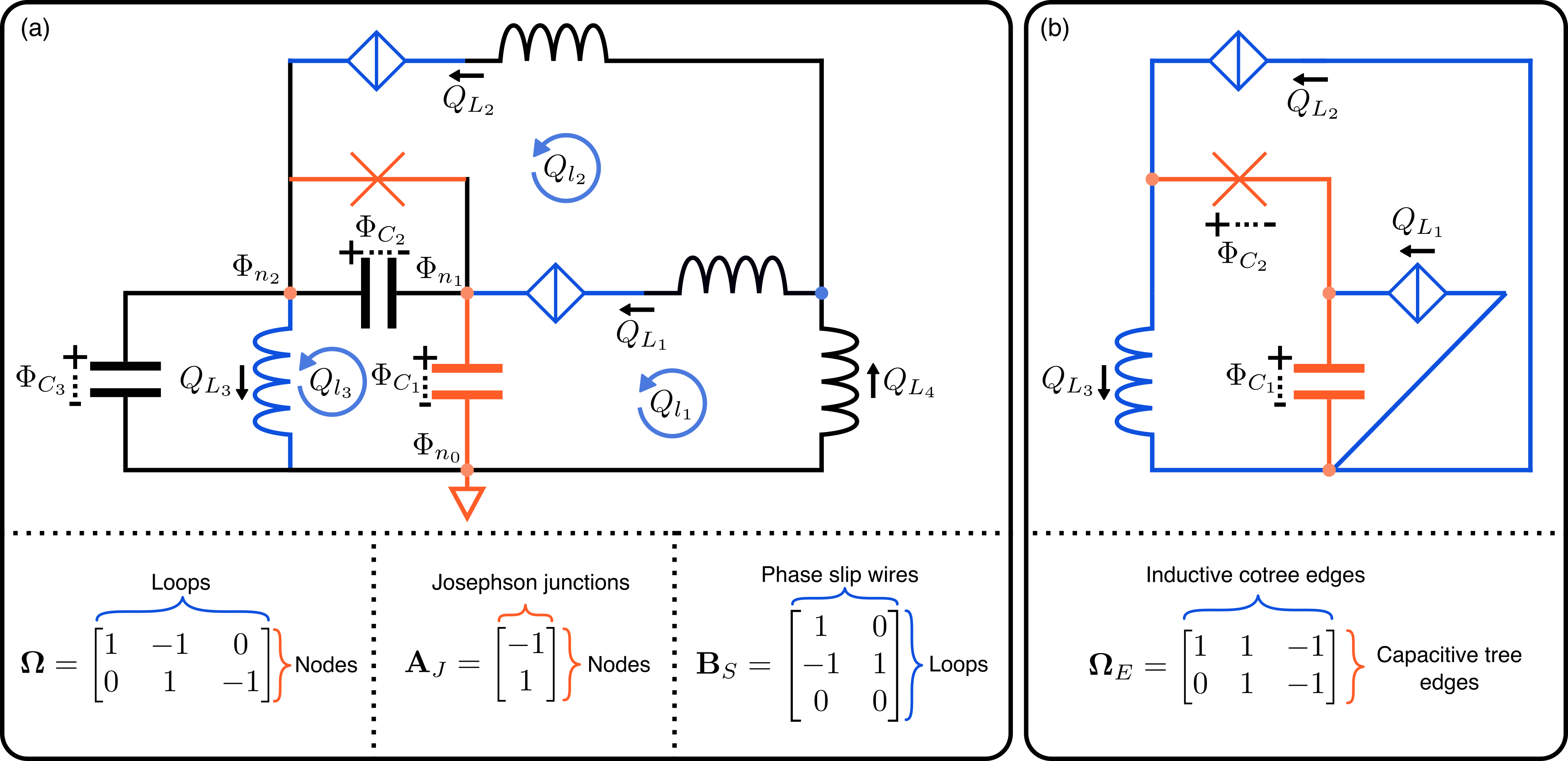}
    \caption{Superconducting circuit models represented in standard and tree-cotree notations. (a) Superconducting circuit in standard notation, with capacitors, inductors, Josephson junctions, and quantum phase slip wires. Node fluxes are specified by $\Phi_n$, with capacitive branch fluxes given by $\Phi_C$. Similarly, loop charges are written as $Q_l$ with inductor branch charges $Q_L$. The system's node-loop network matrix $\mathbf{\Omega}$, Josephson junction incidence matrix $\mathbf{A}_J$, and phase slip loop matrix $\mathbf{B}_S$ are pictured below the circuit drawing. A capacitive spanning tree (orange) and an inductive spanning cotree (blue) are also depicted. (b) The same superconducting circuit in tree-cotree notation, where only the capacitive spanning tree (orange) and inductive cotree edges (blue) are drawn. The cutset/loop topology of the system is captured by the edge network matrix $\mathbf{\Omega}_E$.}
\label{fig:TopologicalMatricesTreeCotree}
\end{figure*}

To better understand and work with our superconducting circuit model, we can visualize it in an alternate tree-cotree notation, seen in Fig. \ref{fig:TopologicalMatricesTreeCotree}. In this shorthand, we only draw the circuit's capacitive spanning tree (orange) and an inductive spanning cotree (blue). Because our physical restrictions specify that there are no Josephson junction-only loops or phase slip-only cutsets (see Appendix \ref{subsec:BranchDecomposition}), we can find (1) a set of capacitive tree branches that span the capacitive nodes and go across every Josephson junction, and (2) a set of inductive cotree branches that span the system's inductive loops and lie along every phase slip inductor (such that every loop current can be written as a linear combination of branch currents).

We use the Josephson junction symbol to denote capacitive edges across a Josephson junction, and the linear capacitor symbol for capacitive edges that are not. Similarly, in tree-cotree notation, the quantum phase slip emblem stands for inductive cotree edges along a phase slip, and the linear inductor for those that are not along one. For the linear part of the capacitive spanning tree (inductive cotree) the choice of edges is arbitrary. We note that tree-cotree notation removes capacitive loops and inductive nodes from the circuit drawing, with the these phenomena already recorded in the nodal capacitance matrix $\mathbf{C}$ and loop inductance matrix $\mathbf{L}$. Our notation differs from others used in the field in that we do not require every edge to lie in either the tree or cotree. \cite{michel_h_devoret_quantum_1995, burkard_multilevel_2004, burkard_circuit_2005, rasmussen_superconducting_2021, ciani_lecture_2024, parra-rodriguez_geometrical_2024}.

Fig. \ref{subsec:EdgeNetworkSpanningTreeCotree}(b) gives an example of a circuit in tree-cotree notation (with that circuit's standard notation shown in Fig. \ref{subsec:EdgeNetworkSpanningTreeCotree}(a)). Later, in Section \ref{subsec:EdgeNetworkSpanningTreeCotree}, we will see how the notation transformation corresponds to a change of basis. By changing basis, the node-loop network matrix $\mathbf{\Omega}$ transforms into the edge network matrix $\mathbf{\Omega}_E$ (Eq. \ref{eq:MainNetworkMatrixLoop}) \cite{schrijver_theory_1998, wolsey_integer_2014}, which encodes the cutset/loop connectivity between the tree and cotree edges of the system \cite{bapat_graphs_2014, burkard_multilevel_2004, burkard_circuit_2005}.

\section{Quantization}\label{sec:Quantization}

\subsection{Quantization algorithm}

Using the equations of motion (Eqs. \ref{eq:ContinuityStandardForm} and \ref{eq:FaradayStandardForm}), we now present a straightforward algorithm to generate a quantized Hamiltonian for circuits containing both Josephson junctions and phase slip wires. The underlying process and formalism bear similarity to the ideas presented in
\cite{egusquiza_algebraic_2022, parra-rodriguez_geometrical_2024, osborne_symplectic_2024, osborne_flux-charge_2024}. The primary novelty of our method is the use of the integrated equations of motion (Eqs. \ref{eq:ContinuityIntegratedForm} and \ref{eq:FaradayIntegratedForm}) to predict whether the quantum operators have continuous or discrete/integer-valued spectra (with compact conjugate variables). Physically, this discreteness arises from the discrete spectra of the charge and flux tunneling operators. An illustrative example is given in Section \ref{subsec:FluxoniumWithPhaseSlips}. More details about the following algorithm can be found in Appendix \ref{sec:AppendixC}.

\begin{figure}
    \centering
    \includegraphics[width=1\linewidth]{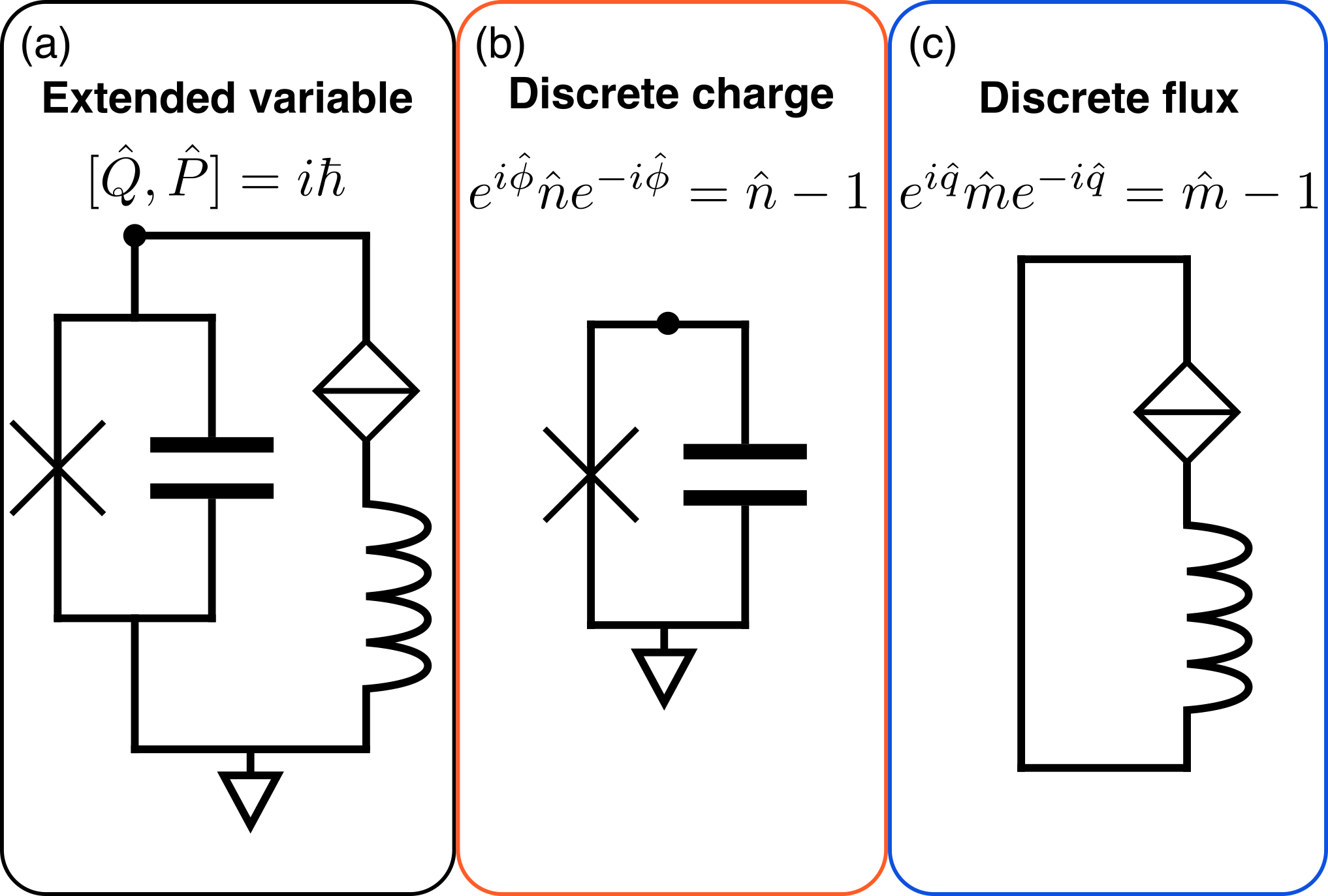}
    \caption{Predicted types superconducting circuit modes. (a) Both operators have extended, continuous spectra (with another doubly-discrete conjugate pair that drops out of Hamiltonian). (b) One operator has discrete charge and the other compact/periodic flux. (c) One operator has discrete flux and the other compact/periodic charge.}
    \label{fig:TypesOfModes}
\end{figure}

In order to quantize the equations of motion, the essential step is to perform basis transformations (as described in section \ref{subsec:ChangeOfBasisMain}) that reduce the network matrix (of rank $k$) to the upper left identity matrix $\mathbf{I}_{kk}$, with zeros everywhere else. This is similar to the process of manipulating the symplectic form, shown in \cite{parra-rodriguez_geometrical_2024, osborne_symplectic_2024}, and gives a result of:
\begin{align}
\mathbf{\Omega} \rightarrow 
\mathbf{U} \mathbf{\Omega} \mathbf{W}^T 
=
\begin{bmatrix}
\mathbf{I}_{kk} & \mathbf{0}_{ks} \\
\mathbf{0}_{jk} & \mathbf{0}_{js}
\end{bmatrix}
\end{align}
Eventually, each index $k$ will produce a pair of charge and flux conjugate operators with continuous spectra over the real numbers. Each $j$ will produce a discrete charge/compact flux pair, arising from capacitive nodes connected by Josephson junctions. Each $s$ will produce a discrete flux/compact charge pair, arising from from inductive loops connected through quantum phase slip/fluxon tunneling wires. These three types of modes are illustrated in Fig. \ref{fig:TypesOfModes}.

The transformations $\mathbf{U}$ (on node flux variables) and $\mathbf{W}$ (on loop charge) are integer-valued and transform the junction incidence and phase slip loop matrices as:
\begin{align}
\mathbf{A}_J \rightarrow 
\mathbf{U} \mathbf{A}_J
&=
\begin{bmatrix}
\mathbf{A}_{kJ} \\
\mathbf{A}_{jJ} 
\end{bmatrix} \\ %newline
\mathbf{B}_S \rightarrow
\mathbf{W} \mathbf{B}_S 
&=
\begin{bmatrix} 
\mathbf{B}_{kS} \\
\mathbf{B}_{sS} 
\end{bmatrix} 
\end{align}
The transformed matrices then remain integer-valued in their entries. We also split them into two different groupings of rows to align their indices with those of the transformed $\mathbf{\Omega}$ matrix, enabling block matrix manipulation. The other transformed quantities can also be divided up into block form, with four total subsets of dynamical variables:
\begin{align}
\vv{\Phi}_{n}
= 
\begin{bmatrix}
\vv{\Phi}_{k}  \\
\vv{\Phi}_{j}
\end{bmatrix} \\
\vv{Q}_{l}
= 
\begin{bmatrix}
\vv{Q}_{k}  \\
\vv{Q}_{s}
\end{bmatrix}
\end{align}

We then begin the quantization procedure by translating our equations of motion into a Hamiltonian. As shown in Appendix \ref{subsec:Quantization1}, we start this process via constructing a Lagrangian whose Euler-Lagrange equations reproduce the standard equations of motion for our system (Eqs. \ref{eq:ContinuityStandardForm} and \ref{eq:FaradayIntegratedForm}) \cite{osborne_symplectic_2024, osborne_flux-charge_2024}.

In the following step, we define a canonical conjugate variable to each coordinate, using the symbol $\Pi$ to denote conjugates to $\Phi$ coordinates and $P$ to indicate those conjugate to $Q$ \cite{egusquiza_algebraic_2022}. There are four sets of conjugate pairs with Poisson Brackets of:
\begin{align}
\{ \vv{\Phi}_k,  \vv{\Pi}_k \} = \mathbf{I}_{kk} \\
\{ \vv{Q}_k,  \vv{P}_k \} = \mathbf{I}_{kk} \\
\{ \vv{\Phi}_j, \ \vv{\Pi}_j \} = \mathbf{I}_{jj} \\
\{ \vv{Q}_s, \  \vv{P}_s \} = \mathbf{I}_{ss}
\end{align}
and brackets of 0 between all other variable pairs. By performing a Legendre transform, we can then write down a Hamiltonian in terms of these conjugate pairs (shown in Eq. \ref{eq:IntermediateHamiltonian}). We now carry out another key step in the process: inserting the conjugate degrees of freedom into the integrated equations of motion (Eqs. \ref{eq:ContinuityIntegratedForm} and \ref{eq:FaradayIntegratedForm}) to assess whether---when quantized---they will have discrete or continuous spectra. The quantities $\vv{\Pi}_j$ and $\vv{P}_s$ take on integer multiples of $2e$ and $\Phi_0$, respectively:
\begin{align}
\vv{\Pi}_j = - 2e \mathbf{A}_{jJ} \vv{N}_J \\
\vv{P}_s = \Phi_0 \mathbf{B}_{sS}  \vv{M}_S
\end{align}
because $\mathbf{A}_{jJ}$, $\vv{N}_J$, $\mathbf{B}_{sS}$, and $\vv{M}_S$ are all integer-valued. Note that the integer-valued nature of $\vv{N}_J$ and $\vv{M}_S$ comes from the quantization of Cooper pair tunneling across a capacitive gap and fluxon tunneling across an inductive loop.

The other two sets of conjugate variables of the Hamiltonian (both with index $k$) can be analyzed in a similar fashion. By performing a canonical transformation (that preserves Poisson brackets) and then inserting the results into the standard equations of motion, we obtain:
\begin{align}
\vv{\Phi}_k &\rightarrow  \vv{\Phi}_k + \vv{P}_k =  \Phi_0\mathbf{B}_{kS}  \vv{M}_S \nonumber \\
\vv{\Pi}_k & \rightarrow  \vv{\Pi}_k = -2e\mathbf{A}_{kJ}  \vv{N}_J \nonumber \\
\vv{Q}_k & \rightarrow \vv{Q}_k + \vv{\Pi}_k = \vv{Q}_k  - 2e \mathbf{A}_{kJ}\vv{N}_J \nonumber \\
\vv{P}_k & \rightarrow  \vv{P}_k = -\vv{\Phi}_k + \Phi_0\mathbf{B}_{kS}  \vv{M}_S \label{eq:CanonicalChangeOfBasisMain}
\end{align}

We have then produced two types of conjugate pairs of variables with index $k$. One of them ($\vv{Q}_k$ and $\vv{P}_k$) will have continuous spectra when quantized. The others ($\vv{\Phi}_k$ and $\vv{\Pi}_k$) represent conjugate pairs where both operators will become discrete. The second type of conjugate pair participates only in the periodic terms of the Hamiltonian, and in fact drops out of the dynamics by: 
\begin{align}
\mathcal{H}_P =&
- \vv{E}_J^T \cos \left( \frac{2\pi}{\Phi_0} 
\begin{bmatrix}
\mathbf{A}_{kJ}^T &  \mathbf{A}_{jJ}^T 
\end{bmatrix}
\begin{bmatrix}
- \vv{P}_k + \Phi_0\mathbf{B}_{kS} \vv{M}_S \\ 
\vv{\Phi}_j
\end{bmatrix}\right) \nonumber \\ %newline
&-
\vv{E}_S^T \cos \left( \frac{2\pi}{2e} 
\begin{bmatrix}
\mathbf{B}_{kS}^T &  \mathbf{B}_{sS}^T 
\end{bmatrix}
\begin{bmatrix}
\vv{Q}_k + 2e\mathbf{A}_{kJ}  \vv{N}_S \\ 
\vv{Q}_s
\end{bmatrix}\right) \nonumber \\ %newequation
 \nonumber \\ %newline 
=& - \vv{E}_J^T \cos \left( \frac{2\pi}{\Phi_0} 
\begin{bmatrix}
\mathbf{A}_{kJ}^T &  \mathbf{A}_{jJ}^T 
\end{bmatrix}
\begin{bmatrix}
- \vv{P}_k  \\ 
\vv{\Phi}_j
\end{bmatrix}\right) \nonumber \\ %newline
&-
\vv{E}_S^T \cos \left( \frac{2\pi}{2e} 
\begin{bmatrix}
\mathbf{B}_{kS}^T &  \mathbf{B}_{sS}^T 
\end{bmatrix}
\begin{bmatrix}
\vv{Q}_k \\ 
\vv{Q}_s
\end{bmatrix}\right) \label{eq:DoublyDiscreteRemoval}
\end{align}
where the (eventually) integer-valued degrees of freedom are removed from the cosines as they appear in multiples of $2\pi$. The presence of these removable, doubly-discrete conjugate variables is the main hypothesis of our quantization procedure.

We perform one more notational change of the $j$ and $s$ variables by defining:
\begin{align}
\vv{n}_j &= \frac{1}{2e} \vv{\Pi}_j  \\
\vv{\phi}_j &= \frac{2 \pi}{\Phi_0} \vv{\Phi}_j \\
\vv{m}_s &=  \frac{1}{\Phi_0} \vv{P}_s \\
\vv{q}_s &= \frac{2 \pi}{2e} \vv{Q}_s 
\end{align}
where $\vv{n}_j$ and $\vv{m}_s$ will become integer-valued operators, while $\vv{\phi}_j$ and $\vv{q}_s$ will become compact on the circle $\mathbb{R} \bmod (2 \pi)$. The notation modification distinguishes these conjugate pairs from from $\vv{Q}_k$ and $\vv{P}_k$, whose spectra extend over the real number line.

After these relabelings, the canonical change of basis in Eq. \ref{eq:CanonicalChangeOfBasisMain}, and the removal of doubly-discrete variable pairs (Eq. \ref{eq:DoublyDiscreteRemoval}), we obtain a final Hamiltonian (in Appendix \ref{subsec:Quantization2}), which we reproduce here:
\begin{align} \label{eq:MainFinalHamiltonian}
\mathcal{H} =
\frac{1}{2}
\begin{bmatrix}
\vv{Q}_k - \vv{Q}_{\text{ext}_k} \\
2e\vv{n}_j - \vv{Q}_{\text{ext}_j}
\end{bmatrix}^T 
\begin{bmatrix}
\mathbf{C}_{kk} & \mathbf{C}_{kj} \\
\mathbf{C}_{jk} & \mathbf{C}_{jj}
\end{bmatrix}^{-1}
\begin{bmatrix}
\vv{Q}_k - \vv{Q}_{\text{ext}_k} \\
2e\vv{n}_j - \vv{Q}_{\text{ext}_j}
\end{bmatrix} \nonumber \\ %newline
+\frac{1}{2}
\begin{bmatrix}
\vv{P}_k -  \vv{\Phi}_{\text{ext}_k} \\
\Phi_0 \vv{m}_s - \vv{\Phi}_{\text{ext}_s}
\end{bmatrix}^T 
\begin{bmatrix}
\mathbf{L}_{kk} & \mathbf{L}_{ks} \\
\mathbf{L}_{sk} & \mathbf{L}_{ss}
\end{bmatrix}^{-1}
\begin{bmatrix}
\vv{P}_k -  \vv{\Phi}_{\text{ext}_k} \\
\Phi_0 \vv{m}_s  - \vv{\Phi}_{\text{ext}_s}
\end{bmatrix} \nonumber \\ %newline 
- \vv{E}_J^T \cos \left( 
-\frac{2\pi}{\Phi_0}\mathbf{A}_{kJ}^T  \vv{P}_k  +  \mathbf{A}_{jJ}^T \vv{\phi}_j
\right) \nonumber \\ %newline
-
\vv{E}_S^T \cos \left(
\frac{2\pi}{2e} \mathbf{B}_{kS}^T\vv{Q}_k  +  \mathbf{B}_{sS}^T \vv{q}_s \right)
\end{align}
This Hamiltonian contains two types of terms: quadratic forms and cosines, as is standard for circuit Hamiltonians \cite{osborne_symplectic_2024, parra-rodriguez_geometrical_2024}. Note that $\vv{P}_k$ has units of flux. 

The physical restrictions imposed in Section \ref{subsec:PhysicalModel} ensure that the capacitance and inductance matrices are invertible. In addition, because we forbid Josephson junction-only loops, external flux is straightforwardly allocated to the linear inductance terms \cite{you_circuit_2019}.

The Hamiltonian can be quantized by promoting the variables to operators, with three resulting types of commutation relations, illustrated in Fig. \ref{fig:TypesOfModes}:
\begin{align}
[\hat{Q},\hat{P}] &= i \hbar \\
e^{i\hat{\phi}} \hat{n} e^{-i\hat{\phi}} &= \hat{n} - 1 \\
e^{i\hat{q}} \hat{m} e^{-i\hat{q}}& = \hat{m} - 1 
\end{align}
The pair $(\hat{Q},\hat{P})$ consists of charge and flux conjugate operators with continuous spectra extended across $\mathbb{R}$. On the other hand, for conjugates $(\hat{n},\hat{\phi})$, the charge operator $\hat{n}$ has spectrum in $\mathbb{Z}$ and the flux operator $\hat{\phi}$ takes values in the compact (circular) manifold $\mathbb{R} \bmod (2 \pi)$. Similarly, for the elements $(\hat{m},\hat{q})$ the flux operator $\hat{m}$ is discrete, while the charge operator $\hat{q}$ is compact/periodic. We note that the compact flux and charge variables appear only in the cosines, and the discrete charge and flux variables only in the quadratic terms. More discussion of these relations can be found in Appendix \ref{subsec:Quantization2}.

These commutation relations line up with orthodox interpretations of operator spectra in circuit quantum electrodynamics, for circuits with either Josephson junctions \cite{devoret_does_2021} or phase slip wires \cite{mooij_superconducting_2006}. However, our methods produce novel hypotheses for circuits containing both types of nonlinear element---where we predict the absence of conjugate pairs $({\hat{\vv{\Phi}}, \hat{\vv{\Pi}}})$ (Eq. \ref{eq:CanonicalChangeOfBasisMain}) from the Hamiltonian. These pairs vanish because they are doubly-discrete and only appear in the cosine portions of the Hamiltonian, as integer multiples of $2 \pi$ (Eq. \ref{eq:DoublyDiscreteRemoval}). This approach contrasts with treatments that predict $\hat{\vv{\Phi}}$ and $\hat{\vv{\Pi}}$ to be conserved quantities with continuous spectra \cite{le_doubly_2019, parra-rodriguez_geometrical_2024}, wherein they can affect the overall Hamiltonian if the qubit begins in a superposition state.

\subsection{Example: fluxonium with phase slips} \label{subsec:FluxoniumWithPhaseSlips} 

An illustrative example of our quantization methodology can be performed on the circuit shown in Fig. \ref{fig:ModelOverview}, which consists of a capacitor and Josephson junction in parallel connected across an inductor and phase slip wire in series. Ideally, this circuit represents a simple model of a fluxonium qubit \cite{manucharyan_fluxonium_2009} with quantum phase slips \cite{manucharyan_evidence_2012, randeria_dephasing_2024} across its inductor. Note, however, that due to charge noise in the inductor, the phase slip rate $E_S$ will often be a fluctuating quantity. In other sources this qubit has been referred to as the realistic dualmon qubit \cite{le_doubly_2019}, and we call it an LC oscillator with quantum tunneling in Appendix \ref{subsec:LCOscillatorQuantumTunneling}. Whatever its label, it is the simplest realistic, nontrivial circuit containing a Josephson junction and a phase slip wire. We demonstrate how to apply our algorithm to quantize this circuit, removing the doubly-discrete degree of freedom, resulting in a single conjugate pair of extended spectra operators.

Since the circuit possesses a single capacitive node connected to a single inductive loop, the junction incidence matrix, phase slip loop matrix, and network matrix are all equal to the one-by-one identity matrix:
\begin{align}
\mathbf{A}_J = \mathbf{B}_S = \mathbf{\Omega} = \begin{bmatrix} 1 \end{bmatrix}
\end{align}

The equations of motion in the standard form (Eqs. \ref{eq:ContinuityStandardForm} and \ref{eq:FaradayStandardForm}) are:
\begin{align}
C \ddot{\Phi}_n+ \dot{Q}_{\text{ext}} + I_0 \sin \left( \frac{2 \pi}{\Phi_0} \Phi_n \right)- \dot{Q}_l  = 0 \\
L \ddot{Q}_l+ \dot{\Phi}_{\text{ext}} + V_0 \sin \left( \frac{2 \pi}{2e} Q_l \right) + \dot{\Phi}_n = 0 
\end{align}

The corresponding integrated equations of motion (Eqs. \ref{eq:ContinuityIntegratedForm} and \ref{eq:ContinuityStandardForm}) can be written as:
\begin{align}
C \dot{\Phi}_n + Q_{\text{ext}} + 2e N_J - Q_l  = 0 \\
L \dot{Q}_l+ \Phi_{\text{ext}} - \Phi_0 M_S  + \Phi_n = 0 
\end{align}
where the constants of integration have been folded into $Q_{\text{ext}}$ and $\Phi_{\text{ext}}$.

We can then write down a Lagrangian for the equations of motion in standard form:
\begin{align}
\mathcal{L} =& \frac{1}{2} C \dot{\Phi}_n^2+ \dot{\Phi}_n Q_\text{ext} +\frac{1}{2} L \dot{Q}_l^2 + \dot{Q}_l \Phi_\text{ext} - \dot{\Phi}_n Q_l \nonumber \\
& +E_J \cos \left( \frac{2 \pi}{\Phi_0} \Phi_n \right) + E_S \cos \left( \frac{2 \pi}{2e} Q_l \right)
\end{align}

Finding the canonically conjugate coordinates gives:
\begin{align}
\Pi_n &= \frac{\partial \mathcal{L}}{\partial \dot{\Phi}_n} = C \dot{\Phi}_n -Q_l + Q_\text{ext}    \\
P_l &= \frac{\partial \mathcal{L}}{\partial \dot{Q}_l} = L \dot{Q}_l + \Phi_\text{ext}
\end{align}

The Legendre transform generates a Hamiltonian of: 
\begin{align}
\mathcal{H} =& \dot{\Phi}_n \Pi_n  + \dot{Q}_l  P_l - \mathcal{L} \nonumber \\
=& \frac{1}{2C} \left(\Pi_n + Q_l -Q_\text{ext} \right)^2 + \frac{1}{2L} \left(P_l - \Phi_\text{ext} \right)^2 \nonumber \\
&- E_J \cos \left( \frac{2 \pi}{\Phi_0} \Phi_n \right) - E_S \cos \left( \frac{2 \pi}{2e} Q_l \right)
\end{align}
with Poisson brackets:
\begin{align}
\{\Phi_n, \Pi_n \}  = 1 \\
\{Q_l, P_l \}  = 1 
\end{align}

Now we perform a canonical transformation and solve for the transformed operators by inserting them into the integrated equation of motion:
\begin{align}
\Phi_n &\rightarrow \Phi_n + P_l = \Phi_0 M_S  \nonumber \\
\Pi_n &\rightarrow \Pi_n =  -2e N_J  \nonumber \\
Q_l &\rightarrow Q_l + \Pi_n =  Q_l - 2e N_J  \nonumber \\
P_l &\rightarrow P_l = - \Phi_n + \Phi_0 M_S \label{eq:FluxoniumPhaseSlipsOperators}
\end{align}

Once we apply this transformation, the doubly-discrete pair of variables $(\Phi_n,\Pi_n)$ drops out of the Hamiltonian:
\begin{align}
\mathcal{H} \rightarrow & \frac{1}{2C} \left(Q_l -Q_\text{ext} \right)^2 + \frac{1}{2L} \left(P_l - \Phi_\text{ext} \right)^2 \nonumber \\
&- E_J \cos \left( \frac{2 \pi}{\Phi_0} (- P_l + \Phi_0 M_S)  \right) \nonumber \\ 
&- E_S \cos \left( \frac{2 \pi}{2e}( Q_l + 2e N_J) \right) \\ %new equation
=& \frac{1}{2C} \left(Q_l -Q_\text{ext} \right)^2 + \frac{1}{2L} \left(P_l - \Phi_\text{ext} \right)^2 \nonumber \\
&- E_J \cos \left( \frac{2 \pi}{\Phi_0}P_l\right) -E_S \cos \left( \frac{2 \pi}{2e} Q_l \right)
\end{align}
where $\{Q_l, P_l \}  = 1 $. To align our variables with the more standard notation for the fluxonium qubit, we redefine: $\Phi = -P_l$ and $Q = Q_l$. We then quantize the Hamiltonian by taking:
\begin{align}
\{\Phi, Q \}  = 1 \rightarrow [\hat{\Phi}, \hat{Q}]  = i \hbar
\end{align}

The quantized Hamiltonian reads:
\begin{align}
\hat{\mathcal{H}} = & \frac{1}{2C} \left(\hat{Q} -Q_\text{ext} \right)^2 + \frac{1}{2L} \left(\hat{\Phi} + \Phi_\text{ext} \right)^2 \nonumber \\
&- E_J \cos \left( \frac{2 \pi}{\Phi_0} \hat{\Phi} \right) -E_S \cos \left( \frac{2 \pi}{2e} \hat{Q} \right)
\end{align}
and is the standard Hamiltonian for the fluxonium qubit, but with DC sensitivity to shifts in external charge, and containing an additional cosine term with prefactor $E_S$. This cosine of charge can be expanded as:
\begin{align}
\cos \left( \frac{2 \pi}{2e} \hat{Q} \right) = \frac{1}{2} \left( e^{\frac{2 \pi}{2e} \hat{Q}} + e^{-\frac{2 \pi}{2e} \hat{Q}} \right)
\end{align}
Since $\hat{Q}$ is the generator of translations in $\hat{\Phi}$:
\begin{align}
\hat{\Phi} e^{ \pm \frac{2 \pi}{2e} \hat{Q}} \ket{\Phi} = (\Phi \pm \Phi_0) e^{\pm \frac{2 \pi}{2e} \hat{Q}} \ket{\Phi} 
\end{align}
we have that this cosine term couples the ket $\ket{\Phi}$ to $\ket{\Phi \pm \Phi_0}$ and represents quantum tunneling of a fluxon through the inductor of the loop. 

With a time-dependent $E_S$, the Hamiltonian aligns with those used to model fluxonium qubits with quantum phase slips in their Josephson junction arrays \cite{manucharyan_evidence_2012, randeria_dephasing_2024}.

This result exemplifies the difference between our method and that presented in \cite{parra-rodriguez_geometrical_2024}. In our approach, one of the system's conjugate variable pairs consists of discrete tunneling fluxes and charges (Eq. \ref{eq:FluxoniumPhaseSlipsOperators}):
\begin{align}
(\hat{\Phi}_n, \hat{\Pi}_n) = (\Phi_0 \hat{M}_S, -2e \hat{N}_J) = (\hat{\Phi}_S, -\hat{Q}_J)
\end{align}
ultimately causing them to drop out of the Hamiltonian. However, in \cite{parra-rodriguez_geometrical_2024}, elements of the conjugate pair $(\hat{\Phi}_S, -\hat{Q}_J)$ would become operators with continuous spectra. In the language of their work, we hypothesize that the pre-canonical manifold of phase space for these variables is $S^1 \times S^1$ instead of $\mathbb{R} \times \mathbb{R}$.

\section{Circuit decomposition}\label{sec:Decomposition}

\subsection{Equivalent circuits}
In our methodology, certain basis changes correspond to transforming the topology of a circuit (containing capacitors, inductors, Josephson junctions, and phase slips) to an alternate layout with an identical Hamiltonian. Here, we discuss how these ``structure-preserving" basis transformations can be used to perform a ``fundamental decomposition," whereby we reduce a circuit into its simplest equivalent form. In this procedure, we manipulate and separate out the linear portions of the circuit, while leaving invariant the nonlinear degrees of freedom: currents and fluxes across Josephson junctions ($I_J$ and $\Phi_J$) and the voltages and charges across phase slips ($V_S$ and $Q_S$).

The decomposition process has complementary visual and mathematical interpretations. We can envision it in tree-cotree notation (Section \ref{subsec:TreeCotreeNotation}) through moving capacitive and inductive edges, while maintaining the tree-cotree structure. Mathematically, we perform the decomposition by applying pivoting operations to the ``edge" network matrix $\mathbf{\Omega}_E$, which is related to the node-loop network matrix $\mathbf{\Omega}$ by a change of basis. We give a more thorough description of these procedures in Appendix \ref{sec:AppendixD}. 

\subsection{Edge network matrix, spanning tree-cotree basis} \label{subsec:EdgeNetworkSpanningTreeCotree}

In this Section (elaborated upon in Appendix \ref{subsec:BranchDecomposition}), we transform our equations of motion to a capacitive tree/inductive cotree basis, which simplifies the decomposition process. The topological information of the system is transferred to the transformed network matrix $\mathbf{\Omega}_E$, which we refer to as the ``edge" network matrix, because it details the cutset/loop connectivity between capacitive and inductive edges \cite{schrijver_theory_1998, wolsey_integer_2014}. This matrix is straightforward to manipulate into equivalent forms, which can be visualized as equivalent circuit transformations in tree-cotree notation (Section \ref{subsec:TreeCotreeNotation}).

To perform the change of basis, we first select a capacitive spanning tree of the capacitive nodes of the graph (orange edges in Fig. \ref{fig:TopologicalMatricesTreeCotree}), whose incidence matrix is $\mathbf{A}_{C_\mathcal{T}}$. We then choose an inductive cotree whose branches span all inductive loops of the system, with loop matrix $\mathbf{B}_{L_\mathcal{T}}$ (blue edges in Fig. \ref{fig:TopologicalMatricesTreeCotree}). We require the capacitive spanning tree to lie across all junctions and the inductive cotree to lie along all phase slip branches.

It is possible to pick such a tree and cotree because of the aforementioned physical constraints we have placed on the circuit, such that there are no junction-only loops (without inductors) or phase slip-only nodes/cutsets (without capacitors). We are thus also guaranteed that $\mathbf{A}_{C_\mathcal{T}}$ and $\mathbf{B}_{L_\mathcal{T}}$ will be of full rank and thus invertible \cite{bapat_graphs_2014}. We can divide each of these matrices into two submatrix blocks:
\begin{align}
    \mathbf{A}_{C_\mathcal{T}} &=
    \begin{bmatrix}
    \mathbf{A}_J &
    \mathbf{A}_{C'} \label{eq:ACSplit}
    \end{bmatrix} \\
    \mathbf{B}_{L_\mathcal{T}} &=
    \begin{bmatrix}
    \mathbf{B}_S &
    \mathbf{B}_{L'} \label{eq:BLSplit}
    \end{bmatrix}
\end{align}
We see the first $J$ columns of $\mathbf{A}_{C_\mathcal{T}}$ form the Josephson junction incidence matrix $\mathbf{A}_J$, while the last $C'$ are the incidence matrix $\mathbf{A}_{C'}$ of the non-junction (linear capacitor) edges. Analogously, the first $S$ columns of $\mathbf{B}_{S_\mathcal{T}}$ represent phase slip loop matrix $\mathbf{B}_S$, while the last $L'$ columns make up the non-phase slip (linear inductor) loop matrix $\mathbf{B}_{L'}$.

We note that junction loops of zero inductance (a common approximation used for the SQUID loop) and phase slip nodes of zero capacitance are treated in Appendix \ref{subsec:ZeroCapInd}. We leave open the possibility of taking these limiting procedures at the end of the decomposition process, in a manner that generalizes the results of \cite{you_circuit_2019}.

Transforming into the edge bases of capacitive tree flux $\vv{\Phi}_{C_\mathcal{T}}$ and inductive cotree charge $\vv{Q}_{{L}_\mathcal{T}}$ is done with the basis change operations (in the notation of Section \ref{subsec:ChangeOfBasisMain}) of $\mathbf{U} = \mathbf{A}_{C_\mathcal{T}}^{-1}$ and $\mathbf{W} = \mathbf{B}_{L_\mathcal{T}}^{-1}$. The corresponding effect on the dynamical capacitive flux and inductive charge variables is to change them to a tree and cotree basis, respectively (Appendix \ref{subsec:NodeFluxLoopChargeTreeCotree}):
\begin{align}
\vv{\Phi}_n &\rightarrow \mathbf{A}_{C_\mathcal{T}}^T  \vv{\Phi}_n  =  \vv{\Phi}_{C_\mathcal{T}} \\
\vv{Q}_l &\rightarrow \mathbf{B}_{L_\mathcal{T}}^T \vv{Q}_l=\vv{Q}_{{L}_\mathcal{T}}
\end{align}

In this basis, the capacitive tree fluxes split into those that lie across junctions ($\vv{\Phi}_{J}$) and those that only lie across linear capacitance ($\vv{\Phi}_{C'}$). In the same way, inductive cotree charges can be split into those that lie along phase slips ($\vv{Q}_{S}$) and those that only lie along linear inductance ($\vv{Q}_{L'}$):
\begin{align}
\vv{\Phi}_{C_\mathcal{T}} &=
\begin{bmatrix}
\vv{\Phi}_{J} \\
\vv{\Phi}_{C'}
\end{bmatrix} \\
\vv{Q}_{{L}_\mathcal{T}} &= 
\begin{bmatrix}
\vv{Q}_{S} \\
\vv{Q}_{L'}
\end{bmatrix}
\end{align}
Thus, the system's dynamical variables now align with the edges of the circuit graph in tree-cotree notation \ref{subsec:TreeCotreeNotation}.

By applying the aforementioned transformations $\mathbf{U}$ and $\mathbf{W}$ to the topological matrices of the system (as in Section \ref{subsec:ChangeOfBasisMain}), we obtain:
\begin{align}
\mathbf{A}_J &\rightarrow \mathbf{A}_{C_\mathcal{T}}^{-1} \mathbf{A}_J =
\begin{bmatrix} \label{eq:IncidenceToIdentity}
\mathbf{I}_{JJ} \\
\mathbf{0}_{C'J}
\end{bmatrix} \\
\mathbf{B}_S &\rightarrow \mathbf{B}_{L_\mathcal{T}}^{-1} \mathbf{B}_S =
\begin{bmatrix}
\mathbf{I}_{SS} \\
\mathbf{0}_{L'S}
\end{bmatrix} \label{eq:LoopToIdentity} \\
\mathbf{\Omega} & \rightarrow \mathbf{A}_{C_\mathcal{T}}^{-1} \mathbf{\Omega} \mathbf{B}_{L_\mathcal{T}}^{T^{-1}} = \mathbf{\Omega}_E =
\begin{bmatrix}
\mathbf{\Omega}_{JS} & \mathbf{\Omega}_{JL'} \\
\mathbf{\Omega}_{C'S} & \mathbf{\Omega}_{C'L'} 
\end{bmatrix} \label{eq:NetworkMatrix}
\end{align}
 
Here, the junction incidence matrix $\mathbf{A}_J$ and the phase slip loop matrix $\mathbf{B}_S$ are transformed to an upper-identity form, and all of their topological information is transferred to the edge form of the network matrix, symbolized by $\mathbf{\Omega}_E$.

This edge network matrix has a straightforward interpretation. Instead of detailing the connectivity between capacitive nodes and inductive loops, it now encodes the connection between capacitive tree edges and inductive cotree edges in the system's fundamental loops. Conceptually, each loop corresponds to taking one inductive cotree edge and following its unique path through the capacitive spanning tree. This gives a definition of:
\begin{align} \label{eq:MainNetworkMatrixLoop}
-\Omega_{E_{ij}} =
\begin{cases} 
      1 & \text{loop of inductive edge $j$ passes} \\
      & \text{forward through capacitive edge i} \\
      -1 & \text{loop of inductive edge $j$ passes} \\
      &\text{backward through capacitive edge i} \\
      0 & \text{loop of inductive edge $j$ does not} \\
      & \text{pass through capacitive edge i}
\end{cases}
\end{align}

The blocks of the edge network matrix in Eq. \ref{eq:NetworkMatrix} detail the loop connectivity between the sub-categories of tree/cotree edges. Equivalently, the edge network matrix can be seen as the fundamental cutset matrix of the system's capacitive edges (Eq. \ref{eq:NetworkMatrixCutset}) \cite{bapat_graphs_2014, burkard_multilevel_2004,burkard_circuit_2005}. The edge network matrix is the standard graph-theoretic network matrix studied in linear programming \cite{schrijver_theory_1998, wolsey_integer_2014}.

We note that after the basis transformation, the capacitive fluxes and inductive charges are still coupled to each other through full capacitance and inductance matrices, respectively---with off-diagonal entries. For the inductors these couplings correspond to mutual inductance, while for capacitors the off-diagonal couplings correspond to the presence of capacitors outside the spanning tree.

\subsection{Transforming between equivalent networks} \label{subsec:TransformingEquivalentNetworks}

Performing certain linear basis transformations on the edge network matrix corresponds to graphically transforming to an equivalent network---as detailed in Appendix \ref{subsec:GraphTheoreticNetworkMatrices} and references \cite{schrijver_theory_1998,wolsey_integer_2014}. Algebraically, these structure-preserving operations involve multiplying rows and columns by $-1$, permuting pairs of rows and columns, and pivoting on nonzero elements of rows and columns. After these transformations, $\mathbf{\Omega}_E$ remains an edge network matrix. Again, we note that these operations will not alter the circuit's nonlinear degrees of freedom.

Graphically, the transformations are most easily envisioned in tree-cotree notation (Section \ref{subsec:TreeCotreeNotation}).
Multiplying rows and columns by $-1$ corresponds to swapping the direction of capacitive spanning tree edges or inductive spanning cotree edges, respectively. Permuting two rows corresponds to swapping the labels of two capacitive edges, while permuting two columns corresponds to relabeling inductive edges. We note that we only allow label swaps between edges of the same type. So, for instance, we do not swap a junction spanning tree edge with that representing a linear capacitor.

The row and column pivoting operations are most crucial to our decomposition procedure, and have a more detailed visual interpretation. Row pivoting (which eliminates all nonzero entries except one from a column) on a nonzero element $\Omega_{E_{ij}}$ removes the $i^{\text{th}}$ capacitive edge from the spanning tree and places it in parallel with the $j^{\text{th}}$ inductive edge, with which it shares a loop. Column pivoting (which removes all nonzero entries except one from a row) can be interpreted as contracting the $j^{\text{th}}$ inductive edge to a single vertex and then re-placing the inductive edge in series with the $i^{\text{th}}$ capacitive edge, whose cutset it lies in. Note that we only allow row pivots using linear capacitive edges (labeled $C'$ in Eq. \ref{eq:NetworkMatrix}), and column pivots using linear inductive edges (labeled $L'$ in Eq. \ref{eq:NetworkMatrix}), such that the nonlinear degrees of freedom $\vv{\Phi}_J$ and $\vv{Q}_S$ are invariant (or equivalently $\mathbf{A}_J$ and $\mathbf{B}_S$ do not change). Also, we observe that these pivot operations are integer valued. The pivoting process and the resulting changes to the edge network matrix are illustrated in Fig. \ref{fig:GMonDecomposition} and Appendix Fig. \ref{fig:PivotIllustration}.

Note that in this tree-cotree basis, we are also free to rearrange edges arbitrarily, as long as we do not change the system's edge network matrix (encoding the cutset/loop topology). For instance, a spanning tree of four capacitive branches with no inductive loops can be represented by any (loop-free) connected 5-node graph (with one node representing ground). In the language of matroid theory, we can transform the tree-cotree graphs up to 2-isomorphism \cite{oxley_matroid_2006}.

After performing a set of structure-preserving transformations to the tree-cotree structure of the network, we can change the equations of motion back to a node/loop basis if desired. To do this, we write down the final capacitive tree incidence matrix $\mathbf{A}_{C_\mathcal{T}}'$ and inductive cotree loop matrix $\mathbf{B}_{L_\mathcal{T}}'$ (which can correspond to the faces of the tree-cotree graph if the circuit is planar). Then, we perform the inverse transformations of those depicted in Eqs. \ref{eq:IncidenceToIdentity}, \ref{eq:LoopToIdentity}, and \ref{eq:NetworkMatrix}, employing change of basis matrices $\mathbf{U}=\mathbf{A}_{C_\mathcal{T}}'^{-1}$, and $\mathbf{W} =\mathbf{B}_{L_\mathcal{T}}'^{-1}$.

\subsection{Fundamental decomposition}
\label{subsec:FundamentalDecompositionMain}
The edge network matrix (and its corresponding circuit) can now be reduced to its fundamental form through a series of structure-preserving transformations, a procedure we discuss further in Appendix \ref{subsec:FundamentalDecomposition}. We begin with the block form of the network matrix given in Eq. \ref{eq:NetworkMatrix}:
\begin{align}
\mathbf{\Omega}_E =
\begin{bmatrix}
\mathbf{\Omega}_{JS} & \mathbf{\Omega}_{JL'} \\
\mathbf{\Omega}_{C'S} & \mathbf{\Omega}_{C'L'} 
\end{bmatrix}
\end{align}

First we perform a maximal number of successive row and column pivots on the nonzero elements of the rank $r$ submatrix $\mathbf{\Omega}_{C'L'}$. Graphically, this operation disentangles the inductive and capacitive branches representing LC oscillators from the rest of the circuit, visually distinguishing the harmonic modes.

We then carry out row pivoting on the nonzero elements of $\mathbf{\Omega}_{C'S}$ and column pivoting on the nonzero elements of $\mathbf{\Omega}_{JL'}$ until the nonzero rows and columns (respectively) of these submatrices are linearly independent. This step separates out the circuit's capacitor-only cutsets and inductor-only loops, which correspond to rows and columns of zeros (respectively) in the edge network matrix. Note that this row pivoting can be thought of as moving linear capacitive tree edges and column pivoting as moving linear inductive cotree edges, as discussed in Section \ref{subsec:TransformingEquivalentNetworks}. We also observe that column pivoting on $\mathbf{\Omega}_{C'S}$ or row pivoting on $\mathbf{\Omega}_{JL'}$ would alter the nonlinear $J$ and $S$ degrees of freedom, and not be structure-preserving. These disallowed transformations correspond to moving Josephson junction or phase slip edges.

Next, we can remove the free modes. These zero rows (in the index range $C'$) and the zero columns (in the index range $L'$), are eliminated through the technique of free mode removal \cite{ding_free-mode_2021}---with further detail in Appendix \ref{subsec:FreeModes}. This process renormalizes some of the system's parameters but does not alter any topological quantities.

Finally, we permute rows and multiply certain rows and columns by $-1$, to obtain a transformed edge network matrix in the fundamental form of:
\begin{align} 
\mathbf{\Omega}_E =
\begin{bmatrix}
\mathbf{\Omega}_{JS} &\mathbf{\Omega}_{Jf} &  \mathbf{0}_{Jr} \\
\mathbf{\Omega}_{pS}  & \mathbf{0}_{pf}  &  \mathbf{0}_{pr} \\
\mathbf{0}_{rS}  & \mathbf{0}_{rf} & \mathbf{I}_{rr}  
\end{bmatrix}
\end{align}

We now comment on the form and interpretation of this matrix. The submatrix $\mathbf{I}_{rr}$ represents the harmonic degrees of freedom, with inductive branches connected across capacitive ones. These LC-oscillators can be mutually uncoupled from each other as shown in \ref{subsec:CapacitanceInductanceDiagonalization}. However, they are still capacitively and inductively coupled to the rest of the circuit. $\mathbf{\Omega}_{JS}$ represents the loop interconnectivity between junction and phase slip branches. The matrix $\mathbf{\Omega}_{Jf}$ enumerates the linear cotree inductors in loops with Junction edges and $\mathbf{\Omega}_{pS}$ the linear tree capacitors in loops with phase slip edges. Importantly, there are no linear inductive cotree edges in loops with linear capacitive edges, outside of the harmonic modes.

We note that the transformed $\mathbf{\Omega}_{JS}$ is generally different from the original submatrix, as is the circuit graph it represents. In this transformed picture there are now $f \leq J$ cotree linear inductive branches and $p \leq S$ linear capacitive spanning tree branches. We use the letter $f$ to denote the eventual presence of fluxonium-like \cite{manucharyan_fluxonium_2009} degrees of freedom in the Hamiltonian, and $p$ to indicate those corresponding to the phase-slip fluxonium analogue.

\subsection{Junction-only decomposition} \label{subsec:JunctionOnlyDecomposition}

\begin{figure}
    \centering
    \includegraphics[width=1\linewidth]{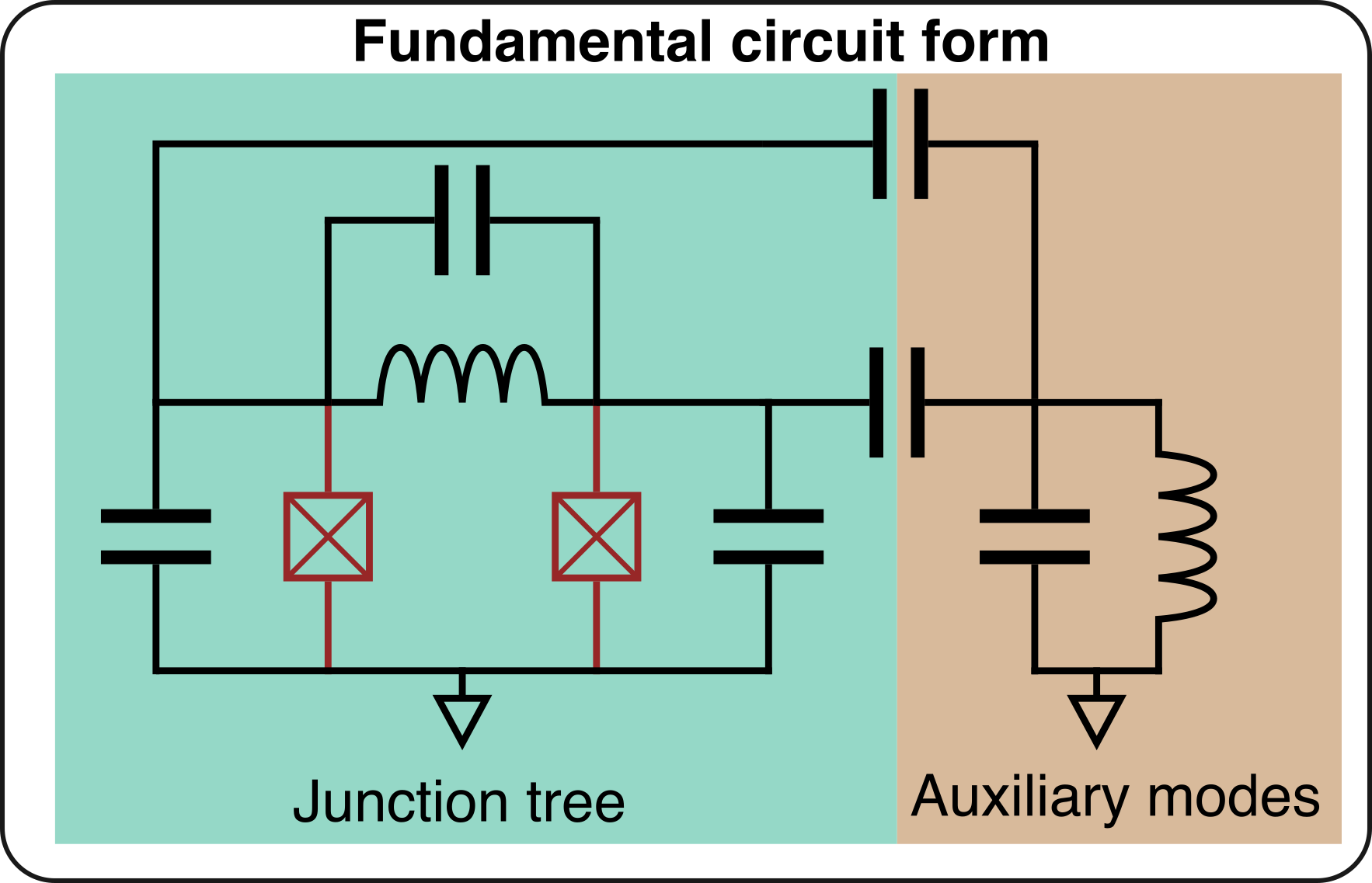}
    \caption{Fundamental decomposition of a Josephson junction circuit. Any such model can be divided into (1) a junction tree (with possible inductors between branches) and (2) a set of auxiliary Harmonic modes, galvanically disconnected from the junction tree.}
    \label{fig:FundamentalCircuitForm}
\end{figure}

For systems without phase slip wires, the possible network geometries take on simplified forms, exemplified in Fig. \ref{fig:FundamentalCircuitForm}. In this regime, once free modes have been removed, the edge network matrix can be written as:
\begin{align} \label{eq:JunctionFundamentalDecomposition}
\mathbf{\Omega}_E 
=
\begin{bmatrix}
\mathbf{\Omega}_{Jf} & \mathbf{0}_{Jr}  \\
\mathbf{0}_{rf} & \mathbf{I}_{rr} 
\end{bmatrix}
\end{align}

Here, the system is described in terms of its number of Josephson junctions $J$, its number of linear inductive branches $f \leq J$, and its number of auxiliary harmonic modes $r$. In this fundamental decomposition, the Josephson junctions form a tree that spans all capacitive nodes (except those corresponding to the harmonic oscillators). The network submatrix $\mathbf{\Omega}_{Jf}$ details the inductive topology of these junction branches, while the submatrix $\mathbf{I}_{rr}$ describes the connectivity (of inductors to ground) of the harmonic branches.

In sum, the form of the network matrix $\mathbf{\Omega}_{Jf}$ and the number of harmonic modes $r$ determine the topology of the circuit. We demonstrate an example decomposition of a Josephson circuit in Section \ref{subsec:ExampleDecomposition}, and show how this fundamental form can be used for circuit classification in Section \ref{subsec:ClassificationEquivalentCircuits}.

Note that for phase slip circuits with no Josephson junctions, the fundamental circuit forms are dual to those with only Josephson junctions \cite{ulrich_dual_2016,osborne_flux-charge_2024}---with phase slips replacing junctions and cutsets replacing loops. The Hamiltonian structure will be identical with charge and flux variables swapped, as well as capacitance and inductance matrices.

\subsection{Example decomposition procedure} \label{subsec:ExampleDecomposition}

\begin{figure*}
    \centering
    \includegraphics[width=1\linewidth]{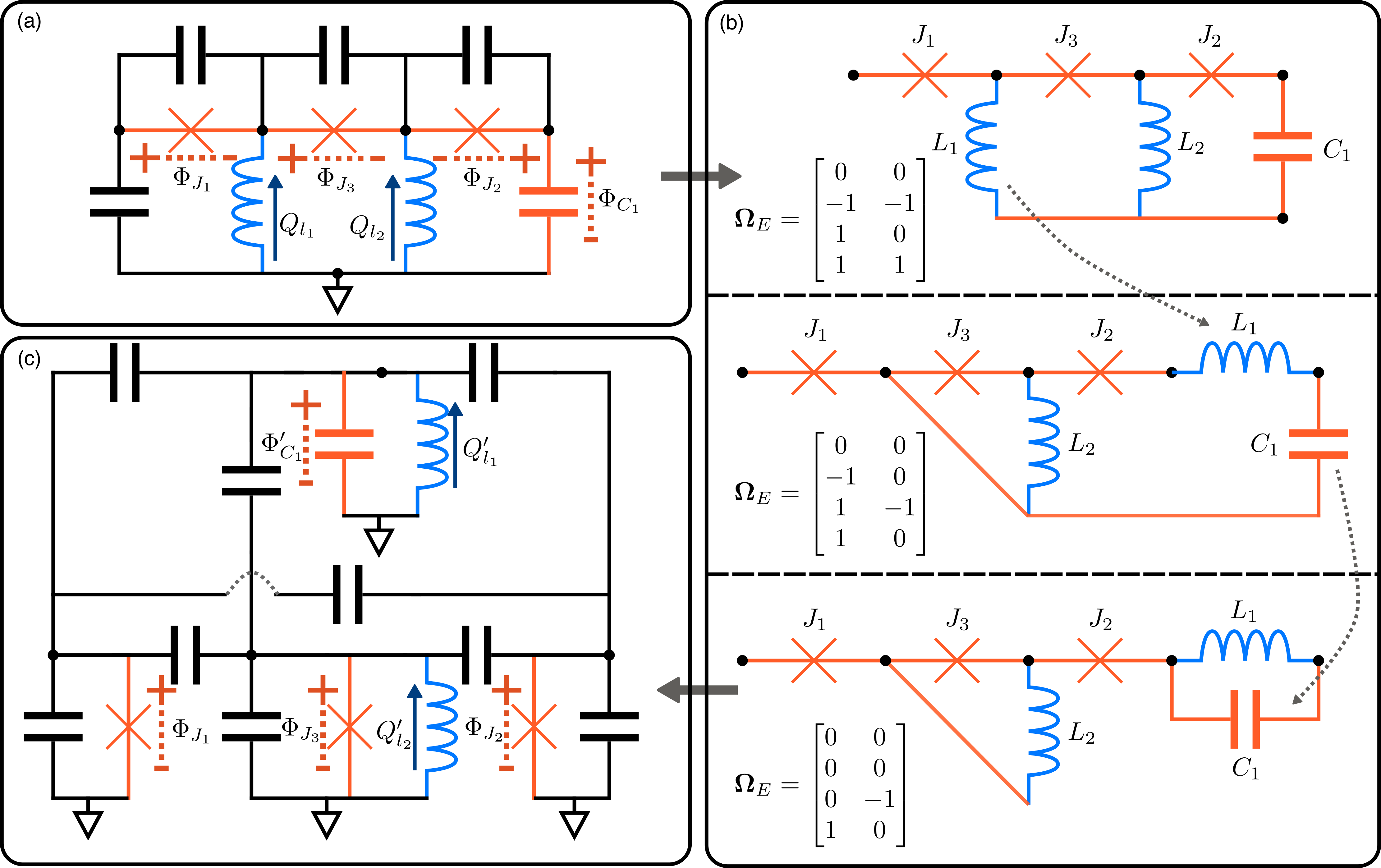}
    \caption{Decomposition applied to an example circuit. (a) Initial four-node circuit with three Josephson junctions and two inductors. A capacitive spanning tree (orange) can be formed by including the bottom right capacitor along with the capacitive branches across the junctions (represented in tree-cotree notation by junction symbols). An inductive spanning cotree (blue) consists of the system's two inductors. (b) Conversion of the circuit to tree-cotree notation, and pivot manipulation into equivalent circuits. Top: initial circuit in tree-cotree notation, and its edge network matrix. The first three rows of the matrix are the Josephson junction cutsets, while the fourth is that of the linear capacitor. Middle: column pivoting of inductor 1 on junction 2. The inductor is contracted to a point and re-placed in series with junction 2 (with which it shares a loop). Bottom: row pivoting of capacitor 1 (the fourth capacitive edge, including junctions) on inductor 1. Capacitor 1 is removed from the circuit and re-placed in parallel with inductor 1. (c) Conversion of spanning tree model back into standard notation. This is done by performing a change of basis, using a capacitive incidence matrix that respects the network's tree-cotree loop/cutset topology.}
    \label{fig:GMonDecomposition}
\end{figure*}

An example of this network matrix decomposition procedure is shown in shown in Fig \ref{fig:GMonDecomposition}. Here, the lumped model may represent inductively-coupled charge qubits, with tunability provided by an external flux offset \cite{chen_qubit_2014}. We show that this circuit model can be transformed into an equivalent circuit of two charge modes coupled through a flux mode and a harmonic mode.

In the initial circuit shown in Fig. \ref{fig:GMonDecomposition}(a) the Josephson junction incidence matrix and the node-loop network matrix (here the incidence matrix of the inductive branches) are given by:
\begin{align}
\mathbf{A}_J &= 
\begin{bmatrix}
1 & 0 & 0 \\
-1 & 1 & 0 \\
0 & -1 & -1 \\
0 & 0  & 1
\end{bmatrix} \\
\mathbf{\Omega} &= 
\begin{bmatrix}
0 & 0 \\
1 & 0 \\
0 & 1 \\
0 & 0
\end{bmatrix}
\end{align}
where the non-ground nodes are ordered from left to right. Since there are no phase slip elements we have no phase slip loop matrix $\mathbf{B}_S$.

To carry out the decomposition we construct a capacitive spanning tree by including the capacitive edge on the bottom right alongside the three capacitive edges across the junctions (denoted for simplicity by the Josephson junction symbol). This adds a fourth column to the incidence matrix, giving:
\begin{align}
\mathbf{A}_{C_\mathcal{T}}
= 
\begin{bmatrix}
1 & 0 & 0 & 0\\
-1 & 1 & 0 & 0 \\
0 & -1 & -1 & 0 \\
0 & 0  & 1 & 1
\end{bmatrix}
\end{align}

At this point, we use Eqs. \ref{eq:IncidenceToIdentity} and \ref{eq:NetworkMatrix} to carry out a change of basis on the flux variables, with basis change operator $\mathbf{U} = \mathbf{A}_{C_\mathcal{T}}^{-1}$. This converts the incidence matrix to the 3-by-3 identity matrix with a row of zeros underneath, and transfers all the topological information into the edge network matrix $\mathbf{\Omega}_E$:
\begin{align}
\mathbf{\Omega}_E = \mathbf{A}_{C_\mathcal{T}}^{-1} \mathbf{\Omega} = 
\begin{bmatrix}
0 & 0 \\
-1 & -1 \\
1 & 0 \\
1 & 1
\end{bmatrix}
\end{align}

Note that the inductive edges already form a cotree to the capacitive spanning tree, and so no loop charge change of basis is performed when transferring the system to the tree-cotree framework.

In Fig. \ref{fig:GMonDecomposition}(b) we give a visual interpretation of the subsequent pivot operations, whose mathematical interpretation we illustrate here. By performing a column pivot of the first column on the second row, we obtain a transformed edge network matrix of:
\begin{align}
\mathbf{\Omega}_E \rightarrow 
\begin{bmatrix}
0 & 0 \\
-1 & 0 \\
1 & -1 \\
1 & 0
\end{bmatrix}
\end{align}

The effect of the above operation on the spanning tree is shown with the first grey arrow of Fig. \ref{fig:GMonDecomposition}. The first inductor is contracted to a point, and is then reinserted in series with the second capacitive element (the second junction).

The second grey arrow indicates the next pivot operation, whereby a row pivot is carried out of the fourth row (capacitor 1) on the first column (inductor 1):
\begin{align}
\mathbf{\Omega}_E \rightarrow 
\begin{bmatrix}
0 & 0 \\
0 & 0 \\
0 & -1 \\
1 & 0
\end{bmatrix}
\end{align}
Visually, this pivot corresponds to removing the capacitor edge from the spanning tree and then reinserting it in parallel with the first inductor. 

The circuit topology has now been placed into its fundamentally decomposed form, in which the harmonic mode has been separated out from the rest of the circuit, and no more simplifying pivots are possible. As mentioned in Section \ref{subsec:TransformingEquivalentNetworks}, we are free to move the tree and cotree edges, as long as we do not alter the fundamental loops of the system.

In Fig. \ref{fig:GMonDecomposition}(c) we exhibit a circuit model in standard notation that corresponds to the decomposed circuit. We first perform a trivial loop basis change that multiplies the second column of $\mathbf{\Omega}_E$ by $-1$ (switching the direction of the second inductor. We then find a new capacitive incidence matrix that obeys the loop topology of the circuit, which in this case will be the identity:
\begin{align}
\mathbf{A}_J' = 
\begin{bmatrix}
1 & 0 & 0 & 0 \\
0 & 1 & 0 & 0 \\
0 & 0 & 1 & 0 \\
0 & 0 & 0 & 1
\end{bmatrix}
\end{align}

This (identity) basis transformation ($\mathbf{U} = \mathbf{A}_J'$) then transforms the edge network matrix into a node-loop network matrix (which here is the incidence matrix of the inductive branches):
\begin{align}
\mathbf{\Omega}' = \mathbf{A}_J' \mathbf{\Omega}_E = 
\begin{bmatrix}
0 & 0 \\
0 & 0 \\
0 & 1 \\
1 & 0
\end{bmatrix}
\end{align}

The reconstructed equivalent circuit in Fig \ref{fig:GMonDecomposition}(c) now takes on the form of two charge modes coupled through a harmonic mode and a flux mode. The detangling of the harmonic mode from the rest of the circuit is the hallmark of our fundamental decomposition procedure.

\subsection{Classification and equivalent circuits} \label{subsec:ClassificationEquivalentCircuits}

\begin{figure}
    \centering
    \includegraphics[width=1\linewidth]{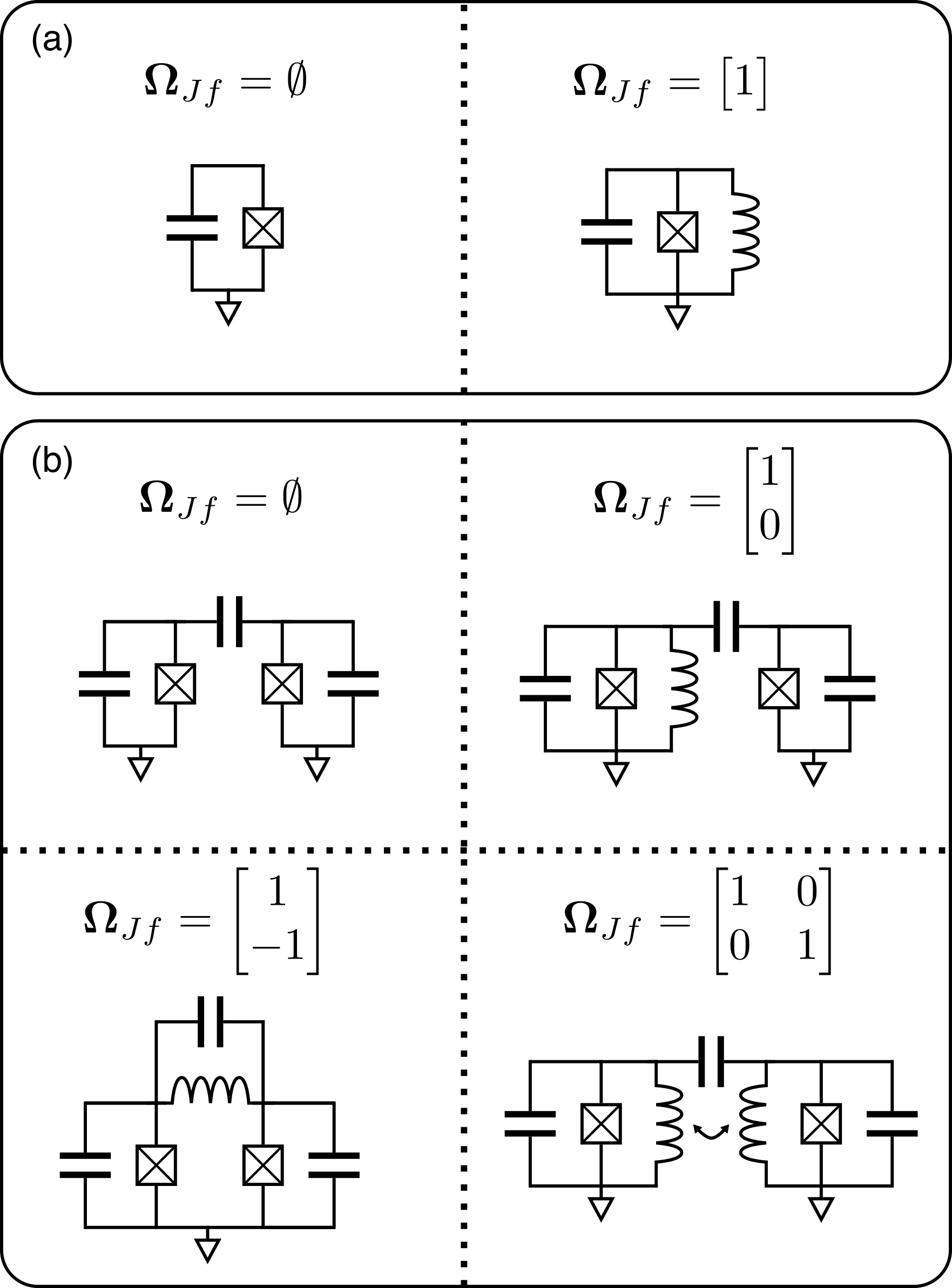}
    \caption{Classification of one and two-Josephson junction circuits, with the harmonic modes removed. (a) Enumeration of one-junction circuits, each with a representative network matrix. (b) Similar enumeration of two-junction circuits.}
    \label{fig:Classification}
\end{figure}

In Section \ref{subsec:JunctionOnlyDecomposition} we discussed how circuits with Josephson junctions (and no phase slip elements) can be decomposed into a fundamental form, encoded by an edge network matrix. In this form, harmonic modes are galvanically separated out from the tree of Josephson junctions that make up the nonlinear portion of the circuit. Thus we can understand the topology of circuits by considering the junction tree and the harmonic modes separately (with more information provided in Appendix \ref{subsec:StructurePreservingTransformations}).

The junction topology is specified by the network submatrix $\mathbf{\Omega}_{Jf}$ given in Eq. \ref{eq:AppendixNetworkMatrix}. Here, there are $J$ junction branches and $f \leq J$ inductive branches, with each representing a loop. Equivalent network matrices can be classified through the set of structure-preserving basis transformations described in Section \ref{subsec:TransformingEquivalentNetworks}, which can move the positions of the inductors, and reverse the direction of circuit elements but still maintain the topology of the junction loops. Note here that we do not include circuits with current or voltage sources embedded into the qubit. 

We illustrate a classification of one and two-junction circuits in Fig. \ref{fig:Classification}. For circuits whose only nonlinear element is a single Josephson junction, there are two possible types of circuit topologies, with network matrices given by:
\begin{align}
 \mathbf{\Omega}_{Jf} =
\begin{cases} 
      \varnothing \\[6pt]
      \begin{bmatrix} 1 \end{bmatrix}
\end{cases}
\end{align}
In the first case there is no inductor across the Josephson junction, resulting in  single charge mode. The second case is that of the flux qubit mode, wherein the two terminals of the Josephson junction have a galvanic connection. These are the potential single-junction circuit types classified up to the presence of harmonic modes.

For two-junction circuits, there are four equivalence classes of fundamental network matrices:
\begin{align}
    \mathbf{\Omega}_{Jf}
    =
\begin{cases} 
      \varnothing  \\[6pt]
    \begin{bmatrix}
        1 \\
        0
    \end{bmatrix} \\[16pt]
    \begin{bmatrix}
        1 \\
        -1
    \end{bmatrix}  \\[16pt]
    \begin{bmatrix}
        1 & 0\\
        0 & 1
    \end{bmatrix} 
\end{cases} 
\end{align}
The first case has no inductors, representing capacitively coupled charge modes. The second case is that of a flux mode capacitively coupled to a charge mode. The third case contains two junctions connected in a single inductive loop (such as in a DC SQUID with its intrinsic inductance included). The fourth case represents two flux modes, coupled both inductively and capacitively. Under the restrictions outlined thus far, the nonlinear portion of any two-junction circuit can be decomposed into one of these forms.

This method of classification based on the network matrix can be used to enumerate more general sets of superconducting circuits, aiding in classification efforts such as that presented in \cite{weissler_enumeration_2024}. 

\section{Circuit model extraction} \label{sec:Synthesis}

\subsection{The network synthesis paradigm}

In this Section, we show how accurate (and transformerless) lumped-element circuit models can be systematically extracted from electromagnetic simulations, by matching the tree-cotree topology of the lumped circuit with that of the simulated device. A central idea is that the linear portion of a lossless, reciprocal superconducting device has the same frequency-domain response function as a particular lumped-element capacitive and inductive network with identical edge network matrix $\mathbf{\Omega}_E$ (introduced in Section \ref{subsec:EdgeNetworkSpanningTreeCotree}). A more complete discussion of the following presentation is found in Appendix \ref{sec:AppendixE}.

Lumped-element circuit models represent an idealized picture of physical reality. However, they can provide a highly accurate model if the nonlinear portions of a device are localized to physically small (compared to the wavelength of light) regions. In this situation, the nonlinear circuit components are well-approximated as true lumped elements, while the linear part of the device may be distributed over a larger volume. In many cases, however, this distributed linear response can be well-approximated with an effective lumped-element model consisting of linear circuit elements.

In simulation, this process is accomplished by replacing the nonlinear parts of the device with electromagnetic ports, and then calculating the multi-port linear response of the system. One then extracts a lumped-element linear circuit model representing the response, in a process known as network synthesis \cite{newcomb_linear_1966, anderson_network_2013}. When the nonlinear elements are reinserted across the port terminals, a full circuit is generated (which can then be quantized).

The necessary linear electromagnetic port simulations are often performed in the frequency domain. Here, at a frequency $\omega$ (in angular units) currents and/or voltages are input and output at these ports. The multi-port response is then calculated and can be written in terms of Laplace variable $s$, which corresponds to the frequency domain for $s = i \omega$. A variety of response functions may be obtained, including the admittance matrix $\mathbf{Y}(s)$ (with voltage input and current output), the impedance matrix $\mathbf{Z}(s)$ (with current input and voltage output), or a hybrid matrix $\mathbf{H}(s)$ (with mixed voltage/current input and output) \cite{david_m_pozar_microwave_2012}:
\begin{align}
\vv{I}(s) &= \mathbf{Y}(s) \vv{V}(s) \\
\vv{V}(s) &= \mathbf{Z}(s) \vv{I}(s) \\
\begin{bmatrix}
\vv{I}_1(s) \\
\vv{V}_2(s)
\end{bmatrix}
&=
\mathbf{H}(s)
\begin{bmatrix}
\vv{V}_1(s) \\
\vv{I}_2(s)
\end{bmatrix}
\end{align}

In this work we utilize a hybrid matrix approach with a specific set of port placements to extract exact circuit models for reciprocal, lossless, superconducting systems.  The hybrid matrix represents a natural flux-charge symmetric framework to analyze devices with Josephson junctions and inductive loops (which can potentially have fluxoid tunneling)---such that each junction is shunted by a capacitor and each phase slip wire is in series with an inductor.

With our placement of ports, we show that the system obeys a zero-frequency constraint, which is expressed in terms of the device's edge network matrix $\mathbf{\Omega}_E$. By manipulating the form of the hybrid response to take advantage of this constraint, the resulting linear response function is easily synthesized with a lumped-element circuit model that only possesses capacitors and inductors---and no multi-port transformers. In essence, the extracted circuit model consists of a core low-frequency component with edge network matrix $\mathbf{\Omega}_E$, coupled capacitively and inductively to LC oscillators that represent the response's poles.

The transformerless nature of our approach differentiates it from other forms of exact, lossless network synthesis \cite{newcomb_linear_1966, labarca_toolbox_2024}. By only using capacitance and inductance for the linear portions of the circuits, our extracted circuits align with the most commonly used and best understood models in the field. In particular, the degrees of freedom of these circuits can be systematically quantized, with certain variables becoming continuous operators and others becoming discrete (as shown in Section \ref{sec:Quantization}). We highlight that this result is accomplished through the combined use of the flux-charge symmetric Hybrid matrix, specific port placements, and the zero-frequency network matrix constraint. We note, however, that unlike in the exact synthesis works mentioned above, our result does not encompass non-reciprocal response. Also, we consider the system to have zero loss, and thus do not include dissipative effects treated in \cite{solgun_blackbox_2014, solgun_multiport_2015}.

Our method resembles a multi-port version (under certain constraints) of the standard single-port Foster synthesis \cite{foster_reactance_1924}, which underlies the original black box approach to superconducting circuit quantization \cite{nigg_black-box_2012}. It provides a generalization of DC capacitance and inductance simulations \cite{minev_circuit_2021} to encompass high frequency effects. Compared to the eigenmode expansion \cite{minev_energy-participation_2021} approach to device simulation, our method has the advantage that it outputs a circuit model, and that it provides a natural framework to model highly anharmonic qubits. Throughout this Section we utilize many similar derivations to those presented in the works of Yarlagadda et al. \cite{yarlagadda_synthesis_1966, yarlagadda_reciprocal_1972}.

\subsection{Hybrid matrix and port placement}

\begin{figure*}
    \centering
    \includegraphics[width=1\linewidth]{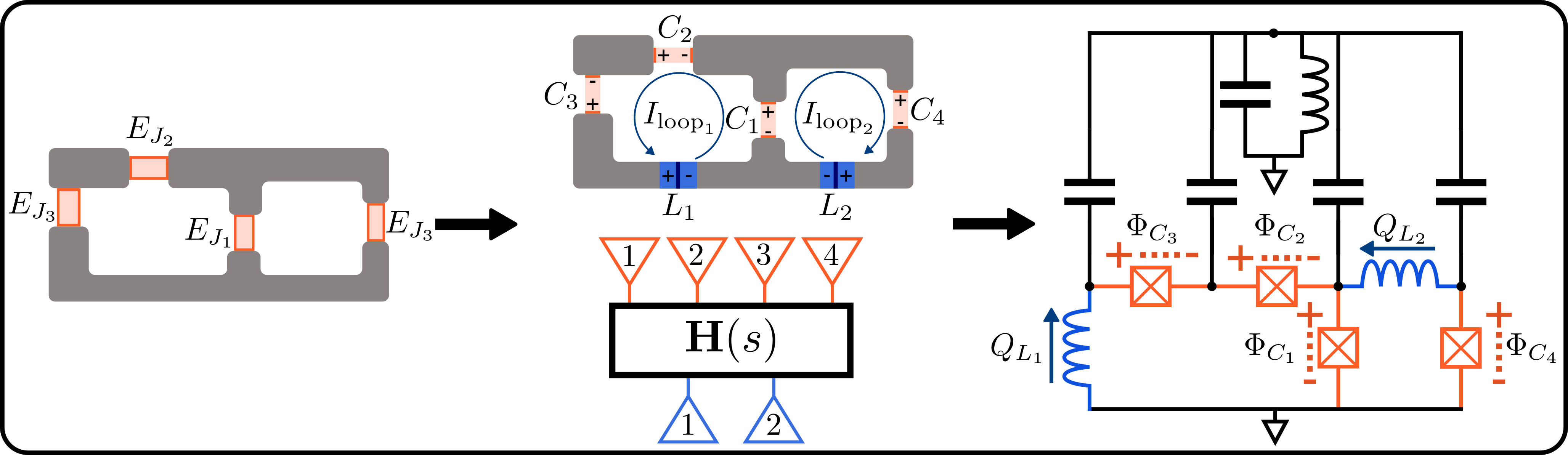}
    \caption{Overview of the model extraction procedure. A layout (with orange rectangles representing regions of Josephson tunneling) is converted into an electromagnetic simulation with capacitive ports replacing the nonlinear junction elements, and inductive ports lying along each loop of the system. The simulation produces a hybrid matrix $\mathbf{H}(s)$, which relates the frequency-dependent vector of junction voltage and inductor current to that of junction current and inductor voltage. This linear response can then be matched by a circuit model whose network topology mirrors that of the layout, with additional coupled resonators representing the high-frequency poles of the system. Josephson junctions are then reinserted across the capacitive ports. We omit most coupling capacitors from the diagram for clarity.}
    \label{fig:SynthesisLayout}
\end{figure*}

In order to simulate the desired hybrid matrix response, we require our electromagnetic ports to mirror the tree-cotree topology of the device (as described in Appendix \ref{subsec:PortPlacementZeroFreq}). We thus prevent the response from having zero-frequency poles, simplifying the synthesis procedure. In our methodology, we divide ports into two categories: capacitive (parallel) ports, which are placed across metal gaps, and inductive (series) ports, which are placed along metal loops. When we extract a circuit model from the simulation, the capacitive ports will become capacitive tree edges, while the inductive ports will represent inductive cotree edges. In general, each Josephson junction and charge drive line will possess a capacitive port and each phase slip element and flux bias line will have an inductive port. We note that the capacitive and inductive ports may be implemented identically to each other in simulation---but they are differentiated in what they represent and how they enter into the hybrid matrix. The general procedure is illustrated in Figs. \ref{fig:SynthesisLayout} and \ref{fig:FluxoniumSimulation}.

As shown in Fig. \ref{fig:SynthesisLayout}, for systems with Josephson junctions, an additional inductive port must be placed along each junction loop, to remove the zero-frequency poles, which represent the loops' nonzero linear inductances. As discussed in Appendix \ref{subsec:PortPlacementZeroFreq} We similarly add extra capacitive ports across cutsets of phase slip elements. Overall, abstracting away regions of superconducting metal as the nodes of a graph, our port placements generate a tree of capacitive ports with a cotree of inductive ports---aligning with eventual the tree-cotree notation of the synthesized circuit model. More accurately, when simulating qubits that are disconnected from the ground plane, the port structure may consist of a forest (collection of trees) of capacitive ports and coforest of inductive ones---which can be turned into an equivalent tree/cotree (up to 2-isomorphism) by assigning one node from each tree in the forest to ground \cite{oxley_matroid_2006}.

We then calculate the device's multi-port response and represent it with the hybrid matrix. We place the capacitive port currents and $\vv{I}_C(s)$ and inductive port voltages $\vv{V}_L(s)$ on the left hand side of the equation, and the capacitive port voltages $\vv{V}_C(s)$ and inductive port currents $\vv{I}_L(s)$ on the right hand side. The hybrid matrix response for such a reciprocal system can then written in block matrix form as \cite{yarlagadda_reciprocal_1972}:
\begin{align}
    \begin{bmatrix}
       \vv{I}_C(s) \\
       \vv{V}_L(s)
    \end{bmatrix}
    =
    \begin{bmatrix}
       \mathbf{H}_{CC}(s) & - \mathbf{H}_{CL}(s) \\
       \mathbf{H}_{CL}^T(s) & \mathbf{H}_{LL}(s)
    \end{bmatrix}
    \begin{bmatrix}
       \vv{V}_C(s) \\
       \vv{I}_L(s)
    \end{bmatrix}
\end{align}
Here, $\mathbf{H}_{CC}(s)$ is a symmetric admittance matrix, while $\mathbf{H}_{LL}(s)$ is a symmetric impedance matrix. Thus, we observe how the hybrid matrix serves as a ``hybrid" of the admittance and impedance response. This property allows us to use the hybrid matrix to extract both capacitive and mutual inductive couplings from the same simulation.

With our port setup, the capacitive branches form a spanning tree of the system, while the inductive branches represent a spanning cotree (one inductive edge per loop). Thus, at zero frequency ($s=0$), the resulting DC voltages and currents obey a constraint in terms the edge network matrix $\mathbf{\Omega}_{E}$:
\begin{align} \label{eq:MainZeroFreqConstraint}
\begin{bmatrix}
       \vv{I}_C(0) \\
       \vv{V}_L(0)
    \end{bmatrix}
    =
    \begin{bmatrix}
       \mathbf{0}_{CC} & - \mathbf{\Omega}_E \\
       \mathbf{\Omega}_E^T & \mathbf{0}_{LL}
    \end{bmatrix}
    \begin{bmatrix}
       \vv{V}_C(0) \\
       \vv{I}_L(0)
    \end{bmatrix}
\end{align}
This is to say that each inductive loop takes a path through the set of capacitive ports and each capacitive branch has a corresponding cutset of inductive edges.

\subsection{Form of hybrid response}

With no zero-frequency poles, the hybrid matrix response will generally possess constant terms, linear terms in $s$ (poles at infinity), and finite-frequency poles. Real-valued matrices representing each of these quantities can be extracted through algorithms such as vector fitting \cite{gustavsen_rational_1999, wassaf_efficiently_2024}. Then, using the zero-frequency constraint and the properties of the poles of a lossless, reciprocal hybrid matrix \cite{yarlagadda_reciprocal_1972}, we show (in Appendices \ref{subsec:LPR}, \ref{subsec:HybridMatrixPoleExpansion}, and \ref{subsec:PoleDecomposition}) that the hybrid matrix can be expanded out into a particular block form:
\begin{align} \label{eq:MainFinalHybridSimulationResponse}
&\mathbf{H}(s) = \nonumber \\ %newline
&\begin{bmatrix}
    s \left[ \mathbf{K}_{{CC}_\infty}+ \sum_r \vv{R}_{C_r} \vv{R}_{C_r}^T \right] & -\mathbf{\Omega}_E \\
    \mathbf{\Omega}_E^T &  s\left[ \mathbf{K}_{{LL}_\infty}+ \sum_r \vv{R}_{L_r} \vv{R}_{L_r}^T \right]
\end{bmatrix} \nonumber 
\\%newline
&+\sum_{r}\frac{s^2}{s^2+\omega_r^2} \begin{bmatrix}
\vv{R}_{C_r} &  \vv{0}_{C}  \\
\vv{0}_{L} & \vv{R}_{L_r}
\end{bmatrix}
\begin{bmatrix}
-s &  -\omega_r  \\
\omega_r & -s
\end{bmatrix}
\begin{bmatrix}
\vv{R}_{C_r} &  \vv{0}_{C}  \\
\vv{0}_{L} & \vv{R}_{L_r}
\end{bmatrix}^T
\end{align}

As previously mentioned, $\mathbf{\Omega}_{E}$ represents the edge network matrix of the system. In addition, $\mathbf{K}_{{CC}_\infty}$ and $\mathbf{K}_{{LL}_\infty}$ denote the residues of the poles at infinity, which can be represented by real, positive definite matrices. The finite-frequency resonant poles of the system are indexed by $r$, and their residues been expanded out as outer products. Here, the participation of each capacitive edge in the resonant mode is captured by $\vv{R}_{C_r}$, while the inductive participations are encapsulated by the $\vv{R}_{L_r}$ term \cite{labarca_toolbox_2024}. 

The behavior of the system is thus divided into two components: the low frequency behavior of the system (linear and constant terms in $s$) and the higher order pole response. Removing the final line of the expansion would produce the standard lowest-order lumped-element response of the system. In general, a cutoff is applied to the number of high-frequency poles we include, such that we accurately model the response up to our operating frequencies..

\subsection{Matching response with lumped circuit model} \label{subsec:MatchingResponseLumpedCircuit}

Now we wish to find a lumped-element linear circuit model that produces this hybrid matrix frequency response. To do this, we construct a lumped circuit whose capacitive tree/inductive cotree structure mirrors that of the simulated device, with identical edge network matrix $\mathbf{\Omega}_E$. We then place a capacitive port in parallel with each edge in the capacitive spanning tree and an inductive port in series with each inductive edge of the cotree. We label this tree/cotree as the ``core circuit."

Next, as shown in we Appendix Fig. \ref{fig:NetworkMatrixSynthesis}, we add a set of auxiliary harmonic modes (galvanically disconnected from the rest of the circuit and from each other), which couple capacitively and inductively to the core circuit of the system. This results in a total capacitance and inductance matrix of:
\begin{align}
\mathbf{C} &= 
\begin{bmatrix}
\mathbf{C}_{CC} & \mathbf{C}_{Cr} \\
\mathbf{C}_{rC} & \mathbf{C}_{rr}
\end{bmatrix} \\
\mathbf{L} &= 
\begin{bmatrix}
\mathbf{L}_{LL} & \mathbf{L}_{Lr} \\
\mathbf{L}_{rL} & \mathbf{L}_{rr}
\end{bmatrix}
\end{align}

The index $r$ refers to the auxiliary harmonic modes, while $C$ and $L$ denote the core circuit capacitive and inductive spanning branches, respectively. We also restrict the auxiliary modes to have no direct coupling to each other, such that $\mathbf{C}_{rr}$ and $\mathbf{L}_{rr}$ are diagonal. The diagonal nature of these matrices allow us to perform a similar outer product expansion (to Eq. \ref{eq:MainFinalHybridSimulationResponse}) of the lumped-circuit's hybrid response (shown in Appendix \ref{subsec:ComparisonEquationsMotion}):
\begin{align}
&\mathbf{H}(s) = 
\begin{bmatrix}
    s \mathbf{C}_{CC} & -\mathbf{\Omega}_E \\
    \mathbf{\Omega}_E^T &  s\mathbf{L}_{LL}
\end{bmatrix}
 \nonumber \\ %newline
&+ \sum_r \frac{s^2}{s^2 + \omega_r^2}
\begin{bmatrix}
\frac{\vv{C}_{C_r}}{\sqrt{C_{rr}}} & \vv{0}_{C} \\
\vv{0}_{L} & \frac{\vv{L}_{L_r}}{\sqrt{L_{rr}}}
\end{bmatrix}
\begin{bmatrix}
-s & -\omega_r \\
\omega_r & -s
\end{bmatrix}
\begin{bmatrix}
\frac{\vv{C}_{C_r}}{\sqrt{C_{rr}}} & \vv{0}_{C} \\
\vv{0}_{L} & \frac{\vv{L}_{L_r}}{\sqrt{L_{rr}}}
\end{bmatrix}^T
\end{align}
Here, $\vv{C}_{C_r}$ and $\vv{L}_{L_r}$ refer to columns of of $\mathbf{C}_{Cr}$ and $\mathbf{L}_{Lr}$, respectively, while $C_{rr}$ and $L_{rr}$ represent the diagonal elements of $\mathbf{C}_{rr}$ and $\mathbf{L}_{rr}$.

We observe that the lumped circuit's hybrid matrix exactly mirrors the hybrid matrix extracted from electromagnetic simulation (Eq. \ref{eq:MainFinalHybridSimulationResponse}). Indeed, they are equivalent under the conditions that:
\begin{align}
{C}_{rr}{L}_{rr} &= \frac{1}{\omega_r^2} \\ %newline
\vv{C}_{C_r} &= \sqrt{C_{rr}} \vv{R}_{C_r} \\ %newline
\vv{L}_{L_r} &= \sqrt{L_{rr}} \vv{R}_{L_r}\\ %newline
\mathbf{C}_{CC} &=  \mathbf{K}_{{CC}_\infty} + \sum_r \vv{R}_{C_r}\vv{R}_{C_r}^T \\ %newline
\mathbf{L}_{LL} &=  \mathbf{K}_{{LL}_\infty} + \sum_r \vv{R}_{L_r}\vv{R}_{L_r}^T 
\end{align}

In essence, a capacitive and inductive lumped circuit can reproduce the response of a simulated system when (1) it matches the system's network tree-cotree topology, (2) its response without resonant poles is the same as the system's low-frequency response, and (3) its auxiliary modes correspond to the resonant poles of the EM response. 

We note that the auxiliary resonances needed to emulate the poles of an electromagnetic response are analogous to the auxiliary modes that appear in the fundamental decomposition of an arbitrary lumped circuit (Section \ref{subsec:FundamentalDecomposition}). In fact, after eliminating free modes, the synthesized lumped circuit will be in a fundamentally decomposed form.

\subsection{Reinserting nonlinear and drive elements}

\begin{figure}
    \centering
    \includegraphics[width=1\linewidth]{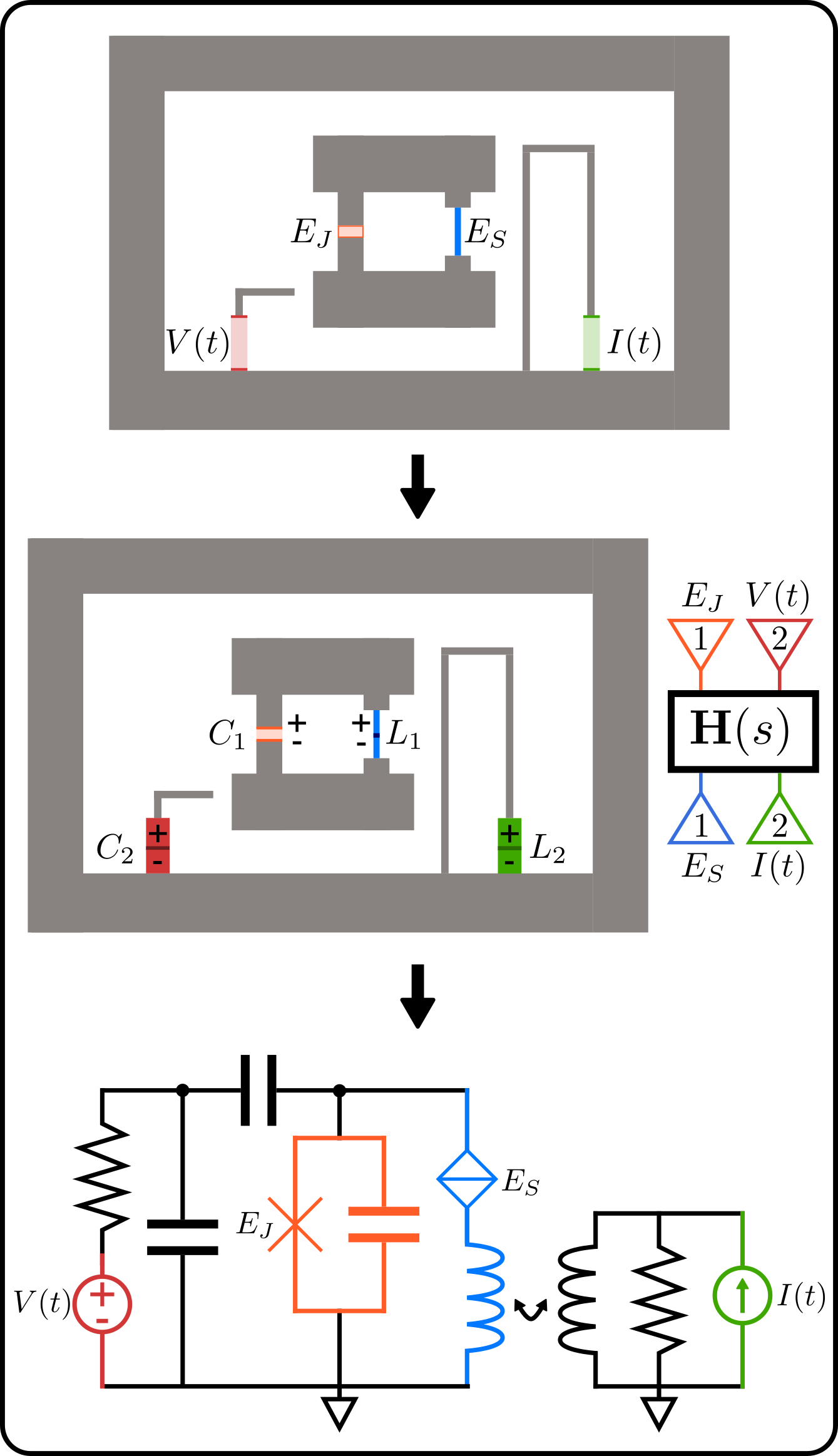}
    \caption{Model extraction for a fluxonium qubit with external charge and flux drives. The nonlinear components of the system consist of Josephson tunneling across the capacitor and fluxon tunneling across the inductor. The drive terms constitute a voltage applied to the charge line and a current to the flux line. This model can be transformed into an electromagnetic simulation (with hybrid matrix $\mathbf{H}(s)$ by replacing the Josephson junction and voltage drive with capacitive ports, and the phase slip wire and flux drive with inductive ports. Finally, from the hybrid matrix response a matching lumped-element model can be synthesiszed. Note that we can systematically add auxiliary resonant modes to increase the model's accuracy.}
    \label{fig:FluxoniumSimulation}
\end{figure}

Once a capacitive/inductive lumped circuit has been generated for the linear part of the device model, the nonlinear and drive elements can be reinserted across the ports. For the ports parallel to capacitors, Josephson tunneling elements can be inserted, just as fluxoid tunneling elements can be placed in series with the inductive ports. This is an advantage of the hybrid matrix synthesis approach: that it allows for Josephson and fluxoid tunneling to be considered in a natural fashion (without placing capacitive shunts across phase slips or inductive series elements across Josephson junctions). Note that we may also want to place additional capacitance across capacitive ports or additional inductance along inductive ports, if we did not fully account for these effects in simulation. For instance, the full parallel-plate structure of a Josephson junction is often added post facto when not included in a full device simulation. The model extraction process for an example Josephson circuit is illustrated in \ref{fig:SynthesisLayout}.

External drives can also be included in circuit models. Usually, voltage sources (with series resistance) are placed on capacitive drive ports and current sources (with shunt resistance) on inductive drive ports. This procedure is illustrated in Fig. \ref{fig:FluxoniumSimulation} for a fluxonium qubit coupled to a charge and flux drive line (with more details following in Section \ref{subsec:ExampleModelExtraction}). Note that we consider the fluxonium's junction array as a linear inductor with phase slip tunneling.  We observe how this method allows for the simultaneous calculation of flux and charge coupling to external lines (and also between qubits, in the multi-qubit case).

\subsection{Example of model extraction} \label{subsec:ExampleModelExtraction}

In Fig. \ref{fig:FluxoniumSimulation} we outline the simulation and model extraction procedure for a multiport circuit. Here, capacitive port 1 ($C_1$) lies across a Josephson junction, with capacitive port 2 ($C_2$) across a charge drive line. Similarly, inductive port 1 ($L_1$) is placed along a phase slip element, while inductive port 2 ($L_2$) is inserted along a flux drive line. We can then perform a frequency domain electromagnetic simulation and extend it to the Laplace domain, with the hybrid matrix response defined as:
\begin{align}
\begin{bmatrix}
I_{C_1}(s) \\
I_{C_2}(s) \\
V_{L_1}(s) \\
V_{L_2}(s)
\end{bmatrix}
=
\mathbf{H}(s)
\begin{bmatrix}
V_{C_1}(s) \\
V_{C_2}(s) \\
I_{L_1}(s) \\
I_{L_2}(s)
\end{bmatrix}
\end{align}

We now expand out the simulated hybrid response as in \ref{eq:MainFinalHybridSimulationResponse},
Though not depicted in the Figure for simplicity, we consider the system to have a single resonant pole, and thus a total response equal to:
\begin{align}
    \mathbf{H} = \mathbf{H}_\text{const}(s) +  \mathbf{H}_\text{lin}(s) +  \mathbf{H}_\text{pole}(s)
\end{align}

The constant term is given by the zero frequency hybrid matrix constraint of:
\begin{align}
\mathbf{H}_\text{const}(s) =
\begin{bmatrix}
0 & 0 & -1 & 0 \\
0 & 0 & 0 & 0 \\
1 & 0 & 0 & 0  \\
0 & 0 & 0 & 0 
\end{bmatrix} = 
\begin{bmatrix}
\mathbf{0}_{CC} & -\mathbf{\Omega}_E \\
\mathbf{\Omega}_E^T & \mathbf{0}_{LL}
\end{bmatrix} 
\end{align}
which encodes the fact that $C_1$ and $L_1$ lie in the same loop.

The linear frequency response is the sum of residues from the pole at infinity plus a contribution from the finite-frequency pole:
\begin{align}
\mathbf{H}_\text{lin}(s) =&
s\begin{bmatrix}
K_{{{C_1} {C_1}}_\infty} & K_{{{C_1} {C_2}}_\infty} & 0 & 0 \\
K_{{{C_2} {C_1}}_\infty}&  K_{{{C_2} {C_2}}_\infty} & 0 & 0 \\
0 & 0 & K_{{{L_1} {L_1}}_\infty}&  K_{{{L_1} {L_2}}_\infty}  \\
0 & 0 & K_{{{L_2} {L_1}}_\infty} & K_{{{L_2} {L_2}}_\infty} 
\end{bmatrix} \nonumber \\ 
&+s\begin{bmatrix}
R_{C_1} & 0 \\
R_{C_2} & 0 \\
0 & R_{L_1} \\
0 & R_{L_2}
\end{bmatrix}
\begin{bmatrix}
R_{C_1} & 0 \\
R_{C_2} & 0 \\
0 & R_{L_1} \\
0 & R_{L_2}
\end{bmatrix}^T
\end{align}

The finite-frequency pole contributes a final term of:
\begin{align}
\mathbf{H}_\text{pole}&(s) = \nonumber \\
&\frac{s^2}{s^2+\omega_r^2} 
\begin{bmatrix}
R_{C_1} & 0 \\
R_{C_2} & 0 \\
0 & R_{L_1} \\
0 & R_{L_2}
\end{bmatrix}
\begin{bmatrix}
-s & -\omega_r \\
\omega_r & -s
\end{bmatrix}
\begin{bmatrix}
R_{C_1} & 0 \\
R_{C_2} & 0 \\
0 & R_{L_1} \\
0 & R_{L_2}
\end{bmatrix}^T
\end{align}

If we now write down a lumped capacitive/inductive circuit the same edge network matrix
\begin{align}
\mathbf{\Omega}_E = 
\begin{bmatrix}
1 & 0 \\
0 & 0
\end{bmatrix}
\end{align}
then using the set of matching criteria from Section \ref{subsec:MatchingResponseLumpedCircuit}, we can construct a linear lumped circuit with parallel and series port terminals across capacitors and along inductors (respectively) that mimics this hybrid response. The nonlinear and drive elements are reinserted across the terminals in the final step.

The capacitance and inductance matrices of the ``core circuit" are equal to:
\begin{align}
\mathbf{C} = 
\begin{bmatrix}
K_{{{C_1} {C_1}}_\infty} & K_{{{C_1} {C_2}}_\infty} \\
K_{{{C_2} {C_1}}_\infty}&  K_{{{C_2} {C_2}}_\infty}
\end{bmatrix} + 
\begin{bmatrix}
R_{C_1} \\
R_{C_2}
\end{bmatrix}
\begin{bmatrix}
R_{C_1} \\
R_{C_2}
\end{bmatrix}^T \\
\mathbf{L} = \begin{bmatrix}
K_{{{L_1} {L_1}}_\infty}&  K_{{{L_1} {L_2}}_\infty}  \\
K_{{{L_2} {L_1}}_\infty} & K_{{{L_2} {L_2}}_\infty} 
\end{bmatrix}
+
\begin{bmatrix}
R_{L_1} \\
R_{L_2}
\end{bmatrix}
\begin{bmatrix}
R_{L_1} \\
R_{L_2}
\end{bmatrix}^T
\end{align}

At this point, we have the information to construct the low-frequency response model shown in Fig. \ref{fig:FluxoniumSimulation}. To also capture the effect of the finite-frequency pole, we augment the network matrix as:
\begin{align}
\mathbf{\Omega}_{E} \rightarrow
\begin{bmatrix}
\mathbf{\Omega}_{E_{11}} & \mathbf{\Omega}_{E_{12}} & 0 \\
\mathbf{\Omega}_{E_{21}} & \mathbf{\Omega}_{E_{22}} & 0 \\
0 & 0 & 1
\end{bmatrix}
\end{align}
where the $1$ in the bottom right corner indicates an auxiliary LC oscillator. Now we set the LC resonator's frequency equal to that of the AC pole by specifying: $C_{rr}L_{rr} = \frac{1}{\omega_r^2}$. To ensure the lumped circuit response accounts for the simulated pole term $\mathbf{H}_\text{pole}(s)$, we augment the capacitance and inductance matrices as:
\begin{align}
\mathbf{C} &\rightarrow
\begin{bmatrix}
C_{11} & C_{12} & \frac{R_{C_1}}{\sqrt{C_{rr}}} \\
C_{21} & C_{22} & \frac{R_{C_2}}{\sqrt{C_{rr}}} \\
\frac{R_{C_1}}{\sqrt{C_{rr}}} & \frac{R_{C_2}}{\sqrt{C_{rr}}} & C_{rr}
\end{bmatrix} \\
\mathbf{L} &\rightarrow
\begin{bmatrix}
L_{11} & L_{12} & \frac{R_{L_1}}{\sqrt{L_{rr}}} \\
L_{21} & L_{22} & \frac{R_{L_2}}{\sqrt{L_{rr}}} \\
\frac{R_{L_1}}{\sqrt{L_{rr}}} & \frac{R_{L_2}}{\sqrt{L_{rr}}} & L_{rr}
\end{bmatrix}
\end{align}

At this point, we have a complete circuit model for the linear response of the system, with port terminals in parallel with capacitive elements and in series with inductive ones---across which nonlinear and drive elements can be inserted (Fig. \ref{fig:FluxoniumSimulation}). While not shown in the Figure, the additional LC oscillator (representing a pole) can be added to the circuit diagram as in Fig. \ref{fig:SynthesisLayout}. The equations of motion for the system are naturally returned in the tree-cotree edge basis (Section \ref{subsec:EdgeNetworkSpanningTreeCotree}), and can be analyzed through techniques presented in Section \ref{sec:LumpedEquationsOfMotion}.

\section{Conclusions}

In this work we have presented an intuitive flux-charge symmetric framework for lumped-element circuit quantum electrodynamics \cite{kerman_fluxcharge_2013, osborne_symplectic_2024, parra-rodriguez_geometrical_2024} and illustrated its applications to quantization, decomposition, and model extraction. We emphasized pivotal ways in which the network matrix $\mathbf{\Omega}$ connects all of these procedures.

In Section \ref{sec:LumpedEquationsOfMotion} we demonstrated a set of flux-charge symmetric equations of motion that are subject to certain physical restrictions, and introduced the network matrix---a key object that encodes the connectivity of the circuit's capacitive and inductive components. We showed how the equations of motion can be written in complementary ``standard" and ``integrated" forms, and how the elements of the equations change under basis transformations. Finally, we gave a visual illustration of our systems in a tree-cotree notational shorthand.

In Section \ref{sec:Quantization} we showed a straightforward algorithm to quantize circuits with capacitors, inductors, Josephson junctions, and phase slip wires. We began by using basis transformations to reduce the network matrix to the identity (with extra rows and columns of zeros). We then utilized the integrated equations of motion to track which of a system's variables become integer-valued when quantized. We generated novel predictions about the Hamiltonians of circuits containing Josephson junctions and phase slip elements.

In Section \ref{sec:Decomposition} we illustrated how to decompose and manipulate circuits, by performing pivoting operations on the circuit's ``edge" network matrix $\Omega_E$ \cite{schrijver_theory_1998, wolsey_integer_2014}. We gave a mathematical and visual interpretation of this ``fundamental decomposition" procedure, which converts circuit models to a maximally simplified equivalent form---where the harmonic and free modes have been separated out from the rest of the circuit. This procedure aids in simplifying and classifying lumped superconducting circuits.

Finally, in Section \ref{sec:Synthesis} we showed how the (edge) network matrix underlies a flux-charge symmetric algorithm to extract lumped-element circuit models from electromagnetic simulation data. Valid for lossless, reciprocal systems with physically small nonlinear components, the procedure synthesizes a lumped-element capacitive/inductive circuit model of a device's distributed linear components. We carried out this process by matching the lumped circuit's edge network matrix to the topology of the physical device, and then setting the capacitance and inductance matrices to reproduce the device's hybrid admittance/impedance matrix response \cite{david_m_pozar_microwave_2012, yarlagadda_reciprocal_1972}. We can use this model extraction technique to generate transformerless circuit models of quantum devices, which are accurate at high frequencies.

In sum, our methods enable powerful and intuitive algorithms for analysis, manipulation, simulation of superconducting quantum devices. The network matrix serves as a unifying concept that reveals the interconnections between each of these topics.

We envision that our methods will aid in the design of superconducting quantum devices. In addition, we aim to extend our transformerless superconducting model extraction techniques to systems with non-reciprocal response. Further, we would like to use our decomposition methods to perform a detailed classification of superconducting circuits.

\section{Acknowledgments}

The authors would like to thank Anjali Premkumar, Jeronimo Martinez, Andrew Osborne, David Schuster, Jens Koch, and Joe Aumentado for useful discussions.

Funding for this project was provided by the National Science Foundation Quantum Leap Challenge Institute for Robust Quantum Simulation (grant number 2120757), and by the U.S. Army Research Office under the Gates on Advanced qubits with Superior Performance program (grant number W911NF2310101).

Princeton University Professor Andrew Houck is also a consultant for Quantum Circuits Incorporated (QCI). Due to his income from QCI, Princeton University has a management plan in place to mitigate a potential conflict of interest that could affect the design, conduct and reporting of this research.

\clearpage

\appendix

\section{Lossless, reciprocal, lumped circuit models without quantum tunneling}\label{sec:AppendixA}

\begin{figure}
    \centering
\includegraphics[width=1\linewidth]{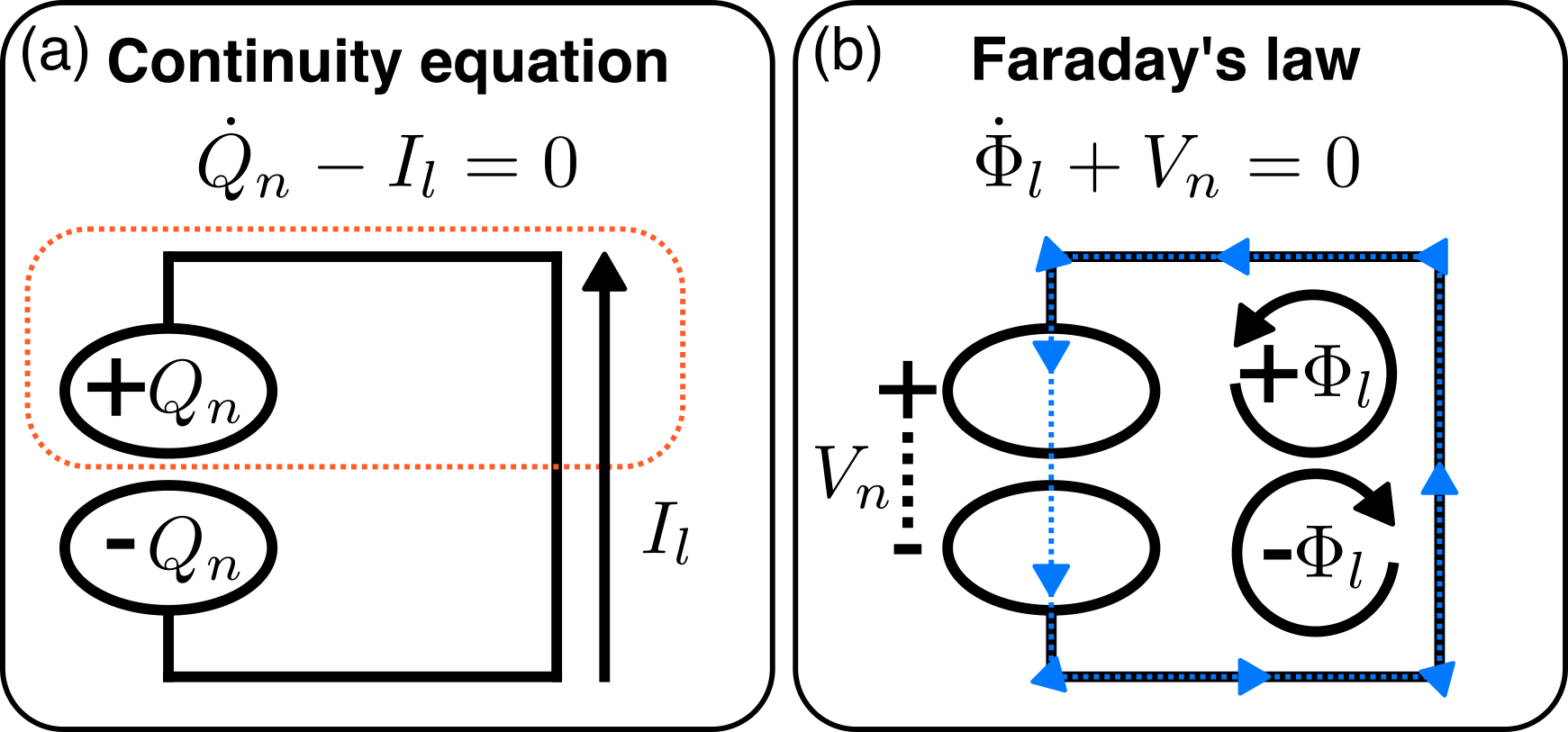}
    \caption{Circuit model illustrations of (a) the electromagnetic continuity equation and (b) Faraday's law.}
    \label{fig:ContinuityFaraday}
\end{figure}

In this Appendix, we provide background on the analysis of lossless, reciprocal lumped-element circuits in the absence of quantum tunneling. These circuit models consist of capacitive nodes/islands (that have node voltage variables) connected by inductive loops (that possess loop current variables), with connectivity encoded by the node-loop network matrix $\mathbf{\Omega}$. Our presentation emphasizes the nodal capacitance matrix and loop inductance matrix as central mathematical objects (and relates them to the standard branch capacitor and inductor symbols). We highlight the physical distinction between the time derivative of charge and current, as well as between the time derivative of flux and voltage, and show how basis transformations of the voltage and current variables can be viewed as transformations of the underlying circuit model. Without nonlinear quantum tunneling, a capacitive/inductive system can always be reduced to an equivalent set of uncoupled harmonic oscillators. This material lays the groundwork for the more complete picture of circuit dynamics described in Appendix \ref{sec:AppendixB}, which includes effects of quantum tunneling.

\subsection{Nodes and loops} 

Lumped circuit models represent a discretized, quasi-static formulation of Maxwell's equations of electromagnetism. For superconducting metals in low-loss, non-magnetic dielectrics, the resulting equations will be lossless and reciprocal (obeying time-reversal symmetry) \cite{david_m_pozar_microwave_2012, newcomb_linear_1966}. Though circuit models are an idealization of physical reality, they can be systematically extracted from electromagnetic simulations as detailed in Appendix \ref{sec:AppendixE}.

We now provide a heuristic explanation for the lumped-element description of linear capacitive/inductive circuits. Fig. \ref{fig:ContinuityFaraday} gives an illustrative example: an electrical system where a wire loop connects two nodes. Current can flow through the loop and cause charge to accumulate on the nodes, and a voltage difference can form between the two nodes, causing magnetic flux to accumulate in the loop. This system is usually depicted as an LC oscillator, with the wire drawn as an inductor, and the nodes connected by a capacitor. We will introduce these branch circuit elements after further discussing the underlying physics.

As depicted in Fig. \ref{fig:ContinuityFaraday}, this system is governed by two electromagnetic equations: the continuity equation and Faraday's law.

In Fig. \ref{fig:ContinuityFaraday}(a), the orange line represents a region of volume that encapsulates a node and intersects with the loop. In general, for a volume region $\mathcal{V}$ with surface $\mathcal{S}$, the volume charge $Q$ and net surface current $I$ are defined by \cite{griffiths_introduction_2013}:
\begin{align}
    Q &=  \iiint_{\mathcal{V}} \rho d { \mathcal{V}} \\
    I &=  \iint_{\mathcal{S}} \vv{J} \cdot d\vv{\mathcal{S}}
\end{align}
Here $\rho$ is the net free charge density and $\vv{J}$ is the net free current density. The only net current flowing into the integration region occurs along the wire of the loop. In this example, the electromagnetic continuity equation states that, for the orange volume region region $\mathcal{V}$ with boundary surface $\partial \mathcal{V}$:
\begin{align} \label{eq:ContinuityEquation}
    \frac{\partial }{\partial t }\iiint_{\mathcal{V}} \rho &  d { \mathcal{V}} -  \oiint_{\partial \mathcal{V}} \vv{J} \cdot d\vv{\mathcal{S}} = 0 \nonumber \\
    \implies & \dot{Q}_n - I_l = 0
\end{align}
We note that the boundary orientation of the surface integral is defined inwards.
We also mention that the lumped-element limit assumes a quasi-static distribution of current flow, such that it can be defined by a single value at all points in the loop---with vanishing net charge buildup inside the wire. We can then define single values of node charge $Q_n$ and loop current $I_l$ that are independent of the precise integration region.

Similar logic applies to derive the equation of motion for loops from Faraday's law. The definitions of net magnetic flux $\Phi$ through a surface $\mathcal{S}$ and the voltage drop $V$ between two points on a path $\Gamma$ are given by:
\begin{align}
    \Phi & =  \iint_\mathcal{S} \vv{B}  \cdot d\vv{\mathcal{S}} \\
    V &= \int_\Gamma \vv{E} \cdot d\vv{\Gamma} 
\end{align}

Then, following Fig. \ref{fig:ContinuityFaraday}(b), we apply Faraday's law to the blue curve. We take the integration line to be deep enough inside the superconductor such that electric field $\vv{E} = \vv{0}$ along the curve everywhere except in the gap between the two islands (where the voltage drop occurs). Faraday's law applied to a surface $\mathcal{S}$ with boundary $\partial \mathcal{S}$ defined by the blue curve implies:
\begin{align} \label{eq:FaradaysLaw}
   \frac{\partial}{\partial t} \iint_S \vv{B}&  \cdot d\vv{\mathcal{S}}  +   \oint_{\partial \mathcal{S}} \vv{E} \cdot d \vv{\Gamma} = 0 \nonumber \\
    \implies & \dot{\Phi}_l +  V_n = 0
\end{align}
Another quasi-static approximation is applied here: that, for the nodes, the voltage drop $V_n$ is independent of the integration path taken between them. This is equivalent to saying that if we combined to inter-node integration paths into a loop, there would be approximately no accumulating net flux through the resulting surface---and thus it is reasonable to speak of a path-independent loop flux $\Phi_l$ and node voltage difference $V_n$.

In the present example, Eq. \ref{eq:ContinuityEquation} represents Kirchhoff's current law and \ref{eq:FaradaysLaw} represents Kirchhoff's voltage law \cite{chen_electrical_2004}. Note that our notation is slightly non-standard. Usually, the quantity $\dot{Q}_n$ is written as a current through the capacitor and $\dot{\Phi}_l$ as a voltage across an inductor. Maintaining these distinctions in notation will aid in conceptual clarity.

\subsection{Capacitance and inductance matrices}
\label{subsec:CapacitanceInductanceMatrices}

To analyze the equations of motion, a key step is expanding the expressions for $Q_n$ in terms of $V_n$ and $\Phi_l$ in terms of $I_l$ using the quasi-static concepts of capacitance and inductance \cite{griffiths_introduction_2013}. 

For capacitance, in a system with $n$ nodes, the vector of all node charges is linearly proportional to the vector of node voltages by the capacitance matrix $\mathbf{C}$:
\begin{align} \label{eq:NodeCapacitance}
\vv{Q}_n &= \mathbf{C} \vv{V}_n \end{align}
Note that the voltage of each node is measured relative to that of an arbitrary ground node.

Similarly, for inductance, the vector of all loop fluxes is linearly proportional to the vector of loop currents by the inductance matrix $\mathbf{L}$:
\begin{align} \label{loopInductance}
\vv{\Phi}_l &= \mathbf{L} \vv{I}_l 
\end{align}
Here too the current through each loop is measured relative to that in a reference loop.

After removing the row and column corresponding to the ground node or loop, the capacitance and inductance matrices are symmetric positive definite (a thorough demonstration for the capacitance matrix is shown in \cite{diaz_positivity_2011}). Their (positive) diagonal elements represent node self-capacitance and loop self-inductance, respectively. The off-diagonal elements of $\mathbf{C}$ represent coupling capacitances between nodes, while the off-diagonal elements of $\mathbf{L}$ embody the mutual inductances between loops. 

Note that the lack of coupling between capacitive and inductive degrees of freedom is a result of electromagnetic reciprocity. Relevant discussion of circuits beyond the reciprocal regime can be found in \cite{tellegen_gyrator_1948, egusquiza_algebraic_2022, parra-rodriguez_geometrical_2024}. Those works also discuss the transformer limit, where the reduced capacitance or inductance matrix has a zero eigenvector, which we do not consider in this manuscript.

\subsection{Node and loop ground}

To perform quantum analysis, we aim to find a Hamiltonian description of the circuit. This requires inverting the reduced node capacitance and loop inductance matrices. In order to make these matrices invertible, the standard process of grounding must take place.

The idea of ground can be illustrated in our example of Fig \ref{fig:ContinuityFaraday}. For the capacitance matrix, we let $V_1$ be the voltage on the upper node and $V_0$ the voltage on the lower node (measured relative to some arbitrary point). Then, assuming the total system to be charge-neutral, using the symmetry of the capacitance matrix gives:
\begin{align}
\begin{bmatrix}
Q_n \\
-Q_n
\end{bmatrix}
=
\begin{bmatrix}
C & -C \\
-C & C
\end{bmatrix}
\begin{bmatrix}
V_1 \\
V_0 
\end{bmatrix}
\end{align}
Though there are two nodes, there is only one linearly independent capacitance equation:
\begin{align}
Q_n = C \left( V_1-V_0 \right) = C V_n 
\end{align}
where $V_n$ is the voltage difference across the nodes. In particular, the capacitance matrix equation is invariant under a uniform offset of both node voltages $V_0$ and $V_1$, which leaves the voltage difference between the two nodes the same. Thus it is typical to perform an offset that sets $V_0= 0$ and $V_1 = V_n$. The node whose voltage is set to 0 is called the ground. The effect of grounding is to remove the row and column of the capacitance matrix corresponding to that node, leaving an invertible reduced capacitance matrix.

More generally, for a system of $N+1$ capacitively connected nodes, the reduced capacitance matrix is of size $N$ by $N$. We note that a circuit model may contain multiple groups of capacitive nodes (that are not interconnected), and in that case each group should have its own ground node. However, we will usually make the physical assumption that we have a single capacitive network.

For inductance the procedure is similar. As an example, we follow Fig. \ref{fig:ContinuityFaraday}(b) and consider $\Phi_l$ to be the flux going through the loop out of the page. If the system is flux neutral then $-\Phi_l$ is the flux out of the page through the exterior loop of the system. Then if $I_1$ represents the counterclockwise current of charge carriers and $I_0$ is the clockwise motion, we have that the inductance matrix equation is:
\begin{align}
\begin{bmatrix}
\Phi_l \\
-\Phi_l
\end{bmatrix}
=
\begin{bmatrix}
L & -L \\
-L & L
\end{bmatrix}
\begin{bmatrix}
I_1 \\
I_0 
\end{bmatrix}
\end{align}
Once again there is only one linearly independent equation of motion: 
\begin{align}
\Phi_l = L \left( I_1-I_0 \right) = L I_l 
\end{align}
This equation is invariant under uniform translations of the clockwise and counterclockwise currents. So, as before, we are free to displace such that $I_0 = 0$ and $I_1 = I_l$, and then eliminate the row and column corresponding to the grounded clockwise loop of the inductance matrix---giving a reduced, invertible inductance matrix.

Once again, for a system of $M+1$ connected loops (including the external loop), there will be $M$ linearly independent inductance equations, and one can set the current through one of the loops to $0$ to reduce the inductance matrix. Also, each disconnected set of loops will have its own ground current reference. However, this process of grounding is often more implicit than in the case of the capacitance matrix. Usually, the location and directions of non-ground loops are specified, while the ground loops are not explicitly drawn on the circuit diagram.

We note that circuits may represent effective models in which nodes or loops of the system have been eliminated through the process of free mode removal (Appendices \ref{subsec:ChangeofBasis1} and \ref{subsec:FreeModes}). In this case, the assumption of overall charge and flux neutrality may not hold. However, the deviations from neutrality result only in constant offset terms that do not enter the equations of motion and can thus be eliminated from the effective circuit model. 

\subsection{External charge and flux}\label{subsec:ExternalChargeAndFlux}

In the equations for capacitance and inductance, we can append a notion of external charge and flux, by adding additional vectors:
\begin{align}
\vv{Q}_n &= \mathbf{C} \vv{V}_n +  \vv{Q}_\text{ext} \label{eq:CapacitiveConstitutive} \\
\vv{\Phi}_l &= \mathbf{L} \vv{I}_l + \vv{\Phi}_\text{ext} \label{eq:InductiveConstitutive}
\end{align}
Conventionally, these external charge and flux quantities can refer to two different phenomena: externally-applied bias fields (where stiff voltage or current sources prevent back-action), and noise from fluctuations in the environment.

In both cases, an external charge can be considered as a sum of the capacitive excitations coming from external node voltages. For a single island under the influence of multiple external charges this can be written:
\begin{align}
Q_{n} & = C V_{n}  -\sum_i \left[ C_c V_\text{ext} \right]_i \nonumber \\
& = C V_{n} + Q_\text{ext}
\end{align}

Similarly, for inductance, external flux is produced by external current loops with mutual inductive coupling to those of the device. A single loop under the influence of external flux can be written:
\begin{align}
\Phi_{l} &= L I_{l} + \sum_i \left[ m I_\text{ext} \right]_i \nonumber \\
&= L I_{l} + \Phi_\text{ext}
\end{align}

Note that this symmetric presentation of external charge and flux is an advantage of considering both capacitive nodes and inductive loops on equal footing.

\subsection{Quantization of an LC-oscillator} \label{subsec:LCQuantization}

Taking the information from the preceding sections, we present the standard quantization procedure for the LC-oscillator circuit given in Fig. \ref{fig:ContinuityFaraday}. As previously noted, the continuity equation (Eq. \ref{eq:ContinuityEquation}) and Faraday's law (Eq. \ref{eq:FaradaysLaw}) result in the equations of motion:
\begin{align}
\dot{Q}_n - I_l = 0 \\
\dot{\Phi}_l + V_n = 0
\end{align}

The capacitive and inductive constitutive relations (Eqs. \ref{eq:CapacitiveConstitutive} and \ref{eq:InductiveConstitutive}) imply that:
\begin{align}
Q_n &= C V_n + Q_\text{ext}  \\
\Phi_l &=  L I_l + \Phi_\text{ext}
\end{align}

All together this means:
\begin{align}
C \dot{V}_n + \dot{Q}_\text{ext} - I_l &= 0 \\
L \dot{I}_l + \dot{\Phi}_\text{ext} + V_n &= 0
\end{align}

By differentiating the continuity equation and eliminating the current variable we get:
\begin{align}
C\ddot{V}_n + \ddot{Q}_\text{ext} + \frac{1}{L} \left( V_n + \Phi_\text{ext} \right) = 0 
\end{align}
This is the Euler-Lagrange equation of motion ($ \frac{d}{dt}\frac{\partial}{\partial \dot{V}_n} \mathcal{L} -\frac{\partial}{\partial V_n} \mathcal{L}=0 $ ) of the Lagrangian:
\begin{align}
\mathcal{L} = \frac{1}{2} C \dot{V}_n^2 + \dot{V}_n \dot{Q}_\text{ext}  - \frac{1}{2L} \left( V_n + \dot{\Phi}_\text{ext} \right)^2
\end{align}

To analyze this circuit quantum mechanically
we convert to a Hamiltonian form by first finding the conjugate current variable to the voltage
\begin{align}
\Pi_n = \frac{\partial\mathcal{L}}{\partial \dot{V}_n} = C\dot{V}_n + \dot{Q}_\text{ext} = I_l
\end{align}
and then taking the Legendre transform:
\begin{align}
\mathcal{H} & = \dot{V}_n I_l - \mathcal{L} \nonumber \\
& = \frac{1}{2 C} \left( I_l - \dot{Q}_\text{ext}  \right)^2+  \frac{1}{2 L} \left( V_n + \dot{\Phi}_\text{ext} \right)^2
\end{align}
This gives the expected Hamiltonian for a harmonic oscillator, which can be quantized by promoting the classical Poisson bracket to the quantum commutation relation:
\begin{align}
\{V_n , I_l \} = 1 \rightarrow [\hat{V}_n , \hat{I}_l] = i \hbar
\end{align}
giving a quantized Hamiltonian of:
\begin{align}
\hat{\mathcal{H}} = \frac{1}{2 C} \left( \hat{I}_l - \dot{Q}_\text{ext}  \right)^2+  \frac{1}{2 L} \left( \hat{V}_n + \dot{\Phi}_\text{ext} \right)^2
\end{align}
where we have followed the standard approximation, which treats external charge and flux variables as classical---instead of quantum---parameters.

Once quantum tunneling circuit elements are added, systems will be quantized in terms of canonically conjugate flux and charge variables, which represent time integrals of voltage and current (respectively) \cite{michel_h_devoret_quantum_1995, vool_introduction_2017}.

\subsection{Node-loop graph circuit model, network matrix}
\label{subsec:NodeLoopGraphCircuitModelNetworkMatrix}

\begin{figure}
    \centering
    \includegraphics[width=1\linewidth]{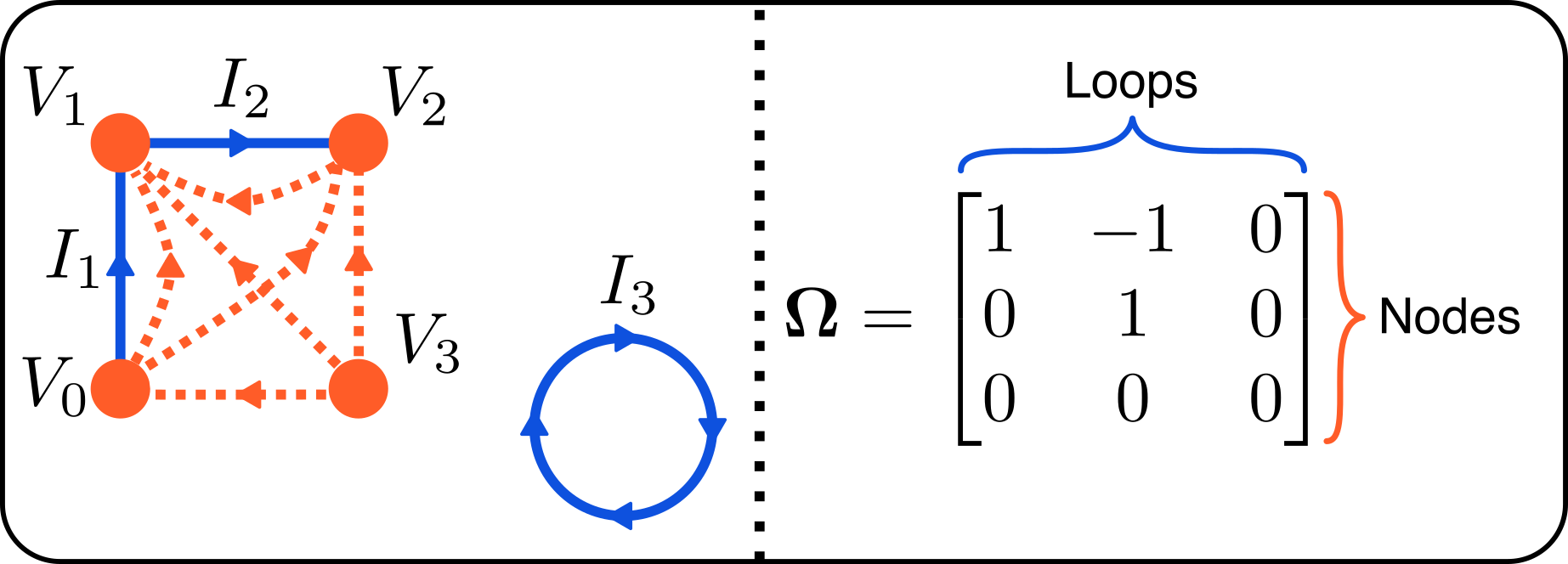}
    \caption{Graph model of electrical circuits with capacitance and inductance. The orange nodes represent capacitive islands and the blue lines the edges of inductive loops. The orange dashed lines are the implicit capacitive edges that may lie in the inductive loops. Here $V_0=0$ is the ground node voltage, and $I_3$ the current through a disconnected self-loop. The network matrix $\mathbf{\Omega}$ encodes the node-loop connectivity of the system, with the row corresponding to the ground node removed.}
\label{fig:GraphsNetworkMatrix}
\end{figure}

As is standard in electrical engineering \cite{chen_electrical_2004}, we employ linear algebraic methods from graph theory to model networks of interconnected nodes and loops. This type of approach forms a cornerstone of superconducting device modeling \cite{michel_h_devoret_quantum_1995, burkard_multilevel_2004, burkard_circuit_2005, parra-rodriguez_canonical_2019, egusquiza_algebraic_2022, parra-rodriguez_geometrical_2024, osborne_symplectic_2024, osborne_flux-charge_2024}.

Electromagnetic circuits can be represented as directed graphs. A directed graph $G$ consists of vertices (or nodes) $V(G)$ and edges (or branches) $E(G)$. Each edge is an ordered pair of vertices, where the order denotes the edge's direction: out of the first vertex and into the second. We use ``branch" as synonym for ``edge," following circuit theory conventions.

A simple representation of capacitor/inductor circuit dynamics treats the graph's vertices as capacitive nodes, and edges as the inductive portions of loops (with return paths through the capacitive sub-network). In Fig. \ref{fig:GraphsNetworkMatrix}, each vertex (orange) represents a node (parameterized with a voltage) and each each directed edge (blue) represents part of an inductive loop (parameterized with a current). We assume that that no island is capacitively isolated from the others, such that the circuit can be written with a single ground node (with voltage $V_0 = 0$). In this picture, there is an implied connected subgraph of ``capacitive" edges spanning the circuit's nodes, depicted with the orange dashed lines. The inductive loops of the system then correspond to simple graph cycles (paths starting and ending on the same vertex that do not self-intersect) that contain at least one inductive edge (and potentially a combination of implicit capacitive edges). We also visually represent inductive loops that do not intersect with any nodes, such as the self-loop shown in the figure with current $I_3$.

The connectivity of the circuit can be encoded in the node-loop network matrix
$\mathbf{\Omega}$. This object represents a more restricted version of the symplectic form discussed in \cite{egusquiza_algebraic_2022,parra-rodriguez_geometrical_2024, osborne_symplectic_2024,osborne_flux-charge_2024}. Intuitively, the matrix denotes both the charge buildup on nodes as current flows through inductive edges of loops, and the flux buildup in loops as voltage accumulates across nodes. The rows of this matrix represent non-ground nodes and the columns represent loop paths, as follows:
\begin{align} \label{eq:AppendixNetworkMatrix}
\Omega_{ij} =
\begin{cases} 
      1 & \text{node $i$ touches $1$ inductive}  \\ & \text{edge in loop $j$ (tip)} \\
      -1 & \text{node $i$ touches $1$ inductive} \\ & \text{edge in loop $j$ (tail)} \\
      0 & \text{node $i$ touches even number} \\ & \text{of inductive edges in loop $j$}
\end{cases}
\end{align}

The loops enumerated in the network matrix represent a maximal linearly independent set. We consider a set of loops to be linearly dependent if the inductive edges in one loop can be written as a symmetric difference of the inductive edges of the other loops. The symmetric difference operation corresponds to taking the union of two sets and then subtracting the intersection \cite{oxley_matroid_2006}. In this framework, two loops are linearly dependent if they contain the same inductive edges, even if they have different capacitive edges---which only enter the model implicitly.

In Fig. \ref{fig:GraphsNetworkMatrix}, and throughout this work, we will often work with a ``standard" basis of loops, where each inductive edge represents a unique loop, with its return path running through the implicit capacitive edges. Using this set of loops, a column of the node-loop network matrix contains at most one value of $1$ and one value of $-1$, with the remaining entries being $0$. Then, in particular, the network matrix is a graph incidence matrix (an object discussed further in Appendix \ref{subsec:IncidenceLoopMatrices}) \cite{bapat_graphs_2014}. Here, the loop current variables correspond to the branch currents across the inductive edges.

Fig. \ref{fig:GraphsNetworkMatrix} demonstrates the standard form of a node-loop network matrix for an example graph network.

Now we can use the properties of this network matrix to produce vector-valued node and loop equations of motion. 

For the continuity equation at each node, we have that the rate of change in net charge is equal to the total incoming current minus the total outgoing---summed over each loop (a generalization of Eq. \ref{eq:ContinuityEquation}). This can be written with the network matrix as:
\begin{align}
\dot{Q}_n - \sum_l \Omega_{n l} I_l = 0
\end{align}

Similarly, the Faraday's law equation of each loop states that the rate of change of flux through the loop plus the net sum of voltage drops across the loop is equal to zero (restating of Eq. \ref{eq:FaradaysLaw}). This is equivalently formulated with the transpose of the network matrix:
\begin{align}
\dot{\Phi}_l + \sum_n \Omega_{n l}^T V_n = 0
\end{align}

Written in a vector form (that employs Eq. \ref{eq:CapacitiveConstitutive} and Eq. \ref{eq:InductiveConstitutive}) we have:
\begin{align}
\mathbf{C} \dot{\vv{V}}_n + \dot{\vv{Q}}_\text{ext} - \mathbf{\Omega} \vv{I}_l &= \vv{0}_n \label{eq:VectorContinuity} \\
\mathbf{L} \dot{\vv{I}}_l + \dot{\vv{\Phi}}_\text{ext} + \mathbf{\Omega}^T \vv{V}_n &= \vv{0}_l \label{eq:VectorFaraday}
\end{align}

Often, in circuit quantum electrodynamics, the current degrees of freedom are eliminated to give a single set of equations in terms of voltage (or rather flux, in the case of Josephson junction tunneling elements):
\begin{align}
\mathbf{C} \ddot{\vv{V}}_n + \ddot{\vv{Q}}_\text{ext} + \mathbf{\Omega} \mathbf{L}^{-1} \left( \mathbf{\Omega}^T \vv{V}_n + \dot{\vv{\Phi}}_\text{ext}  \right) &= \vv{0}_n
\end{align}
Alternatively voltage can be eliminated to produce equations of motion dependent on current:
\begin{align}
\mathbf{L} \ddot{\vv{I}}_l + \ddot{\vv{\Phi}}_\text{ext} + \mathbf{\Omega}^T \mathbf{C}^{-1} \left( \mathbf{\Omega} \vv{I}_l - \dot{\vv{Q}}_\text{ext}  \right) &= \vv{0}_l
\end{align}
These equations can produce a Hamiltonian in a straightforward manner. However, we first simplify them through changes of basis discussed in Appendices \ref{subsec:ChangeofBasis1} and \ref{subsec:CapacitanceInductanceDiagonalization}.

\subsection{Incidence and loop matrices} \label{subsec:IncidenceLoopMatrices}

\begin{figure}
    \centering
    \includegraphics[width=1\linewidth]{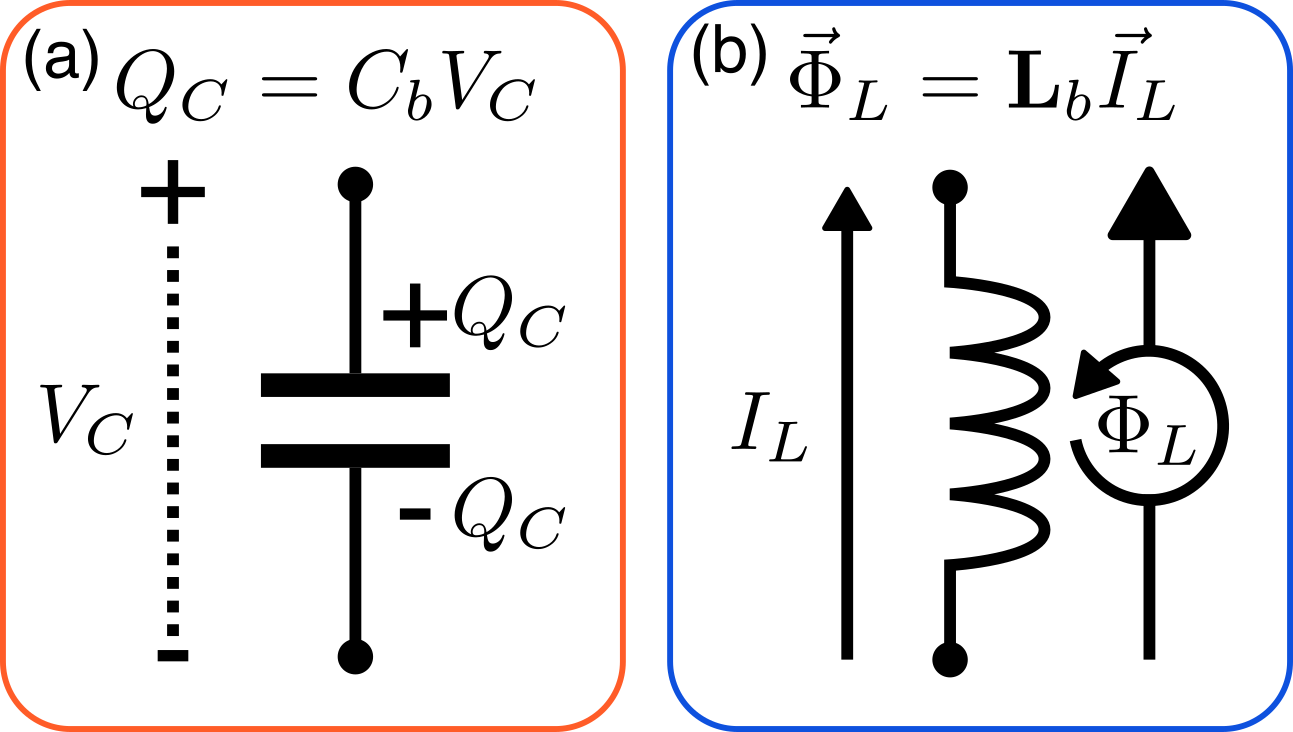}
    \caption{Capacitors and inductors. (a) Branch capacitor circuit element relating branch charge and voltage. (b) Branch inductor circuit element relating magnetic flux and current. Note that the equation is vector-valued: inductors couple to each other through mutual inductance.}
    \label{fig:SingleCapacitorInductor}
\end{figure}

Before analyzing the equations of motion, we introduce the standard branch capacitor and inductor circuit elements (shown in Fig. \ref{fig:BranchCapacitorInductor}). Ultimately, these branch circuit elements represent an alternate method to generate the capacitance and inductance matrices appearing in the equations of motion. Observe that we use these same capacitor and inductor symbols with alternate meanings in our ``tree-cotree" notation (Section \ref{subsec:TreeCotreeNotation} of the Main Text).

First, we make note of a few more graph-theoretic matrices, which are useful when working with capacitors and inductors: the incidence matrix $\mathbf{A}$ and the loop matrix $\mathbf{B}$ \cite{bapat_graphs_2014,strang_introduction_2016}.

An incidence matrix $\mathbf{A}$ quantifies how directed edges connect to nodes:
\begin{align} \label{eq:IncidenceMatrix}
A_{ij} =
\begin{cases} 
      1 & \text{edge $j$ enters node $i$} \\
      -1 & \text{edge $j$ leaves node $i$} \\
      0 & \text{edge $j$ not at node $i$}
\end{cases}
\end{align}

A loop matrix $\mathbf{B}$ is defined for the cycles (loops) of a directed graph, and details which edges belong to the cycle and how they are oriented along the path:
\begin{align} \label{eq:LoopMatrix}
B_{ij} =
\begin{cases} 
      1 & \text{edge $j$ within loop $i$ --- same direction} \\
      -1 & \text{edge $j$ within loop $i$ --- opposite direction} \\
      0 & \text{edge $j$ not in loop $i$}
\end{cases}
\end{align}

We highlight some key conditions on the invertibility of incidence and loop matrices \cite{bapat_graphs_2014,strang_introduction_2016}. Briefly, the reduced incidence matrix (with ground node removed) of a graph is invertible if and only if it represents a tree (a connected set of edges with no cycles). A loop matrix is invertible if and only if a tree can be drawn such that its cotree (edges not in the tree) contains exactly one edge from each loop.

\subsection{Branch capacitors and inductors} \label{sec:BranchCapacitorsAndInductors}

Fig. \ref{fig:SingleCapacitorInductor} shows two types of directed edges in branch circuit models: capacitors and inductors---which represent a common method to encode the capacitance and inductance matrices of a system. However, we again emphasize the resulting capacitance and inductance matrices (Eq. \ref{eq:AppendixNetworkMatrix}) as the fundamental objects. Ultimately this section shows how to write down the equations of motion of the system (Eqs. \ref{eq:VectorContinuity} and \ref{eq:VectorContinuity}) from a lumped circuit written in standard electrical notation.

Branch capacitors and inductors have properties determined by their constitutive relation. That of the branch capacitor (Fig. \ref{fig:SingleCapacitorInductor}(a)) relates the branch charge $Q_C$ across the nodes (of equal magnitude but opposite sign on each) to the branch voltage $V_C$ across them through the capacitance $C_b$:

\begin{figure*}
    \centering
    \includegraphics[width=1\linewidth]{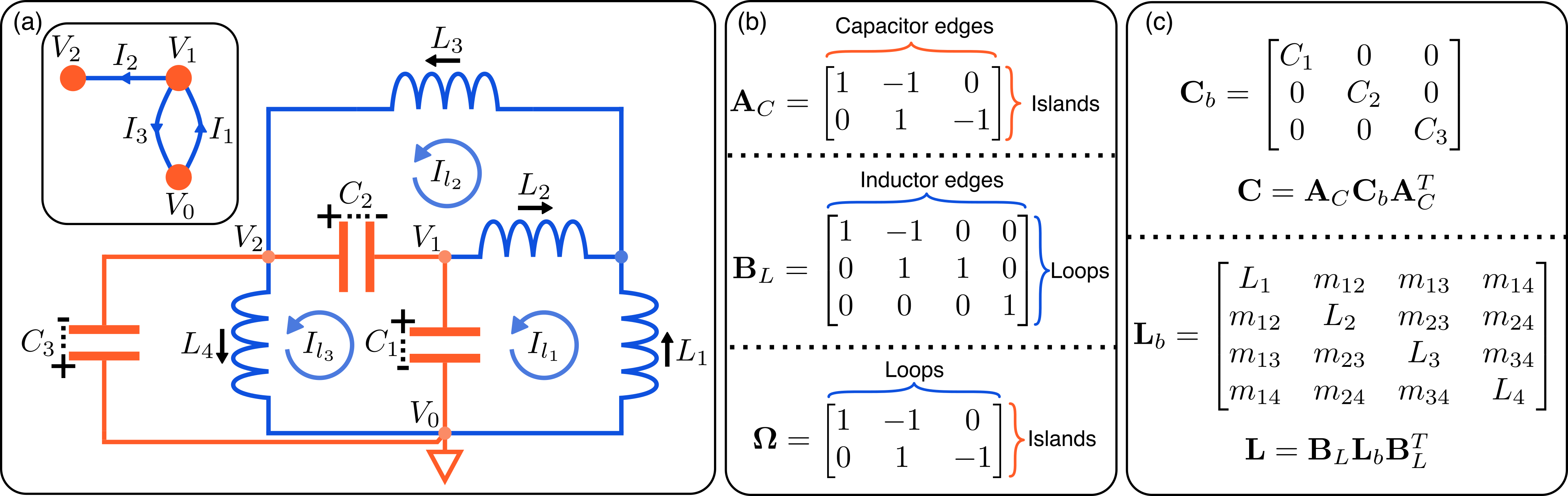}
    \caption{Constructing system matrices from branch circuit models. (a) Circuit diagram with capacitors and inductors. The circuit contains two active capacitive nodes (and a ground node), as well as three linearly independent inductive loops. Inset: equivalent node-loop graph model of the system, with implicit capacitive edges (Appendix \ref{subsec:NodeLoopGraphCircuitModelNetworkMatrix}). (b) Graph matrices of the system. $\mathbf{A}_C$ is the capacitor incidence matrix, $\mathbf{B}_L$ is the inductor loop matrix, and $\mathbf{\Omega}$ is the node-loop network matrix. (c) Calculating the node capacitance matrix from the branch capacitances and the loop inductance matrix from the branch inductances.}
    \label{fig:BranchCapacitorInductor}
\end{figure*}
\begin{align}
Q_C &= C_b V_C
\end{align}
This equation can be made vector-valued by defining $\mathbf{C}_b$ as a diagonal matrix that contains all of the (positive) branch capacitances of the circuit:
\begin{align} \label{eq:CapacitorEq1}
\vv{Q}_C &= \mathbf{C}_b \vv{V}_C
\end{align}

The branch charges can be parameterized in terms of the node charges through the use of an incidence matrix (Eq. \ref{eq:IncidenceMatrix}) $\mathbf{A}_C$, whose rows represent the capacitive islands of the circuit, and whose columns represent the individual branch capacitors incident upon those nodes. Directionality is defined by a $+1$ appearing in the incidence matrix if a capacitor's positive terminal intersects with a given node, and a $-1$ if instead the negative terminal intersects. 

At each node, the total charge is the sum of the charges of all incident capacitors. We also add the standard factor of external charge (Eq. \ref{eq:CapacitiveConstitutive}) to obtain:
\begin{align} \label{eq:CapacitorEq2}
\vv{Q}_n = \mathbf{A}_C \vv{Q}_C + \vv{Q}_\text{ext}
\end{align}
Note that $\mathbf{A}_C$ is the reduced incidence matrix where the row corresponding to the ground node has been eliminated.

For the loops within the capacitive sub-network, we have a form of Kirchhoff's voltage law with no flux term, as there are no inductors in the loop \cite{chen_electrical_2004}:
\begin{align}  \label{eq:CapacitorEq3}
\mathbf{B}_C \vv{V}_C = \vv{0} \iff \vv{V}_C = \mathbf{A}_C^T \vv{V}_n 
\end{align} 
Thus, the sum of voltages around capacitor loops being zero allows for the definition of node voltages, in terms of which the capacitor branch voltages can be expressed.

Now we move on to discussing inductors, whose treatment is more complicated than that of capacitors, as their constitutive relation is fundamentally non-local, with mutual inductances between loops.

An inductive branch of a graph generally represents the flux-current relation of a loop or of part of a loop. Here, we define an inductive loop analogously to our discussion in Appendix \ref{subsec:NodeLoopGraphCircuitModelNetworkMatrix}, where it consists of a simple graph cycle that travels through at least one inductive edge. Ultimately, the properties of the loop do not depend on which capacitive edges are included.

Fig. \ref{fig:SingleCapacitorInductor}(b) lays out the constitutive equation for a network of inductors, relating a vector of branch fluxes to that of branch currents:
\begin{align} \label{eq:InductorEq1}
\vv{\Phi}_L &= \mathbf{L}_b \vv{I}_L 
\end{align}
Note that this equation is necessarily vector-valued, unlike that of the capacitor---and the (positive definite) branch inductance matrix $\mathbf{L}_b$ is usually not diagonal. Its diagonal entries stand for the self-inductances $L$ of inductive branches, while the off-diagonal entries represent mutual inductances $m$ between different inductive branches.

The vector of currents flowing through the inductive edges $\vv{I}_L$ is straightforward to interpret, but the vector of fluxes $\vv{\Phi}_L$ is more complicated. Fundamentally, the branch flux variable represents how flux through an inductive loop containing the branch is incremented as one follows along the loop---positively if branch current is defined to flow in the direction of the loop and negatively for the reverse case. These partial fluxes only acquire a straightforward physical meaning when they are summed up to form the inductance of a full loop \cite{paul_inductance_2011}. Thus, as we consider capacitance to be a quantity defined at capacitive nodes, we similarly view linear inductance as belonging to inductive loops.

All of this can be expressed as an equation by employing a loop matrix (Eq. \ref{eq:LoopMatrix}), with the addition of a loop external flux. Here, we assume that the loops represented by the rows of $\mathbf{B}_L$ form a maximal linearly independent set:
\begin{align} \label{eq:InductorEq2}
\vv{\Phi}_l = \mathbf{B}_L \vv{\Phi}_L + \vv{\Phi}_\text{ext}
\end{align}
In this model, the enclosed flux and circulating current are assigned entirely to the inductors, and the loops are enumerated only by their inductive edges. Also, within this definition of inductance, loops consisting of a single inductor are allowed to exist, which can be equivalently imagined as having no intersection with the nodes of the circuit.

In our framework, nodes connected only to inductors are not treated as islands with a distinct voltage, because there is no physical voltage drop (integral of electric field) through the interior of the perfect conductor wires that these inductors represent. Instead of employing the usual notion of pseudo-voltage across inductors, we account for the Faraday's law effects via the time derivative of the loop flux variable.

The inductor-only nodes serve only as current conservation junctions, where the total net incoming and outgoing currents cancel, with no capacitive charge buildup at the nodes. This is equivalent to saying that the current through each inductor can be written as a sum of (linearly independent) loop currents, in a Kirchhoff's current law for inductors:
\begin{align}  \label{eq:InductorEq3}
\mathbf{A}_L \vv{I}_L = \vv{0} \iff \vv{I}_L = \mathbf{B}_L^T \vv{I}_l 
\end{align}

We have seen that the presence of capacitive loops, which have zero net voltage drop, allows us to write capacitor branch voltages in terms of capacitive node voltages (Eq. \ref{eq:CapacitorEq3}). Similarly, the zero net current accumulation at inductive nodes allows us to write the inductor branch currents as a linear combination of inductive loop currents (Eq. \ref{eq:InductorEq3}). Now, as opposed to these capacitive loops and inductive nodes, we will see that each capacitive node and inductive loop will produce an equation of motion.

The specific interconnection between loops and nodes is treated with the aforementioned node-loop network matrix (Eq. \ref{eq:AppendixNetworkMatrix})---whose columns represent loops and whose rows represent the nodes they are connected to and the direction in which they are connected. As before, the continuity equation for capacitive islands then gives $\dot{\vv{Q}}_n - \mathbf{\Omega} \vv{I}_l = \vv{0}_n$ and the Faraday's law equation for inductive loops implies $\dot{\vv{\Phi}}_l + \mathbf{\Omega}^T \vv{V}_n  = \vv{0}_l$.

Combining the continuity equation with the equations for branch capacitance (Eqs. \ref{eq:CapacitorEq1}, \ref{eq:CapacitorEq2}, and \ref{eq:CapacitorEq3}) gives:
\begin{align} \label{eq:CapacitorContinuityEquation}
\vv{0}_n &= \dot{\vv{Q}}_n - \mathbf{\Omega} \vv{I}_l \nonumber \\
 &= \mathbf{A}_C \dot{\vv{Q}}_C  + \dot{\vv{Q}}_\text{ext} - \mathbf{\Omega} \vv{I}_l \nonumber \\
&= \mathbf{A}_C \mathbf{C}_b \dot{\vv{V}}_C  + \dot{\vv{Q}}_\text{ext} - \mathbf{\Omega} \vv{I}_l \nonumber \\
&= \mathbf{A}_C \mathbf{C}_b \mathbf{A}_C^T \dot{\vv{V}}_n+ \dot{\vv{Q}}_\text{ext} - \mathbf{\Omega} \vv{I}_l 
\end{align}
This is the same as the Eq. \ref{eq:VectorContinuity} but with nodal capacitance matrix $\mathbf{C}$ written in terms of the branch capacitance matrix as:
\begin{align}
\mathbf{C} = \mathbf{A}_C \mathbf{C}_b \mathbf{A}_C^T   
\end{align}
Since each branch capacitance in the diagonal matrix $\mathbf{C}_b$ is positive, the resulting $\mathbf{C}$ is positive definite.

Analogously, combining the Faraday's law equation with branch inductor equations (Eqs. \ref{eq:InductorEq1}, \ref{eq:InductorEq2}, and \ref{eq:InductorEq3}) implies:
\begin{align} \label{eq:InductorFaradaysLaw}
\vv{0}_l &= \dot{\vv{\Phi}}_l + \mathbf{\Omega}^T \vv{V}_n \nonumber \\
&= \mathbf{B}_L \dot{\vv{\Phi}}_L + \dot{\vv{\Phi}}_\text{ext} + \mathbf{\Omega}^T \vv{V}_n \nonumber\\
&= \mathbf{B}_L \mathbf{L}_b \dot{\vv{I}}_L + \dot{\vv{\Phi}}_\text{ext} + \mathbf{\Omega}^T \vv{V}_n \nonumber \\
&= \mathbf{B}_L \mathbf{L}_b \mathbf{B}_L^T \dot{\vv{I}}_l + \dot{\vv{\Phi}}_\text{ext} + \mathbf{\Omega}^T \vv{V}_n 
\end{align}
This is the same as Eq. \ref{eq:VectorContinuity}, but with the loop inductance matrix $\mathbf{L}$ expressed in terms of the branch inductance matrix as:
\begin{align}
\mathbf{L} = \mathbf{B}_L \mathbf{L}_b \mathbf{B}_L^T
\end{align}
The positive definite nature of $\mathbf{L}_b$ means that the derived $\mathbf{L}$ will also be positive definite.

Thus we see that a treatment of circuits with branch capacitors and inductors produces the same equations of motion as a fundamentally node and loop based approach (detailed in Appendix \ref{subsec:NodeLoopGraphCircuitModelNetworkMatrix}). Each of the $n$ capacitive nodes and each of the $l$ inductive loop produces an equation of motion, linked together by the node-loop network matrix. Through our construction, the nodal capacitance matrix and loop inductance matrix should both be invertible. This procedure benefits from notational clarity, which stems from differentiating between capacitive and inductive nodes, and inductive and capacitive loops.

Fig. \ref{fig:BranchCapacitorInductor} provides an example of how branch circuit elements can be used to construct the constituent matrices of a system. It presents a circuit graph whose edges consist of branch capacitors (orange) and inductors (blue). The capacitive nodes (orange) each have a voltage assigned to them, while the purely inductive node (blue) does not. The triangle symbol on the node with voltage $V_0 = 0$ denotes that it is the ground island. The figure further enumerates a set of fundamental loops. It then details how to write down the capacitor incidence matrix, the inductor loop matrix, and the node-loop network matrix, as well as how to derive the nodal capacitance matrix and loop inductance matrix.

\subsection{Equivalent circuit models, change of basis} \label{subsec:ChangeofBasis1}

\begin{figure}
    \centering
    \includegraphics[width=1\linewidth]{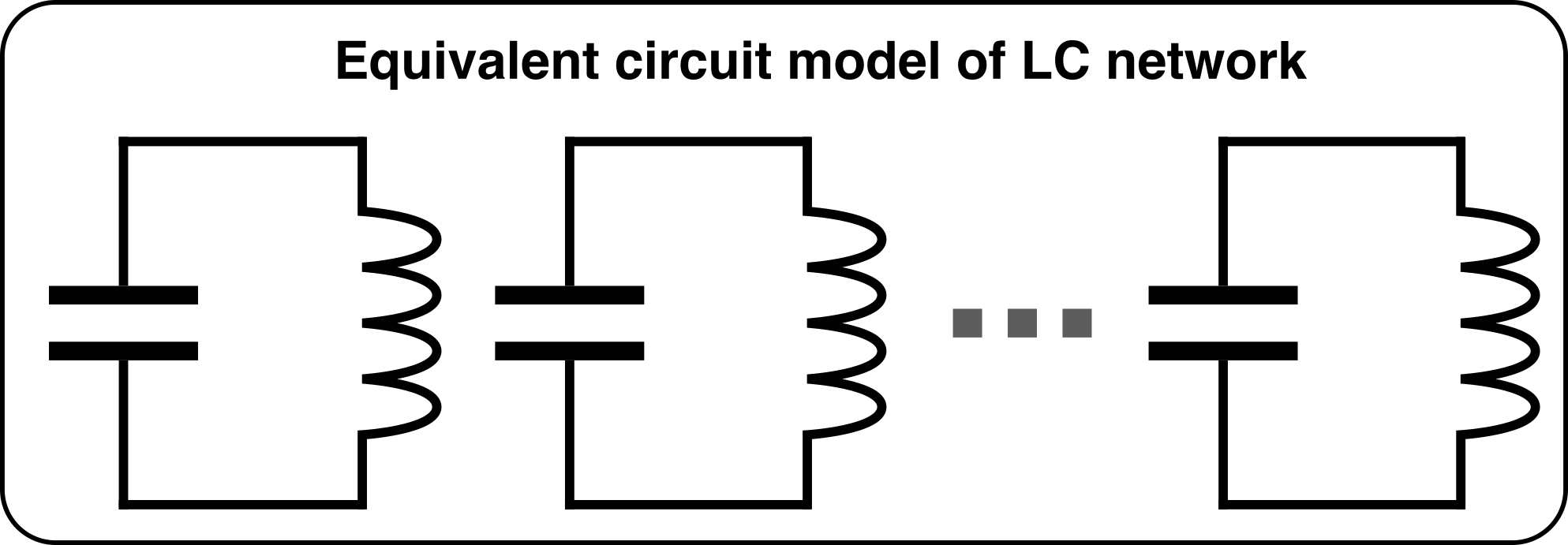}
    \caption{Every LC network can be transformed into an equivalent circuit of uncoupled harmonic oscillators.}
    \label{fig:LCEquivalentModel}
\end{figure}

Further simplification can be performed on the node and loop equations of motion by applying basis transformations to the node voltage and loop current variables. It is useful to view these transformations, when possible, as transformations to an equivalent circuit model. This perspective is valid when the transformed node-loop network matrix $\mathbf{\Omega}$ still encodes a valid circuit network. We now detail how to perform this procedure for this simple case of linear capacitor-inductor networks. Ultimately, as illustrated in Fig. \ref{fig:LCEquivalentModel}, any such circuit can be decomposed into an equivalent representation consisting of uncoupled harmonic oscillators. In Appendix \ref{sec:AppendixD} we generalize this procedure to include nonlinear quantum tunneling elements, resulting in a process we term the ``fundamental decomposition."

Under a change of basis of the node flux or loop current variables, the equations of motion (Eqs. \ref{eq:VectorContinuity} and \ref{eq:VectorFaraday}) alter as:
\begin{align}
\vv{V}_n &\rightarrow \mathbf{U}^{T^{-1}} \vv{V}_n \\
\vv{I}_l &\rightarrow \mathbf{W}^{T^{-1}} \vv{I}_l
\end{align}
We write the transformations in terms of the inverse transpose to simplify the subsequent analysis, which revolves primarily around manipulating the network matrix $\mathbf{\Omega}$. With this change of basis, the equations of motion transform as (multiplying by transformation operators and inserting resolutions of identity):
\begin{align}
[\mathbf{U} \mathbf{C} &\mathbf{U}^T ] [ \mathbf{U}^{T^{-1}} \dot{\vv{V}}_n ]+ [\mathbf{U} \dot{\vv{Q}}_\text{ext}] \nonumber \\
&-[\mathbf{U} \mathbf{\Omega} \mathbf{W}^T] [ \mathbf{W}^{T^{-1}}  \vv{I}_l ]= \vv{0}_n \\
[\mathbf{W}\mathbf{L} &\mathbf{W}^T ][\mathbf{W}^{T^{-1}} \dot{\vv{I}}_l] \nonumber 
+ [\mathbf{W}\dot{\vv{\Phi}}_\text{ext}] \\
&+ [\mathbf{U}\mathbf{\Omega}\mathbf{W}^T]^T [\mathbf{U}^{T^{-1}}\vv{V}_n] = \vv{0}_l 
\end{align}
The change of basis transforms each quantity in the equations of motion as:
\begin{align}
\mathbf{C} &\rightarrow \mathbf{U} \mathbf{C} \mathbf{U}^T \\
\mathbf{L} &\rightarrow \mathbf{W} \mathbf{L} \mathbf{W}^T \\
\vv{Q}_\text{ext} &\rightarrow \mathbf{U} \vv{Q}_\text{ext}\\
\vv{\Phi}_\text{ext} &\rightarrow \mathbf{W} \vv{\Phi}_\text{ext} \\
\mathbf{\Omega} &\rightarrow \mathbf{U} \mathbf{\Omega} \mathbf{W}^T 
\end{align}

We note that the capacitance and inductance matrices maintain their positive definite character under the transformation. We also see that $\mathbf{U}$ (the node voltage change of basis) performs row operations on $\mathbf{\Omega}$ and $\mathbf{W}$ (the loop current change of basis) performs column operations on $\mathbf{\Omega}$. 

In these equations, $\mathbf{\Omega}$ encodes the node and loop topology of the circuit. So, if after transformation the final $\mathbf{\Omega}$ matrix still has an obvious interpretation as a node-loop network matrix, then $\mathbf{U}$ and $\mathbf{W}$ can be seen to transform the system into an equivalent circuit model. Not all basis transformations are necessarily equivalent circuit transformations. In particular, we definitely lose this interpretation if the resultant $\mathbf{\Omega}$ has any elements other than $0$, $-1$, or $+1$.

Since we can perform both row and column operations on $\mathbf{\Omega}$, it has a logical ``reduced" form with the identity as a block submatrix and zeros everywhere else. Let $r$ be the rank of $\mathbf{\Omega}$. Then by performing integer-valued row ($\mathbf{U}$) and column pivoting ($\mathbf{W}$), it is straightforward to transform it into the following form \cite{strang_introduction_2016}:
\begin{align}
\mathbf{\Omega} =
\begin{bmatrix}
\mathbf{I}_{rr} & \mathbf{0}_{r\beta} \\
\mathbf{0}_{\alpha r} & \mathbf{0}_{\alpha\beta}
\end{bmatrix}
\end{align}

This network matrix represents an equivalent circuit with $r$ node, each connected to the ground node through inductive edges (the identity portion of the matrix). The index $r$ is chosen to denote that they serve as LC-resonators. There are also potentially $\alpha$ islands that do not connect to any inductive loops and $\beta$ loops that do not connect to any capacitive nodes (the zero rows and columns, respectively). Thus, LC-circuits can be expressed as a simple equivalent circuit of coupled LC-oscillators, which are in turn coupled to a set of isolated loops and islands.

These disconnected nodes (islands) and loops represent free modes that can now be removed to further simplify the circuit \cite{ding_free-mode_2021}. This is seen by writing the equations of motion in block form as:
\begin{align}
\begin{bmatrix}
\mathbf{C}_{rr} & \mathbf{C}_{r\alpha} \\
\mathbf{C}_{\alpha r} & \mathbf{C}_{\alpha\alpha}
\end{bmatrix}
\begin{bmatrix}
\dot{\vv{V}}_{r} \\
\dot{\vv{V}}_{\alpha} 
\end{bmatrix}
+
\begin{bmatrix}
\dot{\vv{Q}}_{\text{ext}_r} \\
\dot{\vv{Q}}_{\text{ext}_\alpha} 
\end{bmatrix}
-
\begin{bmatrix}
\mathbf{I}_{rr} & \mathbf{0}_{r\beta} \\
\mathbf{0}_{\alpha r} & \mathbf{0}_{\alpha\beta}
\end{bmatrix}
\begin{bmatrix}
\vv{I}_{r} \\
\vv{I}_{\beta} 
\end{bmatrix}
=
\begin{bmatrix}
\vv{0}_{r} \\
\vv{0}_{\alpha} 
\end{bmatrix} \\ %NEXT line
\begin{bmatrix}
\mathbf{L}_{rr} & \mathbf{L}_{r \beta} \\
\mathbf{L}_{\beta r} & \mathbf{L}_{\beta \beta}
\end{bmatrix}
\begin{bmatrix}
\dot{\vv{I}}_{r} \\
\dot{\vv{I}}_{\beta} 
\end{bmatrix}
+
\begin{bmatrix}
\dot{\vv{\Phi}}_{\text{ext}_r} \\
\dot{\vv{\Phi}}_{\text{ext}_\beta} 
\end{bmatrix}
+
\begin{bmatrix}
\mathbf{I}_{rr} & \mathbf{0}_{\alpha r}^T \\
\mathbf{0}_{r\beta}^T & \mathbf{0}_{\alpha\beta}^T
\end{bmatrix}
\begin{bmatrix}
\vv{V}_{r} \\
\vv{V}_{\alpha} 
\end{bmatrix}
=
\begin{bmatrix}
\vv{0}_{r} \\
\vv{0}_{\beta} 
\end{bmatrix}
\end{align}

To eliminate the $\alpha$ and $\beta$ degrees of freedom, we solve for the free variables (the second line of each block system of equations):
\begin{align}
\dot{\vv{V}}_\alpha &= -\mathbf{C}_{\alpha \alpha}^{-1} \left( \mathbf{C}_{\alpha r} \dot{\vv{V}}_r + \dot{\vv{Q}}_{\text{ext}_\alpha}  \right) \\
\dot{\vv{I}}_\beta &= -\mathbf{L}_{\beta \beta}^{-1} \left( \mathbf{L}_{\beta r} \dot{\vv{I}}_r + \dot{\vv{\Phi}}_{\text{ext}_\beta}  \right)
\end{align}
then we reinsert them into the first lines of the block equations. This leads to a reduced set of equations of motion:
\begin{align}
\mathbf{C}_{r r} \dot{\vv{V}}_r + \dot{\vv{Q}}_{\text{ext}_r} -\mathbf{I}_{r r}\vv{I}_r = \vv{0}_r \\
\mathbf{L}_{r r} \dot{\vv{I}}_r + \dot{\vv{\Phi}}_{\text{ext}_r} + \mathbf{I}_{r r}\vv{V}_r = \vv{0}_r 
\end{align}
where the following transformations have been made: 
\begin{align}
\mathbf{C}_{r r} & \rightarrow \mathbf{C}_{r r} - \mathbf{C}_{r \alpha} \mathbf{C}_{\alpha \alpha}^{-1} \mathbf{C}_{\alpha r} \\
\dot{\vv{Q}}_{\text{ext}_r} & \rightarrow \dot{\vv{Q}}_{\text{ext}_r} - \mathbf{C}_{r \alpha} \mathbf{C}_{\alpha \alpha}^{-1} \dot{\vv{Q}}_{\text{ext}_\alpha} \\
\mathbf{L}_{r r} & \rightarrow \mathbf{L}_{r r} - \mathbf{L}_{r \beta} \mathbf{L}_{\beta \beta}^{-1} \mathbf{L}_{\beta r} \\
\dot{\vv{\Phi}}_{\text{ext}_r} & \rightarrow \dot{\vv{\Phi}}_{\text{ext}_r} - \mathbf{L}_{r \beta} \mathbf{L}_{\beta \beta}^{-1} \dot{\vv{\Phi}}_{\text{ext}_\beta}
\end{align}

Note that the resulting $\mathbf{C}_{rr}$ and $\mathbf{L}_{rr}$ matrices remain positive definite, by properties of a matrix's Schur complement. At this point the isolated islands and loops have been removed. Thus, an arbitrary LC network has been transformed into an equivalent system of coupled LC-oscillators, whose topology is encoded in the identity node-loop network matrix $\mathbf{I}_{rr}$.

\subsection{Capacitance and inductance matrix diagonalization} \label{subsec:CapacitanceInductanceDiagonalization}

We can now further reduce our decomposed circuit to remove the capacitive and inductive couplings between the harmonic oscillators. To do this we perform a transformation that simultaneously diagonalizes the capacitance and inductance matrix while leaving the network matrix invariant. This is a standard procedure in the analysis of coupled harmonic oscillators, and we follow treatments given in \cite{eremenko_simultaneous_2019,ciani_lecture_2024}. We start by making a transformation of the form:
\begin{align}
\vv{V}_n &\rightarrow \mathbf{X}^{-1} \vv{V}_n \\
\vv{I}_l &\rightarrow \mathbf{X}^T \vv{I}_l
\end{align}
This has the following effect on the equations of motion:
\begin{align}
[\mathbf{X}^T \mathbf{C} &\mathbf{X}] [ \mathbf{X}^{-1} \dot{\vv{V}}_n ]+ [\mathbf{X}^T \dot{\vv{Q}}_\text{ext}] \nonumber \\
&-[\mathbf{X}^T \mathbf{I}_{rr} \mathbf{X}^{T^{-1}}] [ \mathbf{X}^T \vv{I}_l ]= \vv{0}_n  \\
[\mathbf{X}^{-1} \mathbf{L} &\mathbf{X}^{T^{-1} }][\mathbf{X}^{T} \dot{\vv{I}}_l] 
+ [\mathbf{X}^{-1} \dot{\vv{\Phi}}_\text{ext}] \nonumber  \\
&+ [\mathbf{X}^T \mathbf{I}_{rr}\mathbf{X}^{T^{-1}}]^T [\mathbf{X}^{-1} \vv{V}_n] = \vv{0}_l 
\end{align}
Thus it leaves the network matrix invariant and performs identical transformations on the capacitance and inverse inductance matrix:
\begin{align}
\mathbf{C} & \rightarrow \mathbf{X}^T \mathbf{C} \mathbf{X} \\
\mathbf{L}^{-1} & \rightarrow \mathbf{X}^T \mathbf{L}^{-1} \mathbf{X}
\end{align}

Then we look for a transformation that diagonalizes both $\mathbf{C}$ and $\mathbf{L}^{-1}$ (which will also diagonalize $\mathbf{L} = \left[\mathbf{L}^{-1}\right]^{-1}$). Since both  $\mathbf{C}$ and $\mathbf{L}^{-1}$ are positive definite, then they can be decomposed as $\mathbf{C} = \mathbf{M}^T \mathbf{M}$ and $\mathbf{L}^{-1} = \mathbf{N}^T \mathbf{N}$.

We construct the matrix $\mathbf{M}^{T^{-1}} \mathbf{N}^T \mathbf{N} \mathbf{M}^{-1}$. It is also positive definite, and as such admits an orthogonal eigenbasis, which we take to be the columns of the matrix $\mathbf{Y}$. We let the elements of $\mathbf{Y}$ have units of square root of capacitance, and normalize the columns such that: $\vv{Y}_i \cdot \vv{Y}_i = C_i$. We represent the corresponding positive, diagonal eigenvalue matrix (in units of angular frequency squared) with symbol $\boldsymbol{\omega}^2$. Then the correct transformation $\mathbf{X}$ will be given by the unitless matrix $\mathbf{X} = \mathbf{M}^{-1} \mathbf{Y}$. 

To see this we write down the eigenvalue equation
\begin{align}
\mathbf{M}^{T^{-1}} \mathbf{N}^T \mathbf{N} \mathbf{M}^{-1} \mathbf{Y} =  \mathbf{Y} \boldsymbol{\omega}^2 \nonumber \\ 
\implies \mathbf{N}^T \mathbf{N}\mathbf{X} = \mathbf{M}^{T} \mathbf{M} \mathbf{X} \boldsymbol{\omega}^2 \nonumber \\
\implies \mathbf{X}^T\mathbf{N}^T \mathbf{N}\mathbf{X} =  \mathbf{X}^T \mathbf{M}^{T} \mathbf{M} \mathbf{X} \boldsymbol{\omega}^2
\end{align}
then since
\begin{align}
\mathbf{X}^T\mathbf{C} \mathbf{X} &=
\mathbf{X}^T\mathbf{M}^T \mathbf{M}\mathbf{X} \nonumber \\
&= \mathbf{Y}^T\mathbf{Y} \nonumber \\
&= \mathbf{C}_d 
\end{align}
is a diagonal matrix $\mathbf{C}_d$ (with entries $C_i$) by the orthogonality of the columns of $\mathbf{Y}$, we have that:
\begin{align}
\mathbf{X}^T\mathbf{L}^{-1} \mathbf{X} &= \mathbf{X}^T\mathbf{N}^T \mathbf{N}\mathbf{X} \nonumber \\
& = \mathbf{X}^T \mathbf{M}^{T} \mathbf{M} \mathbf{X} \mathbf{\Omega}^2 \nonumber \\
& = \mathbf{C}_d \boldsymbol{\omega}^2
\end{align}
is also a diagonal matrix whose inverse gives the diagonal form of the inductance matrix after the basis transformation, given by:
\begin{align}
    \mathbf{L}_d = \left[ \mathbf{C}_d \boldsymbol{\omega}^2 \right]^{-1}
\end{align}

Thus, applying this transformation $\mathbf{X}$ produces an equivalent circuit model with diagonal capacitance and inductance matrix and unchanged (identity) network matrix:
\begin{align}
    \mathbf{C} &\rightarrow \mathbf{C}_d \\
    \mathbf{L} &\rightarrow \mathbf{L}_d  \\
    \mathbf{\Omega} &\rightarrow \mathbf{\Omega}
\end{align}
and our transformed equations represent a network of uncoupled harmonic oscillators, as desired.

\subsection{Quantization of vector-valued equations}

At this point, after diagonalizing $\mathbf{C}$ and $\mathbf{L}$, the equations of motion are reduced to an appropriate form for quantization, written in its simplest effective form. These two notions coincide here in the linear case. However, in the presence of tunneling nonlinearity, potentially different transformations are required to prepare a system for quantization (Appendix \ref{sec:AppendixC}) and to decompose a circuit into an equivalent form (Appendix \ref{sec:AppendixD}). 

We can quantize the capacitor-inductor equations of motion in a similar manner to the single-oscillator case shown in Appendix \ref{subsec:LCQuantization}, starting with:
\begin{align}
\mathbf{C} \dot{\vv{V}}_n + \dot{\vv{Q}}_\text{ext} - \vv{I}_l = \vv{0}_r  \\
\mathbf{L} \dot{\vv{I}}_l + \dot{\vv{\Phi}}_\text{ext} + \vv{V}_n = \vv{0}_r
\end{align}

Eliminating the current variable gives:
\begin{align}
\mathbf{C} \ddot{\vv{V}}_n + \ddot{\vv{Q}}_\text{ext} + \mathbf{L}^{-1} \left[ \vv{V}_n + \dot{\vv{\Phi}}_\text{ext} \right]  = \vv{0}_r 
\end{align}

A Lagrangian corresponding to these equations of motion is:
\begin{align}
\mathcal{L} =& \frac{1}{2} \dot{\vv{V}}_n^T \mathbf{C} \dot{\vv{V}}_n + \dot{\vv{V}}_n^T \dot{\vv{Q}}_\text{ext} \nonumber \\
&- \frac{1}{2} \left( \vv{V}_n + \dot{\vv{\Phi}}_\text{ext} \right)^T \mathbf{L}^{-1}\left( \vv{V}_n + \dot{\vv{\Phi}}_\text{ext} \right)
\end{align}

The conjugate momentum variables are:
\begin{align}
\vv{\Pi}_n = \frac{\partial\mathcal{L}}{\partial \dot{\vv{V}}_n} = \mathbf{C} \dot{\vv{V}}_n + \dot{\vv{Q}}_\text{ext} = \vv{I}_l
\end{align}

This gives the following Hamiltonian from taking the Legendre transform:
\begin{align}
\mathcal{H} =& \dot{\vv{V}}_n^T \vv{I}_l - \mathcal{L} \nonumber \\
 =& \frac{1}{2} \left(\vv{I}_l - \dot{\vv{Q}}_\text{ext}  \right)^T \mathbf{C}^{-1}\left(\vv{I}_l - \dot{\vv{Q}}_\text{ext}  \right)\\ 
 +&  \frac{1}{2} \left( \vv{V}_n + \dot{\vv{\Phi}}_\text{ext} \right)^T \mathbf{L}^{-1}\left( \vv{V}_n + \dot{\vv{\Phi}}_\text{ext} \right)
\end{align}

Again we can quantize the equations of motion by taking the Poisson brackets to commutators as:
\begin{align}
\{ \vv{V}_n , \vv{I}_l \} = \mathbf{I}_{rr} \rightarrow [\hat{\vv{V}}_n , \hat{\vv{I}}_l ] = i \hbar \mathbf{I}_{rr}
\end{align}

We note that the above derivation does not depend on the capacitance and inductance matrices being diagonal. Instead, the key step was to transform the node-loop network matrix $\mathbf{\Omega}$ to the identity.

\section{Lumped-element superconducting equations of motion} \label{sec:AppendixB}

To more accurately model lumped-element superconducting circuits, we need to account for quantum mechanical effects. These effects emerge from the quantization of charge and flux carriers, and from their ability to undergo quantum tunneling. Ultimately, the discreteness of charge and flux results in certain operators acquiring a discrete spectrum when they are quantized. The consequence of tunneling is to add nonlinear (sinusoidal) terms to the equations of motion. These equations of motion are most naturally expressed in terms of charge and flux variables---instead of current and voltage (their time derivatives). A key part of the procedure thus involves defining flux variables across capacitors and charge variables along inductors. After applying a set of physical restrictions, the final model can then be expressed in terms of these branch variables in what we call ``tree-cotree" notation, or in an equivalent node flux/loop charge picture.

In this appendix we work with lumped-element models of ideal superconductor. In this framework, charge flows in integer units of $-2e$, as two-electron Cooper pairs. Similarly, magnetic flux through a surface is quantized in integer multiples of the magnetic flux quantum $\Phi_0$ \cite{tinkham_introduction_1996}. Note that we imagine all Cooper pairs to be confined to the superconducting metal and all magnetic flux to be expelled out of the metal due to the Meissner effect. Corrections to this picture resulting from a nonzero magnetic penetration depth (discusssed in Appendix \ref{subsec:AppLondonDepth}) do not qualitatively change the model.

\subsection{Discreteness of charge and flux} \label{subsec:DiscretenessChargeFlux}

In this section, we discuss the effects of charge and flux quantization, giving a heuristic explanation for how to determine whether a quantity takes on a discrete or continuous spectrum when quantized. Essentially, discrete operators result from unweighted integrals of charge density $\rho(\vv{r},t)$ and flux density $\vv{B}(\vv{r},t)$, while continuous operators originate from weighted integrals of these quantities. Our arguments throughout are based on the results presented in \cite{riwar_charge_2021}.

In the unweighted case, volume integrals of charge density and surface integrals of flux density (over any finite region) admit the following charge and flux quantization conditions:
\begin{align}
2eN(t) &= \iiint_V \rho(\vv{r},t) dV \\
\Phi_0 M(t) &= \oiint_S \vv{B}(\vv{r},t) \cdot d\vv{S}
\end{align}

Here $2e$ is the charge of the underlying carrier (we label it with a positive instead of negative charge for notational simplicity) and $N(t)$ is net number of enclosed charge carriers, which takes on an integer-valued spectrum when the system is quantized. Similarly, $\Phi_0$ represents the flux of the underlying quanta, and $M(t)$ is the net number of enclosed flux carriers, which in turn becomes integer-valued when the system is quantized.

Observe that the flux equation is taken in the limit where the London penetration depth of the superconductor goes to zero. When the superconductor admits a nonzero magnetic field, the flux quantization condition is replaced by fluxoid quantization \cite{tinkham_introduction_1996,gross_applied_2016}, which involves an additional line integral of current density. This case is discussed in Appendix \ref{subsec:AppLondonDepth}. The extra term does not qualitatively change the physics, but results in an additional kinetic inductance.

As noted in \cite{riwar_charge_2021}, weighted integrals of charge density will end up corresponding to operators with continuous spectra. Along with the analogous result for flux density this can be written:
\begin{align}
Q(t) &= \iiint_V F_\rho(\vv{r}) \rho(\vv{r},t) dV \\
\Phi(t) &= \oiint_S F_B(\vv{r})\vv{B}(\vv{r},t) \cdot d\vv{S}
\end{align}
Here, $F_\rho(\vv{r})$ and $F_B(\vv{r})$ are non-constant (nor the sums of constant) weighting functions. The charge variable $Q(t)$ and flux variable $\Phi(t)$ will have continuous spectra upon quantization. As we will observe, capacitive gaps in loops and inductive wires between islands result in continuous-spectrum charge and flux variables (respectively) that are defined with weighted integrals.

These charge and flux density operators obey two essential ``continuity equations." The first is the electromagnetic continuity equation (in terms of free charge and current), and the second is Faraday's law---where the magnetic field is interpreted as magnetic flux density:
\begin{align}
\nabla \cdot \vv{J}(\vv{r},t) &=-\frac{ \partial \rho(\vv{r},t)}{\partial t} \label{eq:RhoContinuity} \\
\nabla \times \vv{E}(\vv{r},t) &=-\frac{ \partial \vv{B}(\vv{r},t)}{\partial t} \label{eq:BContinuity}
\end{align}

This then gives that for any volume $V$ of superconductor, the net current flowing in is:
\begin{align} \label{eq:CurrentChargeDiscrete}
I_{\partial V}(t) &= -\oiint_{\partial V} \vv{J}(\vv{r},t) \cdot d\vv{S} \nonumber \\
& = -\oiint_S \nabla \cdot \vv{J}(\vv{r},t) \cdot d\vv{S} \nonumber \\
& =  \frac{\partial}{\partial t} \iiint_V \rho(\vv{r},t) dV \nonumber \\
& = 2e \dot{N}(t)
\end{align}

Similarly the total voltage drop across the boundary of a surface $S$ can be written as:
\begin{align} \label{eq:VoltageFluxDiscrete}
V_{\partial S}(t) &= \oint_{\partial S} \vv{E}(\vv{r},t) \cdot d\vv{\Gamma} \nonumber \\
&= \iint_S \nabla \times \vv{E}(\vv{r},t) \cdot d\vv{S} \nonumber \\
&= -\frac{\partial}{\partial t} \iint_S \vv{B}(\vv{r},t) \cdot d\vv{S} \nonumber \\
&=-\Phi_0 \dot{M}(t)
\end{align}

Thus, just as current is a flow of quantized charge, voltage can be interpreted as the flow of magnetic fluxons into or out of a surface. Similarly to how electric current flows through conducting wires, we can imagine flux current (voltage) flowing through the gaps in superconducting rings. From these equations, the time integrals of currents and voltages will also become discrete-valued when quantized, as is discussed for currents in \cite{riwar_charge_2021}.

\subsection{Inductor charge and capacitor flux} \label{subsec:InductorChargeCapacitorFlux}

\begin{figure}
    \centering
    \includegraphics[width=1\linewidth]{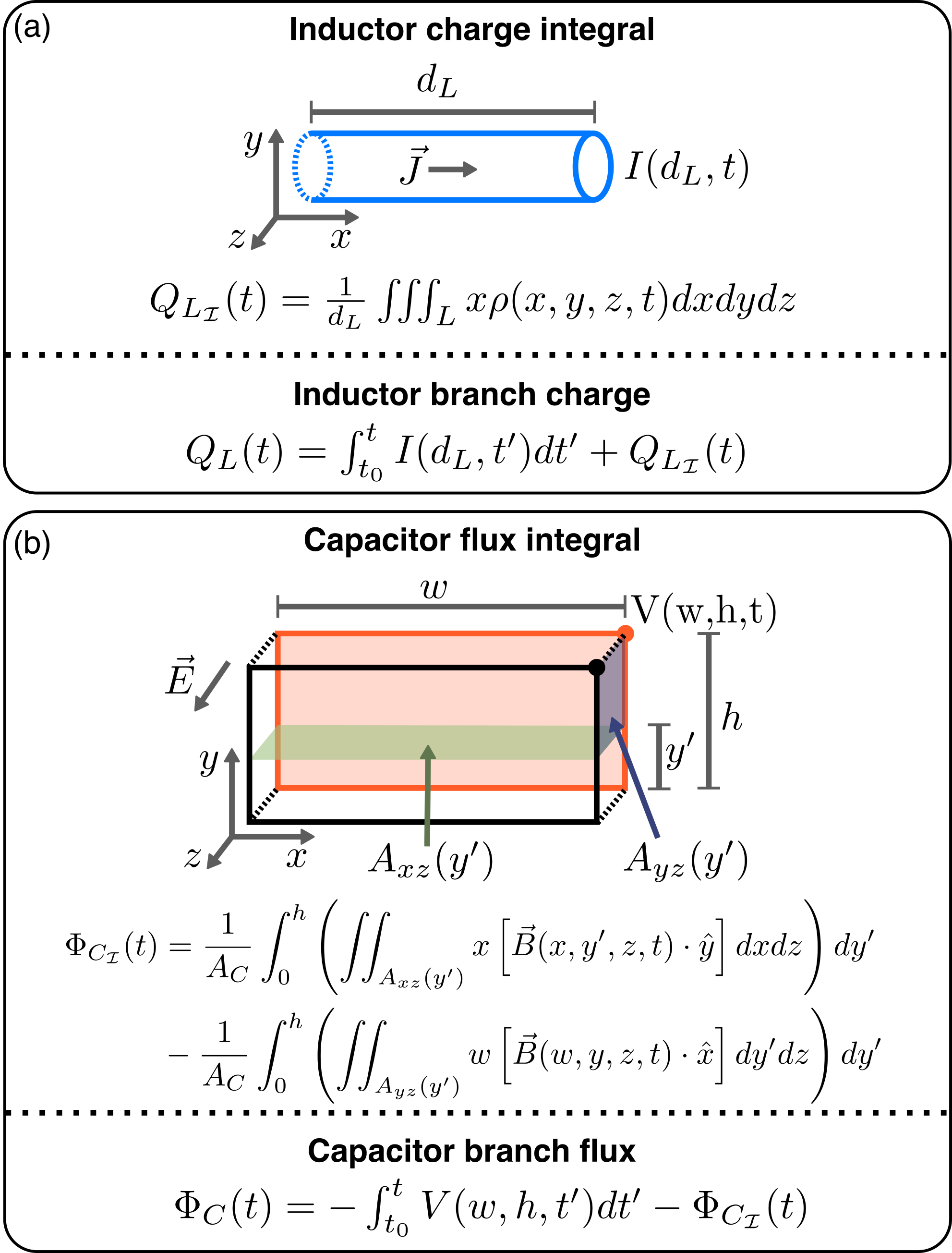}
    \caption{Illustrations of lumped inductor charge and capacitor flux. (a) Inductor charge can be defined as weighted volume of charge density integral plus a time integral of current flowing out of one end. (b) Capacitor flux is an average of weighted flux surface integrals plus a time integral of voltage (fluxon current) at one end.}
\label{fig:InductorChargeCapacitorFlux}
\end{figure}

In the presence of quantum tunneling, lumped circuits are characterized by capacitive flux variables (with time derivative of voltage) and inductive charge variables (with time derivative of current). In essence, the inductor charge variable contains a weighted charge density integral, and the capacitor flux variable possesses a weighted flux density integral. These presence of these weighted averages can break the discreteness of charge on a node (if it connects to another node with an inductor) or flux through a loop (if the loop is interrupted by a capacitor). This presentation gives a heuristic physical picture for lumped superconducting circuits with both node flux \cite{michel_h_devoret_quantum_1995,vool_introduction_2017} and loop charge \cite{mooij_superconducting_2006,kerman_fluxcharge_2013,ulrich_dual_2016} variables.

Models for inductors and capacitors are illustrated in Fig. \ref{fig:InductorChargeCapacitorFlux}. The inductors are represented by straight wires (though they are more realistically seen as wire loops) of length $d_L$ and uniform inductance per unit length of $L/{d_L}$. We model capacitors with two rectangular parallel plates of area $A_C = wh$ and uniform capacitance per unit area of $C/{A_C}$. If $Q_{C}$ is the total charge on the capacitor induced by the (spatially-varying) voltage difference between the plates and $\Phi_{L}$ is the total flux through the inductor induced by the current, then capacitive and inductive constitutive relations can be written:
\begin{align}
Q_{C}(t) & =  \frac{C}{A_C} \iint_C  V(x,y,t)  dx dy \label{eq:DistributedCapacitor} \\
\Phi_{L}(t) & =  \frac{L}{d_L} \int_L I(x,t) dx \label{eq:DistributedInductor}
\end{align}

If we define $V_C(t) = \frac{1}{A_C} \iint_C  V(x,y,t)  dx dy $ and $I_L(t) = \frac{1}{d_L} \int_L I(x,t) dx$, then we have $Q_C(t) = C V_C(t)$ and $\Phi_L(t) = L I_L(t)$, which resemble the standard constitutive relations of a capacitor and an inductor, respectively.
Note that is straightforward to generalize this treatment to include effects of non-uniform capacitance or inductance along the $x$-dimension. 

For an interconnected inductor and capacitor, some of the capacitor charge will lie in the inductive wire and some of the inductor flux will thread through the capacitive cap. We now define a set of weighted charge and flux integrals that allow us to distribute inductor charge between capacitor pads and capacitor flux between inductive loops---and to create node flux and loop charge variables. The equation for the inductor charge integral is:
\begin{align} \label{eq:InductorChargeIntegral}
Q_{L_\mathcal{I}}(t) &= \frac{1}{d_L} \iiint_L x  \rho(x,y,z,t) dx dy dz  \nonumber  \\
&= \frac{1}{d_L}  \int_0^{d_L} x \left[\iint_{A_L(x)} \rho(x,y,z,t) dy dz \right] dx 
\end{align}

As a weighted sum, this variable acquires a continuous spectrum when quantized. In the case that the inductor connects to two capacitors pads, this equation can be interpreted as linearly interpolating the integral of charge density between the ground pad at $x=0$ (with weight 0) and the non-ground pad at $x = d_L$ (with weight 1). In the case where an inductor connects two capacitive islands, and allowing the possibility of some external charge, we can write the charge of the non-ground island as:
\begin{align} \label{eq:IslandCharge}
Q_{C}(t) + Q_\text{ext}(t)  &=  Q_n(t) \nonumber\\ &= \iiint_n \rho(\vv{r},t) dV + Q_{L_\mathcal{I}}(t) \nonumber \\
& = 2e N_n(t) + Q_{L_\mathcal{I}}(t)
\end{align}
where $N_n(t)$ is the total number of net charge carriers inside the capacitor pad. Thus we see that the presence of the inductor breaks the discreteness of charge on the island, turning it into a continuous variable (when quantized).

We can also define a corresponding charge for the ground capacitor pad $Q_g(t)$ (including a contribution from the inductor), by replacing the factor of $x$ in Eq. \ref{eq:InductorChargeIntegral} with $d_L - x$. Then the sum of the two charges would be the integral of net charge over all the superconductor, which is $0$ (assuming overall charge neutrality):
\begin{align}
Q_n(t) + Q_g(t) = \iiint_V \rho(\vv{r},t) dV  = 0
\end{align}
This then gives the expected relation for a capacitor of $Q_g(t) = - Q_n(t)$.

We can use the inductor charge integral to define an inductor branch charge, whose time derivative is the branch current. This will become the loop charge variable needed to quantize phase slip wires:
\begin{align} \label{eq:InductorBranchCharge}
Q_L(t) &= \int_{t_0}^t I(d_L,t') dt' + Q_{L_\mathcal{I}}(t) \nonumber \\
&= 2eN_L(t) + Q_{L_\mathcal{I}}(t)
\end{align}
When the system is quantized, $N_L(t)$ counts the (discrete) net charges that has passed through the end of the wire. Though $N_L(t)$ depends on the start time of integration $t_0$, we will later show that the physics of the quantized model is independent of this choice.

The key property of this inductor branch charge is that its time derivative is the average current through the inductive wire. Taking the time derivative of the inductor charge integral gives:
\begin{align}
\dot{Q}_{L_\mathcal{I}}(t) &=  \frac{1}{d_L} \int_0^{d_L} x \left[\iint_{A_L(x)} \frac{\partial}{\partial t} \rho(x,y,z,t) dy dz \right] dx \nonumber \\
&= - \frac{1}{d_L} \int_0^{d_L} x \left[\iint_{A_L(x)} \frac{\partial }{\partial x} J(x,y,z,t) dy dz \right] dx \nonumber \\
& = - \frac{1}{d_L} \int_0^{d_L} x \frac{\partial I(x,t)  }{\partial x} dx \nonumber \\
& = - I(d_L,t) + \frac{1}{d_L}  \int_0^{d_L} I(x,t) dx 
\end{align}
In this calculation we employ the charge density continuity equation (Eq. \ref{eq:RhoContinuity}), under the assumption that the current density flows only along the $x$-direction, and used integration by parts in the final step. Applying this result when taking the time derivative of Eq. \ref{eq:InductorBranchCharge} implies that:
\begin{align} \label{eq:SecondPhaseSlipRelation}
\dot{Q}_L(t) =\frac{1}{d_L} \int_L I(x,t) dx = I_L(t)
\end{align}
This relation---of the time derivative of inductor branch charge the being current variable from the inductive constitutive relation (Eq. \ref{eq:DistributedInductor})---identifies this branch charge with the loop charge variable used to quantize phase slip wires \cite{ulrich_dual_2016}.

A similar set of derivations is used to construct a weighted capacitor flux integral, which distributes a capacitor's flux between the inductive loops it intersects, and which allows us to define a capacitor branch flux---in terms of which we can express the Josephson effect. The capacitor flux integral is illustrated in Fig. \ref{fig:InductorChargeCapacitorFlux}(b) and is given by:
\begin{align} \label{eq:CapacitorFluxIntegral}
\Phi_{C_\mathcal{I} }(t) &= \frac{1}{A_C} \int_{0}^{h} \left( \iint_{A_{xz}(y')} x \left[ \vv{B}(x,y,z,t) \cdot \hat{y} \right] dx dz \right) dy' \nonumber  \\
& - \frac{1}{A_C} \int_{0}^{h} \left( \iint_{A_{yz}(y')} w \left[ \vv{B}(w,y,z,t) \cdot \hat{x} \right] dy dz \right) dy' 
\end{align}
This quantity fulfills an analogous role to Eq. \ref{eq:InductorChargeIntegral}, in that it is a weighted integral, with weight $0$ at $x=0$ and $1$ at $x = w$. Unlike in the previous case, it is now the sum of two different terms: the integral of magnetic flux density (magnetic field) on a sheet oriented in the $\hat{y}$-direction at $y=y'$ and one oriented in the $-\hat{x}$-direction at $x=w$, going from height $y=y'$ to $y=h$. There is a third implicit $-\hat{x}$-oriented sheet at $x=0$ that is not included, as its weight is zero. Taken together, this set of sheets serves as a staircase that connects between the $\hat{z}$-oriented line  at $(x,y) = (0,0)$ to the $\hat{z}$-oriented line at $(x,y) = (w,h)$. This quantity is then averaged over all $y'$ to produce the total integral.

We can interpret the capacitor flux integral as interpolating the flux distribution between a right hand side flux loop that intersects the capacitor at $(x,y) = (w,h)$ and a ground (external) loop on the left hand side intersecting the capacitor at $(x,y) = (0,0)$. Then for this right hand loop, we have its total flux is:
\begin{align} \label{eq:LoopFlux}
\Phi_L(t) + \Phi_\text{ext}(t)  &= \Phi_l(t) \nonumber \\
&=\oiint_l \vv{B} (\vv{r},t) \cdot d\vv{S} +  \Phi_{C_\mathcal{I}}(t) \nonumber \\
& = \Phi_0 M_l(t) +  \Phi_{C_\mathcal{I}}(t)
\end{align}
Here we see that the total flux in the loop is now the sum of the number of fluxons contained purely in the loop and a weighted integral term of the flux in the capacitive gap. The presence of the capacitor flux integral breaks the discreteness of the total flux through the loop, giving it a continuous spectrum when quantized.

The capacitor flux integral also allows us to define a total branch capacitor flux, which is used to characterize the physics of Cooper pair tunneling (the Josephson effect) through the capacitor. The time derivative of this branch flux is equal the average voltage across the capacitor $V_C(t)$---the second Josephson relation. This branch flux is:
\begin{align} \label{eq:CapacitorBranchFlux}
\Phi_C(t)  &= \int_{t_0}^t V(w,h,t') dt' - \Phi_{C_\mathcal{I}}(t) \nonumber \\
&= -\Phi_0 M_C(t) - \Phi_{C_\mathcal{I}}(t)
\end{align}

Here $M_C(t)$ is the discrete (when quantized) number of fluxons that have passed into the right side loop through the line segment at $(x,y) = (w,h)$, (using Eq. \ref{eq:VoltageFluxDiscrete}). Again this quantity varies as a function of ${t_0}$, but we will later show that the overall physics is independent of this choice.

To show that the time derivative of the capacitor branch flux is equal to the branch voltage, we first demonstrate that:
\begin{align} \label{eq:CapacitorDerivativeDerivation}
&\dot{\Phi}_{C_\mathcal{I}}(t) \nonumber \\ &= \frac{1}{A_C} \int_{0}^{h} \left( \iint_{A_{xz}(y)} x \left[\frac{\partial}{\partial t} \vv{B}(x,y,z,t) \cdot \hat{y} \right] dx dz \right) dy  \nonumber  \\
& - \frac{1}{A_C} \int_{0}^{h} \left( \iint_{A_{y'z}(x)} w \left[ \frac{\partial}{\partial t}  \vv{B}(w,y',z,t) \cdot \hat{x} \right] dy' dz \right) dy \nonumber \\
% &= -\frac{1}{A_C} \int_{0}^{h} \left( \iint_{A_{xz}(y)} x \left[\nabla \times \vv{E}(x,y,z,t) \cdot \hat{y} \right] dx dz \right) dy  \nonumber  \\
% & + \frac{1}{A_C} \int_{0}^{h} \left( \iint_{A_{y'z}(x)} w \left[ \nabla \times \vv{E}(w,y',z,t) \cdot \hat{x} \right] dy' dz \right) dy \nonumber \\
&= \frac{1}{A_C} \int_{0}^{h} \left( \iint_{A_{xz}(y)} x \left[ \frac{ \partial E_z(x,y,z,t)}{\partial x} \right] dx dz \right) dy  \nonumber  \\
& + \frac{1}{A_C} \int_{0}^{h} \left( \iint_{A_{y'z}(x)} w \left[ \frac{\partial E_z(w,y',z,t)}{\partial y'} \right] dy' dz \right) dy \nonumber \\
&= \frac{1}{A_C} \int_{0}^{h} \left( \int_{0}^{w} x \left[ \frac{ \partial V(x,y,t)}{\partial x} \right] dx \right) dy  \nonumber  \\
& + \frac{1}{A_C} \int_{0}^{h} \left( \int_{y}^{h} w \left[ \frac{\partial V(w,y',t)}{\partial y'} \right] dy' \right) dy \nonumber \\
&= - \frac{1}{A_C} \int_{0}^{h}  \int_{0}^{w}  V(x,y,t) dx  dy + \frac{1}{A_C} \int_{0}^{h} w V(w,y,t)dy \nonumber  \\
& - \frac{1}{A_C} \int_{0}^{h} w V(w,y,t)dy  + \frac{wh}{A_C} V(w,h,t)  \nonumber \\
& =  V(w,h,t)  - \frac{1}{A_C} \int_{0}^{h}  \int_{0}^{w}  V(x,y,t) dx  dy
\end{align}
Here we have assumed all of the electric field lies along the $z$-direction, perpendicular to the plates of the capacitor, and we have employed Faraday's law (the magnetic flux continuity equation, Eq. \ref{eq:BContinuity}). We have also used integration by parts and the fact that the capacitor's area is $A_C = wh$.

Combining with Eq. \ref{eq:CapacitorBranchFlux} then gives a relation for the time derivative of capacitive branch flux:
\begin{align} \label{eq:SecondJosephsonRelation}
\dot{\Phi}_C(t) =\frac{1}{A_C} \iint_C V(x,y,t) dx dy = V_C(t)
\end{align}
In the quasi-static limit, this expression for capacitor branch flux in terms of voltage (from Eq. \ref{eq:DistributedCapacitor}) is the second Josephson relation \cite{tinkham_introduction_1996,gross_applied_2016}.

\subsection{LC-oscillator, charge and flux equations, quantum tunneling} \label{subsec:LCOscillatorQuantumTunneling}

\begin{figure}
    \centering
\includegraphics[width=1\linewidth]{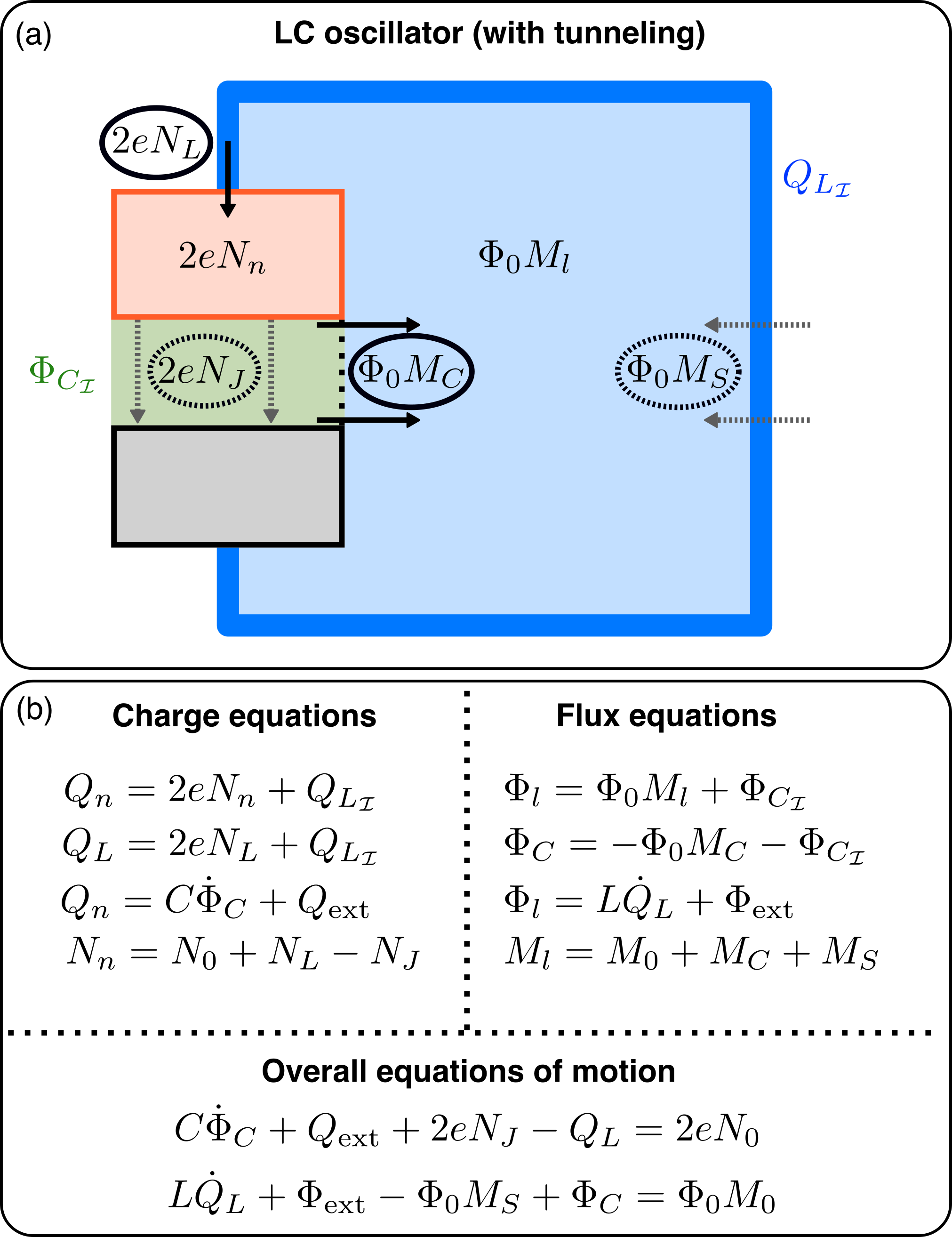}
    \caption{Semiclassical model of LC-oscillator system with quantum tunneling. $N_n$ is the discrete number of Cooper pairs entirely within the capacitor pad, and $M_l$ is the discrete number of fluxons contained entirely within the loop. $N_L$ is the number of charges that have passed from the inductive wire into the capacitor pad, while $M_C$ is the number of fluxons that have passed through the capacitor into the inductive loop. $N_J$ is the the number of charges that have tunneled across the capacitor pads, and $M_S$ is the number of fluxons that have tunneled through the inductor into the loop. $\Phi_{C_\mathcal{I}}$ is the weighted capacitor flux integral, while $Q_{L_\mathcal{I}}$ is the weighted inductor charge integral. The weighted integrals do not take on integer values when quantized. (b) All of these variables can be combined into a set of equations of motion (in ``integrated form"). In the presence of tunneling elements, the ``standard form" equations of motion are found by taking an additional time derivative.}
\label{fig:LCOscillatorSuperconducting}
\end{figure}

We now illustrate the construction of equations of motion for the charge and flux variables of the circuit, in the case of a superconducting LC-oscillator with quantum tunneling, as illustrated in Fig. \ref{fig:LCOscillatorSuperconducting}. This system (whose lumped-element representation is shown in Main Text Fig. \ref{fig:ModelOverview}) can also be conceptualized as a Fluxonium qubit with quantum phase slips \cite{manucharyan_fluxonium_2009} or as a realistic model of the dualmon qubit \cite{le_doubly_2019}.

As shown in the figure, the key addition to the model is the presence of quantum tunneling of Cooper pairs through the capacitor and and fluxons through the wire. The number of Cooper pairs that have tunneled out of the non-ground capacitor pad is labeled $N_J(t)$ (with the index referring to the Josephson effect \cite{tinkham_introduction_1996}), and the number of fluxons that have tunneled into the loop is $M_S(t)$ (through a process known as coherent quantum phase slips \cite{mooij_superconducting_2006}). In total we can relate these quantities to the time integrals of Cooper pair tunneling current $I_J(t)$, and fluxon tunneling voltage $V_S(t)$ as:
\begin{align}
2e N_J(t) = \int_{t_0}^t I_J(t')dt' \\
\Phi_0 M_S(t) = -\int_{t_0}^t V_S(t')dt'
\end{align}

Then, applying the continuity equations for charge density in the capacitor pad (not including the charge contained in the inductor) and flux density (Eqs. \ref{eq:RhoContinuity} and \ref{eq:BContinuity}) in the loop (not including the flux contained inside the capacitor) gives:
\begin{align}
N_n(t) = N_0 + N_L(t) - N_J(t) \label{eq:PadConservation} \\
M_l(t) = M_0 + M_C(t) + M_S(t) \label{eq:LoopConservation} 
\end{align}
where $N_0 = N_n(t_0)$ is the initial number of Cooper pairs inside the capacitor pad and $M_0 = M_l(t_0)$ is the initial number of fluxons in the loop. These equations simply state that the change in the the total number of charges and fluxons is equal to the time integral of the net current of these quantities.

Then combining the equations for capacitor and inductor charge (Eqs. \ref{eq:IslandCharge} and \ref{eq:InductorBranchCharge}), as well as those for inductor and capacitor flux (Eqs. \ref{eq:LoopFlux} and \ref{eq:CapacitorBranchFlux}) gives:
\begin{align}
Q_n(t) + Q_\text{ext}(t) - Q_L(t) &= 2e(N_n(t) - N_L(t)) \\
\Phi_l(t) + \Phi_\text{ext}(t) + \Phi_C(t) &= \Phi_0 (M_l(t) - M_C(t)) 
\end{align}
where the common factors of the inductor charge integral and capacitor flux integral have been eliminated. Now, using the capacitive and inductive constitutive relations (Eqs. \ref{eq:DistributedCapacitor} and \ref{eq:DistributedCapacitor}), as well as the expressions for the time derivatives of capacitor flux and inductor charge (Eqs. \ref{eq:SecondJosephsonRelation} and \ref{eq:SecondPhaseSlipRelation}), and the discrete conservation equations (Eqs. \ref{eq:PadConservation} and \ref{eq:LoopConservation}) results in the total equations of motion:
\begin{align}
C \dot{\Phi}_C(t)+ Q_\text{ext}(t) +2e N_J(t) - Q_L(t) = 2e N_0 \label{eq:1dChargeEquation} \\
L \dot{Q}_L(t)+ \Phi_\text{ext}(t) -\Phi_0 M_S(t) + \Phi_C(t) = \Phi_0 M_0 \label{eq:1dFluxEquation}
\end{align}

Note that we cannot yet quantize and analyze the equations of motion in terms of $\Phi_C(t)$ and $Q_L(t)$, because we still need to express $N_J(t)$ and $M_S(t)$ in terms of these variables. To do this, we need to take a time derivative of these equations and utilize the Josephson and phase slip relations, which characterize the quantum tunneling phenomena. We thus refer to this form of the equations (before taking the time derivative) as the ``integrated" equations of motion.

We also mention that if we write $Q_J(t) = 2e N_J(t)$ and $\Phi_S(t) = \Phi_0 M_S(t)$, then the equations of motion can be expressed as:
\begin{align}
    Q_n(t) + Q_J(t) - Q_L(t) = 2e N_0 \label{eq:1DChargeConservation} \\
    \Phi_l(t) - \Phi_S(t) + \Phi_L(t) \label{eq:1DFluxConservation} = \Phi_0 M_0 
\end{align}
such that they represent charge and flux conservation.

\subsection{Charge and flux tunneling} \label{subsec:ChargeAndFluxTunneling}

\begin{figure}
    \centering
    \includegraphics[width=1\linewidth]{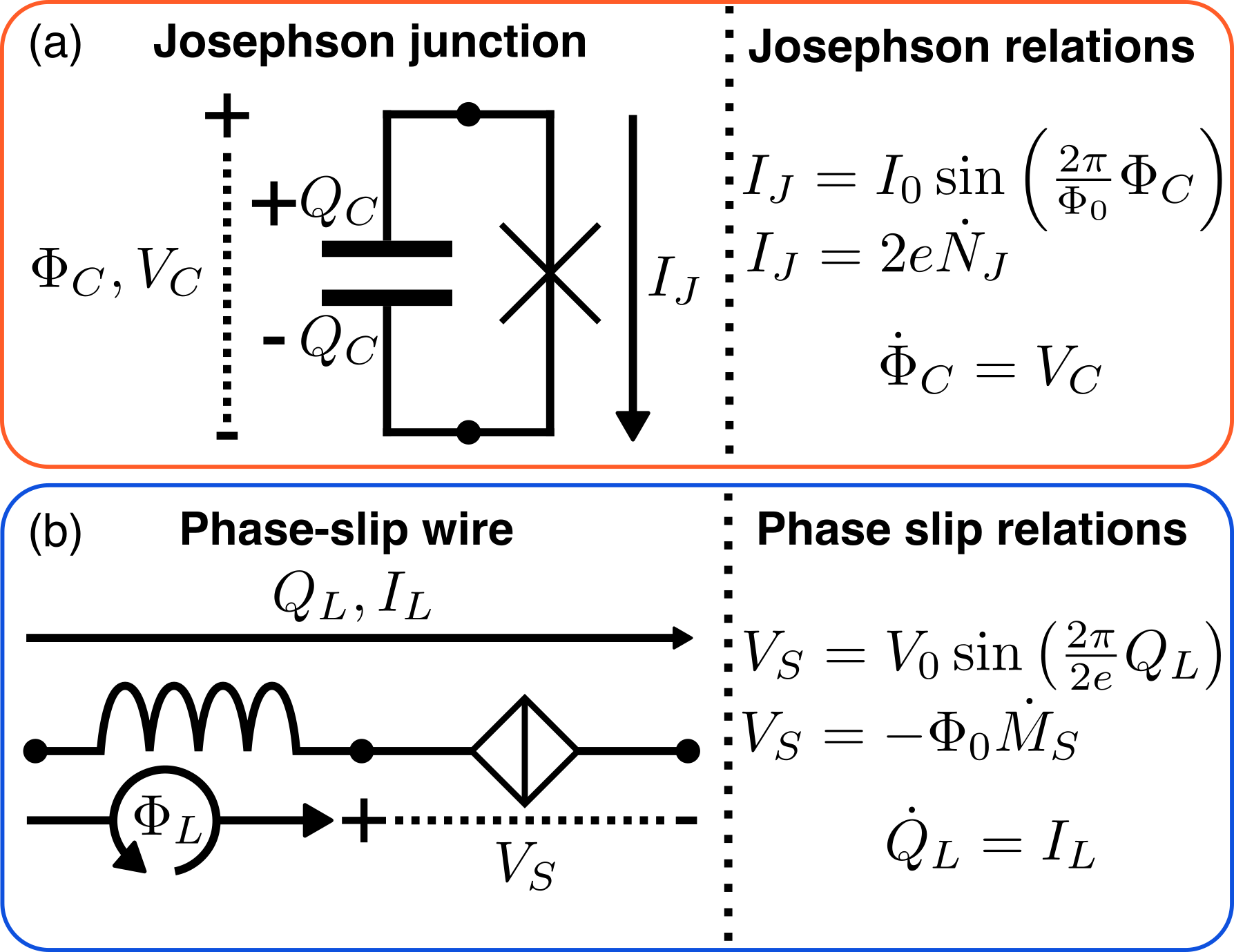}
    \caption{Josephson junctions and quantum phase slips. (a) Constitutive relations for Josephson junction: Cooper pair tunneling element with intrinsic parallel capacitance. Quantized tunneling current is a sinusoidal function of flux across the junction.
    (b) Constitutive relations for phase slip wire with intrinsic series inductance. Quantized tunneling voltage is a sinusoidal function of charge along the wire.}
    \label{fig:JJPS}
\end{figure}

Additional constitutive relations exist for the quantum tunneling elements of superconducting circuits in the quasi-static, lumped-element regime. In particular, the Cooper pair tunneling current can be written in terms of the flux across its corresponding capacitor, and the fluxon tunneling voltage can be written in terms of the charge along its inductor. When quantized, these effects will give rise to hopping terms in the Hamiltonian. Note that we now drop the explicit time-dependence in our notation.

Circuit representations of charge and flux tunneling are given in Fig. \ref{fig:JJPS}. First, we take an additional low-wavelength, lumped-element approximation. In this regime, the tunneling capacitors have a uniform voltage $V_C$ across their areas, and the tunneling inductors have a uniform current $I_L$ along their lengths.

Capacitors in this limit will act as lumped-element Josephson junctions, with $\Phi_C$ proportional to the gauge-invariant phase difference between the superconducting condensates on the capacitor pads. By the first Josephson relation, the tunneling current can then be written \cite{josephson_possible_1962, tinkham_introduction_1996}:
\begin{align}
I_J = 2e \dot{N}_J = I_0 \sin \left(\frac{2 \pi}{\Phi_0} \Phi_C \right)
\end{align}
Here, $I_0$ is the Josephson junction's critical current, which is roughly constant in the case of low magnetic field through the junction. The sine function will generate a cosine hopping term in the Hamiltonian. Note that Josephson tunneling originates from the overlap of the superconducting condensate wavefunctions across the capacitive gap, with higher tunneling rates in thinner capacitors.

The circuit model of a Josephson junction and its relations are illustrated in \ref{fig:JJPS}(a), where we emphasize that junctions are capacitors with Cooper pair tunneling (represented by an $X$-shape in the circuit model). Junctions can also be symbolized with the box around the $X$, to denote additional intrinsic parallel capacitance.

Fluxon tunneling across inductors in the lumped-element limit has a similar sinusoidal constitutive relation, with voltage written in terms of the inductor charge variable \cite{mooij_superconducting_2006,kerman_fluxcharge_2013,ulrich_dual_2016}:
\begin{align}
V_S = - \Phi_0 \dot{M}_S = V_0 \sin \left(\frac{2 \pi}{2e} Q_L \right)
\end{align}
Here, $V_0$ is the inductor's critical voltage (proportional to the fluxon hopping amplitude). Fluxon tunneling occurs when quantum fluctuations cause a suppression of superconductivity across a cross-section of wire, allowing a magnetic fluxon to tunnel through the normal metal gap. This then results in a $2 \pi$ shift in the relative condensate wavefunction on either side of the gap, in what is called a ``phase slip." However, these phase slip wires (as such circuit elements are called) are more commonly realized as Josephson junction arrays \cite{manucharyan_evidence_2012, randeria_dephasing_2024}. Here, a chain of Josephson junctions serves as the discrete analogue of a phase slip wire, with fluxons tunneling through the capacitive gaps between the junctions. These systems are usually modeled with a fluctuating $V_0$ due to charge noise along the inductor.

The phase slip circuit element is shown in Fig \ref{fig:JJPS}(b), and its fluxon tunneling component is represented by a diamond with a line through it. We emphasize that (in this formulation) the phase slip element is a composite object with intrinsic inductance.

\subsection{Corrections from nonzero penetration depth, kinetic inductance} \label{subsec:AppLondonDepth}

The lumped-element picture of superconductivity is modified when the superconductor has a finite London penetration depth (of magnetic field) $\lambda_L$. Written in terms of London constant $\Lambda = \mu_0 \lambda_L^2$, the superconductor is be governed by the linearized London equations 
\cite{tinkham_introduction_1996,gross_applied_2016}: 
\begin{align}
    \frac{ \partial (\Lambda \vv{J})}{\partial t} & = \vv{E}\\ 
    \nabla \times (\Lambda \vv{J}) &=  - \vv{B}
\end{align}

In addition, when the the boundary of a surface lies in a region of superconductor, the flux quantization condition is replaced with the fluxoid quantization condition
\begin{align} \label{eq:KIEquation1}
    \oint_{\partial S} \Lambda \vv{J} \cdot d \vv{\Gamma} +  \iint_{S} \vv{B} \cdot d \vv{S} = \Phi_0 M
\end{align}

For a ring of finite-depth superconductor with a hole in the center, we let $S$ be an interior boundary surface. Then the geometric flux through the loop will consist of an interior flux integral, plus a weighted integral of the magnetic field penetrating the metal of the ring ( labeled as $\Phi_{L_\mathcal{I}}$). The geometric flux constitutive equation $\Phi_l$ (with geometric inductance $L_g$) is then written as:
\begin{align} \label{eq:KIEquation2}
    L_g I_L + \Phi_\text{ext} = \Phi_l &=  \iint_{S} \vv{B} \cdot d \vv{S} + \Phi_{L_\mathcal{I}}
\end{align}
This inductor flux integral is defined exactly like the capacitor flux integral in Eq. \ref{eq:CapacitorFluxIntegral}, interpolating between the exterior boundary of the ring with weight $0$ and the interior boundary with weight $1$. For computational simplicity, we take the integrals with the inductor as a rectangular prism (with length $d_L$ and area $A_L$) instead of a ring. Using the second London equation to write magnetic field in terms of the curl of current density, and performing an analogous derivation to Eq. \ref{eq:CapacitorDerivativeDerivation} then gives:
\begin{align}
    \Phi_{L_\mathcal{I}} &=  \oint_{\partial S} \Lambda \vv{J} \cdot d \vv{\Gamma} - \frac{\Lambda d_L}{A_L} I_L
\end{align}
Again, for simplicity, we have assumed the inductor to have uniform penetration depth. Now we combine this result with that of Eqs. \ref{eq:KIEquation1} and \ref{eq:KIEquation2}:
\begin{align}
    \left( L_g + \frac{\Lambda d_L}{A_L} \right) I_L + \Phi_\text{ext}  = \Phi_0 M
\end{align}
We can thus view $L_k = \frac{\Lambda d_L}{A_L}$ as an additional ``kinetic" inductance \cite{kerman_fluxcharge_2013} that does not substantially change the underlying model. The finite penetration depth will also alter the effective thickness of a Josephson junction, as discussed in \cite{gross_applied_2016}.

\subsection{Node flux and loop charge, capacitive spanning tree/inductive cotree} \label{subsec:NodeFluxLoopChargeTreeCotree}

\begin{figure}
    \centering    \includegraphics[width=1\linewidth]{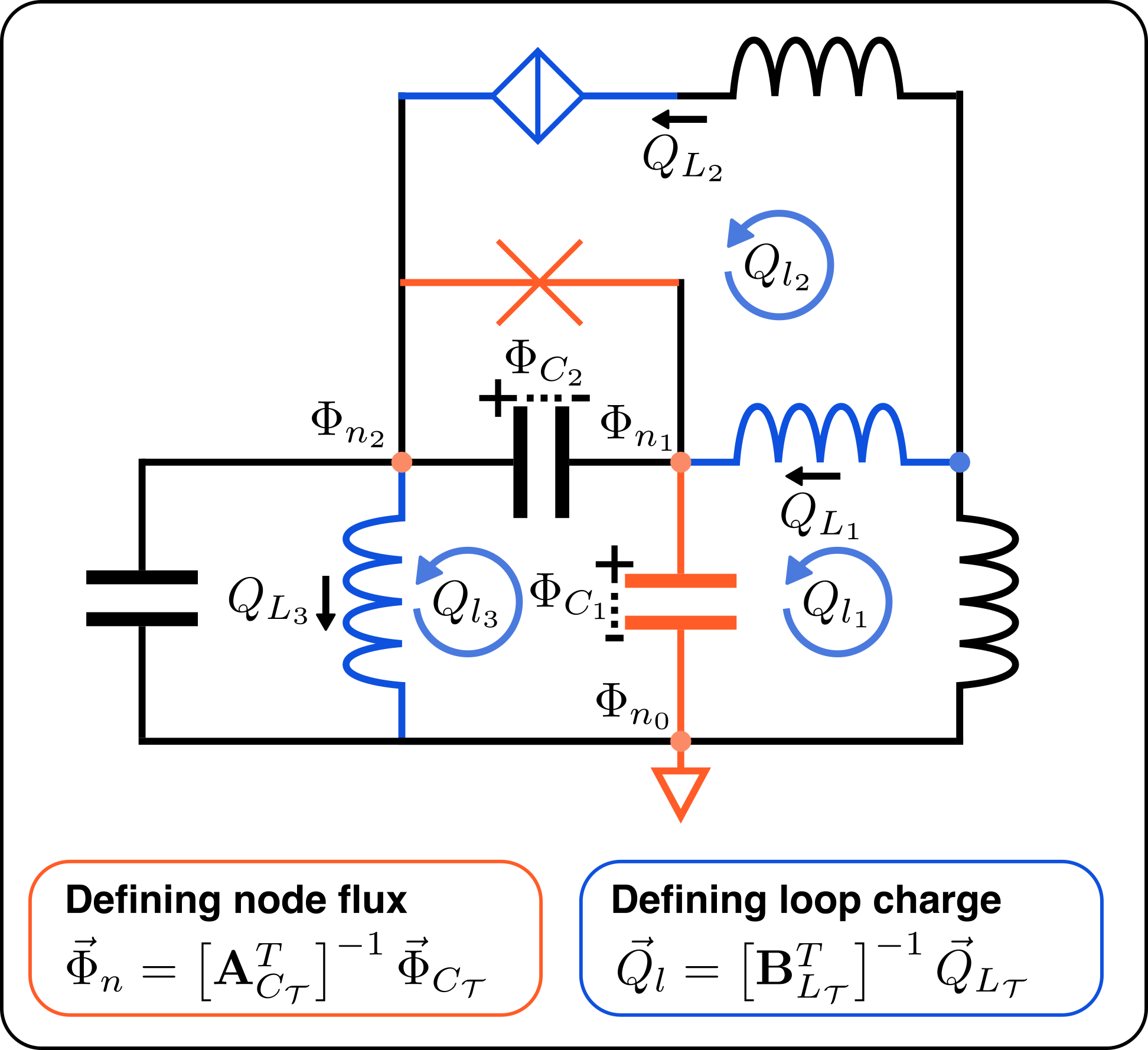}
    \caption{Construction of node flux and loop charge variables. Orange elements represent branches of the capacitive spanning tree, which include each Josephson junction. Node flux is defined by spanning tree relative to ground node with $\Phi_0 = 0$. Blue elements stand for inductive cotree edges, which include all phase slips. Each inductive cotree edge defines a unique loop through the capacitive spanning tree, with the cotree branch charge becoming the loop charge variable.}
\label{fig:DefiningNodeFluxLoopCharge}
\end{figure}

To model the dynamics of more complex networks, we employ vector-valued equations of motion. While doing so, we impose constraints on the circuit connectivity that result from physical considerations. These are in addition to the intrinsic parallel capacitance we include in each Josephson junction element and the intrinsic series inductance in each phase slip wire \ref{fig:JJPS}---and the specification that we have connected capacitive and inductive networks, with invertible capacitance and inductance matrices.

First, as in \cite{burkard_circuit_2005}, we require that there to be no loops consisting entirely of Josephson junctions. This reflects the fact that each physical junction loop will have some intrinsic linear inductance. Next, we mandate that there be no nodes that border only phase slip wires (composite series elements with a phase slip tunneling element and linear inductor). This condition reflects the intrinsic linear capacitance at a node that touches two or more phase slips. These physical restrictions are often relaxed in practice, with junction loop inductances being taken to $0$. Details on how to take these limits are given in \cite{you_circuit_2019}, and further expanded upon in Appendix \ref{subsec:ZeroCapInd}.

Since there are no junction-only loops, we can construct a spanning tree---a loop-free set of $n$ edges that border all $n+1$ capacitive nodes---that includes all of the capacitive Josephson junction fluxes, as well as a minimal set of fluxes across linear capacitors. The full vector of capacitive tree fluxes is labeled $\vv{\Phi}_{C_\mathcal{T}}$, with the first $J$ elements corresponding to the junction fluxes $\vv{\Phi}_{J}$. 

Similarly, the lack of phase slip-only nodes allows us to build an inductive cotree---a set of inductive edges connecting between capacitive tree nodes---that includes the charges along all of the fluxon tunneling/phase slip elements, as well as those along a minimal number of linear inductor segments needed to span all of the system's inductive loops. Each inductive edge of the cotree defines a unique loop through the capacitive spanning tree. The vector of of inductive cotree charges is called $\vv{Q}_{L_\mathcal{T}}$, with the first $S$ elements corresponding to the phase slip charges $\vv{\Phi}_{S}$.
The relation between branch and node/loop variables is illustrated in Fig. \ref{fig:DefiningNodeFluxLoopCharge}. The edges highlighted in orange represent the capacitive spanning tree, while those in blue represent the inductive cotree. A ``tree-cotree" shorthand notation can be obtained by removing each non-tree capacitive edge and contracting each non-cotree inductive edge to a point, as shown in Main Text Fig. \ref{fig:TopologicalMatricesTreeCotree}. The tree-cotree branch basis is central to Appendices \ref{sec:AppendixD} and \ref{sec:AppendixE}. In particular, relevant information can be found in Appendix \ref{subsec:GraphTheoreticNetworkMatrices}.

As is also shown in the figure, node flux and loop charge can be defined from the branch variables through the use of incidence and loop matrices, which are discussed in Appendix \ref{subsec:IncidenceLoopMatrices}. Since $\vv{\Phi}_{C_\mathcal{T}}$ represents the branches of a spanning tree, the incidence matrix $\mathbf{A}_{C_\mathcal{T}}$ (with the ground node's row removed) is invertible \cite{bapat_graphs_2014}. By analogous properties of $\vv{Q}_{{L}_\mathcal{T}}$, loop matrix $\mathbf{B}_{L_\mathcal{T}}$ is also invertible. Node flux and loop charged are then defined by:
\begin{align}
\vv{\Phi}_n &= \left[\mathbf{A}_{C_\mathcal{T}}^T  \right]^{-1} \vv{\Phi}_{C_\mathcal{T}} \label{eq:NodeFluxDefined} \\
\vv{Q}_l &= \left[\mathbf{B}_{L_\mathcal{T}}^T \right]^{-1} \vv{Q}_{{L}_\mathcal{T}} \label{eq:LoopChargeDefined}
\end{align}
Eq. \ref{eq:NodeFluxDefined} defines node fluxes as the sums/differences of branch fluxes along a path in the capacitive spanning tree, starting from the ground node. Eq. \ref{eq:LoopChargeDefined} defines loop charges as the sums/differences of the inductive cotree branch charges in the loop. These loops may correspond to the faces of the tree-cotree graph, when that graph is planar.

If we let $\mathbf{A}_J$ be the column submatrix of $\mathbf{A}_{C_\mathcal{T}}$ that contains all capacitor fluxes across Josephson junctions and $\mathbf{B}_S$ be the the column submatrix of $\mathbf{B}_{L_\mathcal{T}}$ that contains all inductor charges of the phase slip wires, then in particular the junction fluxes and phase slip charges can be written as:
\begin{align}
\vv{\Phi}_J &= \mathbf{A}_J^T\vv{\Phi}_n \\
\vv{Q}_S &= \mathbf{B}_S^T \vv{Q}_l
\end{align}

We also note that taking the derivative of Eqs. \ref{eq:NodeFluxDefined} and \ref{eq:LoopChargeDefined}, and then using Eqs. \ref{eq:SecondPhaseSlipRelation} and \ref{eq:SecondJosephsonRelation} gives:
\begin{align}
\dot{\vv{\Phi}}_n &= \left[\mathbf{A}_{C_\mathcal{T}}^T  \right]^{-1} \vv{V}_{C_\mathcal{T}} = \vv{V}_n \label{eq:NodeFluxDerivative} \\
\dot{\vv{Q}}_l  &= \left[\mathbf{B}_{L_\mathcal{T}}^T \right]^{-1} \vv{I}_{{L}_\mathcal{T}} = \vv{I}_l \label{eq:LoopChargeDerivative}
\end{align}
so the time derivative of node flux is node voltage, and the time derivative of loop charge is loop current.

\subsection{Vector-valued superconducting equations of motion} \label{subsec:Vectorized}

We now illustrate how to construct semiclassical equations of motion for a superconducting network with multiple nodes and loops. The resulting equations are a modification of Eqs. \ref{eq:VectorContinuity} and \ref{eq:VectorFaraday}, with additional terms corresponding to Cooper pair and fluxon tunnelinga. 

We first modify the capacitive (Eq. \ref{eq:CapacitiveConstitutive}) and inductive (Eq. \ref{eq:InductiveConstitutive}) constitutive relations, to express them in terms of node flux and loop charge (using Eqs.
\ref{eq:NodeFluxDerivative} and \ref{eq:LoopChargeDerivative}):
\begin{align}
\vv{Q}_n &= \mathbf{C}\dot{\vv{\Phi}}_n + \vv{Q}_\text{ext} \label{eq:AppCapacitiveConstitutiveCharge}\\
\vv{\Phi}_l &= \mathbf{L}\dot{\vv{Q}}_l + \vv{\Phi}_\text{ext} \label{eq:AppInductiveConstitutiveFlux}
\end{align}
As in Appendix \ref{subsec:CapacitanceInductanceMatrices}, $\mathbf{C}$ is the nodal capacitance matrix and $\mathbf{L}$ is the loop inductance matrix.

We can express charge and flux conservation by writing generalizations of Eqs. \ref{eq:1DChargeConservation} and \ref{eq:1DFluxConservation} to networks of capacitive nodes and inductive loops:
\begin{align}
\vv{Q}_n + \mathbf{A}_J \vv{Q}_J - \mathbf{A}_{L_\mathcal{T}} \vv{Q}_{{L}_\mathcal{T}}  &= 2e \vv{N}_0 \label{eq:SuperconductingChargeVector} \\
\vv{\Phi}_l - \mathbf{B}_S  \vv{\Phi}_S - \mathbf{B}_{C_\mathcal{T} } \vv{\Phi}_{{C}_\mathcal{T}}   &= \Phi_0 \vv{M}_0 \label{eq:SuperconductingFluxVector} 
\end{align}
Here, $\vv{Q}_n$ is the vector of node charges, while $\vv{Q}_J$ represents the charge that has tunneled through the Josephson junction, and $\vv{Q}_L$ the charge that has flowed through cotree inductor branches. The matrix $\mathbf{A}_J$ is the incidence matrix of junction edges at capacitive nodes, and $\mathbf{A}_{L_\mathcal{T}}$ is that of the inductive cotree edges. Similarly, $\vv{\Phi}_l$ is the vector of loop flux, $\vv{\Phi}_S$ is the vector of fluxes along phase slip wires, and $\vv{\Phi}_{C_\mathcal{T}}$ is the vector of fluxes across spanning tree capacitors. The loop matrices $\mathbf{B}_S$ and $\mathbf{B}_{C_\mathcal{T}}$ relate how the phase slip and and capacitive edges lie in inductive loops through the capacitive spanning tree/inductive spanning cotree (Appendix \ref{subsec:NodeFluxLoopChargeTreeCotree}).

To write these equations in terms of node flux and loop charge, we use Eqs. \ref{eq:NodeFluxDefined} and \ref{eq:LoopChargeDefined} to turn the third set of terms in each of Eqs. \ref{eq:SuperconductingChargeVector} and \ref{eq:SuperconductingFluxVector} into:
\begin{align}
- \mathbf{A}_{L_\mathcal{T} } \mathbf{B}_{L_\mathcal{T} }^T \vv{Q}_{l} \\
-\mathbf{B}_{C_\mathcal{T} }  \mathbf{A}_{C_\mathcal{T}}^T \vv{\Phi}_{n} 
\end{align}

We now note that the incidence (or cutset) and loop matrices of a graph satisfy the orthogonality relation $\mathbf{A} \mathbf{B}^T = \mathbf{0}$ \cite{bapat_graphs_2014}. This implies for our graph that:
\begin{align}
\begin{bmatrix}
    \mathbf{A}_{L_\mathcal{T}} & \mathbf{A}_{C_\mathcal{T}}
\end{bmatrix}  
\begin{bmatrix}
    \mathbf{B}_{L_\mathcal{T}}^T\\
    \mathbf{B}_{C_\mathcal{T}}^T
\end{bmatrix} 
= \mathbf{0}_{nl} \\
\implies \mathbf{A}_{L_\mathcal{T}} \mathbf{B}_{L_\mathcal{T}}^T = -\mathbf{A}_{C_\mathcal{T}}\mathbf{B}_{C_\mathcal{T}}^T \label{eq:InductiveCapacitiveIncidenceLoop}
\end{align}
We also point out that:
\begin{align}
 \mathbf{A}_{L_\mathcal{T} } \mathbf{B}_{L_\mathcal{T} }^T = \mathbf{\Omega} \label{eq:NetworkMatrixIncidenceLoop}
\end{align}
where $ \mathbf{\Omega}$ is the node-loop network matrix (Eq. \ref{eq:AppendixNetworkMatrix}).

Then, by utilizing Eqs. \ref{eq:InductiveCapacitiveIncidenceLoop} and \ref{eq:NetworkMatrixIncidenceLoop}, we obtain a modified version of Eqs. \ref{eq:SuperconductingChargeVector} and \ref{eq:SuperconductingFluxVector} (where we also make explicit the discreteness of tunneling charge and flux):
\begin{align}
\vv{Q}_n + \mathbf{A}_J \left[ 2e \vv{N}_J \right]- \mathbf{\Omega} \vv{Q}_l &= 2e \vv{N}_0 \\
\vv{\Phi}_l - \mathbf{B}_S \left[\Phi_0 \vv{M}_S \right]  + \mathbf{\Omega}^T \vv{\Phi}_n  &= \Phi_0 \vv{M}_0 
\end{align}

Inserting the capacitive and inductive constitutive relations (Eqs. \ref{eq:AppCapacitiveConstitutiveCharge} and \ref{eq:AppInductiveConstitutiveFlux}) gives the ``integrated" equations of motion:
\begin{align}
\mathbf{C}\dot{\vv{\Phi}}_n + \vv{Q}_{\text{ext}}+ \mathbf{A}_J \left[ 2e \vv{N}_J \right]- \mathbf{\Omega} \vv{Q}_l &= 2e \vv{N}_0 \label{eq:AppContinuityIntegrated}\\
\mathbf{L}\dot{\vv{Q}}_l + \vv{\Phi}_{\text{ext}}  - \mathbf{B}_S  \left[\Phi_0 \vv{M}_S \right]  + \mathbf{\Omega}^T \vv{\Phi}_n  &= \Phi_0 \vv{M}_0 \label{eq:AppFaradayIntegrated}
\end{align}

Taking the time derivative and using the Josephson and phase slip relations generates the equations of motion in what we call the ``standard" form:
\begin{align}
\mathbf{C}\ddot{\vv{\Phi}}_n + \dot{\vv{Q}}_{\text{ext}}+ \mathbf{A}_J \vv{I}_J - \mathbf{\Omega} \dot{\vv{Q}}_l &= \vv{0}_n \label{eq:AppContinuityStandard} \\
\mathbf{L}\ddot{\vv{Q}}_l + \dot{\vv{\Phi}}_{\text{ext}}  + \mathbf{B}_S \vv{V}_S  + \mathbf{\Omega}^T \dot{\vv{\Phi}}_n  &= \vv{0}_l  \label{eq:AppFaradayStandard}
\end{align}
where
\begin{align}
\vv{I}_J &= \mathbf{I}_0 \sin \left(\frac{2 \pi}{\Phi_0} \mathbf{A}_J^T \vv{\Phi}_n \right) \\
\vv{V}_S &= \mathbf{V}_0 \sin \left(\frac{2 \pi}{2e} \mathbf{B}_S^T \vv{Q}_l \right)
\end{align}
Here, $\mathbf{I}_0 = \text{diag}\left(\vv{I}_0 \right)$ is the diagonal matrix of Josephson critical currents and $\mathbf{V}_0 = \text{diag}\left(\vv{V}_0 \right)$ is the diagonal matrix of fluxoid tunneling critical voltages. The sine functions are applied element-wise to the vectorial arguments $\vv{\Phi}_J = \mathbf{A}_J^T \vv{\Phi}_n$ and $\vv{Q}_S = \mathbf{B}_S^T \vv{\Phi}_l$.

This pair of standard and integrated of equations of motion are a focal point of this work.

Fundamentally, these results are similar to and use notation from \cite{egusquiza_algebraic_2022,parra-rodriguez_geometrical_2024,osborne_symplectic_2024,osborne_flux-charge_2024}. The main difference in our work is that we impose additional physical restrictions on circuit connectivity (Appendix \ref{subsec:NodeFluxLoopChargeTreeCotree}), and that we use the ``integrated" equations of motion to distinguish between discrete and continuous spectrum operators upon quantization. This can result in different predictions for Hamiltonian of circuits containing both Josephson junctions and phase slips. We also note that we do not consider non-reciprocal circuits in this work, unlike in \cite{egusquiza_algebraic_2022,parra-rodriguez_geometrical_2024}.

\subsection{Changes of basis on complete model}  \label{subsec:ChangeOfBasis2}

Basis changes are important when quantizing circuits and manipulating them into equivalent forms. In practice, it is useful to view these transformations as acting on the integer-valued matrices $\mathbf{\Omega}$, $\mathbf{A}_J$, and $\mathbf{B}_S$---which encode the system's topology.

As in Appendix \ref{subsec:ChangeofBasis1}, we perform changes of basis on the node and loop variables:
\begin{align}
\vv{\Phi}_n &\rightarrow \mathbf{U}^{T^{-1}} \vv{\Phi}_n \label{eq:AppPhiU} \\
\vv{Q}_l &\rightarrow \mathbf{W}^{T^{-1}} \vv{Q}_l \label{eq:AppQW}
\end{align}
In this case the integrated equations of motion are transformed as:
\begin{align}
\left[ \mathbf{U} \mathbf{C} \mathbf{U}^T \right] \left[ \mathbf{U}^{T^{-1}} \dot{\vv{\Phi}}_n \right] +  \left[ \mathbf{U}  \vv{Q}_{\text{ext}}\right] +  \left[ 
 \mathbf{U}\mathbf{A}_J  \right] \left[ 2e \vv{N}_J \right] \nonumber \\ %newline
-\left[\mathbf{U} \mathbf{\Omega}\mathbf{W}^T  \right] \left[\mathbf{W}^{T^{-1}} \vv{Q}_l  \right] = 2e \left[ \mathbf{U} \vv{N}_0 \right]  \\ %new equation
\left[ \mathbf{W} \mathbf{L}\mathbf{W}^T \right] \left[ \mathbf{W}^{T^{-1}}\dot{\vv{Q}}_l \right] + \left[ \mathbf{W} \vv{\Phi}_{\text{ext}} \right]  - \mathbf{B}_S  \left[\Phi_0 \vv{M}_S \right] \nonumber \\ %newline
+ \left[\mathbf{U} \mathbf{\Omega}\mathbf{W}^T  \right]^T \left[\mathbf{U} \vv{\Phi}_n  \right] = \Phi_0 \left[ \mathbf{W}\vv{M}_0 \right]
\end{align}

The standard equations of motion transform in an analogous way, wherein a resolution of the identity is inserted into the sinusoidal terms such that:
\begin{align}
\vv{I}_J &= \mathbf{I}_0 \sin \left(\frac{2 \pi}{\Phi_0} \left[\mathbf{U} \mathbf{A}_J \right]^T \left[\mathbf{U}^{T^{-1}} \vv{\Phi}_n \right] \right) \\
\vv{V}_S &= \mathbf{V}_0 \sin \left(\frac{2 \pi}{2e} \left[\mathbf{W} \mathbf{B}_S \right]^T \left[ \mathbf{W}^{T^{-1}} \vv{Q}_l \right] \right)
\end{align}

The system's key matrices are thus transformed as:
\begin{align}
\mathbf{C} &\rightarrow \mathbf{U} \mathbf{C} \mathbf{U}^T \\
\mathbf{L} &\rightarrow \mathbf{W} \mathbf{L} \mathbf{W}^T \\
\mathbf{A}_J &\rightarrow \mathbf{U} \mathbf{A}_J \\
\mathbf{B}_S &\rightarrow \mathbf{W} \mathbf{B}_S \\
\mathbf{\Omega} &\rightarrow \mathbf{U} \mathbf{\Omega} \mathbf{W}^T 
\end{align}
where now $\mathbf{U}$ acts with row operations on $\mathbf{A}_J$ and $\mathbf{\Omega}$, and $\mathbf{W}$ acts with row operations on $\mathbf{B}_S$ and column operations on $\mathbf{\Omega}$.

The vector quantities transform as:
\begin{align}
\vv{Q}_\text{ext} &\rightarrow \mathbf{U} \vv{Q}_\text{ext}\\
\vv{\Phi}_\text{ext} &\rightarrow \mathbf{W} \vv{\Phi}_\text{ext} \\
\vv{N}_0 &\rightarrow \mathbf{U} \vv{N}_0\\
\vv{M}_0 &\rightarrow \mathbf{W} \vv{M}_0 
\end{align}

Thus we see that if $\mathbf{U}$ and $\mathbf{W}$ are integer-valued matrices, then $\mathbf{A}_J$, $\mathbf{B}_S$, $\mathbf{\Omega}$, $\vv{N}_0$, and $\vv{M}_0$ will remain integer-valued.

\subsection{Eliminating free modes} \label{subsec:FreeModes}

To simplify the equations of motion, we would like to remove a system's free modes, which do not affect the overall dynamics. These free modes corresponding to capacitive islands (with no junctions) and inductive loops (with no phase slips). Through this process, we end up renormalizing several of the system's parameters. The following procedure generalizes the one we demonstrated in Appendix \ref{subsec:ChangeofBasis1}, and is similar to those presented in \cite{ding_free-mode_2021,egusquiza_algebraic_2022}.

The presence of free modes is indicated when the flux and charge coordinates can be split up such that:
\begin{align}
\vv{\Phi}_n \rightarrow 
\begin{bmatrix}
\vv{\Phi}_n \\
\vv{\Phi}_\alpha
\end{bmatrix} \\
\vv{Q}_l \rightarrow 
\begin{bmatrix}
\vv{Q}_l \\
\vv{Q}_\alpha
\end{bmatrix}
\end{align}
where the free islands are indexed by $\alpha$ and the free loops indexed by $\beta$. This means that the system's matrices can be broken up as:
\begin{align}
\mathbf{A}_J &= 
\begin{bmatrix}
\mathbf{A}_{nJ} \\
\mathbf{0}_{\alpha J}
\end{bmatrix}\\
\mathbf{B}_{S} &=
\begin{bmatrix}
\mathbf{B}_{l S} \\
\mathbf{0}_{\beta S}
\end{bmatrix}\\
\mathbf{\Omega} &= 
\begin{bmatrix}
\mathbf{\Omega}_{n l} & \mathbf{0}_{n \beta} \\
\mathbf{0}_{\alpha l} & \mathbf{0}_{\alpha \beta}
\end{bmatrix}
\end{align}

The integrated equation of motions (Eqs. \ref{eq:AppContinuityIntegrated} and \ref{eq:AppFaradayIntegrated}) can then be written as:
\begin{align}
\begin{bmatrix}
\mathbf{C}_{nn} & \mathbf{C}_{n \alpha} \\
\mathbf{C}_{\alpha r} & \mathbf{C}_{\alpha \alpha}
\end{bmatrix} 
\begin{bmatrix}
\dot{\vv{\Phi}}_{n}  \\
\dot{\vv{\Phi}}_{\alpha}
\end{bmatrix} 
+
\begin{bmatrix}
\vv{Q}_{\text{ext}_n}  \\
\vv{Q}_{\text{ext}_\alpha}
\end{bmatrix} 
+
\begin{bmatrix}
\mathbf{A}_{nJ} \\
\mathbf{0}_{\alpha J}
\end{bmatrix}
2e \vv{N}_J
\nonumber \\ %new line
-
\begin{bmatrix}
\mathbf{\Omega}_{n l} & \mathbf{0}_{n \beta} \\
\mathbf{0}_{\alpha l} & \mathbf{0}_{\alpha \beta}
\end{bmatrix} 
\begin{bmatrix}
\vv{Q}_{l}  \\
\vv{Q}_{\beta}
\end{bmatrix} 
=
2 e 
\begin{bmatrix}
\vv{N}_{0_n}  \\
\vv{N}_{0_\alpha}
\end{bmatrix} 
\\ %new equation
\begin{bmatrix}
\mathbf{L}_{ll} & \mathbf{L}_{l\beta} \\
\mathbf{L}_{\beta l} & \mathbf{L}_{\beta \beta}
\end{bmatrix} 
\begin{bmatrix}
\dot{\vv{Q}}_{l}  \\
\dot{\vv{Q}}_{\beta}
\end{bmatrix} 
+
\begin{bmatrix}
\vv{\Phi}_{\text{ext}_l} \\
\vv{\Phi}_{\text{ext}_\beta}
\end{bmatrix}
- 
\begin{bmatrix}
\mathbf{B}_{l S} \\
\mathbf{0}_{\beta S}
\end{bmatrix}  \label{eq:AppFreeMode1}
\Phi_0 \vv{M}_S  \\ %new line
+\begin{bmatrix}
\mathbf{\Omega}_{nl}^T & \mathbf{0}_{l \alpha} \\
\mathbf{0}_{\beta n} & \mathbf{0}_{\beta \alpha}
\end{bmatrix} 
\begin{bmatrix}
\vv{\Phi}_{n}  \\
\vv{\Phi}_{\alpha}
\end{bmatrix} 
=
\Phi_0 
\begin{bmatrix}
\vv{M}_{0_l}  \\
\vv{M}_{0_\beta}
\end{bmatrix}  \label{eq:AppFreeMode2}
\end{align}

We can then solve for $\dot{\vv{\Phi}}_\alpha$ and $\dot{\vv{Q}}_\beta$ through the second line of these vector-valued equations:
\begin{align}
\dot{\vv{\Phi}}_\alpha &= -\mathbf{C}_{\alpha \alpha}^{-1} \left( \mathbf{C}_{\alpha n} \dot{\vv{\Phi}}_n + \vv{Q}_{\text{ext}_\alpha} - 2e \vv{N}_{0_\alpha}  \right) \\
\dot{\vv{Q}}_\beta &= -\mathbf{L}_{\beta \beta}^{-1} \left( \mathbf{L}_{\beta l} \dot{\vv{Q}}_l + \dot{\vv{\Phi}}_{\text{ext}_\beta} - \Phi_0 \vv{M}_{0_\beta}  \right)
\end{align}

Substituting these expressions into the first lines of the vector-valued equations Eqs. \ref{eq:AppFreeMode1} and \ref{eq:AppFreeMode2} gives a renormalized set of integrated equations (whose standard form is given by the time derivative):
\begin{align}
\mathbf{C}\dot{\vv{\Phi}}_n + \vv{Q}_{\text{ext}}+ \mathbf{A}_J \left[ 2e \vv{N}_J \right]- \mathbf{\Omega} \vv{Q}_l &= 2e \vv{N}_0 \\
\mathbf{L}\dot{\vv{Q}}_l + \vv{\Phi}_{\text{ext}}  - \mathbf{B}_S  \left[\Phi_0 \vv{M}_S \right]  + \mathbf{\Omega}^T \vv{\Phi}_n  &= \Phi_0 \vv{M}_0
\end{align}
with the system's parameters modified as:
\begin{align}
\mathbf{C} &= \mathbf{C}_{nn} - \mathbf{C}_{n \alpha} \mathbf{C}_{\alpha \alpha}^{-1} \mathbf{C}_{\alpha n} \\
\vv{Q}_{\text{ext}} & = \vv{Q}_{\text{ext}_n} - \mathbf{C}_{n \alpha} \mathbf{C}_{\alpha \alpha}^{-1} \vv{Q}_{\text{ext}_\alpha} \\
\vv{N}_{0} & = \vv{N}_{0_n} - \mathbf{C}_{n \alpha} \mathbf{C}_{\alpha \alpha}^{-1} \vv{N}_{0_\alpha} \\
\mathbf{L} & = \mathbf{L}_{ll} - \mathbf{L}_{l \beta} \mathbf{L}_{\beta \beta}^{-1} \mathbf{L}_{\beta l} \\
\vv{\Phi}_{\text{ext}} & = \vv{\Phi}_{\text{ext}_l} - \mathbf{L}_{l \beta} \mathbf{L}_{\beta \beta}^{-1} \vv{\Phi}_{\text{ext}_\beta} \\
\vv{M}_{0} & = \vv{M}_{0_l} - \mathbf{L}_{l \beta} \mathbf{L}_{\beta \beta}^{-1} \vv{M}_{0_\beta} \\
\end{align}
Importantly this procedure leaves $\mathbf{C}$ and $\mathbf{L}$ as symmetric, positive definite matrices, but alters $\vv{N}_{0}$ and $\vv{M}_{0}$ such that they are no longer necessarily integer-valued. This distinction will become important during the quantization procedure, as non-integer-valued components of $\vv{N}_{0}$ and $\vv{M}_{0}$ will appear as offsets in the final Hamiltonian. Here we can think of $\vv{N}_{0_\alpha}$ and $\vv{M}_{0_\beta}$ as the total charge and flux trapped in each capacitive island and inductive loop of the system, respectively. As these islands and loops are not interconnected through any tunneling elements, the total charge and flux (respectively) that they contain will be constant through time.

In addition, the key integer-valued matrices of the system become submatrices, with some of the rows and columns of zeros removed:
\begin{align}
\mathbf{A}_J = \mathbf{A}_{nJ} \\
\mathbf{B}_S = \mathbf{B}_{lS} \\
\mathbf{\Omega} = \mathbf{\Omega}_{nl} 
\end{align}
Thus, the free mode removal procedure does not substantially alter the topology of the circuit.

\section{Quantization} \label{sec:AppendixC}

We now demonstrate a procedure to construct and quantize a Hamiltonian for these reciprocal-lumped element superconducting circuits. By utilizing our integrated equations of motion, we are able to track which variables represent discrete operators and which represent continuous ones, upon quantization.

\subsection{Superconducting circuit Hamiltonian} \label{subsec:Quantization1}

A key step in the quantization procedure is to reduce the matrix $\mathbf{\Omega}$ (of rank $k$) to a block matrix with the $k$ by $k$ identity matrix $I_{kk}$ in the top left corner and $0$ everywhere else. As detailed in Appendix \ref{subsec:ChangeOfBasis2}, this can be easily accomplished through a series of integer-valued basis transformations $\mathbf{U}$ (performing row operations on $\mathbf{\Omega}$) and $\mathbf{W}$ (performing column operations) on $\mathbf{\Omega}$. These transformations act as:
\begin{align}
\mathbf{\Omega} \rightarrow \mathbf{U} \mathbf{\Omega} \mathbf{W}^T \\
\mathbf{A}_J \rightarrow \mathbf{U} \mathbf{A}_J \\
\mathbf{B}_S \rightarrow \mathbf{W} \mathbf{B}_S
\end{align}
We can expand each of these (still) integer-valued matrices as:
\begin{align}
\mathbf{\Omega} =
\begin{bmatrix}
\mathbf{I}_{kk} & \mathbf{0}_{ks} \\
\mathbf{0}_{jk} & \mathbf{0}_{js}
\end{bmatrix} \\ %newline
\mathbf{A}_J =
\begin{bmatrix}
\mathbf{A}_{kJ} \\
\mathbf{A}_{jJ} 
\end{bmatrix} \\ %newline
\mathbf{B}_S =
\begin{bmatrix} 
\mathbf{B}_{kS} \\
\mathbf{B}_{sS} 
\end{bmatrix} 
\end{align}

Now we make the additional assumption that all the free modes have been removed using the procedure in Appendix \ref{subsec:FreeModes}. This is to say that submatrices $\mathbf{A}_{jJ}$ and $\mathbf{B}_{sJ}$ have rank $j$ and $s$, respectively, and they cannot be row-reduced to have a row of all $0$ (which would represent a free mode). 

We partition the vectors of flux and charge coordinates in terms of their (eventual) quantized mode type:
\begin{align}
\vv{\Phi}_n = 
\begin{bmatrix}
\vv{\Phi}_k \\
\vv{\Phi}_j
\end{bmatrix} \\
\vv{Q}_l = 
\begin{bmatrix}
\vv{Q}_k \\
\vv{Q}_s
\end{bmatrix}
\end{align}

We remove the presence of the $\vv{N}_0$ term by grouping it with $\vv{Q}_\text{ext}$ and do the same for $\vv{M}_0$ by grouping it with $\vv{\Phi}_\text{ext}$:
\begin{align}
\vv{Q}_\text{ext} &\rightarrow \vv{Q}_\text{ext} - 2e \vv{N}_0 \label{eq:QExtTransform} \\
\vv{\Phi}_\text{ext} &\rightarrow \vv{\Phi}_\text{ext} - \Phi_0 \vv{M}_0 \label{eq:PhiExtTransform}
\end{align}

The integrated equations of motion can now be written in block form as:
\begin{align}
\begin{bmatrix}
\mathbf{C}_{kk} & \mathbf{C}_{kj} \\
\mathbf{C}_{jk} & \mathbf{C}_{jj}
\end{bmatrix} 
\begin{bmatrix}
\dot{\vv{\Phi}}_{k}  \\
\dot{\vv{\Phi}}_{j}
\end{bmatrix} 
+
\begin{bmatrix}
\vv{Q}_{\text{ext}_k}  \\
\vv{Q}_{\text{ext}_j}
\end{bmatrix} 
+
\begin{bmatrix}
\mathbf{A}_{kJ} \\
\mathbf{A}_{jJ}
\end{bmatrix}
2e \vv{N}_J
\nonumber \\ %new line
-
\begin{bmatrix}
\mathbf{I}_{kk} & \mathbf{0}_{ks} \\
\mathbf{0}_{jk} & \mathbf{0}_{js}
\end{bmatrix} 
\begin{bmatrix}
\vv{Q}_{k}  \\
\vv{Q}_{s}
\end{bmatrix} 
=
\begin{bmatrix}
\vv{0}_{k}  \\
\vv{0}_{s}
\end{bmatrix} \label{eq:AppBacksolveIntegratedC}
\end{align} %New equations
\begin{align}
\begin{bmatrix}
\mathbf{L}_{kk} & \mathbf{L}_{ks} \\
\mathbf{L}_{sk} & \mathbf{L}_{ss}
\end{bmatrix} 
\begin{bmatrix}
\dot{\vv{Q}}_{k}  \\
\dot{\vv{Q}}_{s}
\end{bmatrix} 
+
\begin{bmatrix}
\vv{\Phi}_{\text{ext}_k } \\
\vv{\Phi}_{\text{ext}_s}
\end{bmatrix}
- 
\begin{bmatrix}
\mathbf{B}_{kS} \\
\mathbf{B}_{sS}
\end{bmatrix}
\Phi_0 \vv{M}_S \nonumber \\ %new line
+\begin{bmatrix}
\mathbf{I}_{kk} & \mathbf{0}_{kj} \\
\mathbf{0}_{sk} & \mathbf{0}_{sj}
\end{bmatrix} 
\begin{bmatrix}
\vv{\Phi}_{k}  \\
\vv{\Phi}_{j}
\end{bmatrix} 
=
\begin{bmatrix}
\vv{0}_{k}  \\
\vv{0}_{j}
\end{bmatrix} \label{eq:AppBacksolveIntegratedL}
\end{align}

The time derivative of these equations gives them in standard form as:
\begin{align}
&\begin{bmatrix}
\mathbf{C}_{kk} & \mathbf{C}_{kj} \\
\mathbf{C}_{jk} & \mathbf{C}_{jj}
\end{bmatrix} 
\begin{bmatrix}
\ddot{\vv{\Phi}}_{k}  \\
\ddot{\vv{\Phi}}_{j}
\end{bmatrix} 
+
\begin{bmatrix}
\dot{\vv{Q}}_{\text{ext}_k}  \\
\dot{\vv{Q}}_{\text{ext}_j}
\end{bmatrix} \nonumber \\ %newline
& + 
\begin{bmatrix}
\mathbf{A}_{kj} \\
\mathbf{A}_{jJ}
\end{bmatrix}
\mathbf{I}_J \sin \left( \frac{2\pi}{\Phi_0} 
\begin{bmatrix}
\mathbf{A}_{kJ}^T &  \mathbf{A}_{jJ}^T 
\end{bmatrix}
\begin{bmatrix}
\vv{\Phi}_k \\ 
\vv{\Phi}_j
\end{bmatrix}\right)
\nonumber \\ %new line
&-
\begin{bmatrix}
\mathbf{I}_{kk} & \mathbf{0}_{ks} \\
\mathbf{0}_{jk} & \mathbf{0}_{js}
\end{bmatrix} 
\begin{bmatrix}
\dot{\vv{Q}}_{k}  \\
\dot{\vv{Q}}_{s}
\end{bmatrix} 
=
\begin{bmatrix}
\vv{0}_{k}  \\
\vv{0}_{s}
\end{bmatrix} 
\end{align} %new align
\begin{align}
&\begin{bmatrix}
\mathbf{L}_{kk} & \mathbf{L}_{ks} \\
\mathbf{L}_{sk} & \mathbf{L}_{ss}
\end{bmatrix} 
\begin{bmatrix}
\ddot{\vv{Q}}_{k} \nonumber \\
\ddot{\vv{Q}}_{s}
\end{bmatrix} 
+
\begin{bmatrix}
\dot{\vv{\Phi}}_{\text{ext}_k } \\
\dot{\vv{\Phi}}_{\text{ext}_s}
\end{bmatrix} \\ %newline
&+ 
\begin{bmatrix}
\mathbf{B}_{ks}  \\
\mathbf{B}_{sS}
\end{bmatrix}
\mathbf{V}_S \sin \left( \frac{2\pi}{2e} 
\begin{bmatrix}
\mathbf{B}_{kS}^T & \mathbf{B}_{sS}^T
\end{bmatrix} 
\begin{bmatrix}
\vv{Q}_k \\
\vv{Q}_s
\end{bmatrix}
\right) \nonumber \\ %new line
&+\begin{bmatrix}
\mathbf{I}_{kk} & \mathbf{0}_{kj} \\
\mathbf{0}_{sk} & \mathbf{0}_{sj}
\end{bmatrix} 
\begin{bmatrix}
\dot{\vv{\Phi}}_{k}  \\
\dot{\vv{\Phi}}_{j}
\end{bmatrix} 
=
\begin{bmatrix}
\vv{0}_{k}  \\
\vv{0}_{s}
\end{bmatrix} 
\end{align}

A Lagrangian that generates the standard equations of motion is: 
\begin{align}
\mathcal{L}
= 
\frac{1}{2}
\begin{bmatrix}
\dot{\vv{\Phi}}_k \\
\dot{\vv{\Phi}}_j
\end{bmatrix}^T 
\begin{bmatrix}
\mathbf{C}_{kk} & \mathbf{C}_{kj} \\
\mathbf{C}_{jk} & \mathbf{C}_{jj}
\end{bmatrix} 
\begin{bmatrix}
\dot{\vv{\Phi}}_k \\
\dot{\vv{\Phi}}_j
\end{bmatrix}
+
\begin{bmatrix}
\dot{\vv{\Phi}}_k \\
\dot{\vv{\Phi}}_j
\end{bmatrix}^T 
\begin{bmatrix}
\vv{Q}_{\text{ext}_k} \\
\vv{Q}_{\text{ext}_j}
\end{bmatrix} \nonumber \\ %newline
+\frac{1}{2}
\begin{bmatrix}
\dot{\vv{Q}}_k \\
\dot{\vv{Q}}_s
\end{bmatrix}^T 
\begin{bmatrix}
\mathbf{L}_{kk} & \mathbf{L}_{ks} \\
\mathbf{L}_{sk} & \mathbf{L}_{ss}
\end{bmatrix} 
\begin{bmatrix}
\dot{\vv{Q}}_k \\
\dot{\vv{Q}}_s
\end{bmatrix}
+
\begin{bmatrix}
\dot{\vv{Q}}_k \\
\dot{\vv{Q}}_s
\end{bmatrix}^T 
\begin{bmatrix}
\vv{\Phi}_{\text{ext}_k} \\
\vv{\Phi}_{\text{ext}_s}
\end{bmatrix} \nonumber \\ %newline 
-
\dot{\vv{\Phi}}_k^T
\vv{Q}_k + \vv{E}_J^T \cos \left( \frac{2\pi}{\Phi_0} 
\begin{bmatrix}
\mathbf{A}_{kJ}^T &  \mathbf{A}_{jJ}^T 
\end{bmatrix}
\begin{bmatrix}
\vv{\Phi}_k \\ 
\vv{\Phi}_j
\end{bmatrix}\right) \nonumber \\ %newline
+
\vv{E}_S^T \cos \left( \frac{2\pi}{2e} 
\begin{bmatrix}
\mathbf{B}_{kS}^T &  \mathbf{B}_{sS}^T 
\end{bmatrix}
\begin{bmatrix}
\vv{Q}_k \\ 
\vv{Q}_s
\end{bmatrix}\right)
\end{align}
where $\vv{E}_J = \frac{\Phi_0}{2 \pi} \vv{I}_0$ and $\vv{E}_S = \frac{2e}{2 \pi} \vv{V}_0$. As noted in \cite{osborne_symplectic_2024, osborne_flux-charge_2024}, the factor $-\dot{\vv{\Phi}}_k^T \vv{Q}_k$ could also be chosen as $\dot{\vv{Q}}_k^T \vv{\Phi}_k$, or as a weighted sum of the two factors.

The Euler-Lagrange equations that generate the equations of motion from this Lagrangian are:
\begin{align}
 \frac{d}{dt}\frac{\partial\mathcal{L}}{\partial \dot{\vv{\Phi}}_n} - \frac{\partial\mathcal{L}}{\partial\vv{\Phi}_n} &= \vv{0}_n \\
 \frac{d}{dt}\frac{\partial\mathcal{L}}{\partial \dot{\vv{Q}}_l} - \frac{\partial\mathcal{L}}{\partial \vv{Q}_l} &= \vv{0}_l
\end{align}

To convert from a Lagrangian to a Hamiltonian picture, we find the conjugate momenta and perform a Legendre transform. At this stage we also use the integrated equations of motion to write these conjugate momenta in terms of $\vv{N}_J$ and $\vv{M}_S$, which become integer-valued operators when quantized.

The conjugate momenta to the entries of $\vv{\Phi}_n$ are denoted $\vv{\Pi}_n$ (with units of charge):
\begin{align}
\begin{bmatrix}
\vv{\Pi}_k \\
\vv{\Pi}_j
\end{bmatrix}
& =
\begin{bmatrix}
\mathbf{C}_{kk} & \mathbf{C}_{kj} \\
\mathbf{C}_{jk} & \mathbf{C}_{jj}
\end{bmatrix} 
\begin{bmatrix}
\dot{\vv{\Phi}}_k \\
\dot{\vv{\Phi}}_j
\end{bmatrix}
+
\begin{bmatrix}
\vv{Q}_{\text{ext}_k} \\
\vv{Q}_{\text{ext}_j}
\end{bmatrix}
+
\begin{bmatrix}
-\vv{Q}_{k} \\
\vv{0}_{j}
\end{bmatrix} 
\nonumber \\ %newline
& =-\begin{bmatrix}
\mathbf{A}_{kJ} \\
\mathbf{A}_{jJ}
\end{bmatrix}
2e \vv{N}_J \label{eq:AppConjPi}
\end{align}
where the second line follows from the integrated equation of motion (Eq. \ref{eq:AppBacksolveIntegratedC}).

The conjugate momenta to the entries of $\vv{Q}_l$ are given by $\vv{P}_l$ (with units of flux):
\begin{align}
\begin{bmatrix}
\vv{P}_k \\
\vv{P}_s
\end{bmatrix}
& = 
\begin{bmatrix}
\mathbf{L}_{kk} & \mathbf{L}_{ks} \\
\mathbf{L}_{sk} & \mathbf{L}_{ss}
\end{bmatrix} 
\begin{bmatrix}
\dot{\vv{Q}}_k \\
\dot{\vv{Q}}_s
\end{bmatrix}
+
\begin{bmatrix}
\vv{\Phi}_{\text{ext}_k} \\
\vv{\Phi}_{\text{ext}_s}
\end{bmatrix}
\nonumber \\ %newline 
& =\begin{bmatrix}
\mathbf{B}_{kS} \\
\mathbf{B}_{sS}
\end{bmatrix}
\Phi_0 \vv{M}_S
+ 
\begin{bmatrix}
-\vv{\Phi}_k \\
\vv{0}_s
\end{bmatrix} \label{eq:AppConjP}
\end{align}
Here, the second equality comes from the other integrated equation of motion (Eq. \ref{eq:AppBacksolveIntegratedL}). 

Using the Legendre transform, the total Hamiltonian of the system is then: 
\begin{align}
&\mathcal{H} = \dot{\vv{\Phi}}_n^T \vv{\Pi}_n + \dot{\vv{Q}}_l^T \vv{P}_l - \mathcal{L} \nonumber =  \\ %newline
&\frac{1}{2}
\begin{bmatrix}
\vv{\Pi}_k + \vv{Q}_k - \vv{Q}_{\text{ext}_k} \\
\vv{\Pi}_j - \vv{Q}_{\text{ext}_j}
\end{bmatrix}^T 
\begin{bmatrix}
\mathbf{C}_{kk} & \mathbf{C}_{kj} \\
\mathbf{C}_{jk} & \mathbf{C}_{jj}
\end{bmatrix}^{-1}
\begin{bmatrix}
\vv{\Pi}_k + \vv{Q}_k - \vv{Q}_{\text{ext}_k} \\
\vv{\Pi}_j - \vv{Q}_{\text{ext}_j}
\end{bmatrix} \nonumber \\ %newline
&+\frac{1}{2}\begin{bmatrix}
\vv{P}_k -  \vv{\Phi}_{\text{ext}_k} \\
\vv{P}_s - \vv{\Phi}_{\text{ext}_s}
\end{bmatrix}^T 
\begin{bmatrix}
\mathbf{L}_{kk} & \mathbf{L}_{ks} \\
\mathbf{L}_{sk} & \mathbf{L}_{ss}
\end{bmatrix}^{-1}
\begin{bmatrix}
\vv{P}_k -  \vv{\Phi}_{\text{ext}_k} \\
\vv{P}_s - \vv{\Phi}_{\text{ext}_s}
\end{bmatrix} \nonumber \\ %newline 
&- \vv{E}_J^T \cos \left( \frac{2\pi}{\Phi_0} 
\begin{bmatrix}
\mathbf{A}_{kJ}^T &  \mathbf{A}_{jJ}^T 
\end{bmatrix}
\begin{bmatrix}
\vv{\Phi}_k \\ 
\vv{\Phi}_j
\end{bmatrix}\right) \nonumber \\ %newline
&-
\vv{E}_S^T \cos \left( \frac{2\pi}{\Phi_0} 
\begin{bmatrix}
\mathbf{B}_{kS}^T &  \mathbf{B}_{sS}^T 
\end{bmatrix}
\begin{bmatrix}
\vv{Q}_k \\ 
\vv{Q}_s
\end{bmatrix}\right)  \label{eq:IntermediateHamiltonian}
\end{align}

We note that Poisson bracket for canonically conjugate variables obeys:
\begin{align}
\{\vv{\Phi}_n , \vv{\Pi}_n \} &= \mathbf{I}_{nn} \label{eq:PhiPiCommutation} \\
\{ \vv{Q}_l, \vv{P}_l \} &= \mathbf{I}_{ll} \label{eq:QPCommutation}
\end{align}
with the Poisson bracket for any other pair of these variables that are not canonically conjugate being $0$.

Now follows another essential step in the procedure: a canonical change of basis \cite{ding_free-mode_2021, parra-rodriguez_canonical_2019,egusquiza_algebraic_2022}---which preserves the Poisson brackets. Performing the following transformation on each of the $k$-coordinates gives:
\begin{align}
\vv{\Phi}_k &\rightarrow  \vv{\Phi}_k + \vv{P}_k = \Phi_0 \mathbf{B}_{kS} \vv{M}_S  \label{eq:DiscretePhi} \\
\vv{\Pi}_k & \rightarrow  \vv{\Pi}_k = -2e\mathbf{A}_{kJ} \vv{N}_J \label{eq:DiscretePi}  \\
\vv{Q}_k & \rightarrow \vv{Q}_k + \vv{\Pi}_k = \vv{Q}_k  - 2e \mathbf{A}_{kJ}\vv{N}_J  \\
\vv{P}_k & \rightarrow  \vv{P}_k = -\vv{\Phi}_k + \Phi_0 \mathbf{B}_{kS}  \vv{M}_S 
\end{align}
The transformation is canonical because we can use the upper $k$-by-$k$ submatrices of Eqs. \ref{eq:PhiPiCommutation} and \ref{eq:QPCommutation} to write:

\begin{align}
\{ \vv{\Phi}_k, \vv{Q}_k \} & \rightarrow \{ \vv{\Phi}_k + \vv{P}_k, \vv{Q}_k + \vv{\Pi}_k\} \nonumber \\
&= \{ \vv{\Phi}_k ,  \vv{\Pi}_k\} + \{ \vv{P}_k, \vv{Q}_k \}  \nonumber \\
& = \mathbf{I}_{kk} - \mathbf{I}_{kk} \nonumber \\
& = \mathbf{0}_{kk}
\end{align}

Since $\mathbf{A}_{kJ}$ and $\mathbf{B}_{kS}$ are integer-valued after quantization, so are $\vv{\Phi}_k$ and $\vv{\Pi}_k$, by Eqs. \ref{eq:DiscretePhi} and \ref{eq:DiscretePi}. On the other hand, $\vv{Q}_k$ and $\vv{P}_k$ will translate in the quantum case to continuous operators. 

Applying the basis change results in a Hamiltonian of:
\begin{align}
\mathcal{H} =
\frac{1}{2}
\begin{bmatrix}
\vv{Q}_k - \vv{Q}_{\text{ext}_k} \\
\vv{\Pi}_j - \vv{Q}_{\text{ext}_j}
\end{bmatrix}^T 
\begin{bmatrix}
\mathbf{C}_{kk} & \mathbf{C}_{kj} \\
\mathbf{C}_{jk} & \mathbf{C}_{jj}
\end{bmatrix}^{-1}
\begin{bmatrix}
\vv{Q}_k - \vv{Q}_{\text{ext}_k} \\
\vv{\Pi}_j - \vv{Q}_{\text{ext}_j}
\end{bmatrix} \nonumber \\ %newline
+\frac{1}{2}
\begin{bmatrix}
\vv{P}_k -  \vv{\Phi}_{\text{ext}_k} \\
\vv{P}_s - \vv{\Phi}_{\text{ext}_s}
\end{bmatrix}^T 
\begin{bmatrix}
\mathbf{L}_{kk} & \mathbf{L}_{ks} \\
\mathbf{L}_{sk} & \mathbf{L}_{ss}
\end{bmatrix}^{-1}
\begin{bmatrix}
\vv{P}_k -  \vv{\Phi}_{\text{ext}_k} \\
\vv{P}_s - \vv{\Phi}_{\text{ext}_s}
\end{bmatrix} \nonumber \\ %newline 
- \vv{E}_J^T \cos \left( \frac{2\pi}{\Phi_0} 
\begin{bmatrix}
\mathbf{A}_{kJ}^T &  \mathbf{A}_{jJ}^T 
\end{bmatrix}
\begin{bmatrix}
- \vv{P}_k + \Phi_0 \mathbf{B}_{kS} \vv{M}_S \\ 
\vv{\Phi}_j
\end{bmatrix}\right) \nonumber \\ %newline
-
\vv{E}_S^T \cos \left( \frac{2\pi}{2e} 
\begin{bmatrix}
\mathbf{B}_{kS}^T &  \mathbf{B}_{sS}^T 
\end{bmatrix}
\begin{bmatrix}
\vv{Q}_k + 2e \mathbf{A}_{kJ}  \vv{N}_J \\ 
\vv{Q}_s
\end{bmatrix}\right) \nonumber \\ %newequation
 =
 \frac{1}{2}
\begin{bmatrix}
\vv{Q}_k - \vv{Q}_{\text{ext}_k} \\
\vv{\Pi}_j - \vv{Q}_{\text{ext}_j}
\end{bmatrix}^T 
\begin{bmatrix}
\mathbf{C}_{kk} & \mathbf{C}_{kj} \\
\mathbf{C}_{jk} & \mathbf{C}_{jj}
\end{bmatrix}^{-1}
\begin{bmatrix}
\vv{Q}_k - \vv{Q}_{\text{ext}_k} \\
\vv{\Pi}_j - \vv{Q}_{\text{ext}_j}
\end{bmatrix} \nonumber \\ %newline
+\frac{1}{2}
\begin{bmatrix}
\vv{P}_k -  \vv{\Phi}_{\text{ext}_k} \\
\vv{P}_s - \vv{\Phi}_{\text{ext}_s}
\end{bmatrix}^T 
\begin{bmatrix}
\mathbf{L}_{kk} & \mathbf{L}_{ks} \\
\mathbf{L}_{sk} & \mathbf{L}_{ss}
\end{bmatrix}^{-1}
\begin{bmatrix}
\vv{P}_k -  \vv{\Phi}_{\text{ext}_k} \\
\vv{P}_s - \vv{\Phi}_{\text{ext}_s}
\end{bmatrix} \nonumber \\ %newline 
- \vv{E}_J^T \cos \left( \frac{2\pi}{\Phi_0} 
\begin{bmatrix}
\mathbf{A}_{kJ}^T &  \mathbf{A}_{jJ}^T 
\end{bmatrix}
\begin{bmatrix}
- \vv{P}_k  \\ 
\vv{\Phi}_j
\end{bmatrix}\right) \nonumber \\ %newline
-
\vv{E}_S^T \cos \left( \frac{2\pi}{2e} 
\begin{bmatrix}
\mathbf{B}_{kS}^T &  \mathbf{B}_{sS}^T 
\end{bmatrix}
\begin{bmatrix}
\vv{Q}_k \\ 
\vv{Q}_s
\end{bmatrix}\right) \label{eq:HamiltonianPenultimate}
\end{align}
where second line follows from the first by eliminating the integer multiples (further multiplied by $2 \pi$) of $\vv{M}_S$ and $\vv{N}_J$ in the cosines, as these variables will become integer-valued operators when quantized. 

We thus obtain a central hypothesis of this work: that the Hamiltonian is independent of the set of integer-spectra canonical coordinates $\vv{\Phi}_k$ and $\vv{\Pi}_k$ (which only appear in the cosines), such that they drop out of the overall equations. This differs from the prediction that they appear as a continuous-spectrum conjugate pair of conserved quantities \cite{le_doubly_2019,parra-rodriguez_geometrical_2024}. We are then left with $k+j+s$ total pairs of canonical coordinates, with each subscript referring to a different kind of mode, as will be described in the subsequent section.

Before proceeding with quantization, we make one further note. In Eqs. \ref{eq:QExtTransform} and \ref{eq:PhiExtTransform}, we lumped the charge offset of $\vv{N}_0$ into $\vv{Q}_\text{ext}$ and the flux offset given by $\vv{M}_0$ into $\vv{\Phi}_\text{ext}$. Now we reverse this procedure and perform an additional set of canonical transformations on the dynamical variables of our system:
\begin{align}
\vv{Q}_\text{ext} \rightarrow \vv{Q}_\text{ext} + 2e \vv{N}_0 \\
\begin{bmatrix}
\vv{Q}_k \\
\vv{\Pi}_j
\end{bmatrix}
\rightarrow 
\begin{bmatrix}
\vv{Q}_k \\
\vv{\Pi}_j
\end{bmatrix}
- 2e \vv{N}_0 \\
\vv{\Phi}_\text{ext} \rightarrow \vv{\Phi}_\text{ext} + \Phi_0 \vv{M}_0 \\
\begin{bmatrix}
\vv{P}_k \\
\vv{P}_s
\end{bmatrix}
\rightarrow 
\begin{bmatrix}
\vv{P}_k \\
\vv{P}_s
\end{bmatrix}
- \Phi_0 \vv{M}_0
\end{align}
This operation takes the constant offsets from the quadratic terms of the Hamiltonian and places them instead in the cosine terms. Thus because cosine is $2\pi$-periodic we can see that any integer-valued part of $\vv{N}_0$ and $\vv{M}_0$ will have no effect on the Hamiltonian. Indeed, these quantities are always integer-valued---except for the dressed portion of these terms that arises from the removal of free modes, as shown in \ref{subsec:FreeModes}.

This is an intuitive result for this heuristic, semiclassical model of the quantum system. Recall for a given island of the circuit (before free modes are removed), $N_0 = N(t_0)$ is the total charge contained in the capacitor pad of the island at time $t_0$, and $M_0 = M(t_0)$ is the total flux contained within a loop at that time. These quantities are only independent of integration starting time $t_0$ for the case of islands with no net current flow or loops with no fluxon flow at their boundaries---corresponding exactly to the free modes. Thus, since the non-integer part of these terms come only from free mode removal, the total Hamiltonian will not depend on $t_0$. Since $t_0$ was arbitrarily chosen, this is a desirable property. It means that for non-free modes of the system, the corresponding values of $N_0$ or $M_0$ can be set to zero.

\subsection{Quantization: extended modes, discrete charge modes, discrete flux modes} \label{subsec:Quantization2}

To continue the quantization procedure, we note that---when quantized---$\vv{\Pi}_j$ and $\vv{P}_s$ both represent discrete/integer-valued quantities (as shown in Eqs. \ref{eq:AppConjPi} and \ref{eq:AppConjP}) and make the following redefinitions:
\begin{align}
\vv{n}_j &= \frac{1}{2e} \vv{\Pi}_j = -\mathbf{A}_{jJ} \vv{N}_J \\
\vv{\phi}_j &= \frac{2 \pi}{\Phi_0} \vv{\Phi}_j \\
\vv{m}_s &=  \frac{1}{\Phi_0} \vv{P}_s =\mathbf{B}_{sS} \vv{M}_S \\
\vv{q}_s &= \frac{2 \pi}{2e} \vv{Q}_s 
\end{align}

Here $\vv{n}_j$ replaces the use of $\vv{\Pi}_j$ and makes clear its discrete nature, as $\vv{m}_s$ does for $\vv{P}_s$. The re-scaling of $\vv{\phi}_J$ to $\vv{\phi}_j$ and $\vv{Q}_s$ to $\vv{q}_s$ are not a necessity, but serve to simplify future notation.

With these redefinitions, the semiclassical Hamiltonian (Eq. \ref{eq:HamiltonianPenultimate}) can be rewritten in its final form:
\begin{align} \label{eq:AppFinalHamiltonian}
\mathcal{H} =
\frac{1}{2}
\begin{bmatrix}
\vv{Q}_k - \vv{Q}_{\text{ext}_k} \\
2e\vv{n}_j - \vv{Q}_{\text{ext}_j}
\end{bmatrix}^T 
\begin{bmatrix}
\mathbf{C}_{kk} & \mathbf{C}_{kj} \\
\mathbf{C}_{jk} & \mathbf{C}_{jj}
\end{bmatrix}^{-1}
\begin{bmatrix}
\vv{Q}_k - \vv{Q}_{\text{ext}_k} \\
2e\vv{n}_j - \vv{Q}_{\text{ext}_j}
\end{bmatrix} \nonumber \\ %newline
+\frac{1}{2}
\begin{bmatrix}
\vv{P}_k -  \vv{\Phi}_{\text{ext}_k} \\
\Phi_0 \vv{m}_s - \vv{\Phi}_{\text{ext}_s}
\end{bmatrix}^T 
\begin{bmatrix}
\mathbf{L}_{kk} & \mathbf{L}_{ks} \\
\mathbf{L}_{sk} & \mathbf{L}_{ss}
\end{bmatrix}^{-1}
\begin{bmatrix}
\vv{P}_k -  \vv{\Phi}_{\text{ext}_k} \\
\Phi_0 \vv{m}_s  - \vv{\Phi}_{\text{ext}_s}
\end{bmatrix} \nonumber \\ %newline 
- \vv{E}_J^T \cos \left( 
-\frac{2\pi}{\Phi_0}\mathbf{A}_{kJ}^T  \vv{P}_k  +  \mathbf{A}_{jJ}^T \vv{\phi}_j
\right) \nonumber \\ %newline
-
\vv{E}_S^T \cos \left(
\frac{2\pi}{2e} \mathbf{B}_{kS}^T\vv{Q}_k  +  \mathbf{B}_{sS}^T \vv{q}_s \right)
\end{align}

In order to quantize these equations, we promote classical variables in the Hamiltonian to quantum operators as:
\begin{align}
\vv{Q}_k \rightarrow \hat{\vv{Q}}_k &, \ \vv{P}_k \rightarrow \hat{\vv{P}}_k \\
\vv{n}_j \rightarrow \hat{\vv{n}}_j &, \ \vv{\phi}_j \rightarrow \hat{\vv{\phi}}_j \\
\vv{m}_s \rightarrow \hat{\vv{m}}_s &, \ \vv{q}_s \rightarrow \hat{\vv{q}}_s
\end{align}
where $\hat{\vv{Q}}_k$ and $\hat{\vv{P}}_k$ have extended continuous spectra defined on all of $\mathbb{R}$, $\hat{\vv{n}}_j$ and $\hat{\vv{m}}_s$ have integer-valued (discrete) spectra, and $\hat{\vv{\phi}}_j$ and $\hat{\vv{q}}_s$ have compact/periodic spectra defined on the interval $\mathbb{R} \bmod{2 \pi}$. 

The defining ``commutation" relations for each set conjugate pairs of operators are then given by: 
\begin{align}
\left[ \hat{\vv{Q}}_k, \hat{\vv{P}}_k \right] &= i \hbar \hat{\mathbf{I}}_{kk} \\
e^{i \hat{\vv{\phi}}_j^T} \text{diag}(\hat{\vv{n}}_j) e^{-i\hat{\vv{\phi}}_j} &= \text{diag}(\hat{\vv{n}}_j) - \hat{\mathbf{I}}_{jj} \\
e^{i \hat{\vv{q}}_s^T} \text{diag}(\hat{\vv{m}}_s)  e^{-i \hat{\vv{q}}_s} &=  \text{diag}(\hat{\vv{m}}_s) - \hat{\mathbf{I}}_{ss}
\end{align}

We have three types of operator pairs. The first promotes the Poisson bracket to the standard canonical commutation relation for extended-spectra modes:
\begin{align}
[\hat{Q},\hat{P}] = i \hbar
\end{align}

The second and third represent discrete charge and discrete flux modes, respectively, and possess modified ``commutation relations" of:
\begin{align}
e^{i\hat{\phi}} \hat{n} e^{-i\hat{\phi}} = \hat{n} - 1 \\
e^{i\hat{q}} \hat{m} e^{-i\hat{q}} = \hat{m} - 1 
\end{align}

These are standard equations that define the inter-relation between a discrete operator and its compact conjugate \cite{devoret_does_2021, mooij_superconducting_2006}, reflecting the fact that the compact variables $\hat{\phi}$ and $\hat{q}$ always appear as the arguments of periodic functions in the Hamiltonian. This periodicity is equivalent to the discreteness of $\hat{n}$ and $\hat{m}$, which are incremented by integer values through the action of $e^{i\hat{\phi}}$ and $e^{i\hat{q}}$, respectively.

Note that we do not write down commutation relations for the conjugate pairs of integer-spectrum operators $({\hat{\vv{\Phi}}_k, \hat{\vv{\Pi}}_k})$, as they drop out of the Hamiltonian (Eq. \ref{eq:HamiltonianPenultimate}).

We demonstrate this quantization procedure with an example circuit in Section \ref{subsec:FluxoniumWithPhaseSlips} of the Main Text.

\section{Circuit decomposition} \label{sec:AppendixD}

In this Appendix, we demonstrate a method to decompose circuits into an equivalent ``fundamental" form, which separates out the harmonic and free modes from the rest of the circuit. For Josephson junction circuits, this decomposition arranges the junctions into a tree, whose nodes and loops are coupled to a set of auxiliary harmonic modes. The topology of the simplified circuit is encoded in the edge network matrix, obtained by performing a change of basis on the node-loop network matrix \cite{schrijver_theory_1998,wolsey_integer_2014}. The decomposition is performed by applying basis transformations to this matrix---for which we give a visual interpretation in our tree-cotree notation. We apply this fundamental decomposition to classify equivalent circuits by their network matrix signature, which can aid in efforts to enumerate superconducting circuits with equivalent Hamiltonians \cite{weissler_enumeration_2024}. In addition, we discuss how to take the zero-inductance limit of junction loops and the zero-capacitance limit of phase slip nodes, providing a generalization of work found in \cite{you_circuit_2019}.

\subsection{Decomposition to the edge network matrix} \label{subsec:BranchDecomposition}

We recall from Appendix \ref{subsec:ChangeOfBasis2} that the one effect of an integer-valued basis transformation $\mathbf{U}$ and $\mathbf{W}$, with coordinate transformations
\begin{align}
\vv{\Phi}_n &\rightarrow \mathbf{U}^{T^{-1}} \vv{\Phi}_n  \\
\vv{Q}_l &\rightarrow \mathbf{W}^{T^{-1}} \vv{Q}_l 
\end{align}
is to perform the following operations on the circuit's constituent incidence, loop, and network matrices:
\begin{align}
\mathbf{A}_J & \rightarrow \mathbf{U} \mathbf{A}_J \\
\mathbf{B}_S & \rightarrow \mathbf{W} \mathbf{B}_S \\
\mathbf{\Omega} & \rightarrow \mathbf{U} \mathbf{\Omega} \mathbf{W}^T 
\end{align}

Thus the matrix $\mathbf{U}$ perform simultaneous row operations on $\mathbf{A}_J$ and $\mathbf{\Omega}$, and the matrix $\mathbf{W}$ applies column operations to $\mathbf{B}_S^T$ and $\mathbf{\Omega}$. Note that (as previously discussed in Appendix \ref{subsec:ChangeOfBasis2}) these operations affect other quantities as well, but if as $\mathbf{U}$ and $\mathbf{W}$ are integer-valued there are no further topological effects.

We now transform the node-loop network matrix to the edge (branch) network matrix by performing particular basis changes: the reverse of those shown in Eqs. \ref{eq:NodeFluxDefined} and \ref{eq:LoopChargeDefined}. This new basis consists of the branch flux variables of the capacitive spanning tree and the branch charge variables of the inductive spanning cotree, with transformation matrices $\mathbf{U} = \mathbf{A}_{C_\mathcal{T}}^T$ and $\mathbf{W} = \mathbf{B}_{L_\mathcal{T}}^T$:
\begin{align}
\vv{\Phi}_n &\rightarrow \mathbf{A}_{C_\mathcal{T}}^T  \vv{\Phi}_n  =  \vv{\Phi}_{C_\mathcal{T}} \\
\vv{Q}_l &\rightarrow \mathbf{B}_{L_\mathcal{T}}^T \vv{Q}_l=\vv{Q}_{{L}_\mathcal{T}}
\end{align}
The interpretation of these quantities and the accompanying tree-cotree notation are elaborated upon in Appendix \ref{subsec:NodeFluxLoopChargeTreeCotree}, as well as in Fig. \ref{fig:PivotIllustration}. In our abbreviated tree-cotree circuit diagrams, we denote spanning tree capacitive edges with the Josephson junction symbol if they lie across one, and with the capacitor symbol otherwise. Similarly, we represent inductive segments with either a phase slip emblem if there is flux tunneling across them, or else we notate them as linear inductors. In this basis, the capacitance matrix now details the coupling between tree branch fluxes, while the inductance matrix does the same for the cotree edge charges---making clearer the duality between capacitance and inductance.

As in Appendix \ref{subsec:NodeFluxLoopChargeTreeCotree}, note that the following procedure is possible because we have required every junction loop to have some inductance and every phase slip island to have some capacitance. $\mathbf{A}_{C_\mathcal{T}}$ is the incidence matrix of a capacitive spanning tree of the graph's capacitive nodes. The first $J$ columns correspond to the junction incidence matrix $\mathbf{A}_J$. The last $C'$ columns correspond to an arbitrarily chosen set of capacitive connections that complete the spanning tree. Similarly $\mathbf{B}_{L_\mathcal{T}}$ represents the loop matrix of a spanning cotree of the graph's inductive edges, and its first $S$ columns represent the loop matrix of the phase slip edges $\mathbf{B}_S$. As before, the last $L'$ columns represent non-phase slip inductors that fill out the spanning cotree. Symbolically, we can express these incidence and loop matrices as:
\begin{align}
    \mathbf{A}_{C_\mathcal{T}} &=
    \begin{bmatrix}
    \mathbf{A}_J &
    \mathbf{A}_{C'} 
    \end{bmatrix} \\
    \mathbf{B}_{L_\mathcal{T}} &=
    \begin{bmatrix}
    \mathbf{B}_S &
    \mathbf{B}_{L'} 
    \end{bmatrix}
\end{align}

Because of the tree-cotree structure, both $\mathbf{A}_{C_\mathcal{T}}^{-1}$ and $\mathbf{B}_{L_\mathcal{T}}^{-1}$ exist, are integer-valued \cite{bapat_graphs_2014}, and apply the following transformations to the topological matrices of the system:
\begin{align}
\mathbf{A}_J &\rightarrow \mathbf{A}_{C_\mathcal{T}}^{-1} \mathbf{A}_J =
\begin{bmatrix}
\mathbf{I}_{JJ} \\
\mathbf{0}_{C'J}
\end{bmatrix} \\
\mathbf{B}_S &\rightarrow \mathbf{B}_{L_\mathcal{T}}^{-1} \mathbf{B}_S =
\begin{bmatrix}
\mathbf{I}_{SS} \\
\mathbf{0}_{L'S}
\end{bmatrix} \\
\mathbf{\Omega} & \rightarrow \mathbf{A}_{C_\mathcal{T}}^{-1} \mathbf{\Omega} \mathbf{B}_{L_\mathcal{T}}^{T^{-1}} = \mathbf{\Omega}_E
\end{align}
Thus in this edge (branch) basis $\mathbf{A}_J$ and $\mathbf{B}_S$ are reduced identity matrices with additional rows of $0$, while $\mathbf{\Omega}$ is transformed into $\mathbf{\Omega}_E$, the fundamental cutset matrix of the system, which we will refer to as the ``edge network matrix."

This can be seen by substituting Eq. \ref{eq:NetworkMatrixIncidenceLoop}:
\begin{align}
\mathbf{\Omega}_E &= \mathbf{A}_{C_\mathcal{T}}^{-1} \mathbf{\Omega} \mathbf{B}_{L_\mathcal{T}}^{T^{-1}} \nonumber \\
&= \mathbf{A}_{C_\mathcal{T}}^{-1} \mathbf{A}_{L_\mathcal{T}} \mathbf{B}_{L_\mathcal{T}}^T \mathbf{B}_{L_\mathcal{T}}^{T^{-1}} \nonumber \\
&=\mathbf{A}_{C_\mathcal{T}}^{-1} \mathbf{A}_{L_\mathcal{T}} 
\end{align}
The final line is a common graph-theoretic expression that is equal to the fundamental cutset matrix, when $\mathbf{A}_{C_\mathcal{T}}$ represents the incidence matrix of a tree and $\mathbf{A}_{L_\mathcal{T}}$ the incidence matrix of a cotree \cite{bapat_graphs_2014, burkard_multilevel_2004}.

We now describe the meaning of this fundamental cutset matrix. Note that for a directed graph $G$ with a spanning tree $\mathcal{T}$, removing each $t_i \in \mathcal{T}$ cuts the tree into two disjoint components, one connected to the negative terminal of $t_i$ and one to its positive terminal. If the graph then contains additional non-tree (cotree) edges $E$, then each edge $e_j \in E$ is in the cutset of $t_i$ if its terminals touch the two disjoint components. 

Here our tree edges $\mathcal{T}$ consist of the capacitive edges and the cotree edges $E$ are the inductive edges. This allows us write the definition of the fundamental cutset matrix as:
\begin{align} \label{eq:NetworkMatrixCutset}
\Omega_{E_{ij}} =
\begin{cases} 
      1 & \text{inductive edge j in fundamental cutset} \\
      & \text{of capacitive edge i (same orientation)} \\
      -1 & \text{inductive edge j in fundamental cutset} \\
      &\text{of capacitive edge i (opposite orientation)} \\
      0 & \text{inductive edge j not in fundamental} \\
      & \text{of capacitive edge i}
\end{cases}
\end{align}

An essential result in graph theory says that if $\Omega_{E}$ is the fundamental cutset matrix of a system, then $-\Omega_{E}^T$ is its fundamental loop matrix \cite{bapat_graphs_2014}. In essence, this statement arises from the orthogonality of loops and cutsets. Thus we can equivalently define $\Omega_{E}$ (up to a minus sign) in the language of loops as:
\begin{align} \label{eq:NetworkMatrixLoop}
-\Omega_{E_{ij}} =
\begin{cases} 
      1 & \text{loop of inductive edge $j$ passes} \\
      & \text{forward through capacitive edge i} \\
      -1 & \text{loop of inductive edge $j$ passes} \\
      &\text{backward through capacitive edge i} \\
      0 & \text{loop of inductive edge $j$ does not} \\
      & \text{pass through capacitive edge i}
\end{cases}
\end{align}

In this dual loop-cutset picture, each inductive cotree edge defines a loop through the capacitive tree edges (passing through no other inductive edge), and each capacitive edge defines a cutset of inductive tree edges (containing no other capacitive edges). Our basis changes then alter the definition of the network matrix: instead of detailing how loops connect to capacitive islands, the edge network matrix now describes how loops (characterized by inductive edges) connect to the spanning tree of capacitive edges. In addition, by reducing the junction incidence and phase slip loop matrices to upper identity matrices, the topological content is now shifted largely from those matrices to the network matrix.

\subsection{Graph-theoretic edge network matrices} \label{subsec:GraphTheoreticNetworkMatrices}

In this basis, the circuit's edge network matrix coincides (up to an inconsequential minus sign) with the definition of a tree-cotree edge network matrix from linear/integer programming. Edge network matrices (just called network matrices in that field) can be understood and manipulated by noting that they thus possess the following properties (which are not necessarily true of node-loop network matrices) \cite{schrijver_theory_1998, wolsey_integer_2014}:

\begin{enumerate}[nosep]
    \item Deleting a row or column from an edge network matrix (or replacing it with all $0$) results in an edge network matrix. \label{item:NetworkProp1}
    \item Multiplying the rows or column of an edge network matrix by $-1$ results in an edge network matrix. \label{item:NetworkProp2}
    \item Taking the edge network matrix $\mathbf{\Omega}_E$ and augmenting it as $[\mathbf{\Omega}_E,\mathbf{I}]$ gives an edge network matrix. \label{item:NetworkProp3}
    \item Pivoting (through row or column operations) on a nonzero entry of an edge network matrix returns an edge network matrix. \label{item:NetworkProp4}
    \item Edge network matrices are totally unimodular (the determinant of any square submatrix is $-1$, $+1$, or $0$). \label{item:NetworkProp5}
\end{enumerate}

We give brief explanations of the intuition behind properties 1, 2, and 4. In Property \ref{item:NetworkProp1}, deleting a row corresponds to contracting the two sides of a capacitive edge into a single node, and deleting a column corresponds to removing an inductive edge (and not contracting its endpoints). With Property \ref{item:NetworkProp2}, multiplying a row by $-1$ changes the orientation of the corresponding capacitive tree edge and multiplying a column by $-1$ alters the direction of the inductive loop edge. 

Row and column pivoting (Property \ref{item:NetworkProp4}) also admits a visual interpretation. Here we define row pivoting on a nonzero element $\mathbf{\Omega}_{E_{ij}}$ to be the set of row operations on $\mathbf{\Omega}_E$ that use the $i^{\text{th}}$ row to Gaussian eliminate all other nonzero elements of the $j^{\text{th}}$ column. Similarly, column pivoting on $\mathbf{\Omega}_{E_{ij}}$ employs column operations, with linear combinations of the $j^{\text{th}}$ column used eliminate all other nonzero elements in the $i^{\text{th}}$ row.

In the tree-cotree picture, performing a row pivot on $\mathbf{\Omega}_{E_{ij}}$ (which must be nonzero) corresponds to removing capacitive edge $i$ from the tree and placing it in parallel with inductive edge $j$. Similarly, performing a column pivot on $\mathbf{\Omega}_{E_{ij}}$ involves contracting inductive cotree edge $j$ to a single vertex and then reinserting the edge in series with capacitive edge $i$.

The visual interpretation of network matrix pivoting is shown in Fig. \ref{fig:PivotIllustration} and corresponds to the symmetric difference operations that are performed during the pivot process. Since edge network matrices are closed under pivoting, the entries of the transformed matrix can only consist of $0$, $1$, and $-1$. So, during the pivoting, we are performing element-wise operations of $0 \pm 1 = \pm 1 $, $0 \pm 0 = 0 $, and  $\pm 1 \mp 1 = 0$. This binary operation is a signed symmetric difference, meaning that if exactly one element of the sum is nonzero then the final element is nonzero ($\pm 1$), while if both elements are either zero or nonzero the final sum is zero \cite{oxley_matroid_2006}.

For instance, by row pivoting a capacitive tree edge $C_i$ on an inductive cotree edge $L_j$, one can change the inductive cutset of another capacitive edges $C_k$. In particular, if both $C_i$ and $C_k$ contain $L_j$ in their cutsets, then we add or subtract a copy of the $i^{\text{th}}$ row to the $k^{\text{th}}$ row, which is a symmetric difference operation. The transformed inductive cutset of $C_k$ consists of the inductive edges that were in only one of the cutsets of $C_k$ and $C_i$. 

Moving capacitive tree edge $C_i$ to be in parallel with inductive edge $L_j$ produces the required symmetric difference operation. This equivalence can be understood by noting that the inductive edges in the cutset of $C_k$ are those that lie across $C_k$ when all other capacitive edges are contracted to single vertices. Then, by leaving $C_i$ uncontracted, we see that moving $C_i$ to lie across edge $L_j$ removes from the cutset of $C_k$ all inductive elements shared with the cutset of $C_i$, and adds in all inductive elements that were only in the cutset of $C_i$---a symmetric difference. Analogous logic shows that column pivoting corresponds to moving an inductive cotree edge $L_j$ to be in series with a capacitive tree edge $C_i$.

We note that in the edge network matrix tree-cotree picture, we can perform additional visual operations that leave the equations of motion invariant. In particular, we can arbitrarily move tree/cotree edges of the system, as long as the cutset/loop topology remains unchanged. We can thus disconnect and reconnect loops that are only connected at one vertex and share no edges, and the capacitive spanning tree can be become a forest. These types of operations fall in the domain of matroid theory \cite{oxley_matroid_2006}.

\subsection{Pivoting, structure-preserving basis transformations} 

\label{subsec:StructurePreservingTransformations}
\begin{figure*}
    \centering
  \includegraphics[width=.8\linewidth]{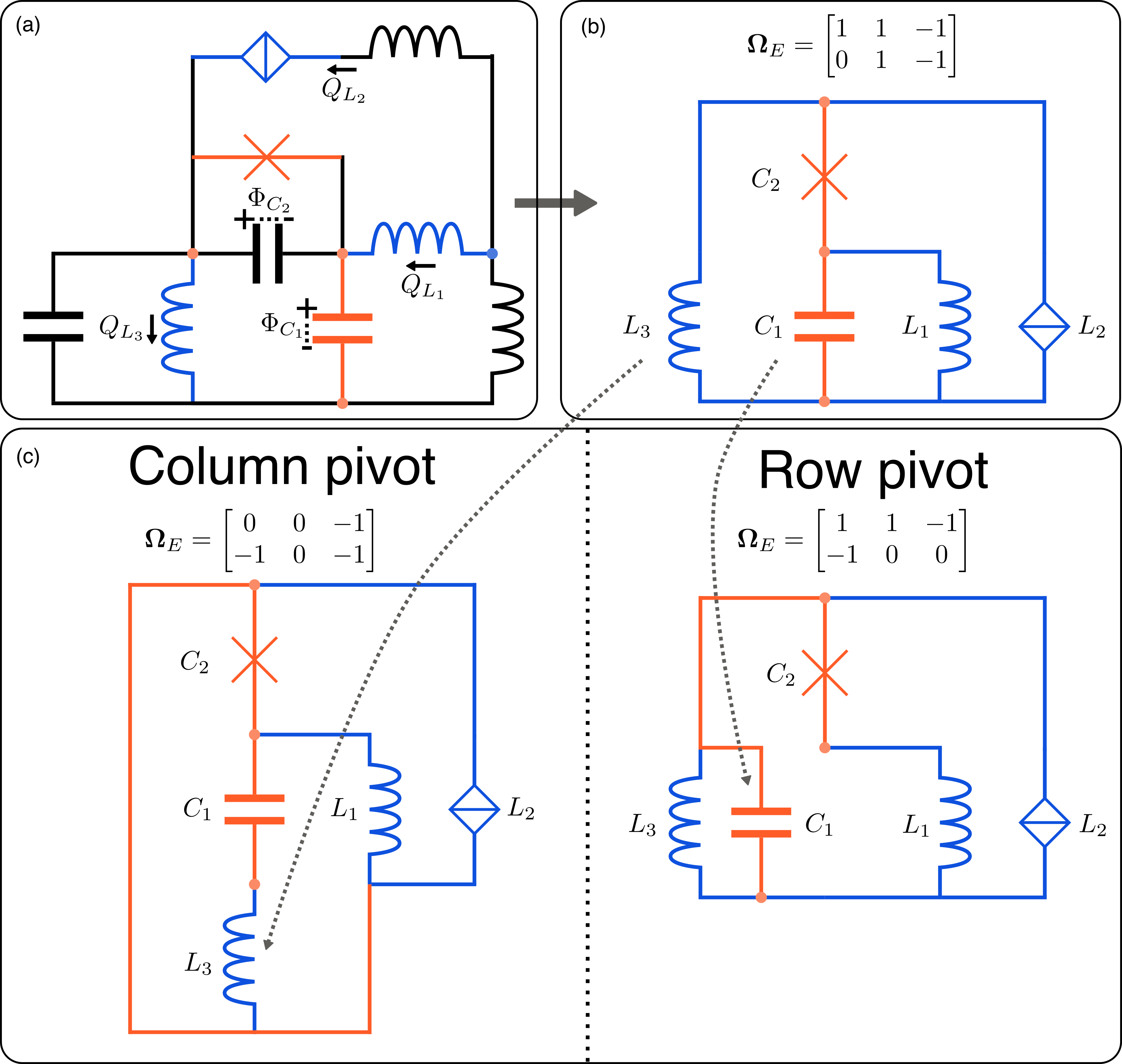}
    \caption{Visual interpretation of the pivoting procedure in the decomposition algorithm. (a) Initial circuit model with eventual spanning tree edges (orange) and cotree edges (blue). (b) Conversion of standard circuit model to tree-cotree notation. (c) Left: pivoting on element 1 of column 3, and its interpretation of moving inductor 3 to be in series with capacitor 1. Right: pivoting on element 3 of row 1, and the corresponding visual interpretation of moving capacitor 1 to be in parallel with inductor 3.}
    \label{fig:PivotIllustration}
\end{figure*}

As discussed in Appendix \ref{subsec:GraphTheoreticNetworkMatrices}, the closure of edge network matrices under linear-algebraic pivoting allows us to manipulate and decompose circuits into equivalent forms. Recall that in this edge basis, the equations of motion for the Josephson fluxes and phase slip charges are enumerated by the incidence and loop matrices (respectively):
\begin{align}
\mathbf{A}_J =
\begin{bmatrix}
\mathbf{I}_{JJ} \\
\mathbf{0}_{C'J}
\end{bmatrix} \\ %newline
\mathbf{B}_S =
\begin{bmatrix}
\mathbf{I}_{SS} \\
\mathbf{0}_{L'S}
\end{bmatrix}
\end{align}

The first $J$ rows of $\mathbf{A}_J$ now correspond to the capacitive fluxes across Josephson junctions and the last $C'$ to the fluxes across non-junction edges. Similarly, the first $S$ rows of $\mathbf{B}_S$ enumerate inductor charges along phase slip wires, and the last $L'$ give inductor charges just along bare inductors. 

So, when performing changes of basis, we are free to change the $C'$ flux variables and the $L'$ charge variables, and still have a valid equivalent circuit with unchanged $\mathbf{A}_J$ and $\mathbf{B}_S$ matrices (or Josephson junction fluxes and phase slip charges), by applying transformations of the form:
\begin{align}
\mathbf{U}  =
\begin{bmatrix}
\mathbf{I}_{JJ} &  \mathbf{U}_{JC'} \\
\mathbf{0}_{C'J} & \mathbf{U}_{C'C'}
\end{bmatrix}  \\ %newline
\mathbf{W} =
\begin{bmatrix}
\mathbf{I}_{SS} &  \mathbf{W}_{SC'} \\
\mathbf{0}_{L'S} & \mathbf{W}_{L'L'}
\end{bmatrix}  \\ %newline
\end{align}

If the transformation also maintains $\mathbf{\Omega}_E$ as an edge network matrix then we refer to it as a structure-preserving transformation. In addition to these operations, we also allow permutation within the first $J$ rows of $\mathbf{A}_J$ (and thus also the first $J$ rows of $\mathbf{\Omega}_E$) and within the first $S$ rows of $\mathbf{B}_S$ (the first $S$ columns of $\mathbf{\Omega}_E$), as well as negative multiplication of any of these rows. After these these transformations, $\mathbf{A}_J$ and $\mathbf{B}_S$ can be returned to their upper-identity form by a simple relabeling of coordinates.

We can use $\mathbf{U}$ to perform row operations (row pivoting) on $\mathbf{\Omega}_E$ and $\mathbf{W}$ to perform column operations (column pivoting) on $\mathbf{\Omega}_E$. In particular, if we break up the network matrix into the form:
\begin{align} \label{eq:EdgeNetworkBlock1}
\mathbf{\Omega}_E =
\begin{bmatrix}
\mathbf{\Omega}_{JS} & \mathbf{\Omega}_{JL'} \\
\mathbf{\Omega}_{C'S} & \mathbf{\Omega}_{C'L'} 
\end{bmatrix}
\end{align}
then with structure-preserving transformations we can perform a row pivot using any nonzero element of $\mathbf{\Omega}_{C'S}$ or $\mathbf{\Omega}_{C'L'}$, or a column pivot using a nonzero elements of $\mathbf{\Omega}_{JL'}$ or $\mathbf{\Omega}_{C'L'}$---leaving $\mathbf{A}_J$ and $\mathbf{B}_S$ invariant, and maintaining $\mathbf{\Omega}_E$ as an edge network matrix due to the closure under pivoting. We note that these pivoting operations $\mathbf{U}$ or $\mathbf{W}$ are integer-valued, containing only $0$, $+1$, and $-1$ among their elements. 

After employing structure-preserving basis transformations, $\mathbf{\Omega}_E$ encodes the topology of an equivalent circuit network. Algorithms exist to extract a circuit graph model from this network matrix \cite{schrijver_theory_1998, wolsey_integer_2014}, if one does not track the basis changes visually as shown in Fig. \ref{fig:PivotIllustration}.

\subsection{Fundamental decomposition} \label{subsec:FundamentalDecomposition}

We now describe the algorithm to reduce a circuit to an equivalent form that we call the ``fundamental decomposition." We start with the edge network matrix in block form (Eq. \ref{eq:EdgeNetworkBlock1}) and carry out a series of pivoting operations. If the submatrix $\mathbf{\Omega}_{C'L'}$ has rank $r$, then by a sequential set of structure-preserving operations, we can pivot on its nonzero rows and columns to reduce it a block matrix with the identity as a submatrix in the lower right corner. This transforms the edge network matrix to:
\begin{align} \label{eq:OmegaEFirstTransformation}
\mathbf{\Omega}_E 
\rightarrow
\begin{bmatrix}
\mathbf{\Omega}_{JS} & \mathbf{\Omega}_{J \lambda} & \mathbf{0}_{Jr} \\
\mathbf{\Omega}_{\kappa S} &  \mathbf{0}_{\kappa \lambda} & \mathbf{0}_{\kappa r}  \\ 
\mathbf{0}_{rS} & \mathbf{0}_{r \lambda}  & \mathbf{I}_{rr} 
\end{bmatrix}
\end{align}
where $\mathbf{\Omega}_{JS}$ has been altered in the pivot operation. Now we wish to reduce $\mathbf{\Omega}_{\kappa S}$ to a matrix of full row rank and $\mathbf{\Omega}_{J \lambda}$ to one of full column rank. There are several ways to do this while still preserving the network matrix structure. 

One option is to successively pivot on nonzero rows of $\mathbf{\Omega}_{\kappa S}$ and columns of $\mathbf{\Omega}_{J \lambda}$. $\mathbf{\Omega}_{JS}$ may be altered by this pivot operation but will remain a network matrix by closure under pivoting. We can repeat the procedure until $\mathbf{\Omega}_{\kappa S}$ is transformed into into nonzero matrix of full row rank, along with an auxiliary set of zero rows. Similarly, $\mathbf{\Omega}_{J \lambda}$ will be transformed into a nonzero matrix of full column rank and a set of zero columns.

Alternatively, we can pick a maximal linearly independent number of rows of $\mathbf{\Omega}_{\kappa S}$ and number of columns of $\mathbf{\Omega}_{J \lambda}$. Then we can express the remaining rows/columns of these matrices in terms of these linearly independent sets, performing row/column operations to set all of the linearly dependent rows/columns to $0$. Since edge network matrices are closed under deletion (or setting to zero) of a set of rows and columns, $\mathbf{\Omega}_E$ will remain an edge network matrix. These operations correspond to contracting the linearly-dependent capacitive edges to single vertices and removing linearly dependent inductive edges. In this case, $\mathbf{\Omega}_{JS}$ remains unchanged.

Using either procedure, the reduced form of $\mathbf{\Omega}_{\kappa S}$ will have a set of $\alpha$ zero rows and $\mathbf{\Omega}_{J \lambda}$ will have a set of $\beta$ zero columns. We then perform a structure-preserving transformation that permutes these zero rows to be the last rows of the network matrix and the zero columns to be the last columns. The network transformed network matrix can be written as:
\begin{align} \label{eq:AppFundamentalFormOmegaFV1}
\mathbf{\Omega}_E 
\rightarrow
\begin{bmatrix}
\mathbf{\Omega}_{JS} &\mathbf{\Omega}_{Jf} &  \mathbf{0}_{Jr}& \mathbf{0}_{J \beta}  \\
\mathbf{\Omega}_{pS}  & \mathbf{0}_{pf}  &  \mathbf{0}_{pr} & \mathbf{0}_{p \beta} \\
\mathbf{0}_{rS}  & \mathbf{0}_{rf} & \mathbf{I}_{rr}  & \mathbf{0}_{r \beta} \\
\mathbf{0}_{\alpha S}  & \mathbf{0}_{\alpha f} & \mathbf{0}_{\alpha r}  & \mathbf{0}_{\alpha \beta} 
\end{bmatrix}
\end{align}

This equation represents what we call the ``fundamental decomposition" of the circuit. Here, the network matrix separates itself into several different sectors. $\mathbf{\Omega}_{JS}$ represents phase slip inductor degrees of freedom connected in loops to Josephson junction capacitors, $\mathbf{\Omega}_{Jf}$ represents standard inductors connected to Josephson junction capacitors, and $\mathbf{\Omega}_{pS}$ represents standard capacitors connected to phase slip inductors. The set of standard inductors connected to standard capacitors is given by: $\mathbf{I}_{rr}$, wherein each inductor lies in a loop containing a unique capacitive spanning tree branch. The $\alpha$ degrees of freedom stand for isolated capacitive islands and the $\beta$ modes for isolated inductive loops. Both of them result in free modes that can be removed as Appendix \ref{subsec:FreeModes}. As described in Appendix \ref{subsec:GraphTheoreticNetworkMatrices}, we can visualize our basis transformations and the connectivity of the transformed circuits. Then, by reading off an incidence matrix for the capacitive spanning tree (compatible with the network topology), we can shift from a branch flux picture to one of node flux using Eq. \ref{eq:NodeFluxDefined}.

The fundamental circuit form represents a maximally simplified equivalent circuit model. In particular, it separates out the harmonic modes (indexed with $r$) and the free modes (indexed with $\alpha$ and $\beta$). Those harmonic LC-oscillators are the only places in the circuit where a standard (non-phase slip) inductor is in the same loop as a standard (non-Josephson junction) capacitive spanning tree branch. In addition, the circuit reduces to a minimal number of additional cotree inductive branches $f$ (that lie in cutsets of Josephson junction edges) and additional tree capacitive branches $p$ (which lie in loops with phase slip edges). We note also that since $\mathbf{\Omega}_{Jf}$ has full column rank we have $f \leq J$ and since $\mathbf{\Omega}_{pS}$ has full row rank we have $p \leq S$.

\subsection{Classifying circuits by equivalent fundamental network matrices}

The fundamental decomposition generates a reduced network matrix $\mathbf{\Omega}_E$, whose signature makes it possible to classify equivalent circuits by performing a set of structure-preserving transformations. 

Recall that a basis transformation takes: $\mathbf{\Omega}_E \rightarrow \mathbf{U} \mathbf{\Omega}_E \mathbf{W}^T$, where $\mathbf{U}$ transforms the capacitive flux variables and $\mathbf{W}$ the inductive charge. As discussed in As discussed in Appendix \ref{subsec:StructurePreservingTransformations}, the transformations that leave the topological structure of the system invariant are a subset of permutation operations (with minus signs allowed for any element, which correspond to switching the direction of the element) and pivot operations. 

The set of applicable permutations can be broken down into several subclasses. Operations labeled by $\mathbf{U}_J^\text{per}$ permute the first $J$ rows of $\mathbf{\Omega}_E$, while those labeled $\mathbf{U}_p^\text{per}$ permute the next $p$. These transformations can be thought of as relabeling the Junctions and capacitive branches of the system, as well as potentially reversing their direction. Similarly, $\mathbf{W}_S^\text{per}$ is a permutation in the first $S$ columns of $\mathbf{\Omega}_E$, while $\mathbf{W}_f^\text{per}$ acts in the next $f$ columns. These operations can thought of as swapping the labels and directions of the phase slip inductors and standard inductors, respectively.

Likewise, there are a set of row and column pivot operations that can be used to convert between equivalent network matrices. We can pivot (as a set of row operations) on a nonzero element in the submatrix $\mathbf{\Omega}_{pS}$ of $\mathbf{\Omega}_E$, as given in Eq. \ref{eq:AppFundamentalFormOmegaFV1}. We can also pivot via column operations on a nonzero element of $\mathbf{\Omega}_{Jf}$. These pivoting operations do not change the labeling of any junction capacitor or phase slip inductor. As referenced in Appendix \ref{subsec:GraphTheoreticNetworkMatrices}, pivoting preserves the character of $\mathbf{\Omega}_E$ as a graph-theoretic edge network matrix. As was also discussed in that Appendix, using our tree-cotree notation, a row pivot corresponds to moving a capacitive tree branch across an inductive cotree edge in its cutset, while a column pivot contracts an inductive cotree edge to a single node and then reinserts it in series with a capacitor that lies in the inductor's loop.

We note that by applying a successive series of these permutation and pivot transformations, Eq. \ref{eq:AppFundamentalFormOmegaFV1} can be written in a more specific form:
\begin{align}  \label{app:FundamentalFormOmegaFV2}
\mathbf{\Omega}_E \rightarrow \begin{bmatrix}
\mathbf{\Omega}_{J'S'} &  \mathbf{0}_{J' p} & \mathbf{\Omega}_{J'f} & \mathbf{0}_{J' r} & \mathbf{0}_{J' \beta}  \\ %newline 
\mathbf{0}_{f S'} & \mathbf{0}_{fp} & \mathbf{I}_{ff} & \mathbf{0}_{f r} & \mathbf{0}_{f \beta}  \\%newline 
\mathbf{\Omega}_{pS'} &  \mathbf{I}_{pp} & \mathbf{0}_{pf} & \mathbf{0}_{pr} &\mathbf{0}_{p \beta} \\ %newline 
\mathbf{0}_{rS'} & \mathbf{0}_{rp} & \mathbf{0}_{rf} &  \mathbf{I}_{rr} & \mathbf{0}_{r \beta} \\ %newline 
\mathbf{0}_{\alpha S'} & \mathbf{0}_{\alpha p} & \mathbf{0}_{\alpha f} & \mathbf{0}_{\alpha r} & \mathbf{0}_{\alpha \beta} 
\end{bmatrix}
\end{align}

In this matrix we have used pivoting and permutation to separate out two identity submatrices, $\mathbf{I}_{pp}$ and $\mathbf{I}_{ff}$. This separation allows for some simplification in the quantization procedure, as well as a aiding us in recognizing equivalent network matrices.

For circuits represented by two fundamentally decomposed network matrices $\mathbf{\Omega}_{E_1}$ and $\mathbf{\Omega}_{E_2}$, we say that they are topologically equivalent if one can be transformed into the other through a sequence of structure-preserving permutation and pivot transformations. We also note that we usually wish to consider network matrices without free modes included, which removes the $\alpha$ rows and $\beta$ columns (of all zeros) from  $\mathbf{\Omega}_E$, as in Appendix \ref{subsec:FreeModes}.

This type of procedure can aid in enumerating possible superconducting circuits, as in \cite{weissler_enumeration_2024}. We give some simple applications of this methodology in Section \ref{subsec:ClassificationEquivalentCircuits} of the Main Text.

\subsection{Fundamental circuit form of Junction-only circuits}

After the elimination of free modes, for circuits with no phase slip inductors, we can write down the fundamental network matrix (Eq. \ref{eq:AppFundamentalFormOmegaFV1}) as:
\begin{align}
\mathbf{\Omega}_E 
=
\begin{bmatrix}
\mathbf{\Omega}_{Jf} & \mathbf{0}_{Jr}  \\
\mathbf{0}_{rf} & \mathbf{I}_{rr} 
\end{bmatrix}
\end{align}

Here, the fundamental form of the circuit is easier to interpret. In standard circuit notation, it corresponds to a tree of $J$ Josephson junctions (so $J+1$ nodes), with $f \leq J$ inductor edges (whose connectivity with the junctions is encapsulated in $\mathbf{\Omega}_{Jf}$) running between nodes of the tree. This tree is then capacitively and inductively coupled to a set of $r$ harmonic oscillators. As in Appendix \ref{subsec:CapacitanceInductanceDiagonalization}, the harmonic oscillators can be mutually decoupled from each other through an additional basis transformation.

We can then to classify superconducting circuits by the number of Josephson junctions in the tree, the loop topology of the tree, and number of auxiliary harmonic modes.

\subsection{Quantization from decomposition}

The fundamental form of the network matrix also provides a straightforward way to carry our quantization procedure (shown in Appendix \ref{subsec:Quantization1} and \ref{subsec:Quantization2}). In addition, it allows us to more specifically classify the system's extended degrees of freedom, by the types of cosine terms in which they appear in the Hamiltonian.

Starting with Eq. \ref{app:FundamentalFormOmegaFV2} and assuming we have removed the free modes of the circuit, we obtain a fundamental network matrix of:
\begin{align}
\mathbf{\Omega}_E = \begin{bmatrix}
\mathbf{\Omega}_{J'S'} &  \mathbf{0}_{J' p} & \mathbf{\Omega}_{J'f} & \mathbf{0}_{J' r} \\ %newline 
\mathbf{0}_{f S'} & \mathbf{0}_{fp} & \mathbf{I}_{ff} & \mathbf{0}_{f r} \\%newline 
\mathbf{\Omega}_{pS'} &  \mathbf{I}_{pp} & \mathbf{0}_{pf} & \mathbf{0}_{pr}  \\ %newline 
\mathbf{0}_{rS'} & \mathbf{0}_{rp} & \mathbf{0}_{rf} &  \mathbf{I}_{rr}
\end{bmatrix}
\end{align}
Its corresponding junction and phase slip matrices will be given by:
\begin{align}
\mathbf{A}_J =
\begin{bmatrix}
\mathbf{I}_{J'J'} & \mathbf{0}_{J'f} \\
\mathbf{0}_{fJ'} & \mathbf{I}_{ff} \\
\mathbf{0}_{pJ'} & \mathbf{0}_{pf} \\
\mathbf{0}_{rJ'} & \mathbf{0}_{rf}
\end{bmatrix} \\
\mathbf{B}_S =
\begin{bmatrix}
\mathbf{I}_{S'S'} & \mathbf{0}_{S'p} \\
\mathbf{0}_{pS'} & \mathbf{I}_{pp} \\
\mathbf{0}_{fS'} & \mathbf{0}_{fp} \\
\mathbf{0}_{rS'} & \mathbf{0}_{rp}
\end{bmatrix}
\end{align}

As before, the key goal of the procedure will be to transform $\mathbf{\Omega}_E$ to have the identity as its upper left submatrix---through a series of basis changes---such that it can be quantized. In this section we only discuss the effects of the basis transformations on the topological matrices of $\mathbf{\Omega}_E$, $\mathbf{A}_J$, and $\mathbf{B}_S$. However, as noted in Appendix \ref{subsec:ChangeOfBasis2}, there are additional effects on the other constituent matrices and vectors of the system. 

First we note that we can perform a row transformation that uses $\mathbf{I}_{ff}$ to eliminate $\mathbf{\Omega}_{J'f}$ in $\mathbf{\Omega}_E$ and a column transformation that uses $\mathbf{I}_{pp}$ to eliminate $\mathbf{\Omega}_{pS'}$. This changes the matrices $\mathbf{\Omega}_E$ $\mathbf{A}_J$, and $\mathbf{B}_S$ to the following forms:
\begin{align} \label{eq:AppSimplifiedNetworkMatrix}
\mathbf{\Omega}_E = \begin{bmatrix}
\mathbf{\Omega}_{J'S'} &  \mathbf{0}_{J' p} & \mathbf{0}_{J'f} & \mathbf{0}_{J' r}   \\ %newline 
\mathbf{0}_{f S'} & \mathbf{0}_{fp} & \mathbf{I}_{ff} & \mathbf{0}_{f r} \\%newline 
\mathbf{0}_{pS'} &  \mathbf{I}_{pp} & \mathbf{0}_{pf} & \mathbf{0}_{pr}  \\ %newline 
\mathbf{0}_{rS'} & \mathbf{0}_{rp} & \mathbf{0}_{rf} &  \mathbf{I}_{rr} 
\end{bmatrix}
\end{align}
\begin{align}
\mathbf{A}_J =
\begin{bmatrix}
\mathbf{I}_{J'J'} & -\mathbf{\Omega}_{J'f} \\
\mathbf{0}_{fJ'} & \mathbf{I}_{ff} \\
\mathbf{0}_{pJ'} & \mathbf{0}_{pf} \\
\mathbf{0}_{rJ'} & \mathbf{0}_{rf} 
\end{bmatrix} \\
\mathbf{B}_S =
\begin{bmatrix}
\mathbf{I}_{S'S'} & -\mathbf{\Omega}_{pS'}^T \\
\mathbf{0}_{pS'} & \mathbf{I}_{pp} \\
\mathbf{0}_{fS'} & \mathbf{0}_{fp} \\
\mathbf{0}_{rS'} & \mathbf{0}_{rp} 
\end{bmatrix}
\end{align}

To simplify $\mathbf{\Omega}_E$ further, we note that there exist invertible transformations $\mathbf{X}$ and $\mathbf{Y}$ such that:
\begin{align}
\mathbf{X} \mathbf{\Omega}_{J'S'} \mathbf{Y}^T = 
\begin{bmatrix}
    \mathbf{I}_{dd} & \mathbf{0}_{ds} \\
    \mathbf{0}_{jd} & \mathbf{0}_{js}
\end{bmatrix}
\end{align}
where $d$ is the rank of $\mathbf{\Omega}_{J'S'}$. In the above equation, $j = J'-d$ and $s = S'-d$. Applying these operations to the entire network matrix
\begin{align}
\mathbf{\Omega}_E \rightarrow
\begin{bmatrix}
\mathbf{X} & \mathbf{0} \\
\mathbf{0} & \mathbf{I}
\end{bmatrix}
\mathbf{\Omega}_E
\begin{bmatrix}
\mathbf{Y} & \mathbf{0} \\
\mathbf{0} & \mathbf{I}
\end{bmatrix}^T
\end{align}
then gives that:
\begin{align}
\mathbf{\Omega}_E = \begin{bmatrix}
\mathbf{I}_{dd} & \mathbf{0}_{d s} &  \mathbf{0}_{d p} & \mathbf{0}_{df} & \mathbf{0}_{d r}   \\ %newline 
\mathbf{0}_{j d} & \mathbf{0}_{j s} & \mathbf{0}_{jp} & \mathbf{0}_{jf} &  \mathbf{0}_{j r} \\ %newline 
\mathbf{0}_{f d} &  \mathbf{0}_{f s} & \mathbf{0}_{fp} & \mathbf{I}_{ff} & \mathbf{0}_{f r} \\%newline 
\mathbf{0}_{pd} & \mathbf{0}_{p s} & \mathbf{I}_{pp} & \mathbf{0}_{pf} & \mathbf{0}_{pr}  \\ %newline 
\mathbf{0}_{rd} & \mathbf{0}_{r s} & \mathbf{0}_{rp} & \mathbf{0}_{rf} &  \mathbf{I}_{rr} 
\end{bmatrix}
\end{align}

If we now label: 
\begin{align}
\mathbf{G} =  -\mathbf{X} \mathbf{\Omega}_{J'f}\\
\mathbf{H} =  -\mathbf{X} \mathbf{\Omega}_{p S'}^T
\end{align}

then the corresponding transformed  $\mathbf{A}_J$ and $\mathbf{B}_S$ are:
\begin{align}
\mathbf{A}_J =
\begin{bmatrix}
\mathbf{X}_{d J'} & \mathbf{G}_{d f} \\
\mathbf{X}_{j J'} & \mathbf{G}_{j f} \\
\mathbf{0}_{fJ'} & \mathbf{I}_{ff} \\
\mathbf{0}_{pJ'} & \mathbf{0}_{pf} \\
\mathbf{0}_{rJ'} & \mathbf{0}_{rf}
\end{bmatrix} \\
\mathbf{B}_S =
\begin{bmatrix}
\mathbf{Y}_{d S'} & \mathbf{H}_{d p} \\
\mathbf{Y}_{s S'} & \mathbf{H}_{s p} \\
\mathbf{0}_{pS'} & \mathbf{I}_{pp} \\
\mathbf{0}_{fS'} & \mathbf{0}_{fp} \\
\mathbf{0}_{rS'} & \mathbf{0}_{rp}
\end{bmatrix}
\end{align}

Performing a set of row and column permutations changes these matrices to the forms:
\begin{align}
\mathbf{\Omega}_E = \begin{bmatrix}
\mathbf{I}_{dd} & \mathbf{0}_{d f} &  \mathbf{0}_{d p} & \mathbf{0}_{dr} & \mathbf{0}_{d s}   \\ %newline 
\mathbf{0}_{f d} & \mathbf{I}_{f f} & \mathbf{0}_{fp} & \mathbf{0}_{fr} &  \mathbf{0}_{f s} \\ %newline 
\mathbf{0}_{p d} &  \mathbf{0}_{p f} & \mathbf{I}_{pp} & \mathbf{0}_{pr} & \mathbf{0}_{p s} \\%newline 
\mathbf{0}_{rd} & \mathbf{0}_{r f} & \mathbf{0}_{rp} & \mathbf{I}_{rr} & \mathbf{0}_{rs}  \\ %newline 
\mathbf{0}_{jd} & \mathbf{0}_{j f} & \mathbf{0}_{jp} & \mathbf{0}_{jr} &  \mathbf{0}_{js} 
\end{bmatrix}
\end{align}
\begin{align}
\mathbf{A}_J =
\begin{bmatrix}
\mathbf{X}_{d J'} & \mathbf{G}_{d f} \\
\mathbf{0}_{fJ'} & \mathbf{I}_{ff} \\
\mathbf{0}_{pJ'} & \mathbf{0}_{pf} \\
\mathbf{0}_{rJ'} & \mathbf{0}_{rf} \\
\mathbf{X}_{j J'} & \mathbf{G}_{j f}
\end{bmatrix} \label{eq:AppAJsimplified} \\
\mathbf{B}_S =
\begin{bmatrix}
\mathbf{Y}_{d S'} & \mathbf{H}_{d p} \\
\mathbf{0}_{fS'} & \mathbf{0}_{fp} \\
\mathbf{0}_{pS'} & \mathbf{I}_{pp} \\
\mathbf{0}_{rS'} & \mathbf{0}_{rp} \\
\mathbf{Y}_{s S'} & \mathbf{H}_{s p} 
\end{bmatrix} \label{app:BSsimplified} 
\end{align}

Now that the network matrix has been transformed into a form that contains the identity in the upper left corner and and zeros elsewhere, we can use it to quantize the equations of motion (Eqs. \ref{eq:AppContinuityStandard} and \ref{eq:AppFaradayStandard}) as detailed in Appendices \ref{subsec:Quantization1} and \ref{subsec:Quantization2}. In those earlier discussions, we noted that there exist three types of conjugate degrees of freedom with their own commutation relations: discrete charge modes (with index $j$), discrete flux modes (with index $s$), and extended variable modes (with index $k$). These operators appear in a Hamiltonian (Eq. \ref{eq:AppFinalHamiltonian}) with two distinct types of terms: quadratic and cosine. For the discrete charge and flux modes, only the conjugate variables (charge and flux, respectively) participate in the quadratic, while for the extended modes both degrees of freedom do.

We note that the earlier Hamiltonian and Eqs. \ref{eq:AppAJsimplified} and \ref{app:BSsimplified} demonstrate how each type of mode participates in the cosine terms of the Hamiltonian. Up to absolute value, the participation of a mode in the Josephson junction cosines is given by the presence of nonzero rows of $\mathbf{A}_J$, and participation in the phase slip cosines by nonzero rows of $\mathbf{B}_S$.

Thus we see that since $\mathbf{X}$ and $\mathbf{Y}$ have full rank, their submatrices $\mathbf{X}_{jJ'}$ and $\mathbf{Y}_{sS'}$ have full row rank, which means that the compact flux of freedom of each discrete charge mode participates in at least one junction cosine and the compact charge of a discrete flux mode in a phase slip cosine---as would be expected \cite{devoret_does_2021,mooij_superconducting_2006}.

We also note that this procedure divides up the extended-spectrum conjugate pairs $k$ into four subclasses: $d$, $f$, $p$, and $r$. These degrees of freedom are indexed by their participation in cosines of the system. The $d$ (or dual) degrees of freedom participate in both junction and phase slip cosines, the $f$ degrees of freedom have only flux in junction cosines, the $p$ degrees of freedom have only charge in phase slip cosines, and the $r$ (or resonator) degrees of freedom participate in neither. We note that we label the $f$ degree of freedom for its similarity to that of an ideal fluxonium qubit the $p$ for the phase slip analogue of that qubit.

\subsection{Limits of zero inductance, capacitance} \label{subsec:ZeroCapInd}

In this work, we have considered every Josephson junction loop to have some finite self-inductance and every phase slip island some self-capacitance. However, superconducting circuits are often analyzed in limits where these quantities go to zero. One example is the tunable transmon qubit, which contains a two-junction SQUID loop, which commonly modeled without an inductor. However, removing these circuit elements creates a question of how to allocate external charges and fluxes to the remaining circuit elements. 

For systems with capacitors, inductors, and Josephson junctions, it was shown in \cite{you_circuit_2019} how to model circuits with vanishing loop inductance. In particular, there is a unique prescription to allocate the external flux of each loop to the fluxes of the Josephson junctions in that loop. In this case, the the sum of all the junction fluxes around the loop will possess a term equal to the externally-applied flux.

Here we give a presentation of this procedure in a flux-charge symmetric framework, generalizing this junction/inductor procedure to include the analogous phase slip/capacitor case.

First we note that we do not consider limits where capacitance vanishes in parallel with a Josephson junction or inductance vanishes in series with a phase slip. These cases are known to produce singular results \cite{rymarz_consistent_2023, egusquiza_consistent_2024}. Instead we only consider situations where capacitance vanishes at a node with no junctions or inductance vanishes in a loop with no phase slips. This would then result in capacitance and inductance matrices of the form:
\begin{align}
\mathbf{C} &= 
\begin{bmatrix}
\mathbf{C}_{nn} & \mathbf{0} \\
\mathbf{0} & \mathbf{0}
\end{bmatrix}  \\
\mathbf{L} &= 
\begin{bmatrix}
\mathbf{L}_{ll} & \mathbf{0} \\
\mathbf{0} & \mathbf{0}
\end{bmatrix} 
\end{align}
where the zero rows and columns correspond to the vanishing capacitances and inductances at those nodes and loops, respectively.

We now assume that the network matrix has been transformed to a form given in \ref{eq:AppSimplifiedNetworkMatrix} by the previously-described procedure. In this case, the modified network matrix divides up into sub-sectors of the form:
\begin{align}
\mathbf{\Omega} \rightarrow \begin{bmatrix}
\mathbf{\Omega}_{12} &  \mathbf{0}_{13} & \mathbf{0}_{14} & \mathbf{0}_{15} \\ %newline 
\mathbf{0}_{4 2} & \mathbf{0}_{43} & \mathbf{I}_{44} & \mathbf{0}_{4 5} \\%newline 
\mathbf{0}_{32} &  \mathbf{I}_{33} & \mathbf{0}_{34} & \mathbf{0}_{35}  \\ %newline 
\mathbf{0}_{52} & \mathbf{0}_{53} & \mathbf{0}_{54} &  \mathbf{I}_{55}
\end{bmatrix}
\end{align}

Unlike in that earlier equation, $\mathbf{\Omega}_{12}$ does not just refer to the connection between Josephson junctions and phase slip elements but also to the connection between any nodes/cutsets of nonzero capacitance and loops of nonzero inductance. The first two sets of row indices represent flux degrees of freedom nonzero capacitance, while the the first two column indices stand for charge degrees of freedom with nonzero inductance. We can write out the system's other matrices as:
\begin{align}
\mathbf{A}_J = 
\begin{bmatrix}
\mathbf{A}_{1J} \\
\mathbf{A}_{4J} \\
\mathbf{0}_{3J} \\
\mathbf{0}_{5J} 
\end{bmatrix}  , \
\mathbf{B}_S = 
\begin{bmatrix}
\mathbf{B}_{2S} \\
\mathbf{B}_{3S} \\
\mathbf{0}_{4S} \\
\mathbf{0}_{5S}  \\
\end{bmatrix} \\
\mathbf{C} = 
\begin{bmatrix}
\mathbf{C}_{11} & \mathbf{C}_{14} & \mathbf{0}_{13} & \mathbf{0}_{15} \\
\mathbf{C}_{41} & \mathbf{C}_{44} & \mathbf{0}_{43} & \mathbf{0}_{45} \\
\mathbf{0}_{31} & \mathbf{0}_{34} & \mathbf{0}_{33} & \mathbf{0}_{35} \\
\mathbf{0}_{51} & \mathbf{0}_{54} & \mathbf{0}_{53} & \mathbf{0}_{55}
\end{bmatrix}  \\
\mathbf{L} = 
\begin{bmatrix}
\mathbf{L}_{22} & \mathbf{L}_{23} & \mathbf{0}_{24} & \mathbf{0}_{25} \\
\mathbf{L}_{32} & \mathbf{L}_{33} & \mathbf{0}_{34} & \mathbf{0}_{35} \\
\mathbf{0}_{42} & \mathbf{0}_{43} & \mathbf{0}_{44} & \mathbf{0}_{45} \\
\mathbf{0}_{52} & \mathbf{0}_{53} & \mathbf{0}_{54} & \mathbf{0}_{55}
\end{bmatrix} 
\end{align}

The flux and charge vectors can then be expanded as:
\begin{align}
\vv{\Phi}_n = 
\begin{bmatrix}
\vv{\Phi}_{n_1}  \\
\vv{\Phi}_{n_4}  \\
\vv{\Phi}_{n_3}  \\
\vv{\Phi}_{n_5} 
\end{bmatrix} , \
\vv{Q}_l = 
\begin{bmatrix}
\vv{\Phi}_{l_2}  \\
\vv{\Phi}_{l_3}  \\
\vv{\Phi}_{l_4}  \\
\vv{\Phi}_{l_5} 
\end{bmatrix} 
\end{align}
and the external charges and flux by:
\begin{align}
\vv{Q}_\text{ext} = 
\begin{bmatrix}
\vv{Q}_{\text{ext}_1}  \\
\vv{Q}_{\text{ext}_4}  \\
\vv{Q}_{\text{ext}_3}  \\
\vv{Q}_{\text{ext}_5} 
\end{bmatrix} , \
\vv{\Phi}_\text{ext} = 
\begin{bmatrix}
\vv{\Phi}_{\text{ext}_2}  \\
\vv{\Phi}_{\text{ext}_3}  \\
\vv{\Phi}_{\text{ext}_4}  \\
\vv{\Phi}_{\text{ext}_5} 
\end{bmatrix} 
\end{align}

Now we examine the equations of motion for this system. The fact that the capacitance and inductance matrix have blocks of zeros allows us to expand the third set of flux equations (corresponding to third block column of $\mathbf{\Omega}$) and the third set of charge equations (corresponding to third block row of $\mathbf{\Omega}$) as:
\begin{align}
\vv{\Phi}_{\text{ext}_4} + \vv{\Phi}_{4} &= \vv{0}_4 \\
\vv{Q}_{\text{ext}_3} - \vv{Q}_{3} &= \vv{0}_3
\end{align}

The effect of zero capacitance and inductance can be seen to form a constraint, which turns $\vv{\Phi}_{4}$ and $\vv{Q}_{3}$ from dynamical variables (which result in operators upon quantization) into external drives (which remain real-valued functions of time). 

We then take take the standard equations of motion for the first sets of flux and charge variables (Eqs. \ref{eq:AppContinuityStandard} and \ref{eq:AppFaradayStandard}), inserting the constraint conditions:
\begin{align}
\mathbf{C}_{11} \ddot{\vv{\Phi}}_1 - \mathbf{C}_{14}\ddot{\vv{\Phi}}_{\text{ext}_4} + \dot{\vv{Q}}_{\text{ext}_1} + \mathbf{A}_{1J} \vv{I}_J - \mathbf{\Omega}_{12} \dot{\vv{Q}}_2  = \vv{0}_1 \\
\mathbf{L}_{22} \ddot{\vv{Q}}_2 + \mathbf{L}_{23}\ddot{\vv{Q}}_{\text{ext}_3} + \dot{\vv{\Phi}}_{\text{ext}_2} + \mathbf{B}_{2S} \vv{V}_S + \mathbf{\Omega}_{12}^T \dot{\vv{\Phi}}_1  = \vv{0}_2
\end{align}
with the Josephson and phase slip terms equal to: 
\begin{align}
\vv{I}_J &= \mathbf{I}_0 \sin \left(\frac{2 \pi}{\Phi_0} \begin{bmatrix} \mathbf{A}_{1J}^T & \mathbf{A}_{4J}^T \end{bmatrix} \begin{bmatrix} \vv{\Phi}_1 \\ -\vv{\Phi}_{\text{ext}_4} \end{bmatrix} \right) \\
\vv{V}_S &= \mathbf{V}_0 \sin \left(\frac{2 \pi}{2e}  \begin{bmatrix} \mathbf{B}_{2S}^T & \mathbf{B}_{3S}^T \end{bmatrix} \begin{bmatrix} \vv{Q}_2 \\ \vv{Q}_{\text{ext}_3} \end{bmatrix} \right)
\end{align}

These equations of motion for flux variables $\vv{\Phi}_1$ and charge variables $\vv{Q}_2$ can be then quantized. However, we note that as described in \cite{you_circuit_2019}, the effect of (for instance) of external flux threading a loop with zero inductance is to add an additional external charge term, equal to $-\mathbf{C}_{14}\dot{\vv{\Phi}}_{\text{ext}_4}$. In the case of time-dependent external flux, this term will participate in the capacitive quadratic portion of the Hamiltonian. An analogous relation holds for external charge on islands of zero capacitance, which generates additions to external flux terms in the inductive quadratic term.

In that previous work, the authors moved to an ``irrotational gauge," wherein the effect of external flux was entirely allocated to the junction cosine terms. We now perform a generalization of that procedure, though we note that it does not always remove the requisite external variables from the quadratic terms. To move to the irrotational guage we make the canonical transformation:
\begin{align}
     \vv{\Phi}_1 \rightarrow \vv{\Phi}_1 -\mathbf{\Gamma} \vv{\Phi}_{\text{ext}_4} \\
     \vv{Q}_2 \rightarrow \vv{Q}_2 + \mathbf{\Lambda} \vv{Q}_{\text{ext}_3}
\end{align}
where $\mathbf{\Gamma} = \mathbf{C}_{11}^{-1}\mathbf{C}_{14}$ and  $\mathbf{\Lambda} = \mathbf{L}_{22}^{-1}\mathbf{L}_{23}$. The equations of motion are then modified as:
\begin{align}
\mathbf{C}_{11} \ddot{\vv{\Phi}}_1  + \dot{\vv{Q}}_{\text{ext}_1} + \mathbf{A}_{1J} \vv{I}_J - \mathbf{\Omega}_{12} ( \dot{\vv{Q}}_2 -\mathbf{\Lambda}\dot{\vv{Q}}_{\text{ext}_3} )   = \vv{0}_1 \\ %new equation
\mathbf{L}_{22}\ddot{\vv{Q}}_2 +  \dot{\vv{\Phi}}_{\text{ext}_2} + \mathbf{B}_{2S} \vv{V}_S +\mathbf{\Omega}_{12}^T (\dot{\vv{\Phi}}_1 + \mathbf{\Gamma} \dot{\vv{\Phi}}_{\text{ext}_4}) = \vv{0}_2
\end{align}
with
\begin{align}
\vv{I}_J &= \mathbf{I}_0 \sin \left(\frac{2 \pi}{\Phi_0} \begin{bmatrix} \mathbf{A}_{1J}^T & \mathbf{A}_{4J}^T \end{bmatrix} \begin{bmatrix} \vv{\Phi}_1 + \mathbf{\Gamma}\vv{\Phi}_{\text{ext}_4} \\ -\vv{\Phi}_{\text{ext}_4} \end{bmatrix} \right) \\
\vv{V}_S &= \mathbf{V}_0 \sin \left(\frac{2 \pi}{2e}  \begin{bmatrix} \mathbf{B}_{2S}^T & \mathbf{B}_{3S}^T \end{bmatrix} \begin{bmatrix} \vv{Q}_2 - \mathbf{\Lambda} \vv{Q}_{\text{ext}_3} \\ \vv{Q}_{\text{ext}_3} \end{bmatrix} \right)
\end{align}

After performing the change of basis, the time-dependent charge (zero capacitance) and flux (zero inductance) terms are in part shifted to the periodic terms. However, if $\mathbf{\Omega}_{12}$ is nonempty and nonzero, then $\vv{Q}_{\text{ext}_3}$ dresses $\vv{Q}_{\text{ext}_1}$ and $\vv{\Phi}_{\text{ext}_4}$ dresses $\vv{\Phi}_{\text{ext}_2}$.  Notationally, we can absorb these effects into $\vv{Q}_{\text{ext}_1}$ and $\vv{\Phi}_{\text{ext}_2}$ with the following transformation:
\begin{align}
\vv{Q}_{\text{ext}_1} &\rightarrow \vv{Q}_{\text{ext}_1} - \mathbf{\Omega}_{12}\mathbf{\Lambda}\vv{Q}_{\text{ext}_3} \\
\vv{\Phi}_{\text{ext}_2} &\rightarrow \vv{\Phi}_{\text{ext}_2} - \mathbf{\Omega}_{12}^T\mathbf{\Gamma}\vv{\Phi}_{\text{ext}_4}
\end{align}
In the limit where all inductances go to zero, $\mathbf{\Omega}_{12}$ is empty and $\vv{\Phi}_{\text{ext}_4}$ only appears in the periodic terms, as in \cite{you_circuit_2019}.

\section{Circuit model extraction} \label{sec:AppendixE}

In this Appendix, we examine the problem of systematically constructing lumped-element circuit models from electromagnetic simulations of device layouts. To perform this procedure, we demonstrate a novel network synthesis approach \cite{newcomb_linear_1966,anderson_network_2013} that extracts a linear lumped circuit model (with no transformers) from a system's simulated hybrid admittance/impedance response matrix. The lumped circuit's network matrix reflects the network topology of the simulated device. The outline of this procedure is that (1) nonlinear and external drive elements are replaced by electromagnetic ports, (2) the (approximately) linear environment is simulated in the frequency domain, (3) a lumped model for the linear environment is constructed, and (4) the nonlinearities and drives are reinserted in place of the ports to construct a complete circuit model.

The keys to the accuracy of this method are that (1) the environment has an approximately linear response and (2) that the nonlinear elements (replaced by electromagnetic ports) are small compared to the operating wavelengths. Note that with frequency-domain synthesis (unlike a purely DC analysis), the linear portion of the circuit does not need to be lumped and sub-wavelength for a lumped model to be accurate. 

The linear environment of a superconducting metal in a low-loss dielectric can be approximated as lossless and reciprocal (not breaking time-reversal symmetric, non-magnetic). With a particular placement of electromagnetic ports, we show that such an environment can be represented using a network of capacitors and inductors---to a desired degree of accuracy.

\subsection{Modeling superconductivity in electromagnetic simulation}

Here, we present background on the electromagnetic simulation of linear superconducting metal in a low-loss dielectric environment. From these simulations, circuit models of a device can be extracted and the tunneling effects of the small, nonlinear ``quantum" elements added in afterwards. 

In most electromagnetic simulations, superconducting metal exists in a distinct geometric region of space, with the rest of the model space occupied by other metals and dielectrics. This does not then account for the presence of the superconducting condensate outside of the boundary of its ion cores. In a similar fashion to normal conducting metal, superconductors are characterized by the presence of mobile charge carriers with current density $\vv{J}$. In addition to Maxwell's equations of electromagnetism, this current density obeys two key constitutive relations: the (linearized) London equations \cite{tinkham_introduction_1996,gross_applied_2016}:
\begin{align}
    \frac{ \partial (\Lambda \vv{J})}{\partial t} & = \vv{E} \label{eq:AppLondon1} \\
    \nabla \times (\Lambda \vv{J}) &=  - \vv{B} \label{eq:London2}
\end{align}

The first London equation (Eq. \ref{eq:AppLondon1}) is the superconducting equivalent of Ohm's law for normal metals, giving a relation between current density and electric field. However, the presence of the time derivative in the current density term indicates that a supercurrent can flow with no electric field. This first London equation therefore enables the modeled material to act as a perfect conductor.

The second London equation (Eq. \ref{eq:London2}) implies the Meissner effect, detailing how magnetic field decays exponentially inside of a superconductor with penetration depth $\lambda_L$ (and London constant $\Lambda = \mu_0 \lambda_L^2$). This phenomenon differentiates superconductors from ideal perfect conductors, which would resist change in interior magnetic field but not expel it.

In practice, we wish to perform electromagnetic simulations in the frequency domain. Here, all quantities and equations are Fourier-transformed and written in terms of frequency $\omega$, whereby time derivatives are converted into multiplications by factors of $i\omega$. In simulation, the superconducting medium is then defined to obey the first London equation. As a result, due to Faraday's law (Eq. \ref{eq:BContinuity}), it will also end up obeying a modified form of the second London equation. The two equations appear in the frequency domain as:
\begin{align}
    i \omega (\Lambda \vv{J}(\omega)) &= \vv{E}(\omega) \label{eq:FirstLondonFreqDomain} \\ 
    \omega \nabla \times (\Lambda \vv{J}(\omega)) &= - \omega \vv{B}(\omega)
\end{align}
We note that for $\omega \neq 0$ (and not necessarily for $\omega = 0$), the second London equation is also obeyed in the frequency domain. Thus, finite-frequency simulations can be performed by just specifying the first London equation, while DC simulations must also specify the second equation, or else be approximated by a low frequency AC calculation.

However, modeling metal in terms of bulk current can prove computationally intensive. Thus, superconductors are often simulated by constraining current to flow only on metal surfaces \cite{gross_applied_2016, kerr_surface_1999}.

Instead of the first London equation, the constitutive relation now links the tangential electric and magnetic fields fields at the metal surface \cite{yuferev_surface_2010}:
\begin{align}
\vv{E}_t(\omega) &= Z_s \hat{n} \times \vv{H}_t(\omega) \\
 & = Z_s \vv{K}_s(\omega)
\end{align}
Here $\vv{K}_s(\omega)$ is the surface current flowing on the metal and $Z_s$ is the metal's surface impedance. We note that 
the equivalence $\hat{n} \times \vv{H}_t(\omega) = \vv{K}_s$ operates under the assumption that magnetic field $\vv{H}_t(\omega)$ goes to zero inside the metal. This presents additional complication when using the electromagnetic method of moments \cite{kerr_surface_1999}. 

If the metal is nearly lossless at operating AC frequencies, the first order (kinetic inductance) expression for surface impedance can be inserted into this equation to give:
\begin{align}
Z_s &\approx i \omega \mu_0 \lambda_L \\
\implies  \vv{E}_t(\omega) &\approx i \omega \mu_0 \lambda_L^2 \frac{\vv{K}_s(\omega)}{\lambda_L} \nonumber \\
& \approx i \omega \left(\Lambda \frac{\vv{K}_s(\omega)}{\lambda_L} \right)
\end{align}

This expression is now the surface equivalent of the frequency domain first London equation (Eq. \ref{eq:FirstLondonFreqDomain}), with volume current $\vv{J}(\omega)$ replaced by $\vv{K}(\omega)/\lambda_L$, the surface current divided by the London penetration depth. Intuitively, most of the current flows within the penetration depth of the metal's surface. Specifying this surface form of the first London equation for a material implies that a corresponding version of the second London equation will be satisfied at finite frequencies, as in the previous volumetric model of superconductor.

Alongside the linear response effects of the superconducting metal, we would like our superconducting circuit models to capture the macroscopic quantum mechanical effects of superconductivity. The most prominent of these phenomena are the quantization of charge and flux carriers and their quantum tunneling across weak links. Tunneling (as mentioned in Appendix \ref{subsec:ElectromagneticPorts}) is modeled by replacing the nonlinear elements with electromagnetic ports and then inserting tunneling elements (Josephson junctions and phase slips) into the final lumped circuit model.

To ensure charge and flux quantization in our final lumped models, the simulated net charge on each superconducting island and the net fluxoid through each superconducting loop should be constant, conserved quantities whose time derivatives equal zero. These constant quantities can then be manually assigned to integer multiples of $2e$ or $\Phi_0$ for the cases of trapped net charge on an island or trapped net flux through a loop, respectively.

Charge conservation is usually ensured in an electromagnetic simulator, while the fluxoid conservation requires that for any superconducting loop:
\begin{align}
    \oint_{\partial S} \Lambda \vv{J} \cdot d \vv{\Gamma} +  \iint_{S} \vv{B} \cdot d \vv{S} &= \Phi_0 M \\
    \implies \frac{\partial}{\partial t} \left( \oint_{\partial S} \Lambda \vv{J} \cdot d \vv{\Gamma} +  \iint_{S} \vv{B} \cdot d \vv{S} \right) &=0
\end{align}
where the quantity in parentheses should be the conserved fluxoid. We observe that if the simulator is set up to obey the first London equation (Eq. \ref{eq:AppLondon1}), then the fluxoid conservation condition will be automatically satisfied through Faraday's law of electromagnetism.

\subsection{Electromagnetic ports} \label{subsec:ElectromagneticPorts}

In order to perform a simulation and extract a circuit model, applied electromagnetic fields are needed, which mimic the excitation profile of the actual device. These excitations are provided by electromagnetic ports, positioned in place of certain device elements (especially those that are nonlinear) and at input and output ports.

In a circuit model, an electromagnetic port consists of a positive and negative pair of terminals \cite{david_m_pozar_microwave_2012}. The net current flowing into the negative terminal equals the current flowing out of the positive terminal. This net flow is known as the port current. The port voltage is then defined as the voltage drop between the positive and negative terminal.

Ports serve as the interface between electromagnetic simulations and circuit models. Usually, they function in simulation by applying an electromagnetic field (and thus a voltage difference) between nearby pieces of metal on the positive and negative terminals, and then solving a discretized form of Maxwell's equations \cite{roth_introduction_2024}. From the solution, the net current through the port can be calculated. The linear relation between the port's voltage and current (as a function of frequency) defines a ``black box" circuit element \cite{nigg_black-box_2012}. In this discussion, we focus on lumped ports, where the size of the port excitation is highly sub-wavelength.

In superconducting quantum devices, ports can be inserted in place of nonlinear elements and drive elements. We now describe the ideal port configuration for a capacitor with Cooper pair tunneling (Josephson junction), and an inductor with fluxoid tunneling (phase slip wire). A central idea is to link lumped ports with our description of ideal lumped tunneling elements shown in Fig. \ref{fig:InductorChargeCapacitorFlux}. In essence, the lumped port current and voltage variables correspond to Josephson and phase slip excitations, connecting between the nonlinear elements and the black-box linear environment. 

To represent the excitation of a capacitor with Josephson tunneling, a port would ideally be placed in the area between the two plates of the capacitor. A uniform current density would be applied over the area of the junction, giving a total current of $I = |\vv{J}| A_C$, where $A_C$ is the area of the capacitor. Uniform current density is a reasonable approximation when the junction is subject to low external magnetic field \cite{gross_applied_2016}. From the resulting simulation, an average voltage across the capacitor $V$ (weighted by the capacitance per unit area) can be calculated. If this port itself is small compared to the characteristic RF wavelengths of the system, then these voltage and current variables match up well with those of the ideal lumped capacitor found in Appendix \ref{subsec:InductorChargeCapacitorFlux}---where voltage is the time derivative of the capacitor flux and current is the time derivative of capacitor charge. To account for Cooper pair tunneling, a Josephson junction circuit element can then be attached in parallel across the black box port model.

Similarly, the excitation of an ideal phase slip inductor would involve applying a uniform electric field along the length of a piece of wire, with total voltage drop $V = |\vv{E}| d_L$, where $d_L$ is the length of the inductor. Then, current $I$ would be measured as an average of the currents along each length of wire, weighted by the inductance per unit length. These definitions align well with the notion of the ideal lumped inductor in Appendix \ref{subsec:InductorChargeCapacitorFlux}---where current is the time derivative of the inductor charge and voltage is the time derivative of inductor flux. To model the voltage drops from phase slips/fluxoid tunneling, a phase slip circuit element can then be attached across the port of the resulting black box model.

Thus, if we can construct a circuit model for the linear black box, we will have a full model of the system. Note that in general the black box simulation may have an arbitrary number of ports, corresponding to multiple nonlinear and drive elements.

In practice, with commercial electromagnetic solvers, we do not have access to this much port customization. Unlike our ideal configuration of the capacitor and the inductor excitations, usually only specific electric field (voltage) excitations are possible. However, as long as the target port region is highly sub-wavelength, the precise geometry of the port excitation should have minimal effect on the calculated response.

In addition, a simulation may omit small features, like the gap capacitor of a Josephson junction, causing further departure of the simulation from the ideal. However, as long as these small omitted regions do not play a key role in the inter-port couplings, they can be corrected after the simulation by inserting additional lumped circuit elements in parallel or series with the port's terminals (such as additional capacitance across the omitted plates of a Josephson junction).

\subsection{Admittance, impedance, and hybrid matrices} \label{subsec:AdmittanceImpedanceHybridMatrices}

To construct a circuit model for linear electromagnetic response of an N-port system, we generally utilize an N-by-N response matrix. This matrix relates some combination input and output currents and voltages. Commonly simulated examples are the admittance response $\mathbf{Y}$ or impedance response $\mathbf{Z}$. For a linear time-invariant (LTI) system \cite{ciani_lecture_2024} the admittance and impedance responses can be written in the time domain as convolutions:
\begin{align}
\vv{I}(t) = \int_{-\infty}^{\infty} \mathbf{Y}(t-\tau)\vv{V}(\tau) d\tau \\
\vv{V}(t) = \int_{-\infty}^{\infty} \mathbf{Z}(t-\tau)\vv{I}(\tau) d\tau \\
\end{align}

Taking the Fourier transform of these equations puts the admittance and impedance matrices in their commonly-used frequency domain form. These responses are the quantities extracted from frequency-domain electromagnetic simulation. Convolution in the time domain becomes multiplication in the frequency domain, as written here:
\begin{align}
\vv{I}(\omega) = \mathbf{Y}(\omega) \vv{V}(\omega) \\
\vv{V}(\omega) = \mathbf{Z}(\omega) \vv{I}(\omega)
\end{align}

For an LTI system, we can also compute the Laplace-domain responses. These are written in terms of the Laplace-domain variable $s = \sigma + i \omega$. When $s = i\omega$, the overall response functions are identical to those in the Fourier domain. In general, the Laplace-domain equations are written:
\begin{align}
\vv{I}(s) = \mathbf{Y}(s) \vv{V}(s) \\
\vv{V}(s) = \mathbf{Z}(s) \vv{I}(s)
\end{align}
When an inverse exists, $\mathbf{Y}(s)^{-1} = \mathbf{Z}(s)$. Note that because we consider the response to be LTI, it represents a zero-state response function in the Laplace domain, where initial voltages and currents are assumed to be zero. As is convention in the field of network synthesis we employ the Laplace-domain response functions throughout this work.

Other types of response functions---beyond admittance and impedance---are commonly employed. In this work, we focus in particular on hybrid matrices \cite{newcomb_linear_1966, anderson_network_2013}. These response matrices are formed by dividing up the ports into two categories. For one category of ports, which we label with $C$, the current variable is on the LHS and voltage is multiplied by the response matrix on the RHS (as in admittance). For another, denoted by $L$, the voltage is on the LHS and the current is multiplied by the response matrix on the RHS (as in impedance). In total, this gives:
\begin{align}
    \begin{bmatrix}
       \vv{I}_C(s) \\
       \vv{V}_L(s)
    \end{bmatrix}
    =
       \mathbf{H}(s)
    \begin{bmatrix}
       \vv{V}_C(s) \\
       \vv{I}_L(s)
    \end{bmatrix}
\end{align}

We note that our choice of index names reflects our future identification of the $C$ ports as spanning across capacitors and the $L$ ports as lying along inductors.

\subsection{Lossless positive real (reciprocal) matrices} \label{subsec:LPR}

Construction of a circuit model is enabled by noting key properties obeyed by the Laplace-domain electromagnetic response matrices, which correspond to the physical attributes of the system being modeled. For superconductors far below their critical temperature embedded in low-loss dielectrics, we make the common approximation that the system is lossless. In that case, the admittance, impedance, or hybrid response function is lossless positive real (LPR). The LPR criteria for a response matrix $\mathbf{H}(s)$ are listed here \cite{newcomb_linear_1966, anderson_network_2013}:

\begin{enumerate}[nosep]
    \item The matrix $\mathbf{H}(s)$ can be written as a  rational function of real coefficients. \label{item:LPRProp1}
    \item Each pole of $\mathbf{H}(s)$ (including the pole at $\infty$) is simple and has positive semi-definite Hermitian residue matrix $\mathbf{K}$. \label{item:LPRProp2}
    \item Poles of $\mathbf{H}(s)$ (except the pole at $\infty$) occur at $s = \pm i \omega$, where $\omega$ represents a real-valued frequency. \label{item:LPRProp3}
    \item $\mathbf{H}(s)+\mathbf{H}^{\dagger}(s) = \mathbf{0}$ for all $s = i \omega$, where $\omega$ is real and does not correspond to a pole of $\mathbf{H}(s)$. \label{item:LPRProp4}
\end{enumerate}

For Property \ref{item:LPRProp1}, the simulation response of a distributed device will not exactly satisfy the condition. However, the true response function is usually approximated to a high degree of accuracy by a real-rational expansion, obtained through an algorithm like vector fitting \cite{gustavsen_rational_1999}.

Property \ref{item:LPRProp2} applies generally to passive systems. Properties \ref{item:LPRProp3} and \ref{item:LPRProp4} reflect the lossless nature of the response. Essentially, they guarantee that finite-frequency poles come in pairs on the imaginary axis $s = \pm i \omega_r$, and do not contain any real component corresponding to loss or gain.

In addition to these LPR criteria, we will restrict ourselves to the discussion of systems obeying electromagnetic reciprocity. This assumption is well-founded for systems that do not contain non-reciprocal elements, such as ferrites in a permanent magnetic field \cite{david_m_pozar_microwave_2012}.

The reciprocity condition implies that admittance matrices $\mathbf{Y}(s)$ and impedance matrices $\mathbf{Z}(s)$ are symmetric. Hybrid matrices obey a more complicated ``hybrid" symmetry in the reciprocal case, with symmetric diagonal blocks and anti-symmetric off-diagonals
\cite{anderson_network_2013, yarlagadda_reciprocal_1972}:
\begin{align} \label{eq:HybridMatrixReciprocal}
    \begin{bmatrix}
       \vv{I}_C(s) \\
       \vv{V}_L(s)
    \end{bmatrix}
    =
    \begin{bmatrix}
       \mathbf{H}_{CC}(s) & - \mathbf{H}_{CL}(s) \\
       \mathbf{H}_{CL}^T(s) & \mathbf{H}_{LL}(s)
    \end{bmatrix}
    \begin{bmatrix}
       \vv{V}_C(s) \\
       \vv{I}_L(s)
    \end{bmatrix}
\end{align}
Here, $\mathbf{H}_{CC}(s)$ and $\mathbf{H}_{LL}(s)$ are symmetric, as they are admittance and impedance matrices, respectively. Note that admittance and impedance matrices form a sub-category of hybrid matrices, such that most results demonstrated for hybrid matrices also apply to them.

\subsection{Port placement, zero-frequency response} \label{subsec:PortPlacementZeroFreq}

Before detailing how to expand hybrid matrices in terms of their poles, we examine the question of port placement on an electromagnetic structure, to further constrain the form of this hybrid matrix. In so doing so, we force the zero-frequency properties of the system to obey a network matrix constraint, enabling the simulated system to be representing by lumped circuit with the same topology.

We first divide the simulation ports into two types, based on their notation in Eq. \ref{eq:HybridMatrixReciprocal}. Ports denoted $C$ are capacitive, and represent parallel excitations across gaps between metals. Ports denoted $L$ are inductive, and provide series excitations along metal segments. In practice the implementation of each type of port in simulation may be similar, differing only by how they enter into the hybrid matrix.

For systems with Josephson junctions (and no phase slips) we place a capacitive port across each junction, and then put one inductive port along a portion of each loop containing at least one junction (as shown in Main Text Fig. \ref{fig:SynthesisLayout}). Thus, when the loop inductive ports are set to open, there is no galvanic connection between the two terminals of the junction. Similarly, when the junction capacitive ports are shorted, there is a unique loop current for each inductive port that flows through a subset of the capacitive ports. In particular, the capacitive branches define a tree, and the inductive edges define a cotree such that, at zero frequency ($s=0$) we have:
\begin{align}
\vv{V}_L &= \mathbf{\Omega}_E^T \vv{V}_C \bigg|_{\vv{I}_L = \vv{0}_L} \\
\vv{I}_C &= -\mathbf{\Omega}_E \vv{I}_L \bigg|_{\vv{V}_C = \vv{0}_C}
\end{align}

The object $\mathbf{\Omega}_E$ is an edge network matrix, which describes the tree-cotree connectivity of circuits, as discussed in Appendices \ref{subsec:BranchDecomposition} and \ref{subsec:GraphTheoreticNetworkMatrices}. An equivalent statement is that the hybrid matrix at zero frequency is equal to
\begin{align} \label{eq:ZeroFreqHybrid}
\mathbf{H}(s=0) = 
    \begin{bmatrix}
       \mathbf{0}_{CC} & -\mathbf{\Omega}_E \\
       \mathbf{\Omega}_E^T & \mathbf{0}_{LL} 
    \end{bmatrix}
\end{align}

This constraint will prove useful in simplifying the form of the hybrid matrix response, and ultimately in matching it to a circuit model with equivalent network topology.

As a whole, this procedure is well-motivated when the inductor port width is small compared to the system's RF wavelengths (the junction capacitive ports should always be small). The tree-cotree port configuration prevents the emergence of zero-frequency poles \cite{newcomb_linear_1966}, and enables charge and flux degrees of freedom to be modeled on an equal footing.

For systems that also contain phase slips along inductive segments, a similar construction should be employed, such that the circuit is represented by a capacitive tree and inductive cotree, described by an edge network matrix. In addition to each junction requiring a capacitive port, each phase slip wire will also need to have an inductive port. However, the addition of these phase slip loop ports may disturb the tree-cotree port structure, necessitating the placement of additional capacitive ports across the phase slip elements to prevent zero-frequency poles.

In simulation, Josephson junction arrays that serve as inductors, such as those in Fluxonium qubits \cite{manucharyan_fluxonium_2009, manucharyan_evidence_2012}, can be approximately modeled with an inductive port along the junction array instead of a capacitive port across each junction. This simplifies the simulations and mirrors the standard approximations made to the system's Hamiltonian. The junction array can be represented in EM simulation with a kinetic inductance material, and can include a phase slip wire in the final circuit model (as in Main Text Fig. \ref{fig:FluxoniumSimulation}).

\subsection{Hybrid matrix pole expansion} \label{subsec:HybridMatrixPoleExpansion}

Now we have a set of conditions and constraints in place that define the simulated hybrid matrix of our lossless, reciprocal superconducting systems. This will allow us to simplify the Laplace-domain response into a form that is easily compared to that of a lumped-element circuit. In this subsection, we provide a derivation of the hybrid matrix pole expansion, using standard methods \cite{newcomb_linear_1966,anderson_network_2013}. This is essentially the same as the result presented in \cite{yarlagadda_reciprocal_1972}, but with our novel zero-frequency network matrix constraint.

Along with that aforementioned zero-frequency constraint, we utilize the properties of the lossless, reciprocal hybrid matrix given in Appendix \ref{subsec:LPR}. Those conditions state that the hybrid matrix $\mathbf{H}(s)$ can be written as a rational polynomial function of $s$ with real coefficients, that all poles of the system are simple, and that their residues are positive semi-definite. In addition, for values of $s$ on the imaginary axis we should have the lossless condition that $\mathbf{H}(i\omega)= -\mathbf{H}^{\dagger}(i\omega)$. We also use hybrid symmetry condition given in Eq. \ref{eq:HybridMatrixReciprocal}.

The simple poles of the system come in two general types: poles on the imaginary axis, and poles at infinity. Before beginning the process, one can first carry out polynomial long division and partial fraction decomposition on the real-rational hybrid matrix. The fact that the system is lossless and all poles are simple means that each rational term with $s$ in the denominator takes the form of $\frac{\mathbf{K}_\mu}{s-i\omega_\mu}$. Here $\omega_\mu$ is the resonance frequency of the pole and $\mathbf{K}_\mu$ is its positive semi-definite residue matrix. 

Then, the remaining rational terms of the matrix without $s$ in the denominator must be polynomial functions of $s$. For degree of $s$ in the polynomial term greater than or equal to 1, these terms represent poles at infinity (zeros of $\mathbf{H}(1/s)$). Since the pole at infinity is simple, only non-constant polynomial terms of degree 1 are allowed, of the form $s \mathbf{K}_\infty$, where $\mathbf{K}_\infty$ is the positive semi-definite residue matrix. The only remaining possible term is a constant matrix $\mathbf{M}_{\text{const}}$, which is not a residue matrix, and will end up being determined the zero-frequency response constraint.

The total hybrid matrix can be written down in its pole decomposition as \cite{newcomb_linear_1966}:
\begin{align}
\mathbf{H}(s) = s \mathbf{K}_\infty +  \mathbf{M}_{\text{const}} +  \sum_\mu \frac{\mathbf{K}_\mu}{s-i\omega_\mu}
\end{align}

The structure of the residue matrices allows for a more specific expansion of these terms. For the poles at infinity, the entries of residue matrix must be real so that $\mathbf{H}(s)$ is real rational. This implies that its residue matrix is block diagonal---since the lossless condition ($i \omega \mathbf{K}_\infty = i\omega \mathbf{K}_\infty^T$) means that the residue matrix is symmetric, while the hybrid symmetry condition states that the off-diagonal blocks are anti-symmetric, and thus zero. The linear $s$ term can then be written as:
\begin{align}
s \mathbf{K}_\infty = 
    \begin{bmatrix}
       s\mathbf{K}_{{CC}_\infty} & \mathbf{0}_{{CL}} \\
       \mathbf{0}_{{LC}} & s\mathbf{K}_{{LL}_\infty}
    \end{bmatrix}
\end{align}
where $\mathbf{K}_{{CC}_\infty}$ and $\mathbf{K}_{{LL}_\infty}$ are symmetric and positive semi-definite. We make one final assumption about the the system's characteristics: that both of these matrices are in fact positive definite. This will occur in simulation when each capacitive port lies across a distinct gap of nonzero capacitance and each each inductive port lies along a distinct loop of nonzero inductance.

We also note that the real-valued constant matrix has its form determined by the lossless and hybrid conditions. These will lead it to be antisymmetric instead of symmetric, implying that:
\begin{align}
\mathbf{M}_{\text{const}} = 
    \begin{bmatrix}
       \mathbf{0}_{CC} & -\mathbf{M}_{CL} \\
       \mathbf{M}_{CL}^T & \mathbf{0}_{LL} 
    \end{bmatrix}
\end{align}

Now we look at the terms generated by imaginary axis poles at $s = i \omega_\mu$. Due to the real-rational nature of the hybrid response, these poles come in conjugate pairs:
\begin{align} \label{eq:SinglePoleExpansion}
\frac{\mathbf{K}_\mu}{s - i \omega_\mu} + \frac{\mathbf{K}_\mu^*}{s + i \omega_\mu} = \frac{s \left[ \omega_\mu^2 \mathbf{R}_\mu \right] + i\omega_\mu \left[ \omega_\mu^2\overline{\mathbf{R}}_\mu \right]}{s^2+\omega_\mu^2}
\end{align}
where we have defined for $\mathbf{K}_\mu$:
\begin{align}
\omega_\mu^2 \mathbf{R}_\mu &=  \mathbf{K}_\mu + \mathbf{K}_\mu^* = \mathbf{K}_\mu + \mathbf{K}_\mu^T \\ %newline
\omega_\mu^2 \overline{\mathbf{R}}_\mu &= \mathbf{K}_\mu - \mathbf{K}_\mu^* = \mathbf{K}_\mu - \mathbf{K}_\mu^T 
\end{align}

Since the residue matrix $\mathbf{K}_\mu$ is Hermitian and also obeys the hybrid symmetry condition, it must be true that its diagonal blocks are real and its off diagonal blocks are imaginary. Then we can write down the following relations:
\begin{align}
\mathbf{K}_\mu &=
    \begin{bmatrix}
       \mathbf{K}_{{CC}_\mu} & -\mathbf{K}_{{CL}_\mu} \\
       \mathbf{K}_{{CL}_\mu}^T & \mathbf{K}_{{LL}_\mu}
    \end{bmatrix} \\
\omega_\mu^2 \mathbf{R}_\mu &=
    \begin{bmatrix}
       2\mathbf{K}_{{CC}_\mu} &  \mathbf{0}_{CL} \\
       \mathbf{0}_{LC}  & 2\mathbf{K}_{{LL}_\mu}
    \end{bmatrix} \\
\omega_\mu^2 \overline{\mathbf{R}}_\mu &=
\begin{bmatrix}
       \mathbf{0}_{CC} &  -2\mathbf{K}_{{CL}_\mu} \\
       2\mathbf{K}_{{CL}_\mu}^T  & \mathbf{0}_{LL}
\end{bmatrix}
\end{align}

Here $\mathbf{K}_{{CC}_\mu}$ and $\mathbf{K}_{{LL}_\mu}$ are real-valued matrices, and $\mathbf{K}_{{CL}_\mu}$ is imaginary-valued, implying that $\mathbf{R}_\mu$ is real-valued and $\overline{\mathbf{R}}_\mu$ is imaginary-valued. As required, this then implies that Eq. \ref{eq:SinglePoleExpansion} has real-valued coefficients.

Now we set $\mathbf{R}_{{CC}_\mu} = 2\mathbf{K}_{{CC}_\mu} $, $\mathbf{R}_{{LL}_\mu} = 2\mathbf{K}_{{LL}_\mu} $, and $\mathbf{R}_{{CL}_\mu} = 2 \mathbf{K}_{{CL}_\mu} $, leading to an expansion of the hybrid matrix terms of:
\begin{align} \label{eq:ResidueMatrixFinitePole}
s\mathbf{R}_\mu + i\omega_\mu\overline{\mathbf{R}}_\mu 
= 
\begin{bmatrix}
   s\mathbf{R}_{{CC}_\mu} & -i \omega_\mu \overline{\mathbf{R}}_{{CL}_\mu} \\
   i \omega_\mu \overline{\mathbf{R}}_{{CL}_\mu}^T & s\mathbf{R}_{{LL}_\mu}
\end{bmatrix}
\end{align}

From here we can write down an equation for each block of the reciprocal hybrid matrix given in Eq. \ref{eq:HybridMatrixReciprocal}. Bearing in mind that $\mathbf{H}_{CL}(s)$ is the negative of the off-diagonal block element, we match terms to obtain:
\begin{align}
\mathbf{H}_{CC}(s) &= s \mathbf{K}_{{CC}_\infty} + s \sum_\mu \frac{\omega_\mu^2 \mathbf{R}_{{CC}_\mu}}{s^2 + \omega_\mu^2} \nonumber \\ %newlines
&= s \left( \mathbf{K}_{{CC}_\infty}+ \sum_\mu \mathbf{R}_{{CC}_\mu}\right) - s^3 \sum_\mu \frac{\mathbf{R}_{{CC}_\mu}}{{s^2 + \omega_\mu^2}} \\ %newequation
\mathbf{H}_{LL}(s) &= s \mathbf{K}_{{LL}_\infty} + s \sum_\mu \frac{\omega_\mu^2 \mathbf{R}_{{LL}_\mu}}{s^2 + \omega_\mu^2} \nonumber \\ %newline
& = s \left( \mathbf{K}_{{LL}_\infty}+ \sum_\mu \mathbf{R}_{{LL}_\mu} \right) - s^3 \sum_k \frac{\mathbf{R}_{{LL}_\mu}}{s^2 + \omega_\mu^2} \\ %newequation
\mathbf{H}_{CL}(s) &= \mathbf{M}_{CL} + i \omega_\mu \sum_\mu  \frac{\omega_\mu^2 \overline{\mathbf{R}}_{{CL}_\mu}}{s^2 + \omega_\mu^2} \nonumber \\ %newline
& =  \left( \mathbf{M}_{CL} + i \omega_\mu \sum_\mu  \overline{\mathbf{R}}_{{CL}_\mu} \right) - s^2 i \omega_\mu  \sum_\mu \frac{\overline{\mathbf{R}}_{{CL}_\mu}}{s^2 + \omega_\mu^2} \nonumber  \\
& =  \left( \mathbf{\Omega}_E \right) - s^2 i \omega_\mu  \sum_\mu \frac{\overline{\mathbf{R}}_{{CL}_\mu}}{s^2 + \omega_\mu^2} \label{eq:HCLHybridResponse}
\end{align}

For each block of the matrix, the first and second lines are equivalent through polynomial long division. The final line of \ref{eq:HCLHybridResponse} follows from the zero-frequency response condition (Eq. \ref{eq:ZeroFreqHybrid}). This expansion, where the finite-frequency pole terms have an additional factor of $s^2$ in their numerator, will also prove useful when comparing this black-box response to that of a lumped circuit. In a matrix form, the expression reads:
\begin{align}
&\mathbf{H}(s) = \nonumber \\ %newline
&\begin{bmatrix}
    s \left[ \mathbf{K}_{{CC}_\infty}+ \sum_\mu \mathbf{R}_{CC_\mu} \right] & -\mathbf{\Omega}_E \\
    \mathbf{\Omega}_E^T &  s\left[ \mathbf{K}_{{LL}_\infty}+ \sum_\mu \mathbf{R}_{LL_\mu} \right]
\end{bmatrix} \nonumber \\ %newline
&-  \sum_\mu \frac{s^2}{s^2 + \omega_\mu^2} \begin{bmatrix}
   s\mathbf{R}_{{CC}_\mu} & -i \omega_\mu \overline{\mathbf{R}}_{{CL}_\mu} \\
   i \omega_\mu \overline{\mathbf{R}}_{{CL}_\mu}^T & s\mathbf{R}_{{LL}_\mu}
\end{bmatrix}
\end{align}

\subsection{Pole decomposition} \label{subsec:PoleDecomposition}

In order to facilitate comparisons to a lumped circuit, we perform one more expansion of each finite-frequency residue term---which relies on the positive semi-definite nature of the residue matrices. We proceed in a fashion most similar to the approach presented in \cite{yarlagadda_reciprocal_1972}.

Each finite-frequency pole contains an expression in the form of Eq. \ref{eq:ResidueMatrixFinitePole}. Recall that the residue matrix $\mathbf{K}_\mu$ is Hermitian positive semi-definite, while $\mathbf{R}_{{CC}_\mu} = 2\mathbf{K}_{{CC}_\mu}$ and $\mathbf{R}_{{LL}_\mu} = 2\mathbf{K}_{{LL}_\mu}$ are real and proportional to diagonal blocks of this matrix. Thus $\mathbf{R}_{{CC}_\mu}$ and $\mathbf{R}_{{LL}_\mu}$ can be decomposed as:
\begin{align}
\mathbf{R}_{{CC}_\mu} &= \mathbf{U}_1 \mathbf{\Sigma}_{CC} \mathbf{U}_1^T \\
\mathbf{R}_{{LL}_\mu} &= \mathbf{W}_1 \mathbf{\Sigma}_{LL} \mathbf{W}_1^T 
\end{align}
Here, $\mathbf{U}_1$ and $\mathbf{W}_1$ are real invertible matrices, while $\mathbf{\Sigma}_{CC}$ and $\mathbf{\Sigma}_{LL}$ are diagonal, with diagonal elements only equal to $1$ or $0$.

Expanding the pole term in the new basis given by $\mathbf{U}_1$ and $\mathbf{W}_1$ will lead to an entry of the form $\mathbf{U}_1^{-1} \overline{\mathbf{R}}_{{CL}_\mu} \mathbf{W}^{T^{-1}}$ in the off-diagonal block of the matrix. Note that this imaginary matrix can be written in its singular value decomposition as: $\mathbf{U}_2 ( i\mathbf{D}_{CL}) \mathbf{W}_2^T$. Here $\mathbf{U}_2$ and $\mathbf{W}_2$ are real orthogonal matrices, and $\mathbf{D}_{CL}$ is real-valued and diagonal.

This procedure then decomposes the the off-diagonal component of the matrix as:
\begin{align}
\overline{\mathbf{R}}_{{CL}_\mu} &= (\mathbf{U}_1\mathbf{U}_2)(i\mathbf{D}_{CL}) ( \mathbf{W}_1\mathbf{W}_2 )^T
\end{align}

We now make the following matrix labels:
\begin{align}
\mathbf{U}_{CC} &= \mathbf{U}_1\mathbf{U}_2 \\
\mathbf{W}_{LL} &= \mathbf{W}_1\mathbf{W}_2 
\end{align}
Using the fact that $\mathbf{U}_2$ and $\mathbf{W}_2$ are unitary and do not affect $\mathbf{\Sigma}_{CC}$ and $\mathbf{\Sigma}_{LL}$, respectively, we see that: 
\begin{align}
&\begin{bmatrix}
   s\mathbf{R}_{{CC}_\mu} & -i \omega_\mu \overline{\mathbf{R}}_{{CL}_\mu} \\
   i \omega_\mu \overline{\mathbf{R}}_{{CL}_\mu}^T & s\mathbf{R}_{{LL}_\mu}
\end{bmatrix} =\nonumber \\
  &\begin{bmatrix}
    \mathbf{U}_{CC} &  \mathbf{0}_{CL}  \\ 
    \mathbf{0}_{LC} & \mathbf{W}_{LL}
    \end{bmatrix}
    \begin{bmatrix}
    s\mathbf{\Sigma}_{CC} &  \omega_\mu \mathbf{D}_{CL}  \\
    -\omega_\mu \mathbf{D}_{CL}^T & s\mathbf{\Sigma}_{LL}
    \end{bmatrix}
    \begin{bmatrix}
    \mathbf{U}_{CC} &  \mathbf{0}_{CL}  \\
    \mathbf{0}_{LC} & \mathbf{W}_{LL}
    \end{bmatrix}^T
\end{align}

Now, as in \cite{yarlagadda_reciprocal_1972} we use the properties residue matrix to constrain the values of diagonal matrix $\mathbf{D}_{CL}$. In particular, for Laplace variable value $s = i \omega_\mu$, the pole term is proportional to the residue matrix as:
\begin{align}
\begin{bmatrix}
   i\omega_\mu \mathbf{R}_{{CC}_\mu} & -i \omega_\mu \overline{\mathbf{R}}_{{CL}_\mu} \\
   i \omega_\mu \overline{\mathbf{R}}_{{CL}_\mu}^T & i \omega_\mu \mathbf{R}_{{LL}_\mu}
\end{bmatrix}
= 2 i \omega_\mu \mathbf{K}_\mu
\end{align}
Since $\mathbf{K}_\mu$ is positive semi-definite, performing this same substitution in our expansion of the pole term implies that:
\begin{align}
\begin{bmatrix}
    \mathbf{\Sigma}_{CC} &  - i\mathbf{D}_{CL}  \\
    i\mathbf{D}_{CL}^T & \mathbf{\Sigma}_{LL}
    \end{bmatrix}
\end{align}
is positive semi-definite as well. This means that the only nonzero elements of $\mathbf{D}_{CL}$ occur in positions where the corresponding diagonal elements of $\mathbf{\Sigma}_{CC}$ have $\mathbf{\Sigma}_{LL}$ values of $1$. In addition, if $d_\nu$ is a nonzero diagonal element of $\mathbf{D}_{CL}$, then $d_\nu^2 \leq 1$.

Because of these properties, we can perform an outer product expansion in terms of the column vectors $\vv{U}_{C_\nu}$ of $\mathbf{U}_{CC}$ and $\vv{W}_{L_\nu}$ of $\mathbf{W}_{LL}$.
\begin{align}
  \begin{bmatrix}
    \mathbf{U}_{CC} &  \mathbf{0}_{CL}  \\ 
    \mathbf{0}_{LC} & \mathbf{W}_{LL}
    \end{bmatrix}
    \begin{bmatrix}
    s\mathbf{\Sigma}_{CC} &  \omega_\mu \mathbf{D}_{CL}  \\
    -\omega_\mu \mathbf{D}_{CL}^T & s\mathbf{\Sigma}_{LL}
    \end{bmatrix}
    \begin{bmatrix}
    \mathbf{U}_{CC} &  \mathbf{0}_{CL}  \\
    \mathbf{0}_{LC} & \mathbf{W}_{LL}
    \end{bmatrix}^T \nonumber \\ %new equation
    =
    \sum_{\nu=1}^N \begin{bmatrix}
    \vv{U}_{C_\nu} &  \vv{0}_{C}  \\
    \vv{0}_{L} & \vv{W}_{L_\nu}
    \end{bmatrix}
    \begin{bmatrix}
    s &  \omega_\mu d_\nu  \\
    -\omega_\mu d_\nu & s
    \end{bmatrix}
    \begin{bmatrix}
    \vv{U}_{C_\nu} &  \vv{0}_{C}  \\
    \vv{0}_{L} & \vv{W}_{L_\nu}
    \end{bmatrix}^T \nonumber \\ %new equation
    = 
    \sum_{\nu=1}^N \begin{bmatrix}
    d_\nu \vv{U}_{C_\nu} &  \vv{0}_{C}  \\
    \vv{0}_{L} & \vv{W}_{L_\nu}
    \end{bmatrix}
    \begin{bmatrix}
    s/d_\nu^2 &  \omega_\mu  \\
    -\omega_\mu & s
    \end{bmatrix}
    \begin{bmatrix}
    d_\nu \vv{U}_{C_\nu} &  \vv{0}_{C}  \\
    \vv{0}_{L} & \vv{W}_{L_\nu}
    \end{bmatrix}^T  \nonumber \\ %new equation
    = \sum_{\nu=1}^N \begin{bmatrix}
    d_\nu \vv{U}_{C_\nu} &  \vv{0}_{C}  \\
    \vv{0}_{L} & \vv{W}_{L_\nu}
    \end{bmatrix}
    \begin{bmatrix}
    s &  \omega_\mu  \\
    -\omega_\mu & s
    \end{bmatrix}
    \begin{bmatrix}
    d_\nu \vv{U}_{C_\nu} &  \vv{0}_{C}  \\
    \vv{0}_{L} & \vv{W}_{L_\nu}
    \end{bmatrix}^T \nonumber \\ %newline
    + \sum_{\nu=1}^N \begin{bmatrix}
    \xi \vv{U}_{C_\nu} &  \vv{0}_{C}  \\
    \vv{0}_{L} & \vv{0}_{L_\nu}
    \end{bmatrix}
    \begin{bmatrix}
    s &  \omega_\mu  \\
    -\omega_\mu & s
    \end{bmatrix}
    \begin{bmatrix}
    \xi \vv{U}_{C_\nu} &  \vv{0}_{C}  \\
    \vv{0}_{L} & \vv{0}_{L_\nu}
    \end{bmatrix}^T
\end{align}
where $\xi = d_\nu \sqrt{\frac{1}{d_\nu^2}-1}$, which is real because $d_\nu^2 \leq 1$. This gives us the vectors of our outer product expansion, which we define as:
\begin{align}
\vv{R}_{C_\nu} &= d_\nu \vv{U}_{C_\nu},  \ \vv{R}_{L_\nu} = \vv{W}_{L_\nu} \\
\vv{R}_{C_{N+\nu}} &= \xi \vv{U}_{C_{\nu}},  \ \vv{R}_{L_{N+\nu}} = \vv{0}
\end{align}
Removing any pairs of zero vectors, and relabeling each vector in terms of index $\nu$, we obtain the pole expansion:
\begin{align}
    &\begin{bmatrix}
    s\mathbf{R}_{{CC}_\mu} &  -i\omega_\mu \mathbf{R}_{{CL}_\mu} \\
    i\omega_\mu \mathbf{R}_{{CL}_\mu}^T & s\mathbf{R}_{{LL}_\mu}
    \end{bmatrix}
    \\
    &=\sum_{\nu} \begin{bmatrix}
    \vv{R}_{C_\nu} &  \vv{0}_{C}  \\
    \vv{0}_{L} & \vv{R}_{L_\nu}
    \end{bmatrix}
    \begin{bmatrix}
    s &  \omega_\mu  \\
    -\omega_\mu & s
    \end{bmatrix}
    \begin{bmatrix}
    \vv{R}_{C_\nu} &  \vv{0}_{C}  \\
    \vv{0}_{L} & \vv{R}_{L_\nu}
    \end{bmatrix}^T
\end{align}
A point of note is that $\vv{R}_{C_\nu}$  and $\vv{R}_{L_\nu}$ are real-valued. Now we expand each pole in this way, using the index $r$ to label the various outer products:
\begin{align} \label{eq:AppFinalHybridSimulationResponse}
&\mathbf{H}(s) = \nonumber \\ %newline
&\begin{bmatrix}
    s \left[ \mathbf{K}_{{CC}_\infty}+ \sum_r \vv{R}_{C_r} \vv{R}_{C_r}^T \right] & -\mathbf{\Omega}_E \\
    \mathbf{\Omega}_E^T &  s\left[ \mathbf{K}_{{LL}_\infty}+ \sum_r \vv{R}_{L_r} \vv{R}_{L_r}^T \right]
\end{bmatrix} \nonumber 
\\%newline
&+\sum_{r}\frac{s^2}{s^2+\omega_r^2} \begin{bmatrix}
\vv{R}_{C_r} &  \vv{0}_{C}  \\
\vv{0}_{L} & \vv{R}_{L_r}
\end{bmatrix}
\begin{bmatrix}
-s &  -\omega_r  \\
\omega_r & -s
\end{bmatrix}
\begin{bmatrix}
\vv{R}_{C_r} &  \vv{0}_{C}  \\
\vv{0}_{L} & \vv{R}_{L_r}
\end{bmatrix}^T
\end{align}

Observe that different values of $r$ may have the same or different resonance frequencies $\omega_r$. The labeling does distinguish which sets of outer products share the same resonance frequency.

\subsection{Comparison to equations of motion of lumped circuit} \label{subsec:ComparisonEquationsMotion}

\begin{figure}
    \centering
    \includegraphics[width=1\linewidth]{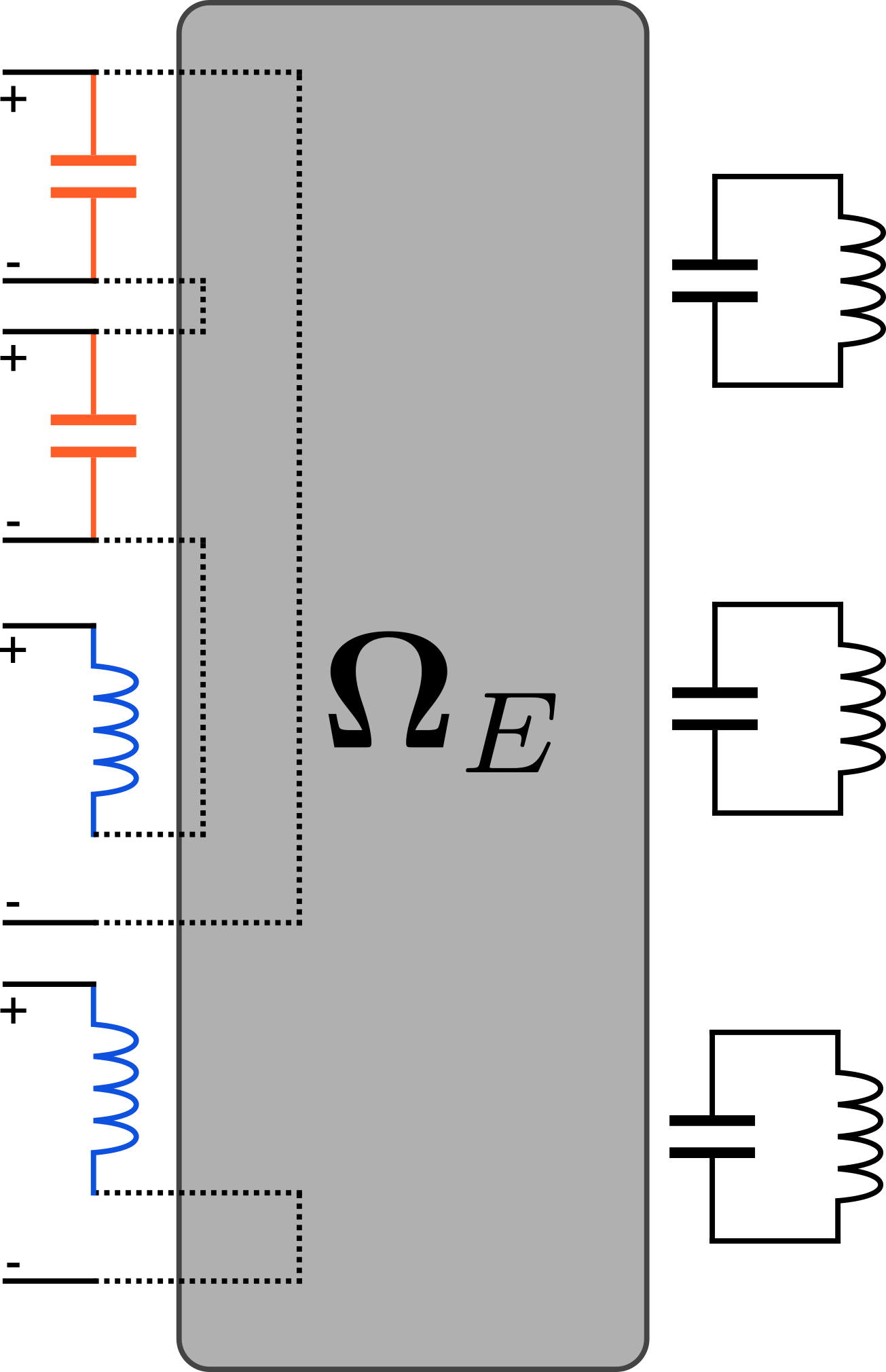}
    \caption{Schematic of model extraction procedure in tree-cotree notation. Capacitive (parallel) and inductive (series) ports are interconnected through the edge network matrix $\mathbf{\Omega}_E$ that matches topology of the simulation. Nonlinear tunneling and drive elements can be inserted across port terminals. In addition, a set of resonant modes (on the right) are used to synthesize the response of the system's finite-frequency poles. In this schematic, capacitive and inductive branch variables are each coupled through the off-diagonal terms of a capacitance and inductance matrix, respectively. This picture can be converted to standard nodal circuit model by finding an incidence matrix for the capacitive branches that respects the circuit's cutset-loop topology.}
\label{fig:NetworkMatrixSynthesis}
\end{figure}

Now we address the question of synthesizing the response function with a lumped circuit model. For a lumped circuit in tree-cotree notation (Appendix \ref{subsec:NodeFluxLoopChargeTreeCotree} and Main Text Fig. \ref{fig:TopologicalMatricesTreeCotree}), with a set of capacitive spanning tree edges and inductive cotree edges, we can place capacitive ports in parallel with the capacitive connections and inductive ports in series with the inductive ones. The lumped circuit will then have terminals into which additional circuit elements (such as nonlinear junctions and phase slips) can be inserted, as outlined in Fig. \ref{fig:NetworkMatrixSynthesis}.

Then in the tree-cotree edge basis of these ports described in Appendix \ref{subsec:BranchDecomposition}, the equations of motion read:
\begin{align}
\vv{I}_C(t) &= \mathbf{C}\dot{\vv{V}}_C(t) - \mathbf{\Omega}_E \vv{I}_L(t)  \\
\vv{V}_L(t) &= \mathbf{L}\dot{\vv{I}}_L(t) + \mathbf{\Omega}_E^T \vv{V}_C(t)
\end{align}
Here, $\mathbf{\Omega}_E$ is again an edge network matrix, $\mathbf{C}$ is the capacitance matrix, and $\mathbf{L}$ is the inductance matrix. These are essentially Eqs. \ref{eq:AppContinuityStandard} and \ref{eq:AppFaradayStandard} expressed in the edge basis of the ports, in the absence of external drives. However, the direction of applied current at capacitive ports $\vv{I}_C(t)$ goes in the opposite direction as the Josephson current $\vv{I}_J(t)$, and the voltage along inductive ports $\vv{V}_L(t)$ has the opposite convention to that of voltage on a phase slip wire $\vv{V}_S(t)$, producing differences in sign.

Considering the system to have linear time-invariant response with zero initial conditions for current and voltage (the zero-state response condition discussed in Appendix \ref{subsec:AdmittanceImpedanceHybridMatrices}), the Laplace transform of these equations looks like:
\begin{align}
\vv{I}_C(s) &= s\mathbf{C}\vv{V}_C(s) - \mathbf{\Omega}_E \vv{I}_L(s)  \\
\vv{V}_L(s) &= s\mathbf{L}\vv{I}_L(s) + \mathbf{\Omega}_E^T \vv{V}_C(s)
\end{align}

This type of model can match the low frequency response of a superconducting device, but in order to synthesize the resonant poles of the system, we require the model to be augmented with auxiliary resonant modes:
\begin{align}
\begin{bmatrix}
    \vv{I}_C(s) \\
    \vv{0}_r
\end{bmatrix}
=
s
\begin{bmatrix}
    \mathbf{C}_{CC} & \mathbf{C}_{Cr} \\
    \mathbf{C}_{rC} & \mathbf{C}_{rr}
\end{bmatrix}
\begin{bmatrix}
\vv{V}_C(s) \\
\vv{V}_r(s)
\end{bmatrix}
-
\begin{bmatrix}
    \mathbf{\Omega}_E & \mathbf{0} \\
    \mathbf{0} & \mathbf{I}_{rr}
\end{bmatrix}
\begin{bmatrix}
    \vv{I}_L(s) \\
    \vv{I}_r(s)
\end{bmatrix} \\ %new equation
\begin{bmatrix}
    \vv{V}_L(s) \\
    \vv{0}_r
\end{bmatrix}
=
s
\begin{bmatrix}
    \mathbf{L}_{LL} & \mathbf{L}_{Lr} \\
    \mathbf{L}_{rL} & \mathbf{L}_{rr}
\end{bmatrix}
\begin{bmatrix}
\vv{I}_L(s) \\
\vv{I}_r(s)
\end{bmatrix}
+
\begin{bmatrix}
    \mathbf{\Omega}_E^T & \mathbf{0} \\
    \mathbf{0} & \mathbf{I}_{rr}
\end{bmatrix}
\begin{bmatrix}
    \vv{V}_C(s) \\
    \vv{V}_r(s)
\end{bmatrix}
\end{align}
Here, $\mathbf{C}_{rr}$ and $\mathbf{L}_{rr}$ are diagonal, with their elements obeying the resonance condition ${C}_{rr}{L}_{rr} = \frac{1}{\omega_r^2}$. The identity submatrix of the network matrix (as discussed in Appendix \ref{subsec:FundamentalDecomposition}) refers to a group of harmonic oscillators, uncoupled from each other, but coupled to the port degrees of freedom through the off-diagonal blocks of the capacitance and inductance matrices, respectively (though galvanically disconnected).

To bring this equation to a more familiar form we write $\vv{V}_r(s)$ and $\vv{I}_r(s)$ in terms of $\vv{V}_C(s)$ and $\vv{I}_L(s)$, giving:
\begin{align}
\begin{bmatrix}
\mathbf{I}_{rr} & - s\mathbf{C}_{rr} \\
- s\mathbf{L}_{rr} & -\mathbf{I}_{rr} 
\end{bmatrix}
\begin{bmatrix}
\vv{I}_{r}(s) \\
\vv{V}_{r}(s)
\end{bmatrix}
=
s
\begin{bmatrix}
\mathbf{C}_{rC} \vv{V}_{C}(s) \\
\mathbf{L}_{rL} \vv{I}_{L}(s)
\end{bmatrix}
\end{align}

The inverse of the matrix on the LHS gives (using the diagonality of the submatrices involved):
\begin{align}
&\begin{bmatrix}
\mathbf{I}_{rr} & - s\mathbf{C}_{rr} \\
- s\mathbf{L}_{rr} & -\mathbf{I}_{rr} 
\end{bmatrix}^{-1} \nonumber \\ %newline
% &=  \begin{bmatrix}
% \left( \mathbf{I}_{rr} + s^2 \mathbf{L}_{rr} \mathbf{C}_{rr} \right)^{-1} & - s \mathbf{C}_{rr}\left( \mathbf{I}_{rr} + s^2 \mathbf{L}_{rr} \mathbf{C}_{rr} \right)^{-1} \\
% - s \mathbf{L}_{rr}\left( \mathbf{I}_{rr} + s^2 \mathbf{L}_{rr} \mathbf{C}_{rr} \right)^{-1} & - \left( \mathbf{I}_{rr} + s^2 \mathbf{L}_{rr} \mathbf{C}_{rr} \right)^{-1}
% \end{bmatrix} \nonumber \\ %newline
&=  \begin{bmatrix}
\boldsymbol{\omega}_{rr}^2 \left( \boldsymbol{\omega}_{rr}^2 + s^2\mathbf{I}_{rr} \right)^{-1} & -s \mathbf{L}_{rr}^{-1}\left( \boldsymbol{\omega}_{rr}^2 + s^2\mathbf{I}_{rr} \right)^{-1} \\
- s \mathbf{C}_{rr}^{-1}\left( \boldsymbol{\omega}_{rr}^2 + s^2\mathbf{I}_{rr} \right)^{-1} & - \boldsymbol{\omega}_{rr}^2 \left( \boldsymbol{\omega}_{rr}^2 + s^2\mathbf{I}_{rr} \right)^{-1}
\end{bmatrix} 
\end{align}

Now we can eliminate $\vv{V}_r(s)$ and $\vv{I}_r(s)$ from the frequency-domain equations of motion, resulting in the expressions:
\begin{align}
    \vv{I}_C(s) &= \left[ s \mathbf{C}_{CC} - s^3 \sum_r \frac{\frac{1}{C_{rr}} 
 \vv{C}_{C_{r}} \vv{C}_{C_{r}}^T }{s^2+\omega_r^2} \right] \vv{V}_C(s) \nonumber \\ %newline
    &-\left[ \mathbf{\Omega}_E   + s^2 \sum_r \frac{ \omega_r^2 \vv{C}_{C_{r}} \vv{L}_{L_{r}}^T}{s^2+\omega_r^2} \right] \vv{I}_L(s) \\ %newline
\vv{V}_L(s) &= \left[ s \mathbf{L}_{LL} - s^3 \sum_r \frac{\frac{1}{L_{rr}} 
 \vv{L}_{L_{r}} \vv{L}_{L_{r}}^T }{s^2+\omega_r^2} \right] \vv{I}_L(s) \nonumber \\
    &+\left[ \mathbf{\Omega}_E^T  + s^2 \sum_r \frac{ \omega_r^2 \vv{L}_{L_{r}} \vv{C}_{C_{r}}^T}{s^2+\omega_r^2} \right] \vv{V}_C(s)
\end{align}
Here $\vv{C}_{C_{r}}$ is the $r$th column of $\mathbf{C}_{Cr}$ and $\vv{L}_{L_{r}}$ is the $r$th column of $\mathbf{L}_{Lr}$. These equations imply the lumped circuit has a hybrid matrix response of the form:
\begin{align}
&\mathbf{H}(s) = 
\begin{bmatrix}
    s \mathbf{C}_{CC} & -\mathbf{\Omega}_E \\
    \mathbf{\Omega}_E^T &  s\mathbf{L}_{LL}
\end{bmatrix}
 \nonumber \\ %newline
&+ \sum_r \frac{s^2}{s^2 + \omega_r^2}
\begin{bmatrix}
\frac{\vv{C}_{C_r}}{\sqrt{C_{rr}}} & \vv{0}_{C} \\
\vv{0}_{L} & \frac{\vv{L}_{L_r}}{\sqrt{L_{rr}}}
\end{bmatrix}
\begin{bmatrix}
-s & -\omega_r \\
\omega_r & -s
\end{bmatrix}
\begin{bmatrix}
\frac{\vv{C}_{C_r}}{\sqrt{C_{rr}}} & \vv{0}_{C} \\
\vv{0}_{L} & \frac{\vv{L}_{L_r}}{\sqrt{L_{rr}}}
\end{bmatrix}^T
\end{align}

This hybrid matrix is identical to that in Eq. \ref{eq:AppFinalHybridSimulationResponse} if the edge network matrix $\mathbf{\Omega}_E$ of the simulation and the lumped circuit are identical, if the lumped model resonance frequencies $\omega_r = 1/{\sqrt{{C}_{rr}{L}_{rr}}}$ equal the frequencies of the simulated resonant poles, and if we set:
\begin{align}
\vv{C}_{C_r} &= \sqrt{C_{rr}} \vv{R}_{C_r} \\ %newline
\vv{L}_{L_r} &= \sqrt{L_{rr}} \vv{R}_{L_r}\\ %newline
\mathbf{C}_{CC} &= \mathbf{K}_{{CC}_\infty} + \sum_r \vv{R}_{C_r}\vv{R}_{C_r}^T \\ %newline
\mathbf{L}_{LL} &= \mathbf{K}_{{LL}_\infty} + \sum_r \vv{R}_{L_r}\vv{R}_{L_r}^T 
\end{align}

Two final properties to check are that the synthesized capacitance and inductance matrices capacitance and inductance matrices are positive definite and thus invertible. For the capacitance matrix, since $\mathbf{C}_{rr}$ is positive definite, then the total capacitance matrix $\mathbf{C}$ is positive definite IFF its Schur complement is positive definite:
\begin{align}
&\mathbf{C}_{CC}-\mathbf{C}_{Cr}\mathbf{C}_{rr}^{-1}
\mathbf{C}_{Cr}^T  \nonumber \\
&= \mathbf{K}_{{CC}_\infty} + \sum_r \vv{R}_{C_r}\vv{R}_{C_r}^T  -\sum_r \vv{R}_{C_r}\vv{R}_{C_r}^T  \nonumber \\
&=\mathbf{K}_{{CC}_\infty}
\end{align}

Since we assumed $\mathbf{K}_{{CC}_\infty}$ is positive definite, the statement holds. Analogous logic shows that $\mathbf{L}$ is positive definite.

In essence, the lumped circuit should have a layout topology that mirrors that of the EM simulation. The capacitance and inductance matrices match the responses of the the simulated infinite and finite-frequency poles. The finite-frequency poles each correspond to an auxiliary resonator in the circuit model, with the $\vv{R}_{C_r}$ and $\vv{R}_{L_r}$ vectors representing participation ratios \cite{labarca_toolbox_2024} of the capacitive and inductive port excitations (respectively) in the modes.

\subsection{Port replacement}

Once a model has been constructed for the lossless linear portion of the circuit, nonlinear and drive elements can inserted across the model's terminals. Josephson tunneling can be placed in parallel with capacitive ports, and fluxoid tunneling can be inserted in series with inductive ports. This hybrid matrix extraction method enables Josephson junctions to be shunted by a capacitor and simultaneously for phase slips to be in series with an inductor, thus giving each of the nonlinear tunneling elements its natural linear component.

In addition, external drives, represented by resistors and voltage or current sources, can be placed at terminals that represent input or output ports. The excitations of these ports (including the zero point fluctuations of the resistive bath) can then be treated to various degrees of approximation.

At this point, the model is expressed in terms of capacitive branch flux variables and inductive branch charges. It can be converted to a nodal capacitive model by finding a spanning tree for the capacitive branches that respects the loop structure of the device, and performing a change of basis (Eq. \ref{eq:NodeFluxDefined}).

\bibliography{references}

\end{document}